\documentclass[10pt,twoside]{article}

\usepackage{fancyheadings}
\usepackage{epsf}
\usepackage{psfig}
\usepackage{lscape}
\usepackage{natbib}
\usepackage{graphicx}
\usepackage{relsize}
\usepackage{amssymb}

\newcommand{\apjs}{ApJS}
\newcommand{\apjl}{ApJL}
\newcommand{\aap}{A\&A}

\usepackage{lineno}
\newcommand{\oversim}[2]{\protect{\mbox{\lower0.5ex\vbox{%
   \baselineskip=0pt\lineskip=0.2ex
   \ialign{$\mathsurround=0pt #1\hfil##\hfil$\crcr#2\crcr\sim\crcr}}}}} 
\newcommand{\simgreat}{\mbox{$\,\mathrel{\mathpalette\oversim>}\,$}} 
\newcommand{\simless} {\mbox{$\,\mathrel{\mathpalette\oversim<}\,$}} 
\begin{document}

\lhead{} \chead{}
\rhead{\small \it Stellar Systems and Galactic Structure, 2012,
  Vol.~5 of
  Planets, Stars \& Stellar Systems,
  Editor-in-chief: Oswalt, T. D., with McLean, I.S.; Bond, H.E.;
  French, L.; Kalas, P.; Barstow, M.A.; Gilmore, G.F.; Keel,
  W.C. (Eds), Springer} 

\cfoot{}

\begin{center}

\vspace{1cm}
{\bf \huge The stellar and sub-stellar IMF
  of\\ \vspace{1ex} simple and composite populations}\\

\vspace{1cm}

  Pavel Kroupa$^{1}$, Carsten Weidner$^{2,3}$, Jan
  Pflamm-Altenburg$^{1}$, Ingo Thies$^{1}$, J{\"o}rg
  Dabringhausen$^{1}$, Michael Marks$^{1}$ \& Thomas Maschberger$^{4,5}$\\

\vspace{1.5cm}

{\it
\begin{minipage}{1.0\textwidth}
\begin{itemize}
  \item[$^1$] Argelander-Institut f\"ur Astronomie (Sternwarte),
  Universit\"at Bonn, Auf dem H\"ugel 71, D-53121~Bonn, Germany\\
    
  \item[$^2$] Scottish Universities Physics Alliance (SUPA), School of Physics and
    Astronomy, University of St. Andrews, North Haugh, St. Andrews, Fife
  KY16 9SS, UK

  \item[$^3$] Instituto de Astrofísica de Canarias, C/ V\'ia L\'actea, s/n,
    E38205 La Laguna (Tenerife), Spain\\

  \item[$^4$] Institute of Astronomy, Madingley Road, Cambridge CB3 0HA, UK\\

  \item[$^5$] Institut de Plan{\'e}tologie et d’Astrophysique de Grenoble, BP 53,
    F-38041 Grenoble C{\'e}dex 9, France\\
\end{itemize}\end{minipage}}

\end{center}

\clearpage
\thispagestyle{empty}
\cleardoublepage

\thispagestyle{empty}
\noindent\begin{center}{\bf \Large Abstract}

\vspace{0.5cm}
\begin{minipage}{\columnwidth} 
  The current knowledge on the stellar IMF is documented. It is
  usually described as being invariant, but evidence to the contrary
  has emerged: it appears to become top-heavy when the star-formation
  rate density surpasses about $0.1\,M_\odot/({\rm yr}\,{\rm pc}^3)$
  on a~pc~scale and it may become increasingly bottom-heavy with
  increasing metallicity and in increasingly massive early-type
  galaxies.  It declines quite steeply below about $0.07\,M_\odot$
  with brown dwarfs (BDs) and very low mass stars having their own
  IMF. The most massive star of mass $m_{\rm max}$ formed in an
  embedded cluster with stellar mass $M_{\rm ecl}$ correlates strongly
  with $M_{\rm ecl}$ being a result of gravitation-driven but
  resource-limited growth and fragmentation induced starvation.  There
  is no convincing evidence whatsoever that massive stars do form in
  isolation. Massive stars form above a density threshold in embedded
  clusters which become {\it saturated} when $m_{\rm max}=m_{\rm max
    *}\approx 150\,M_\odot$ which appears to be the canonical physical
  upper mass limit of stars. Super-canonical massive stars arise
  naturally due to stellar mergers induced by stellar-dynamical
  encounters in binary-rich very young dense clusters.

  Various methods of discretising a stellar population are introduced:
  {\it optimal sampling} leads to a mass distribution that perfectly
  represents the exact form of the desired IMF and the $m_{\rm
    max}$-to-$M_{\rm ecl}$ relation, while {\it random sampling}
  results in statistical variations of the shape of the IMF. The
  observed $m_{\rm max}$-to-$M_{\rm ecl}$ correlation and the small
  spread of IMF power-law indices together suggest that optimally
  sampling the IMF may be the more realistic description of star
  formation than random sampling from a universal IMF with a constant
  upper mass limit.

  Composite populations on galaxy scales, which are formed from
  many~pc scale star formation events, need to be described by the
  integrated galactic IMF. This IGIMF varies systematically from
  top-light to top-heavy in dependence of galaxy type and star
  formation rate, with dramatic implications for theories of galaxy
  formation and evolution.

\end{minipage}
\end{center}
\clearpage
\thispagestyle{empty}
\cleardoublepage
\clearpage
\rhead{}
\chead{\it CONTENTS}
\lhead{}
\cfoot{\thepage}
\tableofcontents 
\cleardoublepage
\rhead{}
\chead{\it LIST OF FIGURES}
\lhead{}
\listoffigures
\clearpage
\thispagestyle{empty}
\cleardoublepage
\rhead{}
\chead{\it LIST OF TABLES}
\lhead{}
\listoftables
\clearpage
\thispagestyle{empty}
\cleardoublepage
\lhead{\leftmark\\\rightmark}
\chead{}
\rhead{}
\cfoot{\thepage}
\section{Introduction and Historical Overview}
\label{sec:intro}

The distribution of stellar masses that form together, the stellar
initial mass function (IMF), is one of the most important
astrophysical distribution functions. The determination of the IMF is
a very difficult problem because stellar masses cannot be measured
directly and because observations usually cannot assess all stars in a
population requiring elaborate bias corrections. Indeed, the stellar
IMF is not measurable (the {\sc IMF Unmeasurability Theorem} on
p.~\pageref{kroupa_theorem:IMFtheorem}).  Nevertheless, impressive
advances have been achieved such that the shape of the IMF is
reasonably well understood from low-mass brown dwarfs (BDs) to very
massive stars.

The IMF is of fundamental importance because it is a mathematical
expression for describing the mass-spectrum of stars born collectively
in ``one event''. Here, {\it one event} means a {\it
  gravitationally-driven collective process of transformation of the
  interstellar gaseous matter into stars on a spatial scale of about
  one~pc and within about one~Myr}. Throughout this text such events
are referred to as {\it embedded star clusters}, but they need not
lead to gravitationally bound long-lived open or globular clusters.
Another astrophysical function of fundamental importance is the
star-formation history (SFH) of a stellar system.

The IMF and the SFH are connected through complex self-regulating
physical processes on galactic scales, whereby it can be summarised
that for late-type galaxies the star-formation rate (SFR) increases
with increasing galaxy mass and the deeper gravitational
potential. For early-type galaxies the same is true except that the
SFH was of short duration ($\simless\;$few$\,$Gyr). Together the IMF
and SFH contain the essential information on the transformation of
dark gas to shining stars and the spectral energy distribution
thereof. They also contain the essential information on the cycle of
matter, which fraction of it is locked up in feeble stars and
sub-stellar objects, and how much of it is returned enriched with
higher chemical elements to the interstellar medium or atmosphere of a
galaxy. Knowing the rate with which matter is converted to stars in
galaxies is essential for understanding the matter cycle and the
matter content in the universe at a fundamental level.

This text is meant to outline essentials in IMF work, to document its
form which appears to be invariant for the vast majority of resolved
star-formation events and to describe modern evidence for IMF
variation and how whole galaxies are to be described as composite or
complex populations. The literature on the IMF is vast, and it is
unfortunately not possible to cover every research paper on this
topic, although some attempt has been made to be as inclusive as
possible.  Sec.~\ref{sec:intro} gives a brief historical review and a
short overview of the topic, and pointers to other reviews are
provided in Sec.~\ref{sec:otherreviews}.

\subsection{Solar neighbourhood}

Given the importance of the IMF a major research effort has been
invested to distill its shape and variability. It began by first
considering the best-known stellar sample, namely that in the
neighbourhood of the Sun.

The seminal contribution by \cite{S55} whilst staying in Canberra
first described the IMF as a power-law,
$dN=\xi(m)\,dm=k\,m^{-\alpha}$, where $dN$ is the number of stars in
the mass interval $m,m+dm$ and $k$ is the normalisation constant.  By
modelling the spatial distribution of the then observed stars with
assumptions on the star-formation rate, Galactic-disk structure and
stellar evolution time-scales, Salpeter arrived at the power-law index
(or ``slope'') $\alpha=2.35$ for $0.4 \simless m/M_\odot \simless 10$,
which today is known as the ``Salpeter IMF''.\footnote{As noted by
  \cite{Zinnecker11}, Salpeter used an age of~6~Gyr for the MW disk;
  had he used the now adopted age of~12~Gyr he would have arrived at a
  ``Salpeter index'' $\alpha\approx 2.05$ instead of~2.35.  }

This IMF form implies a diverging mass density for $m\rightarrow 0$,
which was interesting since dark matter was speculated, until the
early~1990's, to possibly be made-up of faint stars or sub-stellar
objects.  Studies of the stellar velocities in the solar-neighbourhood
also implied a large amount of missing, or dark, mass in the disk of
the Milky Way (MW) \citep{B84}. Careful compilation in Heidelberg of
the Gliese {\it Catalogue of Nearby Stars} beginning in the 1960's
\citep{JW97}\footnote{The latest version of the catalogue can be found
  at http://www.ari.uni-heidelberg.de/datenbanken/aricns/, while
  http://www.nstars.nau.edu/ contains the Nearby Stars (NStars)
  database.}, and the application at the beginning of the 1980's of an
innovative photographic pencil-beam survey-technique reaching deep
into the Galactic field in Edinburgh by \citet{RG82} significantly
improved knowledge of the space density of low-mass stars (LMSs,
$m/M_\odot \simless 0.5$).

Major studies extending Salpeter's work to lower and larger masses
followed, showing that the mass function (MF) of Galactic-field stars
turns over below one solar mass thus avoiding the divergence. Since
stars with masses $m\simless 0.8\,M_\odot$ do not evolve significantly
over the age of the Galactic disk, the MF equals the IMF for these.
While the work of \cite{MS79} relied on using the then-known nearby
stellar sample to define the IMF for $m<1\,M_\odot$, \cite{Sc86}
relied mostly on a more recent deep pencil-beam star-count survey
using photographic plates. \cite{Sc86} stands out as the most thorough
and comprehensive analysis of the IMF in existence, laying down
notation and ideas in use today.  

The form of the IMF for low-mass stars was revised in the early 1990's
in Cambridge, especially through the quantification of significant
non-linearities in the stellar mass--luminosity relation and
evaluation of the bias due to unresolved binary systems
\citep{KTG90,KTG91,KTG93}. This work lead to a detailed understanding
of the shape of the stellar luminosity function (LF) in terms of
stellar physics. It also resolved the difference between the results
obtained by \cite{MS79} and \cite{Sc86} through rigorous modelling of
all biases affecting local trigonometric-based and distant
photometric-parallax-based surveys, such as come from an intrinsic
metallicity scatter, evolution along the main sequence and contraction
to the main sequence.  In doing so this work also included an updated
local stellar sample {\it and} the then best-available deep
pencil-beam survey. As such it stands unique today as being the only
rigorous analysis of the late-type-star MF using simultaneously both
the {\it nearby trigonometric parallax} and the {\it far, pencil-beam}
star-count data to constrain the one underlying MF of stars.  This
study was further extended to an analysis of other ground-based
pencil-beam surveys being in excellent agreement with measurements
with the HST of the LF through the entire thickness of the MW
(Fig.~\ref{fig:MWlf} below).

These results ($\alpha\approx 1.3, 0.1-0.5\,M_\odot$), were confirmed
by \citet{RGH02} using updated local star-counts that included
Hipparcos parallax data.  Indeed, the continued long-term
observational effort on nearby stars by Neill Reid, John Gizis and
collaborators forms one of the very major pillars of modern IMF work;
continued discussion of controversial interpretations have much
improved and sharpened our general understanding of the issues.  A
re-analysis of the nearby mass function of stars in terms of a
log-normal form in the mass range $0.07-1\,M_\odot$ was provided by
Gilles Chabrier, finding agreement to the deep HST star-count data
once unresolved multiple stars and a metal-deficient colour-magnitude
relation for thick-disk M~dwarfs are accounted for \citep{Ch03}. The
log-normal form with a power-law extension to high masses is
indistinguishable from the older canonical two-part power-law form
(see Fig.~\ref{fig:canIMF} below). The immense analysis by
\cite{Bochanski10} of the LF and MF of~$15\times 10^6$ field low-mass
dwarfs ($0.1\simless m/M_\odot \simless 0.8$) derived from Sloan
Digital Sky Survey Data Release~6 photometry using the photometric
parallax method again finds good consistency with the previous work on
the stellar LF and MF.

The Galactic-field stellar IMF for $0.08\simless m/M_\odot\simless 1$
can thus be regarded as being reasonably well-constrained. It
converges to a finite mass density such that low-mass stars cannot be
a dynamically-significant contributor to the MW disk. The dynamical
evidence for significant amounts of dark matter in the disk was
finally eroded by \cite{KG91} and \cite{FF94}.

\subsection{Star clusters}

In contrast to the Galactic-field sample, where stars of many ages and
metallicities are mixed, clusters offer the advantage that the
stars have the same age and metallicity and distance. And so a very
large effort has been invested to try to extract the IMF from open and
embedded clusters, as well as from associations.  

On the theoretical side, the ever-improving modelling of stellar and
BD atmospheres being pushed forward with excellent results notably by
the Lyon group (Isabelle Baraffe and Gilles Chabrier), has allowed
consistently better constraints on the faint-star MF by a wide variety
of observational surveys \footnote{\label{foot:pms} Here it should be
  emphasised and acknowledged that the intensive and highly fruitful
  discourse between Guenther Wuchterl and the Lyon group has led to
  the important understanding that the classical evolution tracks
  computed in Lyon and by others are unreliable for ages less than a
  few~Myr \citep{Tout99,WT03}. This comes about because the emerging
  star's structure retains a memory of its accretion history. In
  particular \cite{WK01} present a SPH computation of the
  gravitational collapse and early evolution of a solar-type star
  documenting the significant difference to a pre-main sequence track
  if the star is instead classically assumed to form in
  isolation.}. Furthermore, the development of high-precision $N$-body
codes that rely on complex mathematical and algorithmic regularisation
of the equations of motions through the work of Sverre Aarseth and
Seppo Mikkola \citep{Aarseth99} and others has led to important
progress on understanding the variation of the dynamical
properties\footnote{\label{foot:dynprop}The {\it dynamical properties}
  of a stellar system are its {\it mass} and, if it is a multiple
  star, its orbital parameters ({\it semi-major axis, mass ratio,
    eccentricity}, both inner and outer if it is a higher-order
  multiple system). See also Sec.~\ref{sec:bins}.} of stellar
populations in individual clusters and the Galactic field.

In general, the MF found for clusters is consistent with the
Galactic-field IMF for $m<1\,M_\odot$, but it is still unclear as to
why open clusters have a significant deficit of white dwarfs
\citep{Fellhauer_etal03}. Important issues are the rapid and violent
early dynamical evolution of clusters due to the expulsion of residual
gas and the associated loss of a large fraction of the cluster
population, and the density-dependent disruption of primordial binary
systems \citep{KAH}.

Young clusters have thus undergone a highly complex dynamical
evolution which continues into old age \citep{BaumMakin03} and are
therefore subject to biases that can only be studied effectively with
full-scale $N$-body methods, thus imposing a complexity of analysis
that surpasses that for the Galactic-field sample. The pronounced
deficit of BDs and low-mass stars in the 600 Myr old Hyades are
excellent observational proof how dynamical evolution affects the MF
\citep{BKM08,Roeseretal11}. While estimating masses of stars younger
than a few~Myr is subject to major uncertainty$^{\rm \ref{foot:pms}}$,
differential reddening is also a complicated problem to handle when
studying the IMF (Sec.~\ref{sec:biases}).

\vspace{2mm} \centerline{ \fbox{\parbox{\columnwidth}{ {\sc Important Note}:
      In order to constrain the shape and variation of the IMF from
      star-count data in clusters high-precision $N$-body modelling must
      be mastered in addition to stellar-evolution theory and
      knowledge of the properties of multiple stars.}}}  \vspace{2mm}

\subsection{Intermediate-mass and massive stars}

Intermediate-mass and particularly massive stars are rare and so
larger spatial volumes need to be surveyed to assess their distribution
by mass.

For intermediate-mass and massive stars, the \citet{Sc86} IMF is based
on a combination of Galactic-field star-counts and OB association data
and has a slope $\alpha\approx 2.7$ for $m \simgreat 2\,M_\odot$ with
much uncertainty for $m\simgreat 10\,M_\odot$.  The previous
determination by \cite{MS79} also implied a relatively steep field-IMF
with $\alpha \approx 2.5, 1 \le m/M_\odot \le 10$ and $\alpha \approx
3.3, m>10\,M_\odot$. \cite{ES06} point out that structure in the MF is
generated at a mass if the SFR of the population under study varies on
a time-scale comparable to the stellar evolution time-scale of its
stars near that mass. Artificial structure in the stellar IMF may be
deduced in this case, and this effect is particularly relevant for the
field-IMF. Perhaps the peculiar structure detected by \cite{GBH05} in
the field-MF of stars in the Large Magellanic Cloud may be due to this
effect.

An interesting finding in this context reported by \citet{M98,
  Massey03} is that the OB stellar population found in the field of
the LMC has a very steep MF, $\alpha \approx 4.5$, which can be
interpreted to be due to the preferred formation of small groups or
even isolated O and B-type stars (e.g. \citealt{Selier11}).  Another
or an additional effect influencing the deduced shape of the IMF via a
changing SFR \citep{ES06} is the dynamical ejection of OB stars from
dynamically unstable cores of young clusters. This may lead to such a
steep IMF because, as dynamical work suggests, preferentially the
less-massive members of a core of massive stars are ejected
\citep{ClarkePringle92, P-AK06a}.  This process would require further
study using fully-consistent and high-precision $N$-body modelling of
young clusters to see if the observed distribution of field-OB stars
can be accounted for with this process alone, or if indeed an exotic
star-formation mode needs to be invoked to explain some of the
observations.

The IMF for intermediate-mass and massive stars deduced from star
counts in the field of a galaxy may therefore yield false information
on the true shape of the stellar IMF, and stellar samples with well
defined formation histories such as OB associations and star clusters
are more useful to constrain the IMF. Photometric surveys of such
regions, while essential for reaching faint stars, do not allow a
reliable assessment of the mass distribution of massive stars since
their spectral energy distribution is largely output at short
wavelengths beyond the optical. Spectroscopic classification is
therefore an essential tool for this purpose.

Phil Massey's work at Tucson, based on extensive spectroscopic
classification of stars in OB associations and in star clusters,
demonstrated that for massive stars $\alpha =2.35\pm 0.1,
m\simgreat10\,M_\odot$ \citep{M98}, for a large variety of physical
environments as found in the MW and the Magellanic Clouds, namely OB
associations and dense clusters and for populations with metallicity
ranging from near-solar-abundance to about $1/10$th metal abundance
(Fig.~\ref{fig:kroupa_figmassey} below).  The significant differences
in the metallicity between the outer and the inner edge of the
Galactic disk seem not to influence star-formation, as \citet{YKT08}
find no apparent difference in the stellar MFs of clusters in the
extreme outer Galaxy compared to the rest of the disk down to
$0.1\,M_\odot$.

\subsection{The invariant IMF and its conflict with theory}
\label{sec:introd_IMF}
The thus empirically constrained stellar IMF can be described well by
a two-part power-law form with $\alpha_1=1.3$ for $m\simless
0.5\,M_\odot$ and with the {\it Salpeter/Massey index},
$\alpha_2=2.35$ for $m\simgreat 0.5\,M_\odot$, being remarkably
invariant.\footnote{Note that \cite{Sc98} emphasises that the IMF
  remains poorly constrained owing to the small number of massive
  stars in any one sample. This is a true albeit conservative
  stand-point, and the present authors prefer to accept Massey's
  result as a working IMF-universality hypothesis.  \label{pk_fn1}.}
This IMF is referred to as the {\it canonical IMF} and is given in
Sec.~\ref{sec:canIMF} as the simpler two-part power-law form by
Eq.~\ref{eq:imf} and as the log-normal plus power-law form by
Eq.~\ref{eq:chabimf}.  The empirical result leads to the statement of
the following hypothesis:

\vspace{2mm} \centerline{ \fbox{\parbox{\columnwidth}{{\sc Invariant IMF
      Hypothesis}: There exists a universal parent distribution
    function which describes the distribution of stellar masses in
    individual star-forming events.  }}}
\label{hyp:InvIMFHyp}
\vspace{2mm} 

\noindent
The {\it Invariant IMF Hypothesis} needs to be tested with star-count
data in galactic fields and in individual clusters and OB associations
for possible significant deviations. But it is mandatory to take into
account the biases listed in Sec.~\ref{sec:biases} when doing so. 

The {\it Invariant IMF Hypothesis} is not consistent with theory (see
also Sec.~\ref{sec:theor} and~\ref{sec:IMFvar}): There are two broad
theoretical ansatzes for the origin of stellar masses: \vspace{0mm}
\begin{description}
\item {\sc The Jeans-Mass Ansatz}: According to the Jeans-mass
  argument (e.g. \citealt{Jeans02}, \citealt{L98},
  \citealt{Bate2005a}, \citealt{BCB06}, Eq.~\ref{eq:MJeans} below)
  star-formation at lower metallicity ought to produce stars of, on
  average, heavier mass and thus an effectively top-heavy IMF
  (i.e. the ratio between the number of massive stars and low mass
  stars ought to increase).  At lower metallicity the cooling is less
  efficient causing larger Jeans masses as a requirement for
  gravitational collapse to a proto-star and thus larger stellar
  masses.  That warmer gas produces an IMF shifted to larger masses
  has been demonstrated with state-of-the art SPH simulations
  (e.g. \citealt{KSJ07}).
\item {\sc The Self-Regulatory Ansatz}: Another approach is formulated
  by \cite{AF96} who argue that the {\sc Jeans-Mass Ansatz} is invalid
  since there is no preferred Jeans mass in a turbulent molecular
  cloud. Instead they invoke the central-limit
  theorem\footnote{\label{foot:cenlimth} Citing from \cite{BJ04}:
    ``According to the central limit theorem of statistics, if the
    mass of a protostellar condensation $M_c = f_1 \times f_2 \times
    \dots \times f_N$, then the distribution of $M_c$ tends to a
    lognormal regardless of the distributions of the individual
    physical parameters $f_i (i = 1, \dots N)$, if $N$ is large.
    Depending on the specific distributions of the $f_i$, a
    convergence to a lognormal may even occur for moderate $N$.''  The
    central limit theorem was invoked for the first time by
    \cite{Zinnecker84} to study the form of the IMF from hierarchical
    fragmentation of collapsing cloud cores.  } together with
  self-regulated assembly and they suggest that the final stellar
  masses are given by the balance between feedback energy from the
  forming star (accretion luminosity, outflows) and the rate of
  accretion from the proto-stellar envelope and circum-stellar
  disk. As the proto-star builds-up its luminosity increases until the
  accretion is shut off. When shut-off occurs depends on the accretion
  rate.  Indeed, \cite{BJ04} explain the observed power-law extension
  of the IMF at large stellar masses as being due to a distribution of
  different accretion rates.  The self-regulating character of star
  formation has been studied profusely by the group around Christopher
  McKee (e.g. \citealt{Tan06}) and has been shown to lead to
  decreasing star-formation efficiencies with increasing metallicity
  \citep{Dibetal10,Dib11, Dibetal11}.  In low-metallicity gas the
  coupling of the photons to the gas is less efficient causing a less
  effective opposition against the accreting material. And, at lower
  metallicity the cooling is reduced causing a higher temperature of
  the gas and thus a higher speed of sound with a larger accretion
  rate. The final stellar IMF is expected to be populated by more
  massive stars in metal-poor environments.
\end{description}
Both approaches can be refined by studying a distribution of physical
conditions in a given star-forming cloud, but both lead to the same
conclusion, namely that low-metallicity and high-temperature ought to
produce top-heavy stellar IMFs. This leads to the following robust
theoretical IMF result:

\vspace{2mm} \centerline{ \fbox{\parbox{\columnwidth}{{\sc The Variable IMF
        Prediciton}: Both the Jeans-mass and the self-regulation
      arguments invoke very different physical principles and yet they
      lead to the same result: The IMF ought to become top-heavy under
      low-metallicity and high-temperature star-forming conditions.
\label{box:varIMFpred}}}}
\vspace{2mm} 

\noindent
Star-formation in the very early universe must have therefore produced
top-heavy IMFs \citep{L98,Bromm,Clark11}. But the samples of
simple-stellar populations spanning all cosmological epochs (globular
clusters to current star-formation in embedded clusters) available in
the Local Group of galaxies have until recently not shown convincing
evidence supporting {\sc The Variable-IMF Prediction}. This issue is
addressed in more detail in Sec.~\ref{sec:currentstate}.

\subsection{A philosophical note}
\label{sec:philosophy}

Much of the current discussion on star-formation, from the smallest to
the largest (galaxy-wide) scales, can be categorised into two broad
conceptual approaches which are related to the {\sc Jeans-Mass Ansatz}
{\it vs} the {\sc Self-Regulatory Ansatz} of
Sec.~\ref{sec:introd_IMF}:

\begin{description}
\vspace{-3mm}
\item {\sc Approach~A} is related to the notion that star formation is
  inherently stochastic such that the IMF is a probabilistic
  distribution function only. This is a natural notion under the
  argument that the processes governing star-formation are so many and
  complex that the outcome is essentially stochastic in nature.
  Followers of this line of reasoning argue for example that massive
  stars can form in isolation and that the mass of the most massive
  star cluster forming in a galaxy depends on the time-scale over
  which an ensemble of star clusters is considered (the {\it
    size-of-sample effect}, even at very low SFRs a galaxy would
  produce a very massive star cluster if one waits long enough,
  i.e. if the sample of clusters is large enough). Approach~A can be
  formulated concisely as {\sc nature plays dice when stars form}.

\item {\sc Approach~B} is related to the notion that nature is
  inherently self-regulated and deterministic. This is a natural
  notion given that physical processes must always depend on the
  boundary conditions which are a result of the physical processes at
  hand. An example of such would be gravitationally-driven growth
  processes with feedback in media with limited resources.  The
  emerging phenomena such as the distribution of stellar masses, of
  star-cluster masses and/or of how phase-space is populated to make
  binary stellar systems are, at least in principle, computable. They
  are computable in the sense that statistical mathematics provides
  the required tools such that the distribution functions used to
  describe the outcomes are subject to constrains. For example, a
  young stellar population of mass $M_{\rm ecl}$ is excellently
  described by $\xi(m)$ with the condition $m\le m_{\rm max}(M_{\rm
    ecl})$. However, purely random sampling from $\xi(m)$ even under
  this constraint will not reproduce a realistic population if nature
  follows {\sc Optimal Sampling} (p.~\pageref{box:optdistr}). This is
  because {\sc Optimal Sampling} will never allow a cluster to be made
  up of $M_{\rm ecl}/m_{\rm max}$ stars of mass $m_{\rm max}$, while
  constrained random sampling would.  Approach~B can be formulated
  concisely as {\sc nature does not play dice when stars form}.
\end{description}

\noindent Depending on which of the two notions is applied, the
resulting astrophysical description of galaxies leads to very
diverging results. Either a galaxy can be described as an object in
which stars form purely stochastically such that the galaxy-wide IMF
is equal to the stellar IMF (Approach~A). In this case a thousand
small groups of 20~pre-main sequence stars will have the same stellar
IMF as one very young star-cluster containing 20000 stars.  Or an
embedded cluster or a galaxy are understood to be highly
self-regulated systems such that the galaxy-wide IMF differs from the
stellar IMF (Approach~B, Sec.~\ref{sec:comppop}). According to this
notion, a thousand small groups of 20~pre-main sequence stars would
not contain a single star with $m>5\,M_\odot$, while a very young star
cluster of~20000 stars would contain many such stars.  The different
{\sc Approaches} have very different implications for understanding
the matter cycle in the universe.

\subsection{Hypothesis testing}
\label{sec:hyptest}

The studies aimed at constraining the stellar IMF observationally
typically have the goal of testing the {\sc Invariant IMF Hypothesis}
(p.~\pageref{hyp:InvIMFHyp}) either in individual star-forming events
such as in a star cluster, or on galaxy-wide scales. Here, it is
important to be reminded of the following: 

\vspace{2mm} \centerline{ \fbox{\parbox{\columnwidth}{{\sc Elementary
        Logics of Hypothesis Testing}: Negation of a hypothesis~I does
      not imply that the alternative hypothesis~II is correct
\label{box:hyptest}}}}
\vspace{2mm} 

\noindent
By showing that hypothesis~I is consistent with some data does not
imply that an alternative hypothesis~II is therewith ruled out.  A
case in point is the discussion about dark matter and dark energy:
Ruling out the standard cosmological (LCDM) model does not imply that
any particular alternative is correct \citep{Kroupa10,
  Kroupa12}. Conversely, ruling out a particular alternative does not
imply that the LCDM model is correct \citep{Wojtaketal11}.

Concerning the IMF, if a purely stochastic model ({\sc Approach~A},
Sec.~\ref{sec:philosophy}) is consistent with some observational data,
then this does not imply that the alternative ({\sc Optimal Sampling},
which is related to {\sc Approach~B}) is falsified.

A case in point is provided by the following example relevant for the
tests of the IGIMF theory on p.~\pageref{box:igimfpred}): The masses
of an ensemble of observed dwarf galaxies are calculated from spectral
energy distribution modelling using the universal canonical IMF. These
masses are then applied in testing a possible variation of the
galaxy-wide IMF in terms of the UV and H$\alpha$ flux ratios. If the
universal IMF calculations allow for fluctuating SFRs whereas the
variable IMF calculations do not, then the (wrong) conclusion of such
an approach would plausibly be that nature appears to play dice,
because observational data naturally contain measurement uncertainties
which act as randomisation agents.

The consistent approach would instead be to compute the galaxy masses
assuming a variable galaxy-wide IMF to test whether the hypothesis
that the IMF varies systematically with galaxy mass can be
discarded. The logically consistent procedure would be to calculate
all fluxes within both scenarios independently of each other and
assuming in both that the SFR can fluctuate, and to test these
calculations against the observed fluxes. The result of this
consistent procedure are opposite to those of the above inconsistent
procedure in that the data are in better agreement with the
systematically variable IMF, i.e. that nature does not play dice. 

A final point to consider is is when a hypothesis ought to be finally
discarded. Two examples illustrate this: Consider the Taurus-Auriga
and Orion star forming clouds. Here the number of stars with
$m\simgreat 1\,M_\odot$ is significantly below the expectation from
the purely stochastic model (see~box {\sc IGIMF Predictions/Tests} on
p.~\pageref{box:igimfpred}).  This unambiguously falsifies the
stochastic model.  But the data are in excellent agreement with the
expectation from the IGIMF theory.  Is it then meaningful to
nevertheless keep adopting the stochastic model on cluster and galaxy
problems?  Another example, being perhaps more relevant to
Sec.~\ref{sec:philosophy}, is the issue with the current standard
cosmological model. It is ruled out by \cite{PN10, Kroupa12}. Should
it nevertheless be adopted in further cosmological and related
research?

\subsection{About this text}

As is evident from the above introduction, the IMF may be well defined
and is quite universal in each star-forming event out of which comes a
spatially and temporally well correlated set of stars. But many such
events will produce a summed IMF which may be different because the
individual IMFs need to be added whereby the distribution of the
star-formation events in mass, space and time become an issue. It thus
emerges that it is necessary to distinguish between simple stellar
populations and composite populations. Some definitions are useful:

\vspace{4mm} \centerline{ \fbox{\parbox{\columnwidth}{ {\sc Definitions}:
      \\[0.5mm] $\bullet$ A {\it simple population} obtains from a
      spatially ($\simless\,$few pc) and temporarily ($\simless\,$Myr)
      correlated star formation event (CSFE, also referred to as a
      {\it collective star formation event}, being essentially an
      embedded star cluster). The mass of a CSFE may range from a
      few~$M_\odot$ upwards.\\[0.5mm] $\bullet$ A {\it composite} or
      {\it complex population} consists of more than one simple
      population.\\[0.5mm] $\bullet$ The {\it stellar IMF} refers to
      the IMF of stars in a simple population. \\[0.5mm] $\bullet$ A
      {\it composite} or {\it integrated IMF} is the IMF of a
      composite or complex population, i.e. a population composed of
      many CSFEs, most of which may be gravitationally unbound. The
      galaxy-wide version is the IGIMF (Sec.~\ref{sec:comppop}).  \\[0.5mm]
      $\bullet$ The PDMF is the present-day MF of a stellar population
      not corrected for stellar evolution nor for losses through
      stellar deaths. Note that a {\it canonical PDMF} is a PDMF
      derived from a canonical IMF.  \\[0.5mm] $\bullet$ A {\it
        stellar system} can be a multiple star or a single star. It
      has {\it dynamical properties} (Footnote~\ref{foot:dynprop} on
      p.~\pageref{foot:dynprop}).  \\[0.5mm] $\bullet$ The {\it
        system} luminosity or mass function is the LF or MF obtained
      by counting all stellar systems.  The {\it stellar} LF or {\it
        stellar} MF is the true distribution of all stars in the
      sample, thereby counting all individual companions in multiple
      systems. This is also referred to as the {\it individual}-star
      LF/MF. \\[0.5mm] $\bullet$ Note 1$\,$km/s$=1.0227\,$pc/Myr,
      1$\,$g$\,$cm$^{-3}=1.478\times 10^{22}\,M_\odot\,$pc$^{-3}$ and
      1$\,$g$\,$cm$^{-2}=4788.4\,M_\odot\,$pc$^{-2}$ for a
      solar-metallicity gas.  \\[0.5mm] $\bullet$ The following
      additional abbreviations are used: SFR$=$star formation rate in
      units of $M_\odot\,$yr$^{-1}$; SFH$=$star formation history$=$SFR
      as a function of time; SFRD$=$star formation rate density in
      units of $M_\odot\,$yr$^{-1}\,$pc$^{-3}$.
      \label{box:definitions} }}} \vspace{4mm}

\noindent This treatise provides an overview of the general methods
used to derive the IMF with special attention on the pitfalls that are
typically encountered. The binary properties of stars and of brown
dwarfs are discussed as well because they are essential to understand
the true shape of the stellar IMF. While the stellar IMF appears to
have emerged as being universal in star-formation events as currently
found in the Local Group of galaxies, the recent realisation that
star-clusters limit the mass spectrum of their stars is one form of
IMF variation and has interesting implications for the formation of
stars in a cluster and leads to the insight that composite populations
must show IMFs that differ from the stellar IMF in each cluster.  With
this finale, this treatise reaches the cosmological arena.

\subsection{Other IMF reviews}
\label{sec:otherreviews}

The seminal contribution by \cite{Sc86} on the IMF remains a necessary
source for consultation on the fundamentals of the IMF problem.  The
landmark review by \cite{Massey03} on massive stars in the Local Group
of galaxies is an essential read, as is the review by \cite{ZH07} on
massive-star formation.  Other reviews of the IMF are by \cite{Sc98},
\cite{K02}, \cite{Chrev03}, \cite{BLZ07}, \cite{El09} and
\cite{Bastian2010}. The proceedings of the ``38th Herstmonceux
Conference on the Stellar Initial Mass Function'' \citep{GilHow98} and
the proceedings of the ``IMF\@50'' meeting in celebration of Ed
Salpeter's 80th birthday \citep{IMF50} contain a wealth of important
contributions to the field. A recent major but also somewhat exclusive
conference on the IMF was held from June~20th to 25th~2010 in Sedona,
Arizona, for researchers to discuss the recently accumulating evidence
for IMF variations: ``UP2010: Have Observations Revealed a Variable
Upper End of the Initial Mass Function?''. The published contributions
are a unique source of information on this problem \citep{UP2010}. A
comprehensive review of extreme star formation is available by
\cite{turner09}.

\section{Some Essentials}
Assuming the relevant biases listed in Sec.~\ref{sec:biases} have been
corrected for such that all binary and higher-order stellar systems
can be resolved into individual stars in some complete population such
as the solar neighbourhood and that only main-sequence stars are
selected for, then the number of single stars per pc$^3$ in the mass
interval $m$ to $m+dm$ is $dN=\Xi(m)\,dm$, where $\Xi(m)$ is the {\it
  present-day mass function} (PDMF). The number of single stars per
pc$^3$ in the absolute P-band magnitude interval $M_P$ to $M_P+dM_P$
is $dN=-\Psi(M_P)\,dM_P$, where $\Psi(M_P)$ is the stellar luminosity
function (LF) which is constructed by counting the number of stars in
the survey volume per magnitude interval, and $P$ signifies an
observational photometric pass-band such as the $V$- or $I$-band.
Thus
\begin{equation}
\Xi(m) = -\Psi(M_P)\,\left({dm \over dM_P}\right)^{-1}.
\label{eq:mf_lf}
\end{equation}
Note that the the minus sign comes-in because increasing mass leads to
decreasing magnitudes, and that the LF constructed in one photometric
pass band $P$ can be transformed into another band $P'$ by
\begin{equation}
\Psi(M_P) = {dN \over dM_{P'}} {dM_{P'} \over dM_P} = 
\Psi(M_{P'}) {dM_{P'} \over dM_P}
\label{eq:passband}
\end{equation}
if the function $M_{P'} = {\rm fn}(M_P)$ is known.  Such functions are
equivalent to colour--magnitude relations.

Since the derivative of the stellar mass--luminosity relation (MLR),
$m(M_P)=m(M_P,Z,\tau,\mathbf{s})$, enters the calculation of the MF,
any uncertainties in stellar structure and evolution theory on the one
hand (if a theoretical MLR is relied upon) or in observational ML-data
on the other hand, will be magnified accordingly. This problem cannot
be avoided if the mass function is constructed by converting the
observed stellar luminosities one-by-one to stellar masses using the
MLR and then binning the masses, because the derivative of the MLR
nevertheless creeps-in through the binning process, because {\it equal
luminosity intervals are not mapped into equal mass intervals}.  The
dependence of the MLR on the star's chemical composition, $Z$, it's
age, $\tau$, and spin vector $\mathbf{s}$, is explicitly stated here,
since stars with fewer metals than the Sun are brighter (lower
opacity), main-sequence stars brighten with time and loose mass, and
rotating stars are dimmer because of the reduced internal
pressure. Mass loss and rotation are significant factors for
intermediate and especially high-mass stars \citep{MPV01}.

The IMF, or synonymously here the IGIMF (Sect.~\ref{sec:comppop}),
follows by correcting the observed number of main sequence stars for
the number of stars that have evolved off the main sequence.  Defining
$t=0$ to be the time when the Galaxy that now has an age $t=\tau_{\rm
  G}$ began forming, the number of stars per pc$^3$ in the mass
interval $m,m+dm$ that form in the time interval $t,t+dt$ is
$dN=\xi(m;t)\,dm\times b'(t)\,dt$, where the expected time-dependence
of the IMF is explicitly stated (Sec.~\ref{sec:comppop}), and where
$b'(t)=b(t)/\tau_{\rm G}$ is the normalised star-formation history,
$(1/\tau_{\rm G}) \int_0^{\tau_{\rm G}}b(t)\,dt = 1$.  Stars that have
main-sequence life-times $\tau(m) < \tau_{\rm G}$ leave the stellar
population unless they were born during the most recent time interval
$[\tau_{\rm G}-\tau(m), \tau_{\rm G}]$. The number density of such
stars still on the main sequence with initial masses computed from
their present-day masses and their ages in the range $m,m+dm$ and the
total number density of stars with $\tau(m) \ge \tau_{\rm G}$, are,
respectively,
\begin{equation}
\Xi(m) = 
   \xi(m){1\over \tau_{\rm G}} \times
   \left\{ 
   \begin{array}{l@{\quad\quad,\quad}l}
   \int_{\tau_{\rm G}-\tau(m)}^{\tau_{\rm G}} b(t)dt &
   \tau(m) < \tau_{\rm G},\\
   \int_0^{\tau_{\rm G}} b(t)\,dt & \tau(m) \ge \tau_{\rm G},
   \end{array}\right.
\label{eq:imf_pdmf}
\end{equation}
where the time-averaged IMF, $\xi(m)$, has now been defined. Thus, for
low-mass stars $\Xi=\xi$, while for a sub-population of massive stars
that has an age $\Delta t \ll \tau_{\rm G}$, $\Xi=(\Delta t/\tau_{\rm
  G})\, \xi$ for those stars of mass $m$ for which the main-sequence
life-time $\tau(m)>\Delta t$, indicating how an observed high-mass IMF
in an OB association, for example, has to be scaled to the
Galactic-field IMF for low-mass stars, assuming continuity of the
IMF. In this case the different spatial distribution via different
disk-scale heights of old and young stars also needs to be taken into
account, which is done globally by calculating the stellar surface
density in the MW disk \citep{MS79,Sc86}.  Thus we can see that
joining the cumulative low-mass star counts to the snap-shot view of
the massive-star IMF is non-trivial and affects the shape of the IMF
in the notorious mass range $\approx 0.8-3\,M_\odot$, where the
main-sequence life-times are comparable to the age of the MW disk
(Fig.~\ref{fig:apl}, bottom panel). For a population in a star cluster
or association with an age $\tau_{\rm cl}\ll\tau_{\rm G}$, $\tau_{\rm
  cl}$ replaces $\tau_{\rm G}$ in Eq.~\ref{eq:imf_pdmf}.  Examples of
the time-modulation of the IMF are $b(t)=1$ (constant star-formation
rate) or a Dirac-delta function, $b(t)=\delta(t-t_0)$ (all stars
formed at the same time $t_0$).

The stellar IMF can conveniently be written as an arbitrary number of
power-law segements, 
\begin{equation}
\xi_{\rm BD}=k\,k_{\rm BD}\left({m\over m_1}\right)^{-\alpha_0},
\label{eq:imf_bd}
\end{equation}
\begin{equation}
\xi_\mathrm{star} (m) = k\left\{
          \begin{array}{l@{\quad,\quad}l@{\quad,\quad}l}
   \left({m\over m_1}\right)^{-\alpha_1}  &m_1 < m \le m_2 &n=1,\\
   \left[
       \prod\limits^{n\ge2}_{i=2}\left({m_i\over
          m_{i-1}}\right)^{-\alpha_{i-1}}
       \right] 
        \left({m\over m_n}\right)^{-\alpha_n} 
        &m_n < m \le m_{n+1}  &n\ge2,\\
          \end{array}\right.
\label{eq:imf_mult}
\end{equation}
where $k_{\rm BD}$ and $k$ contain the desired scaling,
$0.01\,M_\odot$ is about the minimum mass of a BD (see
footnote~\ref{foot:fragmlimit} on p.~\pageref{foot:fragmlimit}), and
the mass-ratios ensure continuity. Here the separation into the IMF of
BDs and of stars has already been explicitly stated (see
Sec.~\ref{sec:bds}).

Often used is the ``logarithmic mass function'' (Table~\ref{tab:imfs} below),
\begin{equation}
\xi_{\rm L}(m) = \left( m\,{\rm ln}10\right)\, \xi(m),
\end{equation}
where $dN=\xi_{\rm L}(m)\,dlm$ is the number of stars with mass in the
interval $lm,lm+dlm$ ($lm \equiv {\rm log}_{10}m$)\footnote{Note that
\citet{Sc86} calls $\xi_{\rm L}(m)$ the {\it mass function} and
$\xi(m)$ the {\it mass spectrum}.}.

The stellar mass of a an embedded cluster, $M_\mathrm{ecl}$, can be
used to investigate the expected number of stars above a certain mass
$m$,
\begin{equation}
N(>m) = \int_{m}^{m_{\rm max*}} \xi(m')\,dm',
\label{eq:Nm}
\end{equation}
with the mass in stars of the whole (originally embedded) cluster,
$M_{\rm ecl}$, being calculated from
\begin{equation}
M_{1,2} = \int_{m_1}^{m_2} m'\,\xi(m')\,dm',
\label{eq:Mecl1}
\end{equation}
with $M_{\rm ecl}=M_{1,2}$ for $m_1=m_{\rm low}\approx 0.07\,M_\odot$
(about the hydrogen burning mass limit) and $m_2=m_{\rm max*}=\infty$
(the {\it Massey assertion}, p.~\pageref{p:massey}; but see
Sec.~\ref{sec:stmass_clmass}).  There are two unknowns ($N(>m)$ and
$k$) that can be solved for by using the two equations above.

It should be noted that the IMF is not a measurable quantity: Given
that we are never likely to learn the exact dynamical history of a
particular cluster or population, it follows that we can {\it never}
ascertain the IMF for any individual cluster or population. This can
be summarised concisely with the following theorem:

\vspace{2mm}
\centerline{
\fbox{\parbox{\columnwidth}{
{\sc The IMF Unmeasurability Theorem}: The IMF cannot be extracted directly 
for any individual stellar population. \label{kroupa_theorem:IMFtheorem}}}}
\vspace{2mm}

\noindent {\sc Proof:} For clusters younger than about 1~Myr star
formation has not ceased and the IMF is therefore not assembled yet
and the cluster cores consisting of massive stars have already
dynamically ejected members \citep{P-AK06a}. Massive stars
($m\simgreat 30\,M_\odot$) leave the main sequence before they are
fully assembled \citep{MB01}.  For clusters with an age between~0.5
and a few~Myr the expulsion of residual gas has lead to a loss of
stars \citep{KAH}. Older clusters are either still loosing stars due
to residual gas expulsion or are evolving secularly through
evaporation driven by energy equipartition
\citep{VH97,BaumMakin03}. There exists thus no time when all stars
are assembled in an observationally accessible volume (i.e. a star
cluster). An observer is never able to access all phase-space
variables of all potential members of an OB association. The field
population is a mixture of many star-formation events whereby it can
practically not be proven that a complete population has been
documented.  End of proof.

Note that {\sc The IMF Unmeasurability Theorem} implies that individual
clusters cannot be used to make deductions on the similarity or not of
their IMFs, unless a complete dynamical history of each cluster is
available.

Notwithstanding this pessimistic theorem, it is nevertheless necessary
to observe and study star clusters of any age. Combined with thorough
and realistic $N$-body modelling the data do lead to essential {\it
  statistical} constraints on the {\sc IMF Universality Hypothesis}
(p.~\pageref{kroupa_hyp:univ}, see also p.~\pageref{hyp:InvIMFHyp}).

\subsection{Unavoidable biases affecting IMF studies}
\label{sec:biases}

Past research has uncovered a long list of biases that affect the
conversion of the observed distribution of stellar brightnesses to the
underlying stellar IMF. These are just as valid today, and in
particular analysis of the GAIA-space-mission data will need to take
the relevant ones into account before the stellar IMF can be
constrained anew. The list of all unavoidable biases affecting stellar
IMF studies is provided here with key references addressing these:

\begin{description}

\item {\it Malmquist bias} (affects MW-field star counts): Stars of the
  same mass but with different ages, metallicities and spin vectors
  have different colours and luminosities which leads to errors in
  distance measurements in flux-limited field star counts using
  photometric parallax \citep{SIP}.

\item {\it Colour-magnitude relation} (affects MW-field star counts):
  Distance measurements through photometric parallax are
  systematically affected if the true colour-magnitude relation of
  stars deviates from the assumed relation \citep{RG97} (but see
  Footnote~\ref{fn:cmr} on p.~\pageref{fn:cmr}).

\item {\it Lutz-Kelker bias} (affects MW-field star counts): A
  distance-limited survey is affected by parallax measurement
  uncertainties such that the spatial stellar densities are
  estimated wrongly \citep{LK73}. Correcting for this bias will be
  required when analysing GAIA-space mission star-count data.

\item {\it Unresolved multiple stars} (affects all star counts):
  Companions to stars can be missed because their separation is below
  the resolution limit, or because the companion's luminosity is below
  the flux limit \citep{KTG91}. When using photometric parallax to
  determine distances and space densities unresolved multiple systems
  appear nearer and redder. This affects the measured disk scale
  height as a function of stellar spectral type \citep{KTG93}. Missed
  companions have a significant effect on the deduced shape of the IMF
  for $m\simless 1\,M_\odot$ \citep{KTG91, KTG93, MZ01} but do not
  significantly affect the shape of the stellar IMF for more massive
  stars \citep{MU05, WKM09}.

\item {\it Stellar mass-luminosity relation} (MLR, affects all star
  counts): Main sequence stars of precisely the same chemical
  composition, age and spins follow one perfect mass--luminosity
  relation.  Its non-linearities map a featureless stellar IMF to a
  structured LF but theoretical MLRs are not reliable \citep{KTG90}.
  An ensemble of field stars do not follow one stellar mass-luminosity
  relation such that the non-linearities in it that map to structure
  in the stellar LF are smeared out \citep{KTG93}. Correcting for this
  bias will be required when analysing GAIA-space mission star-count
  data. Pre-main sequence stars have a complicated and time-varying
  mass-luminosity relation \citep{Piskunov_etal04}.

\item {\it Varying SFH} (affects all star counts):
  Variations of the SFH of a population under study over a
  characteristic time-scale leads to structure in the deduced IMF at a
  mass-scale at which stars evolve on that time-scale, if the observer
  wrongly assumes a constant SFH \citep{ES06}.

\item {\it Stellar evolution} (affects all star counts): Present-day
  stellar luminosities must be transformed to initial stellar
  masses. This relies on stellar-evolution theory \citep{Sc86}.

\item {\it Binary-stellar evolution} (affects all star counts):
  Present-day stellar luminosities must be transformed to initial
  stellar masses, but this may be wrong if the star is derived from an
  interacting binary. If important, then this only affects the
  massive-star IMF (F. Schneider and R. Izzard, private
  communication).

\item {\it Pre-main sequence evolution} (affects all populations with
  stars younger than a few~$10^8\,$yr): In young star clusters the
  late-type stellar luminosities need to be corrected for the stars
  not yet being on the main sequence (e.g. \citealt{HC00}). Pre-main
  sequence evolution tracks are highly uncertain for ages $\simless
  1\;$Myr \citep{Tout99,WT03}.  Field star-count data contain an
  admixture of young stars which bias the star-counts \citep{KTG93}.

\item {\it Differential reddening} (affects embedded star clusters):
  Patchily distributed gas and dust affects mass estimation. Variable
  extinction necessitates the introduction of an extinction limit
  which increases the lower mass limit to which the survey is complete
  \citep{Andersen09}.

\item {\it Binning}: Deriving the IMF power-law index from a binned
  set of data is prone to significant bias caused by the correlation
  between the assigned weights and the number of stars per bin. Two
  solutions have been proposed: variable-sized binning \citep{MU05}
  and newly developed (effectively) bias-free estimators for the
  exponent and the upper stellar mass limit
  \citep{maschberger+kroupa2009}.

\item {\it Crowding} (affects star-counts in star clusters): A compact
  far-away star cluster can lead to crowding and superpositions of
  stars which affects the determination of the IMF systematically
  \citep{MA08}.

\item {\it Early and late star-cluster evolution} (affects star-counts
  in star clusters): A large fraction of massive stars are ejected
  from the cluster core skewing the MF in the cluster downwards at the
  high-mass end \citep{P-AK06a, BKO11}.  When the residual gas is
  blown out of initially mass-segregated young clusters they
  preferentially loose low-mass stars within a few~Myr
  \citep{Marks08}. Old star clusters evolve through energy
  equipartition driven evaporation of low-mass stars \citep{VH97,
    BaumMakin03}.

\end{description}


\subsection{Discretising an IMF: optimal sampling and the $m_{\rm
    max}$-to-$M_{\rm ecl}$ relation} 
\label{sec:optsamp}

In view of Sec.~\ref{sec:introd_IMF} and~\ref{sec:philosophy} it is
clearly necessary to be able to set up and to test various hypothesis
as to how a stellar population emanates from a star formation
event. Two extreme hypotheses are: (1) The stars born together are
always perfectly distributed according to the stellar IMF. (2) The
stars born together represent a random draw of masses from the
IMF. Here one method of perfectly distributing the stellar masses
according to the form of the IMF is discussed. Sec.~\ref{sec:massgen}
describes how to generate a random population of stars highly
efficiently.

Note that throughout this text the relevant physical quantity of a
population is taken to be its mass, and {\it never} the number of
stars, $N$, which is not a physical quantity.

It is useful to consider the concept of optimally sampling a
distribution function. The problem to be addressed is that there is a
mass reservoir, $M_{\rm ecl}$, which is to be distributed according to
the IMF such that no gaps arise.

\vspace{2mm} \centerline{ \fbox{\parbox{\columnwidth}{{\sc Ansatz: Optimal
        Sampling}: Given a pre-defined form of a continuous
      distribution function, $\xi(m)$, of the variable $m\in [m_{\rm
        L}, m_{\rm U}]$ (Note: $m_{\rm U}=m_{\rm max*}$
        is adopted here for brevity of notation)  such that $m_2>m_1
      \Longrightarrow \xi(m_1) > \xi(m_2) > 0$, then the physical
      reservoir $M_{\rm ecl}$ is {\it optimally distributed} over
      $\xi(m)$ if the maximum available range accessible to $m$ is
      covered with the condition that a $m$ occurs once above a
      certain limit $m_{\rm max} \in [m_{\rm L}, m_{\rm U}]$,
      $\int_{m_{\rm max}}^{m_{\rm U}}\,\xi(m)\,dm = 1$. 
\label{box:optdistr}}}}
\vspace{2mm} 

\noindent We define $\xi(m)=k\,p(m)$, where $p(m)$ is the density
distribution function of stellar masses.  The last statement in the
above {\sc Ansatz} implies $k=1/(\int_{m_{\rm max}}^{m_{\rm
    U}}\,p(m)\,dm)$. Since the total mass in stars,
$M_*=k\,\int_{m_{\rm L}}^{m_{\rm max}}\,m\,p(m)\,dm$, one obtains
\[
M_* = {\int_{m_{\rm L}}^{m_{\rm max}}\,m\,p(m)\,dm \over 
\int_{m_{\rm max}}^{m_{\rm U}}\,p(m)\,dm}.
\]
It thus follows immediately that $m_{\rm max}' > m_{\rm max}
\Longrightarrow M_* > M_{\rm ecl}$ and also $m_{\rm max}' < m_{\rm
  max} \Longrightarrow M_* < M_{\rm ecl}$. Thus, only $m_{\rm max}' =
m_{\rm max} \Longrightarrow M_* = M_{\rm ecl}$. The concept of optimal
sampling appears to be naturally related to how $M_{\rm ecl}$ is
divided up among the stars: the largest chunk goes to $m_{\rm max}$
and the rest is divided up hierarchically among the lesser stars (see
{\sc Open Question II} on p.~\pageref{quest:openII}).

The above ansatz can be extended to a discretised optimal distribution
of stellar masses: Given the mass, $M_{\rm ecl}$, of the population
the following sequence of individual stellar masses yields a
distribution function which exactly follows $\xi(m)$,
\begin{equation}
  m_{i+1} = \int_{m_{i+1}}^{m_{i}} m\,\xi(m) dm, \quad m_{\rm L} \le  m_{i+1} < m_i,
  \quad m_1 \equiv m_{\rm max}.
\label{eq:optsamp}
\end{equation}
The normalisation and the most massive star in the sequence are set by
the following two equations,
\begin{equation}
1 = \int_{m_{\rm max}}^{m_{\rm max*}} \xi(m)\,dm,
\label{eq:mm}
\end{equation}
with
\begin{equation}
M_{\rm ecl}(m_{\rm max}) - m_{\rm max} = \int_{m_{\rm L}}^{m_{\rm max}}
m\,\xi(m)\,dm
\label{eq:Mecl}
\end{equation}
as the closing condition. These two equations need to be solved
iteratively. An excellent approximation is given by the following
formula (eq.~10 in \citealt{PWK07} assuming $m_{\rm max
  *}=150\,M_\odot$):
\begin{equation} {\rm log}_{10} \left( {m_{\rm max} \over M_\odot}
  \right) = 2.56\;{\rm log}_{10} \left( {M_{\rm ecl} \over M_\odot}
  \right) \, \left( 3.82^{9.17} + \left[ {\rm log}_{10}\left( {M_{\rm ecl}
        \over M_\odot} \right) \right]^{9.17} \right)^{-{1\over 9.17}} - 0.38
\label{eq:mmax}
\end{equation}

Note that Eq.~\ref{eq:Mecl} contains a correction term $m_{\rm max}$:
the mass, $M_{\rm ecl}$, between $m_{\rm L}$ and $m_{\rm max}$ does
not include the star with $m_{\rm max}$ as this star lies between
$m_{\rm max}$ and $m_{\rm max *}$. The semi-analytical calculation of
the $m_{\rm max}-M_{\rm ecl}$ relation by \cite{KW03} and \cite{WKB09}
is less accurate by not including the correction term $m_{\rm
  max}$. The correction turns out to be insignificant as both
semi-analytical relations are next to identical
(Fig.~\ref{fig:mmax_model}) and are a surprisingly good description of
the observational data (Fig.~\ref{fig:mmaxf}).

Eq.~\ref{eq:optsamp} defines, here for the first time, how to sample
the IMF perfectly in the sense that the stellar masses are spaced
ideally such that no gaps arise and the whole accessible range $m_{\rm
  L}$ to $m_{\rm max}$ is fully filled with stars. This is referred to
as {\sc Optimal Sampling} (see also eq.~8.2 in \citealt{Aarseth03} for
a related concept). The disadvantage of this method is that the target
mass $M_{\rm ecl}$ cannot be achieved exactly because it needs to be
distributed into a discrete number of stars.  The mass of the
generated stellar population is $M_{\rm ecl}$ to within $\pm\,m_{\rm
  L}$ ($\equiv m_{\rm low}\approx 0.07\,M_\odot$ for most
applications), because the integral (Eq.~\ref{eq:optsamp}) is
integrated from $m_{\rm max}$ downwards.\footnote{The publicly
  available C programme {\sc Optimf} allowing the generation of a
  stellar population of mass $M_{\rm ecl}$ is available at
  http://www.astro.uni-bonn.de/en/download/software/ $\;$.}
\begin{figure}
\begin{center}
  \rotatebox{0}{\resizebox{0.85
      \textwidth}{!}{\includegraphics{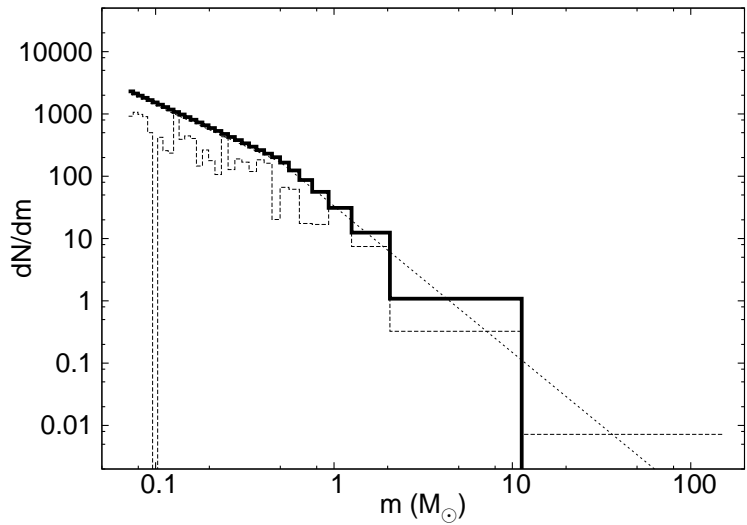}}}
  \vskip -4mm
  \caption[Optimal vs. random sampling: IMF]{ \small{Stellar IMF
      ($dN/dm$) vs. stellar mass ($m$) for a $M_{\rm
        ecl}=149.97\,M_\odot$ cluster.  The thin dashed line is the
      analytical canonical IMF (Eq.~\ref{eq:imf}).  The thick
      histogramm is a population of individual stars generated
      according to {\sc Optimal Sampling} (Eq.~\ref{eq:optsamp}),
      starting with $m_{\rm max}=12\,M_\odot$ (Eq.~\ref{eq:mmax}). The
      optimally sampled population contains 322 stars.  The
      mass-dependent bin width is chosen to ensure that ten stars are
      in each bin.  The thin dashed histogramm is a population of
      stars chosen randomly from the IMF using the mass-generating
      function (Sec.~\ref{sec:massgen}) with $M_{\rm
        ecl}=150.22\,M_\odot$ being reached without an upper limit
      ($m_{\rm max}=\infty$). This population contains 140~stars. The
      same bins are used as in the thick solid histogramm. Note how
      {\sc Optimal Sampling} perfectly reproduces the IMF while random
      sampling shows deviations in the form of gaps. Which is closer
      to reality (remembering that observational data contain
      uncertainties that act as randomisation agents)?  }}
\label{fig:optIMF}
\end{center}
\end{figure}
Fig.~\ref{fig:optIMF} demonstrates how an IMF constructed using {\sc
  Optimal Sampling} compares to one generated with random sampling
from the IMF for a population with $M_{\rm ecl}=150\,M_\odot$.
Fig.~\ref{fig:mmax_model} and Fig.~\ref{fig:avmass}, respectively,
show $m_{\rm max}-M_{\rm ecl}$ and average stellar mass model data
using both sampling techniques.

\begin{figure}
\begin{center}
  \rotatebox{0}{\resizebox{0.85
      \textwidth}{!}{\includegraphics{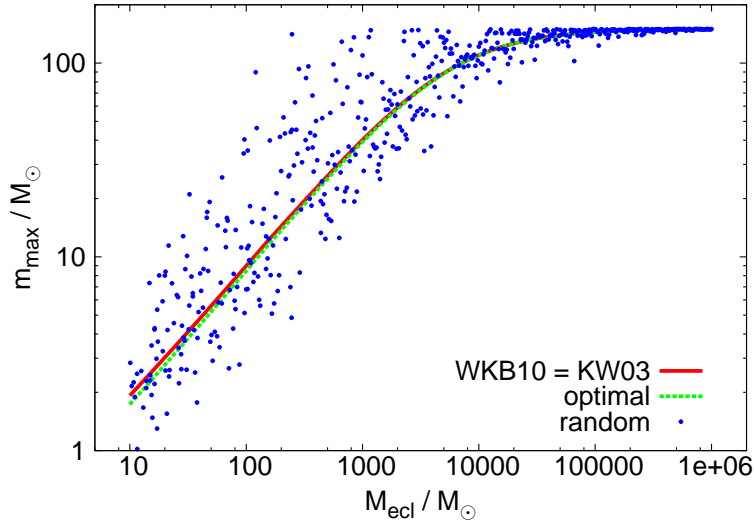}}}
  \vskip -4mm
  \caption[Optimal vs. random sampling:
  $m_\mathrm{max}$-to-$M_\mathrm{ecl}$]{\small{The maximum stellar
      mass, $m_{\rm max}$, is plotted against the stellar mass of the
      population, $M_{\rm ecl}$, for {\sc Optimal Sampling}
      (short-dashed green line) and for randomly drawing a population
      of stars from the IMF (filled blue circles). For each sampling
      method 100~populations are generated per dex in $M_{\rm ecl}$.
      The canonical physical maximum stellar mass, $m_{\rm max
        *}=150\,M_\odot$ is assumed.  The semi-analytical $m_{\rm
        max}-M_{\rm ecl}$ relation from \cite{KW03} and \cite{WKB09}
      (red solid curve) and the corrected version from
      Eq.~\ref{eq:mmax} (green dashed curve) are nearly
      identical. Which is closer to reality, random or optimal
      sampling from the IMF? The small scatter in the observational
      $m_{\rm max}-M_{\rm ecl}$ relation (Fig.~\ref{fig:mmaxf}) and
      the small scatter in observationally derived IMF power-law
      indices around the Salpeter value (Fig.~\ref{fig:ahist}, {\sc
        Open Question~III} on p.~\pageref{quest:openIII}) suggest that
      {\sc Optimal Sampling} may be a more realistic approach to
      nature than purely random sampling.  }}
\label{fig:mmax_model}
\end{center}
\end{figure}

\begin{figure}
\begin{center}
  \rotatebox{0}{\resizebox{0.85
      \textwidth}{!}{\includegraphics{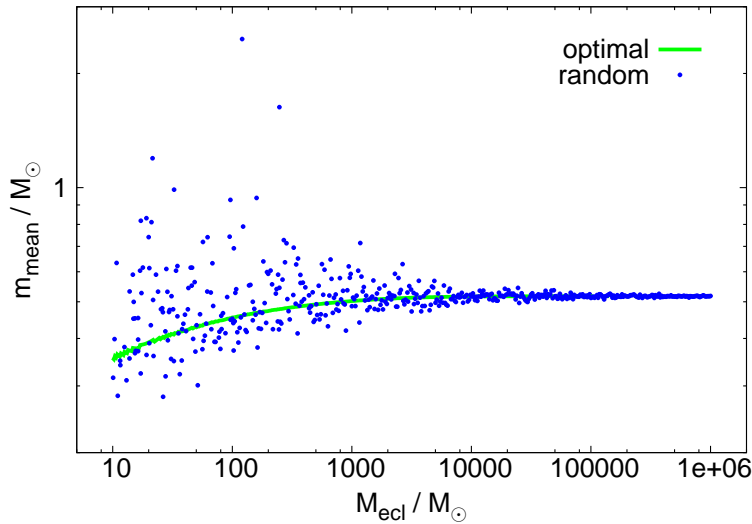}}}
  \vskip -4mm
  \caption[Optimal vs. random sampling: $\bar
  m$--$M_\mathrm{ecl}$]{\small{ The average stellar mass of stellar
      populations of mass $M_{\rm ecl}$ generated with {\sc Optimal
        Sampling} and random sampling from the IMF. Otherwise as
      Fig.~\ref{fig:mmax_model}. Note the extreme random deviations
      from the canonical IMF if it is interpreted to be a
      probabilistic distribution function. }}
\label{fig:avmass}
\end{center}
\end{figure}

\vspace{2mm} \centerline{ \fbox{\parbox{\columnwidth}{{\sc Important
        Hint:} Nature may be prefering {\sc Optimal Sampling}. This is
      evident from the quite tight observational $m_{\rm max}-M_{\rm
        ecl}$ data (Fig.~\ref{fig:mmaxf}) and from {\sc Open Question
        III} (p.~\pageref{quest:openIII}) and the {\sc Sociological
        Hypopthesis} (p.~\pageref{hyp:soc}). Theoretical support that
      nature may not be playing dice comes from the emergence of the
      $m_{\rm max}-M_{\rm ecl}$ relation in SPH and FLASH computations
      of star formation in turbulent molecular clouds (see the {\sc
        BVB Conjecture} on p.~\pageref{quote:bvb}).  }}} \vspace{2mm}

\subsection{Discretising an IMF:  random sampling and the
  mass-generating function} 
\label{sec:massgen}

In order to randomly discretise a stellar population we need to be
able to generate stellar masses independently of each other. This can
be done with constraints (e.g. ensuring that the $m_{\rm max}-M_{\rm
  ecl}$ relation is fulfilled by applying $m\le m_{\rm max}(M_{\rm
  ecl})$, or that stellar masses $m\leq m_{\rm max *}\approx
150\,M_\odot$) or without constraints ($m_{\rm max}=\infty$).  This is
achieved via the mass generating function.  A {\it mass-generating
  function} is a mapping from a uniformly distributed random variable
$X$ to the stellar mass which is distributed according to the
IMF. Generating functions allow efficient random discretisation of
continuous distributions (see \citealt{Kroupa08} for more details).

A generating function can be written in the following way. Assume the
probability distribution function depends on the variable
$\zeta_\mathrm{min} \le \zeta \le \zeta_\mathrm{max}$ (in this case
the stellar mass, $m$). Consider the probability, $X(\zeta)$, of
encountering a value for the variable in the range
$\zeta_\mathrm{min}$ to $\zeta$,
\begin{equation}
X(\zeta) = \int_{\zeta_\mathrm{min}}^{\zeta} p(\zeta')d\zeta',
\end{equation}
with $X(\zeta_\mathrm{min}) = 0 \le X(\zeta) \le X(\zeta_\mathrm{max})
= 1$, and $p(\zeta)$ is the probability distribution function, or
probability density, normalised such that the latter equal sign holds
($X = 1$). For the two-part power-law IMF the corresponding
probability density is
\begin{eqnarray}
p_{1}(m) = k_{p,1} m^{-\alpha_{1}},& 0.07 \le m \le 0.5 M_\odot \nonumber\\
p_{2}(m) = k_{p,2} m^{-\alpha_{2}},& 0.5 \le m \le m_\mathrm{max},
\end{eqnarray}
where $k_{p,i}$ are normalisation constants ensuring continuity at
$0.5\,M_\odot$ and
\begin{equation}
\int_{0.07\,M_\odot}^{0.5\,M_\odot} p_{1}\,dm' + \int_{0.5\,M_\odot}^{m_\mathrm{max}}p_{2}\,dm' = 1,
\end{equation}
whereby $m_\mathrm{max}$ may follow from the mass of the population,
$M_{\rm ecl}$. Defining
\begin{equation}
X_{1}^{'} = \int_{0.07\,M_\odot}^{0.5\,M_\odot} p_{1}(m')\,dm',
\end{equation}
it follows that 
\begin{equation}
X_{1}(m) = \int_{0.07\,M_\odot}^{m} p_{1}(m')\,dm', \quad\quad \mathrm{if}\,m \le 0.5 M_\odot,
\end{equation}
or 
\begin{equation}
X_{2}(m) = X_{1}^{'}+\int_{0.5\,M_\odot}^{m} p_{2}(m')\,dm', \quad\quad \mathrm{if}\,m > 0.5 M_\odot.
\end{equation}
The generating function for stellar masses follows by inverting the
above two equations $X_{i}(m)$. 

The procedure is then to choose a uniformly distributed random variate
$X \in [0,1]$ and to select the generating function $m(X_1 = X)$ if 0
$\le X \le X_1^{'}$, or $m(X_2 = X)$ if $X_1^{'} \le X \le 1$.
Stellar masses are generated until $M_{\rm ecl}$ is reached to within
some pre-set tolerance.  This algorithm is readily generalised to any
number of power-law segments (Eq.~\ref{eq:imf_mult},
Sect.~\ref{sec:canIMF}), such as including a third segment for brown
dwarfs and allowing the IMF to be discontinuous near $0.07\,M_\odot$
(Sec.~\ref{sec:BD_IMF}). Such a form has been incorporated into
Aarseth's {\sc $N$-body4/6/7} programmes.

For a general $\xi(m)$ and if $X(m)$ cannot be inverted stellar masses
may be generated by constructing a table of $X(m), m$ values,
\begin{equation}
M(m)=\int_{0.07\,M_\odot}^m\,m' \, \xi(m')\, dm', \quad X(m) = {M(m)
  \over M_{\rm ecl}},
\end{equation}
such that $X(m_{\rm max}) = 1$.  For a random variate $X$ the
corresponding $m$ is obtained by interpolating the table, whereby the
$X$ are distributed uniformly and the procedure is repeated until
$M_{\rm ecl}$ is reached to some pre-set tolerance.

Another highly efficient method for generating stellar masses randomly from
arbitrary distribution functions is discussed in
Sec.~\ref{sec:num_imf}. 

\subsection{A practical numerical formulation of the IMF}
\label{sec:num_imf}

Assuming the stellar IMF to be a probability density distribution such
that stellar masses can be generated randomly (Sec.~\ref{sec:massgen})
from the IMF with ($m\leq m_{\rm max}$) or without ($m_{\rm
  max}=\infty$) constraints remains a popular approach.  The two-part
description can be straightforwardly expanded to a multi-part power
law. However, the direct implementation of this description requires
multiple {\tt IF}-statements.  In the following a handy numerical
formulation is presented for randomly generating stellar masses which
replaces complicated {\tt IF}-constructions by two straight-forward
loops \citep{P-AK06a}. 

Historically, multi-power-law IMFs start indexing intervals and slopes
at zero instead of one. For simplicity we here index $n$ intervals
from 1 up to $n$.  We now consider an arbitrary IMF with $n$ intervals
fixed by the mass array $[m_{\mathrm{0}},\ldots,m_{\mathrm{n}}]$ and
the array of functions $f_1$,\ldots,$f_\mathrm{n}$. On the i-th
interval $[m_{\mathrm{i-1}},m_{\mathrm{i}}]$ the IMF is described by
the function $f_\mathrm{i}$.  The segment functions refer to the
''linear'' IMF, $\xi(m)=dN/dm$, and not to the logarithmic IMF,
$\xi_{\rm L}(m)=dN/d\log_{10} m$.  At this point it is not required
that the segment functions $f_\mathrm{i}$ are scaled by a constant
such that continuity is ensured on the interval boundaries. They only
need to describe the functional form.

For the case of a multi-power law these functions are 
\begin{equation}
f_{\mathrm{i}}(m)= m^{-\alpha_\mathrm{i}}\;.
\end{equation}

The segment functions may also be log-normal distributions, as for
example in the IMFs of \citet{MS79} or \citet{Chrev03}, but in general
they can be arbitrary.

We first define the two $\Theta$-mappings ($\Theta$-closed and $\Theta$-open)
\begin{equation}
\Theta_{[\phantom{i}]}(x)=\left\{
\begin{array}{cc}
  1&x\ge0\\
  0&x<0
\end{array}\right.\;,
\end{equation}
\begin{equation}
\Theta_{]\phantom{i}[}(x)=\left\{
\begin{array}{cc}
  1&x>0\\
  0&x\le0
\end{array}\right.\;,
\end{equation}
and the function
\begin{equation}
\Gamma_{[\mathrm{i}]}(m) = \Theta_{[\phantom{i}]}(m-m_{\mathrm{i}-1})
\Theta_{[\phantom{i}]}(m_\mathrm{i}-m)\;.
\end{equation}
The $\Gamma_{[\mathrm{i}]}(m)$ function is unity on the interval
$[m_{\mathrm{i}-1},m_{\mathrm{i}}]$ and zero otherwise.

The complete IMF can now be conveniently formulated by
\begin{equation}
\xi(m)=
k\;
\prod_{\mathrm{j}=1}^{\mathrm{n}-1}\;\Delta(m-m_\mathrm{j})\;
\sum_{\mathrm{i}=1}^{\mathrm{n}}\;\Gamma_{[\mathrm{i}]}(m)\;
{\cal C}_{\mathrm{i}}\;f_{\mathrm{i}}(m)\;,
\end{equation}
where $k$ is a normalisation constant and the array
(${\cal C}_{1}$,\ldots,${\cal C}_{\mathrm{n}}$) is to ensure continuity at the
interval boundaries. They are defined recursively by
\begin{equation}
{\cal C}_{1}=1
\;\;\;,\;\;\;
{\cal C}_{\mathrm{i}} = {\cal C}_\mathrm{i-1}\;
\frac{f_\mathrm{i-1}(m_{\mathrm{i}-1})}
{f_{\mathrm{i}}(m_{\mathrm{i}-1})}\;.
\end{equation}
For a given mass $m$ the $\Gamma_{[\mathrm{i}]}$ makes all summands
zero except the one in which $m$ lies.  Only on the inner
interval-boundaries do both adjoined intervals give the same
contribution to the total value.  The product over
\begin{equation}
\Delta(x)=\left\{
\begin{array}{cc}
0.5& x=0\\ 
1& x\not=0
\end{array}
\right.
\end{equation}
halves the value due to this double counting at the interval-boundaries.
In the case of $n$ equals one (one single power law), 
the empty product has, by convention, 
the value of unity.

An arbitrary integral over the IMF is evaluated by
\begin{equation}
\int_{a}^{b}\;\xi(m)\;\mathrm{d}m=
\int_{m_{0}}^{b}\;\xi(m)\;\mathrm{d}m-
\int_{m_0}^{a}\;\xi(m)\;\mathrm{d}m\;,
\end{equation}
where the primitive of the IMF is given by
\[
\int_{m_{0}}^{a}\;\xi(m)\;\mathrm{d}m=k
\sum_{\mathrm{i}=1}^{\mathrm{n}}\;\Theta_{]\phantom{i}[}(a-m_{\mathrm{i}})
\;{\cal C}_{\mathrm{i}}\;\int_{m_{\mathrm{i}-1}}^{m_{\mathrm{i}}}\;
f_{\mathrm{i}}(m)
\;\mathrm{d}m
\]
\begin{equation}
\label{eq_num_imf_int}
\rule{2.0cm}{0pt}+k\sum_{\mathrm{i}=1}^{\mathrm{n}}\;\Gamma_{[\mathrm{i}]}(a)
\;{\cal C}_{\mathrm{i}}\;\int_{m_{\mathrm{i}-1}}^{a}\;f_{\mathrm{i}}(m)
\;\mathrm{d}m\;.
\end{equation}

The expressions for the mass content, i.e. $m\;\xi(m)$, and its
primitive are obtained by multiplying the above expressions in the
integrals by $m$ and one has to find the primitives of
$mf_\mathrm{i}(m)$.

Stars can now be diced from an IMF, $\xi(m)$, based on the above
formulation and the concept of the generating function
(Sect.~\ref{sec:massgen}) in the following way: A random number $X$ is
drawn from a uniform distribution and then transformed into a mass
$m$. The mass segments transformed into the $X$-space are fixed by the
array $\lambda_{0}$,\ldots,$\lambda_{n}$ defined by
\begin{equation}
\lambda_{i}=\int_{m_0}^{m_{\mathrm{i}}}\xi(m)\;\mathrm{d}m.
\end{equation}
If $P(X)$ denotes the uniform distribution with $P(X)=1$ between 0 and
$\lambda_\mathrm{n}$, both functions are related by
\begin{equation}
\int_{m_0}^{m(X)}\xi(m^\prime)\;\mathrm{d}m^\prime = 
\int_0^XP(X^\prime)\;\mathrm{d}X^\prime = X\;.
\end{equation}
If a given $X$ lies between $\lambda_\mathrm{i-1}$ and $\lambda_\mathrm{i}$ the
corresponding mass $m$ lies in the i-th interval $[m_\mathrm{i-1},m_\mathrm{i}]$
and it follows,
\begin{equation}
  X(m) = \lambda_\mathrm{i-1} + k\,{\cal C}_\mathrm{i}
  \left(F_\mathrm{i}(m)-F_\mathrm{i}(m_\mathrm{i-1})\right)\;,
\end{equation}
or 
\begin{equation}
  m(X) = F_\mathrm{i}^{-1}\left(
  \frac{X-\lambda_\mathrm{i-1}}{k\,{\cal C}_\mathrm{i}}+F_\mathrm{i}(m_\mathrm{i-1})
  \right)\;.
\end{equation}
where $F_{\mathrm{i}}$ is a primitive of $f_{\mathrm{i}}$ and
$F^{-1}_{\mathrm{i}}$ is the primitive's inverse mapping.  The
complete expression for the solution for $m$ is given by
\[
m(X)=\sum_{\mathrm{i}=1}^{\mathrm{n}}
{_\lambda}\Gamma_{[\mathrm{i}]}F_{\mathrm{i}}^{-1}\left(
\frac{X-\lambda_{i-1}}{k\,{\cal C}_{\mathrm{i}}}+F_{\mathrm{i}}(m_{\mathrm{i}-1})
\right)
\]
\begin{equation}
  \label{eq_num_imf_dice}
\rule{0.5cm}{0pt}\cdot
\prod_{\mathrm{j}=1}^{\mathrm{n}-1}\;\Delta(X-\lambda_{\mathrm{i}})
\end{equation}
where $_\lambda\Gamma_{\mathrm{i}}$ are mappings which are unity
between $\lambda_{\mathrm{i}-1}$ and $\lambda_{\mathrm{i}}$ and zero
otherwise.  Note that the primitives are determined except for an
additive constant, but it is canceled out in the relevant expressions
in Eq.~\ref{eq_num_imf_int} and \ref{eq_num_imf_dice}.

The most used segment function for the IMF is a power-law,
\begin{equation}
  f(m) = m^{-\alpha}\;.
\end{equation}
The corresponding primitive and its inverse mapping is
\begin{equation}
  F(m) = \left\{
  \begin{array}{cc}
    \frac{m^{1-\alpha}}{1-\alpha}& \alpha\not= 1\\ 
    &\\
    \ln(m) & \alpha=1
  \end{array}
  \right.\;,
\end{equation}
and
\begin{equation}
  F^{-1}(X) = \left\{
  \begin{array}{cc}
    \left((1-\alpha)X\right)^{\frac{1}{1-\alpha}}& \alpha\not= 1\\ 
    &\\
    \exp(X) & \alpha=1
  \end{array}
  \right.\;,
\end{equation}

The other segment function used is a log-normal distribution, i.e. a Gaussian  
distribution of the logarithmic mass,
\begin{equation}
  \xi(lm) \propto \exp\left(
  -\frac{(lm-lm_\mathrm{c}))^2}{2\sigma^2}
  \right)\;,
\end{equation}
where $lm\equiv {\rm log}_{10}m$.  The corresponding segment function
  is
\begin{equation}
  f(m) = \frac{1}{m}\;\exp\left(
  -\frac{(lm-lm_\mathrm{c})^2}{2\sigma^2}
  \right),
\end{equation}
with  the primitive 
\begin{equation}
  F(m)=\sqrt{\frac{\pi}{2}}\sigma \ln 10 \;\mathrm{erf}\left(
  \frac{lm-lm_\mathrm{c}}{\sqrt{2}\sigma}
  \right)\;,
\end{equation}
and the inverse of the primitive
\begin{equation}
  F^{-1}(X) =
  10^{\sqrt{2}\sigma\mathrm{erf}^{-1}\left(\sqrt{\frac{2}{\pi}}\frac{X}{\sigma\ln
      10}\right)+lm_\mathrm{c}}\;,
\end{equation}
where $\mathrm{erf}$ and $\mathrm{erf}^{-1}$ are the Gaussian error function,
\begin{equation}
  \mathrm{erf}(Y) = \frac{2}{\sqrt{\pi}}\int_0^Y e^{-y^2}\;dy\;,
\end{equation}
and its inverse, respectively.

Several accurate numerical approximations of the Gaussian error
function exist but approximations of its inverse are quite rare.  One
such handy numerical approximation of the Gaussian error function
which allows an approximation of its inverse, too, has been presented
by Sergei
Winitzki\footnote{homepages.physik.uni-muenchen.de/\textasciitilde
  Winitzki/erf-approx.pdf} based on a method explained in
\citet{WINI03}:

The approximation of the error function for $Y\ge0$ is
\begin{equation}
  \mathrm{erf(Y)}\approx \left(
  1-\exp\left(
  -Y^2\frac{\frac{4}{\pi}+a\,Y^2}{1+a\,Y^2}
  \right)
  \right)^{\frac{1}{2}}\;,
\end{equation}
with
\begin{equation}
  a = \frac{8}{3\pi}\frac{\pi-3}{4-\pi}\;.
\end{equation}
Values for negative $Y$ can be calculated with 
\begin{equation}
  \mathrm{erf}(Y)=-\mathrm{erf}(-Y)\;.
\end{equation}
The approximation for the inverse of the error function follows 
directly,
\begin{equation}
  \mathrm{erf}^{-1}(Y)\approx \left(
  -\frac{2}{\pi a}- \frac{\ln(1-Y^2)}{2}+
  \sqrt{\left(\frac{2}{\pi a}+\frac{\ln(1-Y^2)}{2}\right)^2
  -\frac{1}{a}\ln(1-Y^2)}
  \right)^{\frac{1}{2}}\;.
\end{equation}

\vspace{2mm} \centerline{ \fbox{\parbox{\columnwidth}{ {\sc Important result}:
      The above algorithm for dicing stars from an IMF supporting
      power-law and log-normal segment functions has been coded in the
      publicly available software package {\sc libimf} available
      at\\http://www.astro.uni-bonn.de/download/software/
}}}
\vspace{2mm}

\subsection{Statistical treatment of the data}
\label{sec:stats}

Whichever is a better description of nature, optimal or random
sampling from the IMF, a set of observationally derived stellar masses
will appear randomised because of uncorrelated measurement
uncertainties. Statistical tools are therefore required to help
analyse the observed set of masses in the context of their possible
parent distribution function and upper limit.

We concentrate here in particular on analysing the high-mass end of
the IMF, which follows a power-law probability density.  Estimating
the exponent via binning (using constant-size bins in logarithmic
space) can introduce significant bias
\citep{MU05,maschberger+kroupa2009}, especially for meager data sets.
This can be remedied by using bins containing approximately the same
number of data points \citep{MU05}.  However, binning does not allow
one to estimate the upper limit of the mass function.  A more suitable
approach to estimate both exponent and upper limit simultaneously is
to use the maximum likelihood method.  The estimate there is given
just by the largest data point, and is consequently also naturally
biased to too small values.  \citet{maschberger+kroupa2009} give a
correction factor which leads to unbiased results for the upper limit.

Besides biases due to the statistical method observational limitations
can also introduce biases in the exponent.  The influence of
unresolved binaries for the high-mass IMF slope is less than $\pm$ 0.1
dex \citep{MA08,WKM09}.  Random superpositions, in contrast, can cause
significant biases, as found by \citet{MA08}. Differential reddening
can affect the deduced shape of the IMF significantly
\citep{Andersen09}. A further point in data analysis besides the
estimation of the parameters is to validate the assumed power-law form
for the IMF, and in particular to decide whether e.g. a universal
upper limit ($m_{max*}=150 M_\odot$) is in agreement with the data.

\begin{figure}
\begin{center}
\rotatebox{0}{\resizebox{0.75 \textwidth}{!}{\includegraphics{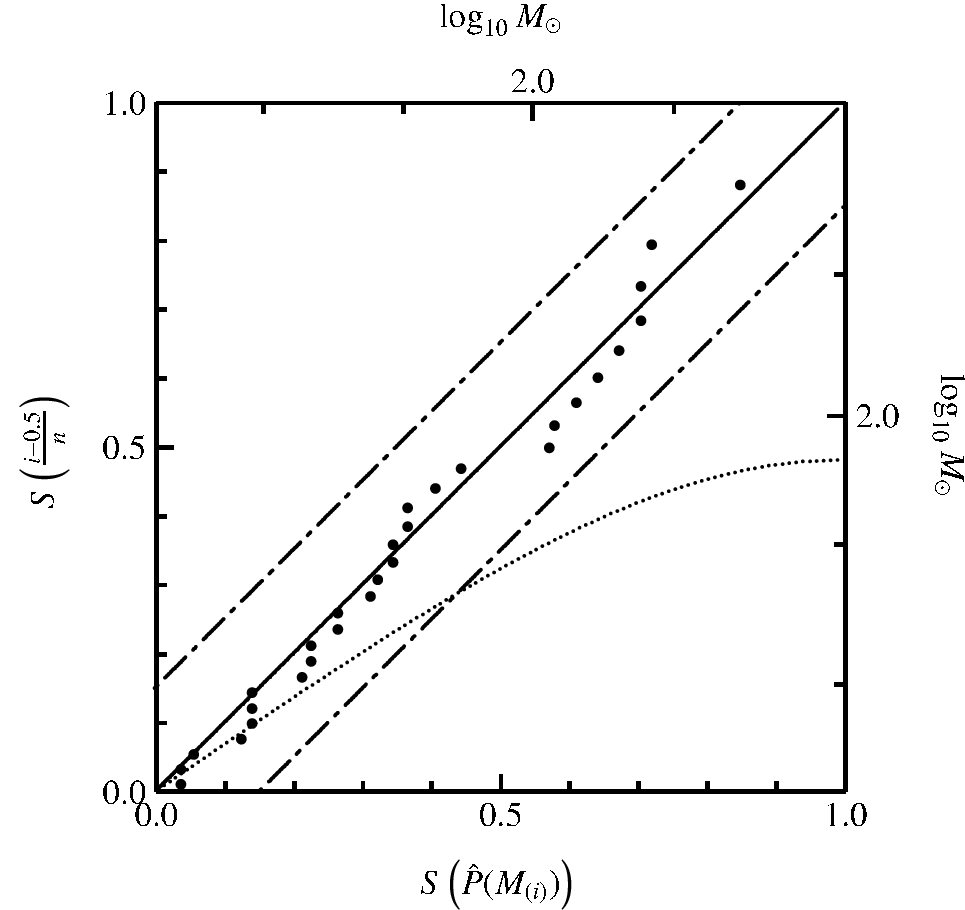}}}
\vskip -4mm
\caption[SPP method to measure IMF]{\small{ SPP plot of the massive
    stars in R136 using a power law truncated at 143 $M_\odot$(from
    \citealp{maschberger+kroupa2009}).  The data are following this
    hypothesis and lie along the diagonal within the 95\% acceptance
    region of the stabilised Kolmogorov-Smirnov test (limited by
    dashed lines).  An infinite power law as the parent distribution
    function, the dotted line bending away from the diagonal, can be
    ruled out. The stellar MF in R136 is thus a power-law with index
    $\alpha=2.2$ truncated at~$143\,M_\odot$ }}
\label{fig:maschberger_spp}
\end{center}
\end{figure}

For this purpose standard statistical tests, as e.g. the
Kolmogorov-Smirnov test, can be utilised, but their deciding powers
are not very high.  They can be significantly improved by making a
stabilising transformation, $S$ \citep{maschberger+kroupa2009}.  A
graphical goodness-of-fit assessment can then be made using the
stabilised probability-probability (SPP) plot,
e.g.~Fig.~\ref{fig:maschberger_spp}.  This plot has been constructed
using a truncated power law ($m_{max*}= 143 M_\odot$), and is aimed to
help to decide whether a truncated or an infinite power law fits the
stellar masses of the 29 most massive stars in R136 (masses taken from
\citealp{MH98}, using the isochrones of
\citealp{chlebowski+garmany1991}).  The points are constructed from
the ordered sample of the masses, with the value of the stabilised
cumulative probability, $S(\hat{P}(m_{(i)}))$ as x coordinate (using
the estimated parameters) and the stabilised empirical cumulative
probability, $S(\frac{i-0.5}{n})$, as the y~coordinate.  The data
follow the null hypothesis of a truncated power law and lie therefore
along the diagonal.  The null hypothesis would be challenged
significantly if the data would leave the region enclosed by the
dashed parallels to the diagonal, which is the 95\% acceptance region
of the stabilised Kolmogorov-Smirnov test.  An alternative hypothesis
of an infinite power law (with the same exponent) is shown as the
dotted line which strongly bends away from the diagonal and is in
significant disagreement with the data.  Plots like
Fig.~\ref{fig:maschberger_spp} can be constructed for any combination
of null hypothesis and alternative hypothesis and have great potential
to improve the statistical analysis of the upper mass end of the IMF.

\vspace{2mm} \centerline{ \fbox{\parbox{\columnwidth}{ {\sc Important result}:
      A program to estimate the parameter of power-law distributed
      data and to calculate goodness-of-fit tests for them is available
      in the publicly available software package {\sc statpl}
      available
      at\\http://www.astro.uni-bonn.de/download/software/
}}}

\vspace{2mm}

\subsection{Binary systems}
\label{sec:bins}

In order to infer the stellar IMF it is necessary to account for all
stars in the population, including the companions of multiple systems.
Indeed, the vast majority of stars are observed to form in multiple
systems. However, since dynamically not evolved star-forming regions
of an age around about 1~Myr such as the Taurus-Auriga sub-clusters
have a high multiplicity fraction of nearly 100~per cent, this
immediately implies that most stars by far must form as binaries. This
is because if they were to form as triple or higher-order systems,
then they would decay on their system crossing time-scale which is far
shorter than 1~Myr leading, by 0.5--1$\,$Myr, to a substantial
population of single stars which is not observed \citep{GK05, GKGB07}.

The IMF appearing largely invariant to star-formation conditions (but
see Sec.~\ref{sec:currentstate}) constitutes a statistical statement
on one of the birth dynamical properties of stars (namely their
distribution of masses). So, since both the IMF and the birth binary
population (BBP) are the result of the same (star-formation) process,
and since the IMF is a result of this process ``one level deeper
down'' than the BBP, it is quite natural to suggest that the formal
mathematical distribution function of all of the birth dynamical
properties (see Footnote~\ref{foot:dynprop} on
p.~\pageref{foot:dynprop}) of stars are also quite invariant.  It
follows that the star-formation outcome in terms of stellar masses
(the IMF) and multiple systems (the birth binary population - BBP) can
be formulated by the {\sc Star Formation Universality Hypothesis}.

\vspace{2mm}
\centerline{ \fbox{\parbox{\columnwidth}{
{\sc 
The Star-Formation Universality
        Hypothesis:}\\[-6mm]
\center {IMF universality $\Longleftrightarrow$
      BBP universality.}
\label{hyp_pk:univ}}}}
\vspace{2mm}

\noindent For stars with $m\simless 5\,M_\odot$ the {\sc Birth Binary
  Population} is deduced from an elaborate analysis:

\vspace{2mm}

\centerline{ \fbox{\parbox{\columnwidth}{\vspace{1mm}
{\sc The birth binary population (BBP):\vspace{1mm}}
\begin{itemize}
\item random pairing from the canonical IMF (Eq.~\ref{eq:imf}) for
  $0.1\simless m/M_\odot \simless 5$;
\item thermal eccentricity distribution of eccentricities, $f_e(e)=2\,e$;
\item the period distribution function
\begin{equation}
f_{P,birth}=\eta \, {lP-lP_{\rm min} \over \delta + \left(lP -
  lP_{\rm min}\right)^2},
\label{eq:fPbirth}
\end{equation}
where $\eta=2.5, \delta=45, lP_{\rm min}=1$ and $\int_{lP_{\rm
    min}}^{lP_{\rm max}} f_{P,birth} \, dlP = 1$ such that the birth
binary fraction is unity ($lP_{\rm max}=8.43; lP\equiv {\rm
  log}_{10}P$ and $P$ is in days). Here $f_e\,de$ and $f_P\,dlP$ are
the fraction of all orbits with eccentricity in the range $e, e+de$
and log-period in the range $lP, lP+dlP$, respectively. Note that
conversion between $P$ and the semi-major axis is convenient through
Kepler's third law: $a^3/P_{\rm yr}^2 = m_1 + m_2$, where $a$ is
in~AU, $P_{\rm yr} = P/365.25$ is the period in years and $m_1, m_2$
are the primary- and secondary-star masses in $M_\odot$.
\end{itemize}
\label{def_pk:bbp}}}}

\vspace{2mm}

\noindent
The following are to be noted: 
\begin{description}
\item (I) The BBP was derived by using observations of pre-main
  sequence and main-sequence binary populations as initial and final
  boundary conditions, respectively. \cite{K95a} postulated that there
  exist two stellar-dynamical operators, $\Omega_P$ and $\Omega_q$,
  which independently transform the period- and mass-ratio
  distribution functions (that are independent products of the
  star-formation process for the majority of binaries) between the
  initial and final states. It is possible to demonstrate that both
  $\Omega_P$ and $\Omega_q$ exist.  Furthermore they are equal and are
  given by a characteristic star-cluster consisting of~200 binary
  systems with a characteristic radius of about 0.8~pc
  (\citealt{K95b}, \citealt{MKO11}),
\begin{equation}
  \Omega_P = \Omega_q \equiv \Omega = 
  (200\,{\rm binaries}, R_{\rm ch}\approx 0.8\,{\rm pc}).
\label{eq:domode}
\end{equation}

\vspace{2mm} \centerline{ \fbox{\parbox{\columnwidth}{ {\sc Important
        result}: The interpretation of this result is that the BBP is,
      like the IMF, a fundamental outcome of star formation and that
      most stars in the MW disk stem from star-formation events that
      are dynamically equivalent to the characteristic, or
      dominant-mode, cluster.  }}}
\vspace{2mm}

\item (II) The deduced maximum binary period in the BBP of $10^{8.43}\,$d
  corresponds to a spatial scale of $\approx 10^4\,$AU which is the
  typical dimension of a pre-stellar cloud core \citep{Kirk05}.
\item (III) The evolved BBP, which matches the Galactic field stellar
  and binary population, also accounts simultaneously for the
  individual-star and the system LFs (Fig.~\ref{fig:lfmods}).
\end{description}

\noindent The BBP is the deduced \citep{K95a,K95b} outcome of star
formation in low to intermediate density ($\rho\simless
10^5\,M_\odot$/pc$^3$) cloud regions (e.g. embedded clusters) but the
rules layed out in Eq.~\ref{eq:fPbirth} may well be formally
applicable to higher density regions as well, whereby wide binaries
are naturally truncated due to close-packing. 

\vspace{2mm} \centerline{ \fbox{\parbox{\columnwidth}{ {\sc Important
        result}: In this sense the BBP is a formal mathematical
      description of the outcome of star formation. Just like the
      formal stellar IMF (Eq.~\ref{eq:imf}) it may never be accessible
      to observations (the {\sc IMF Unmeasurability Theorem},
      p.~\pageref{kroupa_theorem:IMFtheorem}). But, just like the
      stellar IMF it is extractable from the observations.  }}}
\vspace{2mm}

The binary fraction
\begin{equation}
f = {N_{\rm bin} \over N_{\rm bin} + N_{\rm sing}},
\label{eq:fbin}
\end{equation}
where $N_{\rm bin}, N_{\rm sing}$ are the number of binary- and
single-stellar systems in the survey, respectively, is high ($f_{\rm
  bin}>0.8$) in dynamically unevolved populations, whereas $f_{\rm
  bin}\approx 0.5$ for typically open clusters and the Galactic field
as follows from applying $\Omega$ on the BBP (i.e. by performing $N$-body
integrations of dissolving star clusters, \citealt{K95d,MKO11}).

Note that the BBP needs to be transformed to the {\it initial binary
  population} by the process of pre-main sequence eigenevolution
\citep{K95c}, which introduces the observed correlations between
mass-ratio, eccentricity and period for short-period binaries, while
the dynamically evolved initial binary population yields the observed
mass-ratio and period distribution functions with $f\approx 0.5$
\citep{MK11}.

For $m\simgreat 5\,M_\odot$ stars the pairing rules change perhaps
reflecting the outcome of star formation in dense regions such as in
the cores of embedded clusters ($\rho\simgreat
10^5\,M_\odot/$pc$^3$). Using a large sample of young clusters (for a
review see \citealt{SE10}, and also \citealt{SJG11}) it is found that
at least 45--55~per~cent of O stars are spectroscopic binaries: The
mass-ratios for these are larger in comparison with the late-type
stars above: massive binaries have a flat mass-ratio distribution and
$0.2\le q\le 1$.  These systems have short periods, typically less
than about 10~d, but extend from 0.3~d to $10^{3.5}\,$d. The measured
distribution function is provided by eq.~5.2 in \cite{SE10}.  The
overall binary fraction among O~stars is at least~85~per cent
\citep{GM01} as the spectroscopic fraction is augmented by wider
visual binaries with separations between~40 and~200~AU
\citep{Sanaetal11}.  The vast spectroscopic survey by \cite{Chini12}
of about 800~O and~B-type stars affirms such results and establishes,
even for runaway stars, the very high binary fraction and q about 1
pairing.

This leads to the following question:

\vspace{2mm} \centerline{ \label{quest:openI}\fbox{\parbox{\columnwidth}{ {\sc
        Open Question I}: Why do the differing BBP properties between
      $0.1\,M_\odot$ and a~few$\;M_\odot$ on the one hand side, and
      above a few~$M_\odot$ on the other hand side not correspond to
      the structure evident in the IMF, which is a featureless
      power-law above about~$0.5\,M_\odot$ with a flattening below
      this mass?  (Sec.~\ref{sec:canIMF}).  }}} \vspace{2mm}

\noindent 
Below about $0.1\,M_\odot$ very low mass stars and brown dwarfs, with
$f_{\rm bin, BD} \approx 0.15-0.2$, follow entirely separate rules
(Sec.~\ref{sec:BDbins}) being an accompanying but distinct population
to stars.

It has so far not been possible to predict nor to fully understand the
distribution of binary-star birth properties from theory. The
currently most advanced self-consistent gravo-hydrodynamical
simulation without feedback of star formation using the SPH technique
\citep{MB10} leads to too compact clusters of about a 1000 stars and
brown dwarfs from which a binary population emerges which does not
quite have the observed distribution of periods and mass
ratios. However, this may be due to the currently unavoidable omission
of feedback which would limit the depth of the gravitational collapse
perhaps alleviating the binary-star problem \citep{Kroupa11}.

\section{The Maximum Stellar Mass}
\label{sec:maxlim}

While the stellar IMF appears to have a universal two-part power-law
form (Eq.~\ref{eq:imf} below), the existence of a physical truncation
mass as a function of embedded star cluster mass would suggest a form
of IMF variation (Sec.~\ref{sec:mmaxIMFvar}). Here the evidence for
such a truncation is presented.

\subsection{On the existence of a maximum stellar mass:}
The empirically determined range of stellar masses poses important
constraints on the physics of stellar formation, structure and stellar
evolution, as well as on the feedback energy injected into a galaxy's
atmosphere by a population of brand-new stars. The physical limit at
low masses is now well established (Sec.~\ref{sec:bds}), and an upper
mass limit appears to have been found recently.

A theoretical physical limitation to stellar masses has been known
since many decades.  \citet{Edd26} calculated the limit which is
required to balance radiation pressure and gravity, the {\it Eddington
  limit}: $L_{\rm Edd}/L_{\odot}\,\approx\,3.5\,\times\,10^{4}\,m /
M_{\odot}$. Hydrostatic equilibrium will fail if a star of a certain
mass $m$ has a luminosity that exceeds this limit, which is the case
for $m\simgreat 60\,M_{\odot}$. It is not clear if stars above this
limit cannot exist, as massive stars are not fully radiative but have
convective cores. But more massive stars will loose material rapidly
due to strong stellar winds.  \citet{SH59} inferred a limit of
$\approx\,60\,M_{\odot}$ beyond which stars should be destroyed due to
pulsations. But later studies suggested that these may be damped
\citep{BM94}.  \cite{St92} showed that the limit increases to $m_{\rm
  max*}\approx 120-150\,M_\odot$ for more recent Rogers-Iglesia
opacities and for metallicities [Fe/H]$\approx0$. For [Fe/H]$\approx
-1$, $m_{\rm max*}\approx 90\,M_\odot$. A larger physical mass limit
at higher metallicity comes about because the stellar core is more
compact, the pulsations driven by the core having a smaller amplitude,
and because the opacities near the stellar boundary can change by
larger factors than for more metal-poor stars during the heating and
cooling phases of the pulsations thus damping the oscillations. Larger
physical mass limits are thus allowed to reach pulsational
instability.

Related to the pulsational instability limit is the problem that
radiation pressure also opposes accretion for proto-stars that are
shining above the Eddington luminosity.  Therefore the question
remains how stars more massive than $60\,M_{\odot}$ may be formed.
Stellar formation models lead to a mass limit near $40-100\,M_\odot$
imposed by feedback on a spherical accretion envelope
(\citealt{Kahn74}; \citealt{Wolf87}). Some observations suggest that
stars may be accreting material in disks and not in spheres
(e.g. \citealt{CHK04}). The higher density of the disk-material may be
able to overcome the radiation at the equator of the proto-star. But
it is unclear if the accretion-rate can be boosted above the mass-loss
rate from stellar winds by this mechanism.  Theoretical work on the
formation of massive stars through disk-accretion with high accretion
rates thereby allowing thermal radiation to escape pole-wards
(e.g. \citealt{JA96}) indeed lessen the problem and allow stars with
larger masses to form.

Another solution proposed is the merging scenario. In this case
massive stars form through the merging of intermediate-mass
proto-stars in the cores of dense stellar clusters driven by
core-contraction due to very rapid accretion of gas with low specific
angular momentum, thus again avoiding the theoretical feedback-induced
mass limit (\citealt{Bonn98, SPH00}, and the review by
\citealt{ZH07}).  It is unclear though if the very large central
densities required for this process to act are achieved in reality,
but it should be kept in mind that an observable young cluster is, by
necessity, exposed from its natal cloud and is therefore likely to be
always observed in an expanding phase such that the true maximally
reached central density may be very high for massive clusters,
$\approx 10^8\,M_\odot/$pc$^3$ \citep{Dab10, MK10, Conroy2011}.

The search for a possible maximal stellar mass can only be performed
in massive, star-burst clusters that contain sufficiently many stars
to sample the stellar IMF beyond $100\,M_\odot$.  Observationally, the
existence of a finite physical stellar mass limit was not evident
until very recently. Indeed, observations in the 1980's of R136 in the
Large Magellanic Cloud (LMC) suggested this object to be one single
star with a mass of about $2000-3000\,M_\odot$.  \citet{WB85} for the
first time resolved the object into eight components using digital
speckle interferometry, therewith proving that R136 is a massive star
cluster rather than one single super-massive star.  The evidence for
any physical upper mass limit was very uncertain, and \citet{Elm97}
stated that ``observational data on an upper mass cutoff are scarce,
and it is not included in our models [of the IMF from random sampling
in a turbulent fractal cloud]''. Although \citet{MH98} found stars in
R136 with masses ranging up to $140-155\,M_\odot$, \citet{Massey03}
explained that the observed limitation is statistical rather than
physical.  We refer to this as the {\it Massey
  assertion}\label{p:massey}, i.e. that $m_{\rm
  max*}=\infty$. Meanwhile, \citet{Setal99} found, from their
observations, a probable upper mass limit in the LMC near about
$130\,M_\odot$, but they did not evaluate the statistical significance
of this suggestion. \citet{Figer02} discussed the apparent cut-off of
the stellar mass-spectrum near $150\,M_\odot$ in the Arches cluster
near the Galactic centre, but again did not attach a statistical
analysis of the significance of this observation.  \citet{Elm00} also
noted that random sampling from an unlimited IMF for all star-forming
regions in the Milky Way (MW) would lead to the prediction of stars
with masses $\simgreat 1000\,M_\odot$, unless there is a rapid
turn-down in the IMF beyond several hundred~$M_\odot$. However, he
also stated that no upper mass limit to star formation has ever been
observed, a view also emphasised by \citet{Larson03}.

Thus, while theory clearly expected a physical stellar upper mass
limit, the observational evidence in support of this was very unclear.
This, however, changed in~2004.

\subsection{The upper physical stellar mass limit}

Given the observed sharp drop-off of the IMF in R136 near
$150\,M_\odot$, i.e. that the R136 stellar population is observed to
be saturated (p.~\pageref{box:saturated}), \citet{WK04} studied the
above {\it Massey assertion} in some detail. R136 has an age
$\le$~2.5~Myr \citep{MH98} which is young enough such that stellar
evolution will not have removed stars through supernova explosions. It
has a metallicity of [Fe/H]$\approx -0.5$~dex \citep{deBoer_etal85}.
From the radial surface density profile \citet{Setal99} estimated
there to be 1350~stars with masses between 10~and $40\,M_\odot$ within
20~pc of the 30~Doradus region, within the centre of which lies R136.
\citet{MH98} and \citet{Setal99} found that the IMF can be
well-approximated by a Salpeter power-law with exponent $\alpha=2.35$
for stars in the mass range 3~to $120~M_\odot$ (see also
Fig.~\ref{fig:maschberger_spp}).  This corresponds to 8000 stars with
a total mass of $0.68\times 10^5\,M_\odot$.  Extrapolating down to
$0.1\,M_\odot$ the cluster would contain $8\times 10^5$~stars with a
total mass of $2.8\times10^5\,M_\odot$. Using a canonical IMF with a
slope of $\alpha=1.3$ (instead of the Salpeter value of 2.35) between
0.1 and $0.5\,M_\odot$ this would change to $3.4\times 10^5$~stars
with a combined mass of $2\times 10^5\,M_\odot$, for an average mass
of $0.61\,M_\odot$ over the mass range $0.1-120\,M_\odot$.  Based on
the observations by \citet{Setal99}, \citet{WK04} assumed that R136
has a mass in the range $5 \times 10^{4} \le M_{\rm R136}/M_{\odot}
\le 2.5 \times 10^{5}$.  Using the {\it canonical stellar IMF}
(Eq.~\ref{eq:imf} below), They found that $N(>150\,M_\odot)=40$~stars
are missing if $M_{\rm ecl}=2.5\times 10^5\,M_\odot$, while
$N(>150\,M_\odot)= 10$~stars are missing if $M_{\rm ecl}=5\times
10^4\,M_\odot$. The probability that no stars are observed although 10
are expected, assuming $m_{\rm max*}=\infty$, is $P=4.5\times
10^{-5}$. Thus the observations of the massive stellar content of R136
suggest a physical stellar mass limit near $m_{\rm
  max*}=150\,M_\odot$.

A re-analysis of the stellar spectra plus new stellar modelling
suggests however $m_{\rm max*} \approx 300\,M_\odot$ for R136
\citep{Crowther+10}.  But \cite{BKO11} demonstrate that
$m>150\,M_\odot$ stars from readily from merging binaries in
star-burst clusters. Their high-precision Aarseth-$N$-body models of
binary-rich initially mass-segregated R136-type clusters demonstrate
that stars much more massive than the $m_{\rm max *}=150\,M_\odot$
limit appear from massive binaries that merge after becoming eccentric
and hard through a stellar-dynamical encounter near the cluster
core. Such binaries may be ejected from the cluster before merging
thus appearing to an observer as free-floating single stars of mass up
to $300\,M_\odot$.  The fundamental upper mass limit may thus
nevertheless be $m_{\rm max *}\approx150\,M_\odot$.

Results similar to those of \cite{WK04} were obtained by \cite{Fi05}
for the Arches cluster.  The Arches is a star-burst cluster within
25~pc in projected distance from the Galactic centre. It has a mass
$M\approx 1\times 10^5\,M_\odot$ \citep{Bosch_etal01}, age $2-2.5$~Myr
and [Fe/H]$\approx 0$ \citep{Najarro_etal04}. It is thus a counterpart
to R136 in that the Arches is metal rich and was born in a very
different tidal environment to R136.  Using his HST observations of
the Arches, \citet{Fi05} performed the same analysis as \citet{WK04}
did for R136. The Arches appears to be dynamically evolved, with
substantial stellar loss through the strong tidal forces
\citep{Port_etal02} and the stellar mass function with $\alpha=1.9$ is
thus flatter than the Salpeter IMF. Using his updated IMF measurement,
Figer calculated the expected number of stars above $150\,M_\odot$ to
be 33, while a Salpeter IMF would predict there to be 18
stars. Observing no stars but expecting to see~18 has a probability of
$P=10^{-8}$, again strongly suggesting $m_{\rm max*}\approx
150\,M_\odot$. The Arches cluster is thus another example of a
saturated stellar population.

Given the importance of knowing if a finite physical upper mass limit
exists and how it varies with metallicity, \citet{OC05} studied the
massive-star content in 9~clusters and OB associations in the MW, the
LMC and the SMC.  They predicted the expected masses of the most
massive stars in these clusters for different upper mass limits ($120,
150, 200, 1000\,{\rm and}\,10000\,M_{\odot}$). For all populations
they found that the observed number of massive stars supports with
high statistical significance the existence of a general upper mass
cutoff in the range $m_{\rm max*}\in
(120,\,200\,M_{\odot})$.\footnote{More recent work on the physical
upper mass limit can be found in \cite{Koen06,Jesus07,Jesus08}.}

The general indication thus is that a physical stellar mass limit near
$150\,M_\odot$ seems to exist. While biases due to unresolved
multiples that may reduce the true maximal mass need to be studied
further, the absence of variations of $m_{\rm max*}$ with metallicity
poses a problem.  A constant $m_{\rm max*}$ would only be apparent for
a true variation as proposed by the theoretical models, {\it if
  metal-poor environments have a larger stellar multiplicity}, the
effects of which would have to compensate the true increase of $m_{\rm
  max*}$ with metallicity. Interestingly, in a recent hydrodynamical
calculation \citet{MOM09} find a higher binary fraction for low
metallicities.

\subsection{The maximal stellar mass in a cluster, optimal sampling
  and saturated populations}
\label{sec:stmass_clmass}

Above we have seen that there seems to exist a universal physical
stellar mass limit, $m_{\rm max*}$. However, an elementary argument
suggests that star-clusters must additionally limit the masses of
their constituent stars: A pre-star-cluster gas core with a mass
$M_{\rm core}$ can, obviously, not form stars with masses
$m>\epsilon\, M_{\rm core}$, where $\epsilon\approx 0.33$ is the
star-formation efficiency \citep{Lada_Lada03}. Thus, given a freshly
hatched cluster with stellar mass $M_{\rm ecl}$, stars in that cluster
cannot surpass masses $m_{\rm max}=M_{\rm ecl}$, which is the identity
relation corresponding to a ``cluster'' consisting of one massive
star. Note that if we were to construct a (unphysical) model in which
$N$ stars are stochastically chosen from the IMF then such a
constraint would not appear.

\subsubsection{Theory}

As discussed by \cite{SLB09} there are two main theories of massive
star formation: The {\sc First Theory} is essentially a scaled up version of
low-mass star formation, where massive stars form from well-defined
massive cores supported by turbulence. This model requires the
existence of massive pre-stellar cores that manage to evade
fragmentation during their formation stages. Perhaps radiative
feedback can limit the fragmentation, but \cite{SLB09} demonstrate
that radiative feedback does not lead to the formation of massive
pre-stellar cores in isolation. 

The {\sc Second Theory} is based on the {\it competitive accretion
  scenario} or on a refined version thereof, the {\it fragmentation
  limited starvation model}.  Cores are the seeds of the formation of
stars and the most massive of these have a larger gravitational radius
of influence and are therewith more successful at accreting additional
mass. They can thus grow into massive stars.  The massive seeds
typically tend to form and stay at the centre of the gravitational
potential of the forming star cluster which they contribute to because
the gas densities and thus the accretion rates are largest there.
They accrete material via Bondi-Hoyle accretion, but when the velocity
relative to the system is low the accretion is mainly regulated by the
tidal field.  There is no requirement for stellar mergers, which,
however can occur in dense regions (Eq.~\ref{eq:collapset}) and
contribute to the build-up of the IMF.

The first theory would imply that the formation of massive stars can
occur in isolation, i.e. without an accompanying star cluster. The
existing data on the spatial distribution of massive stars do,
however, not support this possibility (Sec.~\ref{sec:isolated}).  The
second theory requires massive stars to be associated with star
clusters.  This is demonstrated by \cite{SLB09} using SPH simulations
of a gas cloud with a changing equation of state as a result of the
cooling process shifting from line emission to dust emission with
increasing density, and with radiative heating as a model of the
feedback process. \cite{Peters10, Peters11a} study this issue
independently with three-dimensional, radiation-hydrodynamical
simulations that include heating by ionizing and non-ionizing
radiation using the adaptive-mesh code FLASH and verify that massive
star formation is associated with low-mass star formation as
fragmentation cannot be suppressed, even including radiative feedback.
Simpler isothermal SPH computations without feedback by \cite{BVB04}
and the FLASH simulations by \cite{Peters10, Peters11a} show that the
most massive star in the forming cluster evolves with time, $t$,
according to the following relation
\begin{equation}
m_{\rm max}(t) = 0.39\,M_{\rm ecl}(t)^{2/3}.
\label{eq:mmax_pred}
\end{equation}
The general form of the observed IMF (Eq.~\ref{eq:imf}) is also
obtained.  This is therefore a prediction of the second theory whereby
it is important to note that Eq.~\ref{eq:mmax_pred} is a result of
purely gravitationally driven star formation calculations with and
without feedback. Interestingly, \cite{Peters10, Peters11a} conclude
that computations with feedback lead to a closer agreement with the
observed $m_{\rm max}-M_{\rm ecl}$ data. The following result emerges:

\vspace{2mm} \centerline{
  \fbox{\parbox{\columnwidth}{ \label{quote:bvb} {\sc The
        Bonnell-Vine-Bate (BVB) Conjecture:} ``Thus an individual
      cluster grows in numbers of stars as the most massive star
      increases in mass. This results in a direct correlation ..., and
      provides a physical alternative to a probabilistic sampling
      from an IMF'' (\citealt{BVB04}).\\[1mm]
      {\sc Main Result}: The self-consistent gravo-hydrodynamical
      simulations of star formation thus yield the result that the
      growth of the most massive star is intimately connected with the
      growth of its hosting cluster, thereby populating the stellar
      IMF.  }}}  \vspace{2mm}

\noindent How does nature arrange the mass of the star-forming
material over the emerging stellar masses? That massive stars can form
in isolation can be discarded statistically
(Sec.~\ref{sec:isolated}). But how regulated, or rather deterministic
is the formation of massive stars and their star clusters?

\begin{figure}
\begin{center}
\rotatebox{0}{\resizebox{0.75 \textwidth}{!}{\includegraphics[width=8cm]
{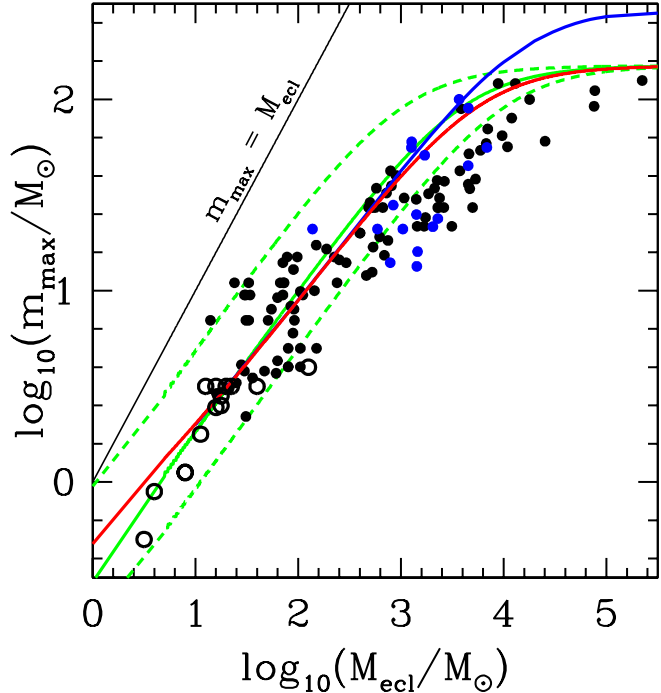}}}
\vspace*{-20mm}
\caption[The $m_{\rm max}-M_{\rm ecl}$ relation]{The $m_{\rm
    max}-M_{\rm ecl}$ relation.  The black solid {\it dots} are
  observed clusters \citep{WKB09}, with newly compiled data being
  shown as blue solid dots. The data are compiled subject to an age
  constraint (no cluster older than 4~Myr is accepted) and to the
  constraint that the cluster still be embedded.  The open circles are
  the most-massive star vs cluster mass in~14 young stellar groups in
  Taurus, Lupus3, ChaI, and IC348 \citep{KM11}. These are dynamically
  unevolved \citep{KB03a} and can be taken to represent pristine
  configurations.  The lower (red) {\it thick solid line} is
  Eq.~\ref{eq:mmax} with $m_{\rm max*}=150\,M_\odot$ and the blue {\it
    thick solid line} is the same but for $m_{\rm max*}=300\,M_\odot$.
  The {\it thin solid line} shows the identity relation, where a
  ``cluster'' consists only of one star.  Assuming unconstrained
  random picking from the canonical IMF with $m_{\rm
    max*}=150\,M_\odot$, 1/6th of all models would lie below the lower
  (green) dashed curve, while 5/6th would lie below the upper (green)
  dashed curve, that is, 2/3rd of all data ought to lie between the
  dashed curves. The median, below which would lie 50~per cent of all
  data for random sampling from the IMF, is plotted as the {\it thick
    (green) solid curve}. Significant deviations from random sampling
  from the stellar IMF are evident in that the scatter of the
  observational data is smaller and the median lies below the
  random-sampling median. The theoretical prediction
  (Eq.~\ref{eq:mmax_pred}) follows nearly precisely the solid lines
  for $M_{\rm ecl} < 10^4\,M_\odot$.  Note also that the horizontal
  axis corresponds to a density. For example, if all clusters form
  with a half-mass radius of 0.5~pc, then the density scale becomes
  log$_{10}\rho_* = $log$_{10}M_{\rm ecl}/M_\odot + 0.28$, where
  $\rho_*$ is in units of $M_\odot/{\rm pc}^3$. Thus, stars more
  massive than $10\,M_\odot$ appear only when $\rho_{\rm gas} >
  10^3\,M_\odot/{\rm pc}^3$ for a star-formation efficiency of~33~per
  cent. Note that the clusters {\it saturate}
  (p.~\pageref{box:saturated}) for $M_{\rm ecl}>10^4\,M_\odot$. }
\label{fig:mmaxf}
\end{center}
\end{figure}

Here it is useful to return to the concept of {\sc Optimal Sampling}
(Sec.~\ref{sec:optsamp}): if nature distributes the available mass
$M_{\rm ecl}$ optimally over the IMF then the $m_{\rm max}-M_{\rm
  ecl}$ relation emerges.  It is plotted in Fig.~\ref{fig:mmaxf} as
the thick-solid curves for two values of $m_{\rm max*}$.

For a given $M_{\rm ecl}$ the observationally derived $m_{\rm max}$
values show a spread rather than one value: Can stars with masses
larger than the optimal $m_{\rm max}$, or even with masses beyond the
canonical upper mass limit of $m_{\rm max *}$, occur?  In discussing
these issues it proves useful to define the concept of a {\it
  saturated population}:

\vspace{2mm} \centerline{ \fbox{\parbox{\columnwidth}{{\sc Definitions}:\\[1mm]
      {\sc Saturated Population}: A population for which $m_{\rm
        max}=m_{\rm max*}$ and $\int_{m_{\rm
          max}}^{\infty}\,\xi(m)\,dm\ge 1$.
      Only a simple population can be saturated. \\[1mm]
      {\sc Un-Saturated Population}: A
      population for which $m_{\rm max} < m_{\rm max *}$.\\[1mm]
      {\sc Super-Saturated Population}: a simple population containing
      super-canonical ($m>m_{\rm max*}$) stars.
     \label{box:saturated}}}}
\vspace{2mm} 

\noindent Thus, a simple stellar population is saturated if its
most-massive star has the physically allowed maximum mass (e.g. R136
and Arches).  An un-saturated population is a simple population in
which $m_{\rm max}$ does not reach $m_{\rm max*}$ (e.g.  $\rho$~Oph
and ONC). The star-burst cluster R136 is associated with
super-canonical stars \citep{Crowther+10} and is therefore a
super-saturated population.  This occurs naturally through merged
massive binaries \citep{BKO11}. Indeed, as an example that such stars
are known to exist is discussed by \cite{Vanbev11}.

\subsubsection{Observational data}

A compilation of clusters only subject to a constraint in age and gas
content (to ensure youth and dynamical virginity) for which the
cluster mass and the initial mass of the heaviest star can be
estimated observationally demonstrates that there exists a strong
correlation between the embedded-cluster mass and the most massive
stellar mass within it. The observational data are plotted in
Fig.~\ref{fig:mmaxf}.

By performing statistical tests \citet{WK05a} and \citet{WKB09} show
with very high confidence that the observational data are not
consistent with the most-massive star being randomly drawn from the
IMF\footnote{A study by \citet{MC08} of the most-massive-star data in
  young star clusters concluded that ''the data are not indicating any
  striking deviation from the expectations of random drawing''.  This
  statement has been frequently misinterpreted that other sampling
  mechanisms are ruled out.  However, the Maschberger \& Clarke
  analysis focusses on low-mass clusters were the data were
  insufficient to decide whether star clusters are populated purely
  randomly from an IMF with constant upper mass limit or for example
  in a sorted fashion. The differences appear clearly at higher
  cluster masses, not included in their analysis but in \citet{WK05a}
  and \cite{WKB09}. \cite{MC08} adapt their data set according to the
  requested result and so their study does not constitute an
  acceptable scientific standard.}.  In addition, if the process of
star formation were equivalent to pure random sampling of stars from
the IMF then this would predict the existence of star clusters
dominated by O~stars or even the formation of massive stars in
isolation \citep[e.g.][]{HA10}. But this hypothesis is ruled out with
high confidence because the theoretically {\it expected} fraction of
massive stars that ought to appear to be isolated (1--4~per cent)
although they were formed in clusters is already larger than the {\it
  observed} fraction of candidate isolated massive stars (see
Sec.~\ref{sec:isolated}).

The newest additions of observational data enhance the empirical
evidence for the existence of a physical $m_{\rm max}-M_{\rm ecl}$
relation, even at masses $M_{\rm ecl}\simless 15\,M_\odot$: The
low-mass data by \cite{KM11} show a remarkably small spread implying
that even the lowest-mass ``clusters'' limit the mass of their most
massive star in a non-trivial way. Indeed, the distribution of all
data shows that random sampling from the IMF is ruled out since the
spread is smaller than the expected spread given the 1/6th and 5/6th
quantiles. The data in Fig.~\ref{fig:mmaxf} also show that for $M_{\rm
  ecl}\simless 10^3\,M_\odot$ the semi-analytical model
(Eq.~\ref{eq:mmax}) is an excellent description.  At larger cluster
masses this model is a fairly good description as well although some
systematic deviation is evident.

It thus appears that the process of star formation ends up close to
optimally sampling the IMF, and that it does not correspond to purely
randomly generating stars from the IMF in support of the {\sc BVB
  Conjecture} (p.~\pageref{quote:bvb}). But why?

\vspace{2mm}
\centerline{ \label{quest:openII}\fbox{\parbox{\columnwidth}{ {\sc
        Open Question II}: Why is it that the star formation process
      samples the IMF close to if not optimally?}}}  \vspace{2mm}

\subsubsection{Interpretation}
\label{sec:mmaxInterp}

The observational data suggest that the dominant physical process
responsible for the $m_{\rm max}-M_{\rm ecl}$ relation is a
competitive resource-limited growth process.  This would be natural
since the proto-stars begin as a distribution of low-mass seed masses
and accrete at various rates thereby depleting the surrounding
interstellar medium, as is in fact evident in self-consistent
gravo-hydrodynamical simulations of star formation with and without
feedback as discussed above. 

For $M_{\rm ecl} > 10^{2.2}\,M_\odot$ the small spread persists in the
observational data, but the data fall below the semi-analytical model
(Eq.~\ref{eq:mmax}). This may hint at additional processes becoming
important perhaps related to the ability for $m\simgreat 10\,M_\odot$
stars to continue to grow through accretion alone.  Another physical
process of possible relevance leading to a reduction of $dm_{\rm max}
/ dM_{\rm ecl}$ for $m_{\rm max}\simgreat 10\,M_\odot$ may be due to
an instability in the cold inter-stellar medium (ISM) developing at a
mass around $M_{\rm ecl}=10^2\,M_\odot$ similar to the ISM
instabilities discussed in \cite{PAK09}.  Such an
accretion-instability may lead to enhanced accretion of gas onto the
pre-cluster cloud core from the surrounding molecular cloud. If
$m_{\rm max} \simgreat 10\,M_\odot$ stars have a reduced accretion
efficiency then this may explain the flattening, since a smaller
fraction of the newly accreted gas adds to the growth of $m_{\rm max}$
and is instead used up by the formation of less-massive cluster stars
({\it fragmentation induced starvation} of \citealt{Peters10,
  Peters11a}).  Furthermore, it may be possible that for $M_{\rm
  ecl}>10^2\,M_\odot$ sub-cluster merging may be becoming an important
physical process: each sub-cluster with $M_{\rm ecl}\simless
100\,M_\odot$ follows the $m_{\rm max}-M_{\rm ecl}$ relation such that
upon amalgamation of the sub-clusters $m_{\rm max}$ changes less than
$M_{\rm ecl}$. The steepening of the $m_{\rm max}-M_{\rm ecl}$
relation for $M_{\rm ecl}\simgreat 10^3\,M_\odot$ may be affected by
the coalescence of massive proto-stars in the dense centres of forming
embedded clusters.

\cite{Peters10, Peters11a} discuss the physics driving the $m_{\rm
  max}-M_{\rm ecl}$ relation (for $M_{\rm ecl} \simless 10^2
\,M_\odot$) and find that $m_{\rm max}$-growth-curves flatten with
increasing $M_{\rm ecl}$ because infalling gas is accreted by the
other stars in the emerging cluster. In particular the appearance of
close companions to the most massive star reduces its growth, while
the star cluster continues to form.  Feedback allows the growth of the
most massive star to be sustained for longer, essentially by heating
the gas such that it is less susceptible to fall into the potentials
of lower-mass companions and stars and is therewith forced to follow
the main potential towards the centre, thereby leading to better
agreement with the observed $m_{\rm max}-M_{\rm ecl}$ relation. If no
cluster of low-mass stars were to form such that none of the gas is
accreted by the other low-mass stars, then $m_{\rm max}\propto M_{\rm
  ecl}$, which is a dependency which is too steep compared to the
data.

\subsubsection{Stochastic or regulated star formation?}

The existence of an observed $m_\mathrm{max} \propto
M_\mathrm{ecl}^{2/3}$ relation (Eq.~\ref{eq:mmax_pred}) different to
the one expected from random sampling from the IMF thus implies that
the formation of massive stars is associated with surrounding low-mass
star formation.  This suggests that the formation of stars within the
cloud cores is mostly governed by gravitationally-driven growth
processes in a medium with limited resources.

If the outcome of star formation were to be inherently stochastic, as
is often assumed to be the case, in the sense that stars are randomly
selected from the full IMF, then this would imply that stellar
feedback would have to be, by stringent logical implication, the
randomisation agent. In other words, the well-ordered process of stars
arising from a molecular cloud core by pure gravitationally driven
accretion as shown to be the case by self-consistent
gravo-hydrodynamical simulations would have to be upset completely
through feedback processes. As there is no physically acceptable way
that this might arise, and since in fact the work of \cite{Peters10,
  Peters11a} has demonstrated that feedback actually helps establish
the $m_{\rm max}-M_{\rm ecl}$ relation, it is concluded that star
formation cannot be a random process, and its outcome cannot be
described by randomly choosing stars from an IMF.\footnote{Choosing
  stars randomly from the IMF is, however, a good first approximation
  for many purposes of study.}

Returning to Sec.~\ref{sec:philosophy}, the following two alternative
hypotheses can thus be stated:

\vspace{2mm} \centerline{ \fbox{\parbox{\columnwidth}{{\sc IMF Random Sampling
        Hypothesis}: A star formation event always produces a
      probabilistically sampled IMF. }}} \vspace{2mm}

\noindent and

\vspace{2mm} \centerline{ \fbox{\parbox{\columnwidth}{{\sc IMF Optimal
        Sampling Hypothesis}: A star formation event always produces
      an optimally sampled IMF such that the $m_{\rm max}-M_{\rm ecl}$
      relation holds true.  }}} \vspace{2mm}

\noindent As stated above, the {\sc IMF Random Sampling Hypothesis}
can already be discarded on the basis of the existing data and
simulations. But can the current data discard the {\sc IMF Optimal
  Sampling Hypothesis}?  The scatter evident in Fig.~\ref{fig:mmaxf}
may suggest that optimal sampling is ruled out, since if it were true
for every cluster then a one-to-one relation between $m_{\rm max}$ and
$M_{\rm ecl}$ would exist. However, the following effects contribute
to introducing a dispersion of $m_{\rm max}$ values for a given
$M_{\rm ecl}$, even if
an exact $m_{\rm max}-M_{\rm ecl}$ relation exists:\\[-5mm]
\begin{itemize}
\itemsep=-1mm
\item The measurement of $M_{\rm ecl}$ and $m_{\rm max}$ are very
  difficult and prone to uncertainty. 
\item An ensemble of embedded clusters of a given stellar mass $M_{\rm
    ecl}$ is likely to end up with a range of $m_{\rm max}$ because
  the pre-cluster cloud cores are likely to be subject to different
  boundary conditions (internal distribution of angular momentum,
  different thermodynamic state, different external radiation field,
  different metallicity, etc.). The details of self-regulation depend
  on such quantities, and this may be compared to the natural
  dispersion of stellar luminosities at a given stellar mass due to
  different metallicity, stellar spins and orientations relative to
  the observer and different stellar ages.\\[-5mm]
\end{itemize}
At present, the {\sc IMF Optimal Sampling Hypothesis} can thus not be
discarded, but its statement here may be conducive to further research
to investigate how exactly valid it is.

\subsubsection{A historical note}

\citet{Larson82} had pointed out that more massive and dense clouds
correlate with the mass of the most massive stars within them and he
estimated that $m_{\rm max}=0.33\,M_{\rm cloud}^{0.43}$ (masses are in
$M_\odot$). An updated relation was derived by \citet{Larson03} by
comparing $m_{\rm max}$ with the stellar mass in a few clusters,
$m_{\rm max}\approx 1.2\,M_{\rm cluster}^{0.45}$. Both are flatter
than the semi-analytical relation, and therefore do not fit the data
in Fig.~\ref{fig:mmaxf} as well \citep{WK05a}. \citet{Elm83}
constructed a relation between cluster mass and its most massive star
based on an assumed equivalence between the luminosity of the cluster
population and its binding energy for a Miller-Scalo IMF (a
self-regulation model). This function is even shallower than the one
estimated by \citet{Larson03}.

\subsection{Caveats}

Unanswered questions regarding the formation and evolution of massive
stars remain. There may be stars forming with $m > m_{\rm max *}$
which implode ``invisibly'' after 1 or 2 Myr.  The explosion mechanism
sensitively depends on the presently still rather uncertain mechanism
for shock revival after core collapse (e.g. \citealt{Janka01}).  Since
such stars would not be apparent in massive clusters older than 2~Myr
they would not affect the empirical maximal stellar mass, and $m_{\rm
  max*, true}$ would be unknown at present.

Furthermore, stars are often in multiple systems. Especially massive
stars seem to have a binary fraction of 80\% or even larger and
apparently tend to be in binary systems with a preferred mass-ratio
$q\simgreat 0.2$ (Sec.~\ref{sec:bins}). Thus, if all O~stars would be
in equal-mass binaries, then $m_{\rm max *\,true} \approx m_{\rm max
  *}/2$.

Finally, it is noteworthy that $m_{\rm max *}\, \approx\,
150\,M_{\odot}$ appears to be the same for low-metallicity
environments ([Fe/H]$ = -0.5$, R136) and metal-rich environments
([Fe/H]$ = $0, Arches), in apparent contradiction to the theoretical
values \citep{St92}. Clearly, this issue needs further study.

\vspace{2mm} \centerline{ \fbox{\parbox{\columnwidth}{ {\sc Main results}: A
      fundamental upper mass limit for stars appears to exist, $m_{\rm
        max*}\approx 150\,M_\odot$.  The mass of a cluster defines the
      most-massive star in it and leads to the existence of a $m_{\rm
        max}-M_\mathrm{ecl}$ relation, which results from competitive
      resource-limited growth and self-regulation processes. The
      outcome of a star formation event appears to be close to an
      optimally sampled IMF.}}}  \vspace{2mm}

\section{The Isolated Formation of Massive Stars?}
\label{sec:isolated}

An interesting problem relevant for the discussion in
Sec.~\ref{sec:stmass_clmass} with major implications for
star-formation theory and the IMF in whole galaxies is whether massive
stars can form alone without a star cluster, i.e. in isolation
(e.g. \citealt{Liklessen03, Krumholz08}). 

Related to this is one of the most important issues in star-formation
theory, namely the still incomplete understanding of how massive stars
($m \simgreat 10\,M_\odot$) form.  From Fig.~\ref{fig:mmaxf} a
gas-density of $\rho_{\rm gas}\simgreat 10^3\,M_\odot/{\rm pc}^3$ for
the formation of $m>10\,M_\odot$ stars can be deduced.  At least four
competing theories have been developed: the competitive accretion
scenario \citep{Bonn98,BVB04,BB06}, collisional merging
\citep{BonnellBate02}, the single core collapse model allowing
isolated massive star formation \citep{Krumholz09}, the
fragmentation-induced starvation model \citep{Peters10, Peters11a},
and the outflow-regulated clump collapse model \citep{WangLi10}, where
massive stars result from the collapse of turbulent cluster-forming
clumps, whose internal dynamics are regulated by proto-stellar outflows.

To help advance this topic it is necessary to find conclusive
constrains for the formation of massive stars from observations. One
important piece of evidence can be deduced from the formation sites of
massive stars. While competitive accretion, collisional merging,
fragmentation-induced starvation descriptions and outflow-regulated
clump collapse predict the formation of massive stars within star
clusters, the core collapse model only needs a sufficiently massive
and dense cloud core and allows for an isolated origin of O stars.

This latter model appears to be inconsistent with the data
(Sec.~\ref{sec:stmass_clmass}). And, even the currently most advanced
radiation-magnetohydrodynamical simulations including ionization
feedback of a $1000\,M_\odot$ rotating cloud lead to suppression of
fragmentation by merely a factor of about two \citep{Peters11}, so
that massive star formation cannot be separated from the formation of
embedded clusters.

Nevertheless, the isolated formation of massive stars remains a
popular option in the research community. Discussing field O~stars,
\cite{M98} writes (his p.~34) ``One is tempted to conclude that these
O3 stars formed pretty much where we see them today, as part of very
modest star-formation events, ones that perhaps produce ``a single O
star plus some change'', as Jay Gallagher aptly put it.''
\cite{Selier11} write ``There is, however, a statistically small
percentage of massive stars ($\approx$5~per~cent) that form in
isolation \citep{DEWIT05, PG07}'' while \cite{Camargo10} confess ``On
the other hand, \cite{DEWIT05} estimate that nearly 95~per cent of the
Galactic O star population is located in clusters or OB associations,
or can be kinematically linked with them''.  \cite{Lamb10} state ``In
a study of Galactic field O stars, \cite{DEWIT04, DEWIT05} find that
$4\pm2$~per cent of all Galactic~O stars appear to have formed in
isolation, without the presence of a nearby cluster or evidence of a
large space velocity indicative of a runaway star.'', while
\cite{Krumholz10} explain ``\cite{DEWIT04, DEWIT05} find that
$4\pm2$~per cent of galactic~O stars formed outside of a cluster of
significant mass, which is consistent with the models presented here
(for example, runs M and H form effectively isolated massive single
stars or binaries), but not with the proposed cluster-stellar mass
correlation.''  Such events of isolated massive-star formation are
conceivable if the equation of state of the inter-stellar medium can
become stiff \citep{Liklessen03}.  Hence, the search for isolated O
stars and the deduction whether or not these stars have formed in situ
can be vital in narrowing down theories and advancing the research
field. Generally, in order to propose that massive stars form in
isolation rather contrived initial conditions in the cloud core are
required such as a strong magnetic field, no turbulence and a highly
peaked density profile \citep{Girichidis11}.

The existence of massive stars formed in isolation would be required
if the stellar distribution within a galaxy were a result of a purely
stochastic process in contradiction to the IGIMF-theory
(Sec.~\ref{sec:comppop}), and would not be in agreement with the
existence of the $m_{\rm max}-M_{\rm ecl}$ relation
(Eq.~\ref{eq:mmax_pred}).

This discussion does not have an easy observational solution because
massive stars have been observed to be quite far away from sites where
late-type stars or star clusters are forming and it can never be
proven with absolute certainty that a given star comes from a cluster.

\begin{figure}
\begin{center}
  \rotatebox{0}{\resizebox{1.2
      \textwidth}{!}{\includegraphics{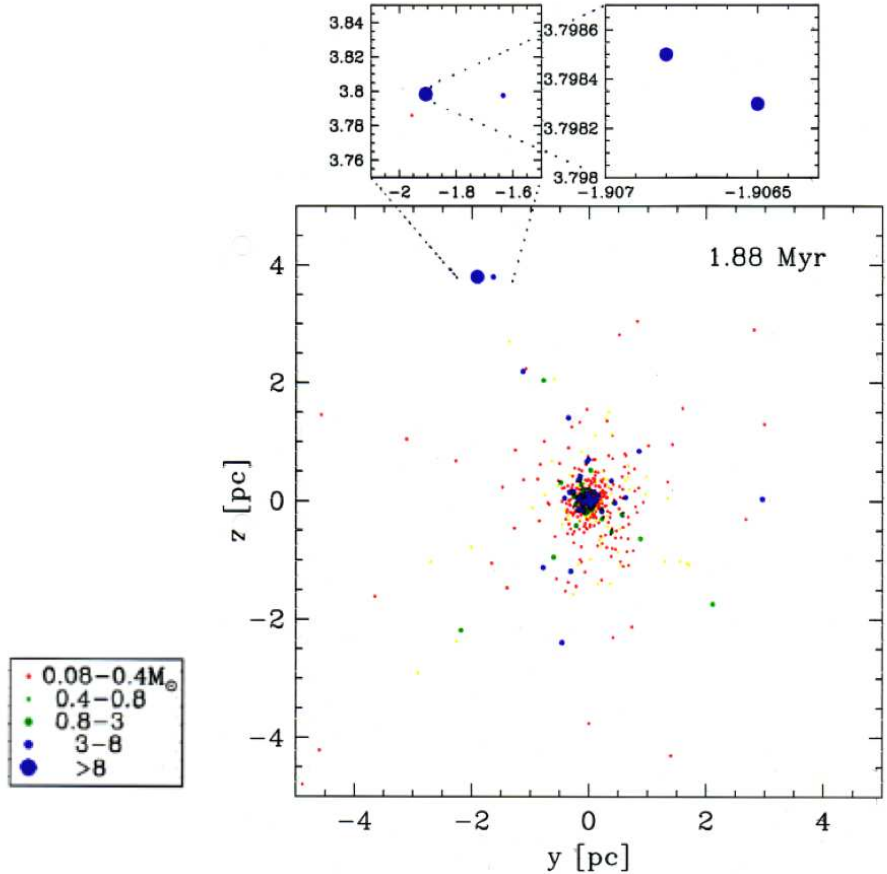}}}
  \vskip 0mm
  \caption[A quadruple system ejected from a star cluster]{\small{An $N$-body5 \citep{Aarseth99} model of a star
      cluster initially not mass segregated, in virial equilibrium,
      with a randomly drawn IMF with stellar masses between
      0.08~and~$30\,M_\odot$ and with a binary fraction of~100~per
      cent (Eq.~\ref{eq:fPbirth}). The cluster consists of~400
      binaries distributed in a Plummer model with a half-mass radius
      of $r_{0.5}=0.1$~pc.  The stellar masses are denoted by
      different symbols as defined in the key. Mass segregation
      develops within the energy-equipartition time-scale $t_{\rm eq}
      < 0.4\,$Myr \citep{Kroupa08} such that the low-mass stars expand
      outwards. The snap-shot shows the system at a time
      of~1.88~Myr. By this time a massive star system has been
      expelled from the cluster by a relatively gentle (about~2~km/s)
      cluster-potential--star recoil. The expelled massive star, which
      is an equal-mass binary formed after an exchange encounter
      within the cluster core, has a very wide intermediate-mass
      companion as well as a very wide M dwarf companion. It is easy
      to misinterpret such a hierarchical multiple system, located
      more than~4pc away from a compact young cluster, to be an ideal
      candidate for isolated massive star formation.  }}
\label{fig:expulsedOstar}
\end{center}
\end{figure}

One possible way to prove that a massive star formed in isolation,
i.e. with at most a few companions, would be to discover an isolated
massive star with wide companions. This would invalidate the star
having been ejected from a cluster. Unfortunately even this possible
criterion is not fool proof evidence for the occurrence of isolated
massive star formation, as the example in Fig.~\ref{fig:expulsedOstar}
demonstrates.

Thus, a decision on whether star formation is random enough to allow
the formation of massive stars by themselves in isolation can only be
reached through statistical arguments, since it can never be proven
beyond doubt that some particular massive star did not form in
isolation. It is thus essential to understand all possible physical
mechanisms that contribute to massive stars being distributed widely
throughout a galaxy.

The ``OB field-star'' MF has $\alpha\approx4.5$, which has been
interpreted to be the result of isolated high-mass star-formation in
small clouds \citep{M98}. Precise proper-motion measurements have
however shown that a substantial number of even the best candidates
for such an isolated population have high space motions \citep{Ram01}
which are best understood as the result of energetic stellar-dynamical
ejections when massive binary systems interact in the cores of
star-clusters in star-forming regions. This interpretation poses
important constraints on the initial properties of OB binary systems
\citep{ClarkePringle92,K01c,P-AK06a, Gvaramadze11}.  Still,
\citet{DEWIT04,DEWIT05} found that $4\pm2$ per cent of all O~stars may
have formed outside any cluster environment.

\begin{figure}
\begin{center}
\rotatebox{0}{\resizebox{0.7
\textwidth}{!}{\includegraphics{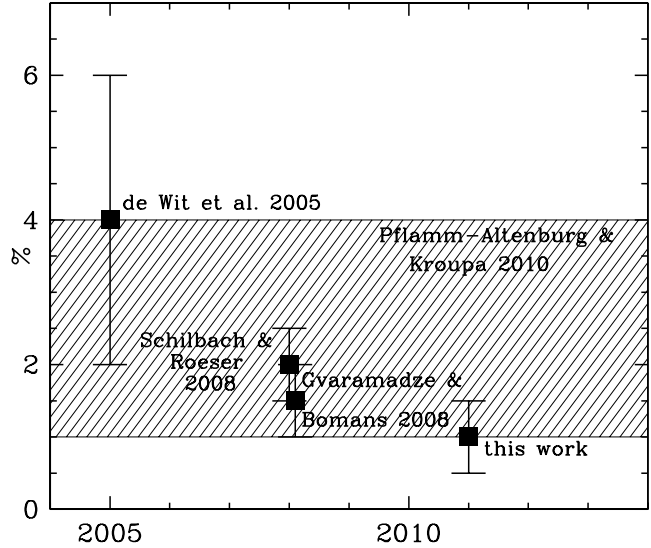}}}
\vskip 0mm
\caption[Evidence history for massive stars formed in
\emph{isolation}]{\small{Estimates of the percentage of massive stars
    ’formed in isolation’ as a function of publication year. The
    present-day (2011) estimate of the fraction of O~stars which can
    not be traced back to their birth clusters is at the lower limit
    of what is expected from the two-step mechanism (shaded
    area). From \cite{Gvaramadzeetal12}.}}
\label{fig:isolated}
\end{center}
\end{figure}

This percentage, however, had to be reduced twice
(Fig.~\ref{fig:isolated}) because \cite{SR08} showed that 6 out of 11
stars that apparently formed in isolation can in fact be back-traced
to their parent clusters. Moreover, \cite{GB08} demonstrated that one
of ”the best examples for isolated Galactic high-mass star formation”
\citep{DEWIT05}, the star HD$\,$165319, has a bow shock and is thus a
runaway star, most likely ejected from the young massive star cluster
NGC$\,$6611. This further reduces the percentage of massive stars
possibly formed in isolation, bringing it to $1.5 \pm 0.5$~per
cent. And finally, using the WISE data, \cite{Gvaramadzeetal12} discovered
a bow shock generated by one more star (HD$\,$48279) from the sample
of ”the best examples for isolated Galactic high mass star
formation”. Correspondingly, the percentage of massive stars which may
have formed in isolation is reduced to $1.0\pm 0.5$ per cent.  Another
example of a possible very massive star having formed in isolation is
VFTS$\,$682 which is an about $150\,M_\odot$ heavy star located about
30~pc in projection from the star burst cluster R136 in the Large
Magellanic Cloud \citep{Bestenlehner11}. However, realistic
binary-rich $N$-body models of initially mass-segregated R136-type
clusters show that such massive stars are ejected with the observed
velocities in all computations therewith readily allowing VFTS$\,$682
to be interpreted as a slow-runaway from R136 \citep{BKO11}

Massive stars may perfectly appear to have formed in isolation despite
originating in clusters: The ``two-step-ejection scenario'', which
places massive stars outside their parent cluster such that they may
fake isolated formation, has been presented by \citet{PAK10}. This is
based on massive binaries first being dynamically ejected from their
parent star cluster. The subsequent type~II supernova explosion then
places the remaining massive star on a random trajectory such that it
can nearly never be traced back to its parent star cluster. This is
necessarily the case for 1--4~per cent of all O~stars assuming all
massive stars to form in star clusters which obey the $m_{\rm
  max}-M_{\rm ecl}$ relation. Further, if the ejected O~star system
consists of a tight inner binary with an outer companion the Kozai
mechanism is likely to force the inner binary into coalescence leading
to a rejuvenated massive star (a massive blue straggler). When the
outer companion explodes the massive blue straggler would be released
in a random direction such that its age nor its motion would allow it
to be traced to its cluster of origin. 

Thus, as shown above, after excluding all observed O~stars that can be
traced back to young star clusters as well as those with bow-shocks,
the current observational evidence for the possible existence of
isolated O~star formation amounts to 1~per cent of all known
O~stars. This is at the lower limit of the expected two-step ejection
fraction of O~stars that cannot be traced back to their cluster of
origin.  There is therefore no meaningful evidence for the formation
of massive stars in isolation. The hypothesis that the isolated
formation of massive stars can be a significant contributor to the
population of massive stars such that the IMF of a whole galaxy
effectively becomes a purely probabilistic invariant distribution
function can formally be negated by noting that the observed IMF is
too invariant (Fig.~\ref{fig:ahist}). That is, an isolated massive star
would correspond to an IMF that significantly differs, by chance, from
the theoretical parent distribution function. But for each such
extreme case there would be many more cases of coeval stellar
populations that by chance have ``strange'' IMFs. There is not
observational evidence whatsoever for this: all known resolved stellar
populations are canonical.

\vspace{2mm} \centerline{ \fbox{\parbox{\columnwidth}{ {\sc Main result}: The
      observed field massive stars are the necessary outcome of the
      formation of massive stars as multiple systems in the cores of
      embedded clusters.  There is no convincing evidence for the
      formation of isolated massive stars. }}} \vspace{2mm}

\section{The IMF of Massive Stars}
\label{sec:massst}

In what follows, the IMF power law indices are $\alpha_1$ for $0.07
\le m/M_\odot < 0.5$, $\alpha_2$ for $0.5 \le m/M_\odot < 1$,
$\alpha_3$ for $1 \le m/M_\odot$.  

Studying the distribution of massive stars ($\simgreat 10\,M_\odot$)
is complicated because they radiate most of their energy at far-UV
wavelengths that are not accessible from Earth, and through their
short main-sequence life-times, $\tau$, that remove them from
star-count surveys.  For example, a $85\,M_\odot$ star cannot be
distinguished from a $40\,M_\odot$ star on the basis of $M_V$ alone
\citep{Massey03}.  Constructing $\Psi(M_V)$ in order to arrive at
$\Xi(m)$ for a mixed-age population does not lead to success if
optical or even UV-bands are used.  Instead, spectral classification
and broad-band photometry for estimation of the reddening on a
star-by-star basis has to be performed to measure the effective
temperature, $T_{\rm eff}$, and the bolometric magnitude, $M_{\rm
  bol}$, from which $m$ is obtained allowing the construction of
$\Xi(m)$ directly (whereby $\Psi(M_{\rm bol})$ and $\Xi(m)$ are
related by Eq.~\ref{eq:mf_lf}).  Having obtained $\Xi(m)$ for a
population under study, the IMF follows by applying
Eq.~\ref{eq:imf_pdmf} after evolving each measured stellar mass to its
initial value using theoretical stellar evolution tracks.

\begin{figure}
\begin{center}
\rotatebox{0}{\resizebox{0.75 \textwidth}{!}{\includegraphics{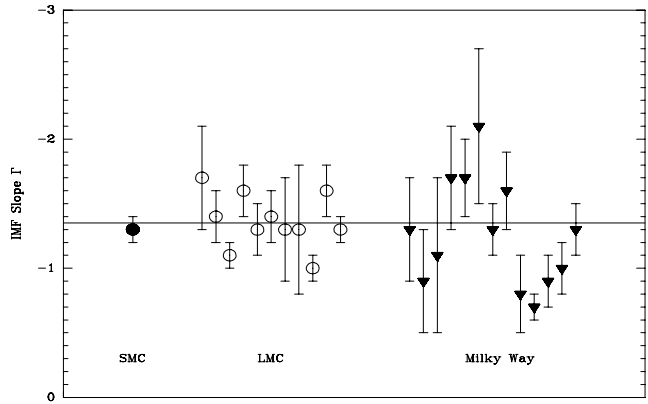}}}
\vskip -4mm
\caption[IMF slopes in the SMC, LMC and MW]{\small{ The IMF slope $\Gamma=1-\alpha$ determined in a
    homogeneous manner for OB associations and clusters in the MW, LMC
    and SMC.  The Small Magellanic Cloud (SMC) has a metallicity
    $Z=0.005$ ([Fe/H]$\approx-0.6$), the Large Magellanic Cloud (LMC)
    has $Z=0.008$ ([Fe/H]$\approx-0.4$) and the Milky Way (MW) has
    $Z=0.018$ ([Fe/H]$\approx-0.05$) within a distance of 3~kpc around
    the Sun.  With kind permission from \cite{Massey03}.  }}
\label{fig:kroupa_figmassey}
\end{center}
\end{figure}

\cite{Massey03} stresses that studies that only rely on broad-band
optical photometry consistently arrive at IMFs that are significantly
steeper with $\alpha_3\approx3$, rather than $\alpha_3=2.2\pm0.1$
found for a wide range of stellar populations using spectroscopic
classification.  Indeed, the application of the same methodology by
Massey on a number of young populations of different metallicity and
density shows a remarkable uniformity of the IMF above about
$10\,M_\odot$ (Fig.~\ref{fig:kroupa_figmassey}).

The available IMF measurements do not take into account the bias
through unresolved systems which may, in principle, be substantial
since the proportion of multiple stars is higher for massive stars
than for low-mass Galactic-field stars (e.g. \citealt{Duchene01}).
For example, in the Orion Nebula Cluster (ONC) each massive star has,
on average, 1.5 companions \citep{Preibisch99}, while in the cluster
NGC~6231, \cite{GM01} find that 80~\% of all O~stars are
radial-velocity binaries.

However, \cite{MA08} and \cite{WKM09} demonstrate that the observed
IMF for massive stars is not affected significantly by unresolved
multiple systems: The models, where initial masses are derived from
the luminosity and colour of unresolved multiple systems, show that
even under extreme circumstances (100 per cent binaries or higher
order multiples), the difference between the power-law index of the
mass function (MF) of all stars and the observed MF is small
($\Delta\alpha \simless 0.1$). Thus, if the observed IMF has the
Salpeter index $\alpha_3 = 2.35$, then the true stellar IMF has an index
not flatter than $\alpha_3 = 2.25$.

Massive main-sequence stars have substantial winds flowing outwards
with velocities of a few~100 to a few~1000~km/s \citep{KP00}.  For
example, $10^{-6.5}<\dot{m}<10^{-6}\,M_\odot$/yr for $m=35\,M_\odot$
with main-sequence life-time $\tau=4.5$~Myr \citep{Garcia96a}, and
$10^{-5.6}<\dot{m}<10^{-5.8}\,M_\odot$/yr for $m=60\,M_\odot$ with
$\tau=3.45$~Myr \citep{Garcia96b}.  More problematical is that stars
form rapidly rotating and are sub-luminous as a result of reduced
internal pressure. But they decelerate significantly during their
main-sequence life-time owing to the angular-momentum loss through
their winds and become more luminous more rapidly than non-rotating
stars \citep{MM00}.  For ages less than 2.5~Myr the models deviate
only by 5--13~\% from each other in mass, luminosity or temperature
\citep{WK05a}. Large deviations are evident for advanced stages of
evolution though because of the sensitivity to the different treatment
of the stellar physics.

The mass--luminosity relation for a population of stars that have a
range of ages is broadened, making mass estimates from $M_{\rm bol}$
uncertain by up to 50~per cent \citep{MPV01}, a bias that probably
needs to be taken into account more thoroughly in the derivations of
the IMF. Another problem is that $m\simgreat40\,M_\odot$ stars may
finish their assembly after burning a significant proportion of their
central~H so that a zero-age-main sequence may not exist for massive
stars \citep{MB01}. However, the agreement between slowly-rotating
tidally-locked massive O-type binaries with standard non-rotating
theoretical stellar models is very good \citep{MPV01}.

\vspace{2mm} \centerline{ \fbox{\parbox{\columnwidth}{ {\sc Main results}: The
      IMF of massive stars is well described by a {\it
        Salpeter/Massey} slope, $\alpha_3=2.3$, independent of
      environment as deduced from resolved stellar populations in the
      Local Group of galaxies.  Unresolved multiple stars do not
      significantly affect the massive-star power-law index.}}}
\vspace{2mm}

\noindent A note to the statement that $\alpha_3=2.3$ is independent
of environment: this is strictly only valid for star formation with
densities less than about $10^5\,M_\odot$/pc$^3$ and for metallicities
of [Fe/H]$\simgreat -2$ as are observed in the Local Group of
galaxies. Evidence has emerged that star formation at higher densities
leads to top-heavy IMFs (see Fig.~\ref{fig:topheavy} below).

\section{The IMF of Intermediate-Mass Stars}
\label{sec:imst}

Intermediate mass ($\approx 1-8\,M_\odot$) stars constitute a
complicated sample to deal with in terms of inferring their mass
function from star-counts in the Galactic field as well as in star
clusters. Their life-times are comparable to the age of the Galactic
disk down to the life-time of typical open clusters (a few~$10^8$~yr,
Fig.~\ref{fig:apl} below). Also, the distribution function of multiple
systems change in this mass range (Sec.~\ref{sec:bins}).  Corrections
of their luminosities for stellar evolution and binarity are thus more
challenging than in the other mass ranges. Also, the diffusion of
stellar orbits within the MW disk away from the birth orbits has a
comparable time-scale (the dynamical period of the Milky Way) such
that an ensemble of intermediate-mass stars does not have an as well
constrained Galactic-disk thickness as the massive ($\approx 50\,$pc)
or low-mass ($\approx 500\,$pc) stars.  This affects the combination
of the essentially two-dimensional (in the Galactic disk plane) star
counts of massive stars with the three-dimensional (solar
neighbourhood within a few hundred~pc) star counts of late-type main
sequence stars to a common density of stars in dependence of stellar
luminosity.  This issue is dealt with excellently by \cite{Sc86}.  The
resulting constraints on the IMF in this mass range are rather
uncertain. Indeed, in Fig.~\ref{fig:apl} it is evident that the
scatter of deduced $\alpha$-indices is very large in this mass range,
and the analysis by \cite{Sc86} may even suggest a discontinuity in
the Galactic-disk IMF in this mass range.

The stellar IMF has $\alpha_2\approx2.3$ for stars with $0.5 <
m/M_\odot < 1$ ({\sc Main Results} on p.~\pageref{box:resonimf}) and
$\alpha_3\approx2.3$ for $m\simgreat 8\,M_\odot$
(Sec.~\ref{sec:massst}) such that $\alpha_3 \approx2.3$ for $1\simless
m/M_\odot \simless 8$ appears natural. However, as noted in
Sec.~\ref{sec:bins}, the initial binary properties appear to be
different for $m<\,{\rm few}\,M_\odot$ in comparison to those for
$m>\,{\rm few}\,M_\odot$ implying {\sc Open Question I} on
p.~\pageref{quest:openI}.

Here it is assumed that the IMF is continuous across this mass range
and thus attention is given to the massive (Sect.~\ref{sec:massst})
and low-mass stars (Sect.~\ref{sec:lmst}).  But in view of the
discontinuity issue uncovered on p.~\pageref{sec:BDbins} for stars and
brown dwarfs, the continuity assumption made here needs to be kept in
mind.
 
\section{The IMF of Low-Mass Stars (LMSs)}
\label{sec:lmst}

Here stars with $0.1\simless m/M_\odot \simless 1$ (the LMSs) are
discussed. They are late-type main sequence stars which constitute the
vast majority of all stars in any known stellar population
(Table~\ref{tab:frac} below). Also, their initial binary properties
follow rather simple rules (Sect.~\ref{sec:bins}).  Very low-mass
stars (VLMSs) with $m\simless 0.15\,M_\odot$ are the subject of
Sec.~\ref{sec:bds}.

There are three well-tried approaches to determine $\Psi(M_V)$ in
Eq.~\ref{eq:mf_lf}.  The first two are applied to Galactic-field stars
(parallax- and flux-limited star-counts), and the third to star
clusters (establishment of members). The sample of Galactic-field
stars close to the Sun (typically within~5 to~20~pc distance depending
on $m$) is especially important because it is the most complete and
well-studied stellar sample at our disposal.

\subsection{Galactic-field stars and the stellar luminosity function} 

Galactic-field stars have an average age of about 5~Gyr and represent
a mixture of many star-formation events. The IMF deduced for these is
therefore a time-cumulated composite IMF (i.e. the IGIMF,
Sec.~\ref{sec:comppop}). For $m\simless 1.3\,M_\odot$ the composite
IMF equals the stellar IMF according to the presently available
analysis, and so it is an interesting quantity for at least two
reasons: For the mass-budget of the Milky-Way disk, and as a
bench-mark against which the IMFs measured in presently- and
past-occurring star-formation events can be compared to distill
possible variations about the mean.

The {\it first and most straightforward method} consists of creating a
local volume-limited catalogue of stars yielding the nearby LF,
$\Psi_{\rm near}(M_V)$.  Completeness of the modern {\it
  Jahreiss--Gliese Catalogue of Nearby Stars} extends to about 25~pc
for $m\simgreat0.6\,M_\odot$, trigonometric distances having been
measured using the Hipparcos satellite, and only to about 5~pc for
less massive stars for which we still rely on ground-based
trigonometric parallax measurements\footnote{\label{fn:compl}Owing to
  the poor statistical definition of $\Psi_{\rm near}(M_V)$ for
  $m\simless 0.5\,M_\odot, M_V\simgreat 10$, it is important to
  increase the sample of nearby stars, but controversy exists as to
  the maximum distance to which the LMS census is complete. Using
  spectroscopic parallax it has been suggested that the local census
  of LMSs is complete to within about 15~\% to distances of 8~pc and
  beyond \citep{RG97}.  However, Malmquist bias allows stars and
  unresolved binaries to enter such a flux-limited sample from much
  larger distances \citep{K01b}.  The increase of the number of stars
  with distance using trigonometric distance measurements shows that
  the nearby sample becomes incomplete for distances larger than 5~pc
  and for $M_V>12$ \citep{J94,Henry97}. The incompleteness in the
  northern stellar census beyond 5~pc and within 10~pc amounts to
  about 35~\% \citep{Hetal03}, and further discovered companions
  (e.g. \citealt{Detal99,Beuzit01}) to known primaries in the distance
  range $5<d<12$~pc indeed suggest that the extended sample may not
  yet be complete.  Based on the work by \cite{Retal03b, Retal03a},
  \cite{Luhman04} however argues that the incompleteness is only about
  15~\%.}.  The advantage of the LF, $\Psi_{\rm near}(M_V)$, created
using this catalogue is that virtually all companion stars are known,
that it is truly distance limited and that the distance measurements
are direct.

The {\it second method} is to make deep pencil-beam surveys using
photographic plates or CCD cameras to extract a few hundred low-mass
stars from a hundred-thousand stellar and galactic images.  This
approach, pioneered by Gerry Gilmore and Neill Reid, leads to larger
stellar samples, especially so since many lines-of-sight into the
Galactic field ranging to distances of a few~100~pc to a~few~kpc are
possible. The disadvantage of the LF, $\Psi_{\rm phot}(M_V)$, created
using this technique is that the distance measurements are indirect
relying on photometric parallax. Such surveys are flux limited rather
than volume limited and pencil-beam surveys which do not pass through
virtually the entire stellar disk are prone to Malmquist bias
\citep{SIP}. This bias results from a spread of luminosities of stars
that have the same colour because of their dispersion of metallicities
and ages, leading to intrinsically more luminous stars entering the
flux-limited sample and thus biasing the inferred absolute
luminosities and the inferred stellar space densities.  Furthermore,
binary systems are not resolved in the deep surveys, or if formally
resolved the secondary is likely to be below the flux limit.

\begin{figure}
\begin{center}
\rotatebox{0}{\resizebox{0.8
\textwidth}{!}{\includegraphics{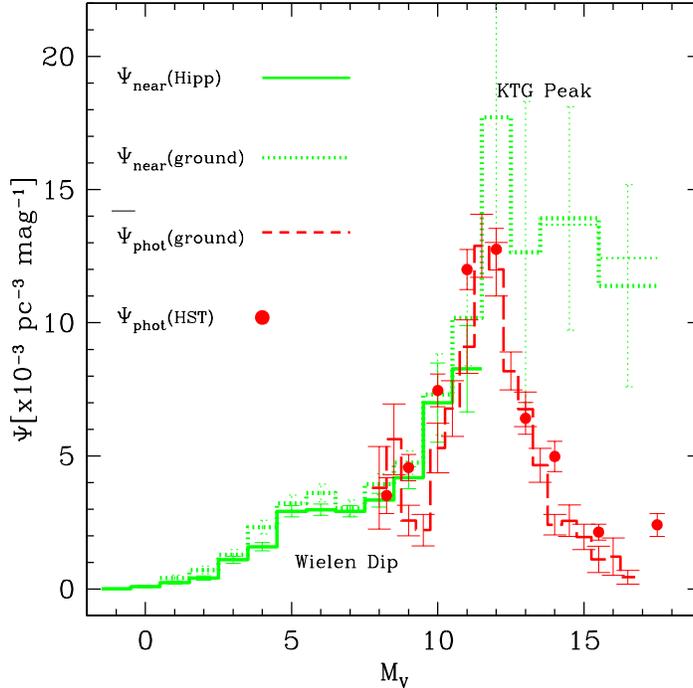}}}
\vskip -28mm
\caption[Luminosity function (LF) of solar-neighbourhood stars]{\small{Stellar luminosity functions (LFs) for
    solar-neighbourhood stars. The photometric LF corrected for
    Malmquist bias and at the mid-plane of the Galactic disk
    ($\Psi_{\rm phot}$) is compared with the nearby LF ($\Psi_{\rm
      near}$). The average, ground-based $\overline{\Psi}_{\rm phot}$
    (dashed histogram, data pre-dating 1995, is confirmed \citep{K95a}
    by Hubble-Space-Telescope (HST) star-count data which pass through
    the entire Galactic disk and are thus less prone to Malmquist bias
    (solid dots).  The ground-based volume-limited
    trigonometric-parallax sample (dotted histogram) systematically
    overestimates $\Psi_{\rm near}$ due to the Lutz-Kelker bias, thus
    lying above the improved estimate provided by the
    Hipparcos-satellite data (solid histogram,
    \citealt{JW97,K01b}). The Lutz-Kelker bias \citep{LK73} arises in
    trigonometric-parallax-limited surveys because the uncertainties
    in parallax measurements combined with the non-linear increase of
    the number of stars with reducing parallax (increasing distance)
    lead to a bias in the deduced number density of stars when using
    trigonometric-parallax limited surveys. The depression/plateau
    near $M_V=7$ is the {\it Wielen dip}.  The maximum near
    $M_V\approx 12, M_I\approx 9$ is the {\it KTG peak}. The thin
    dotted histogram at the faint end indicates the level of
    refinement provided by additional stellar additions \citep{K01b}
    demonstrating that even the immediate neighbourhood within 5.2~pc
    of the Sun probably remains incomplete at the faintest stellar
    luminosities. Which LF is the relevant one for constraining the
    MF? \cite{KTG93} uniquely used both LFs simultaneously to enhance
    the constraints. See text.}}
\label{fig:MWlf}
\end{center}
\end{figure}

\begin{figure}
\begin{center}
\rotatebox{0}{\resizebox{0.8
\textwidth}{!}{\includegraphics{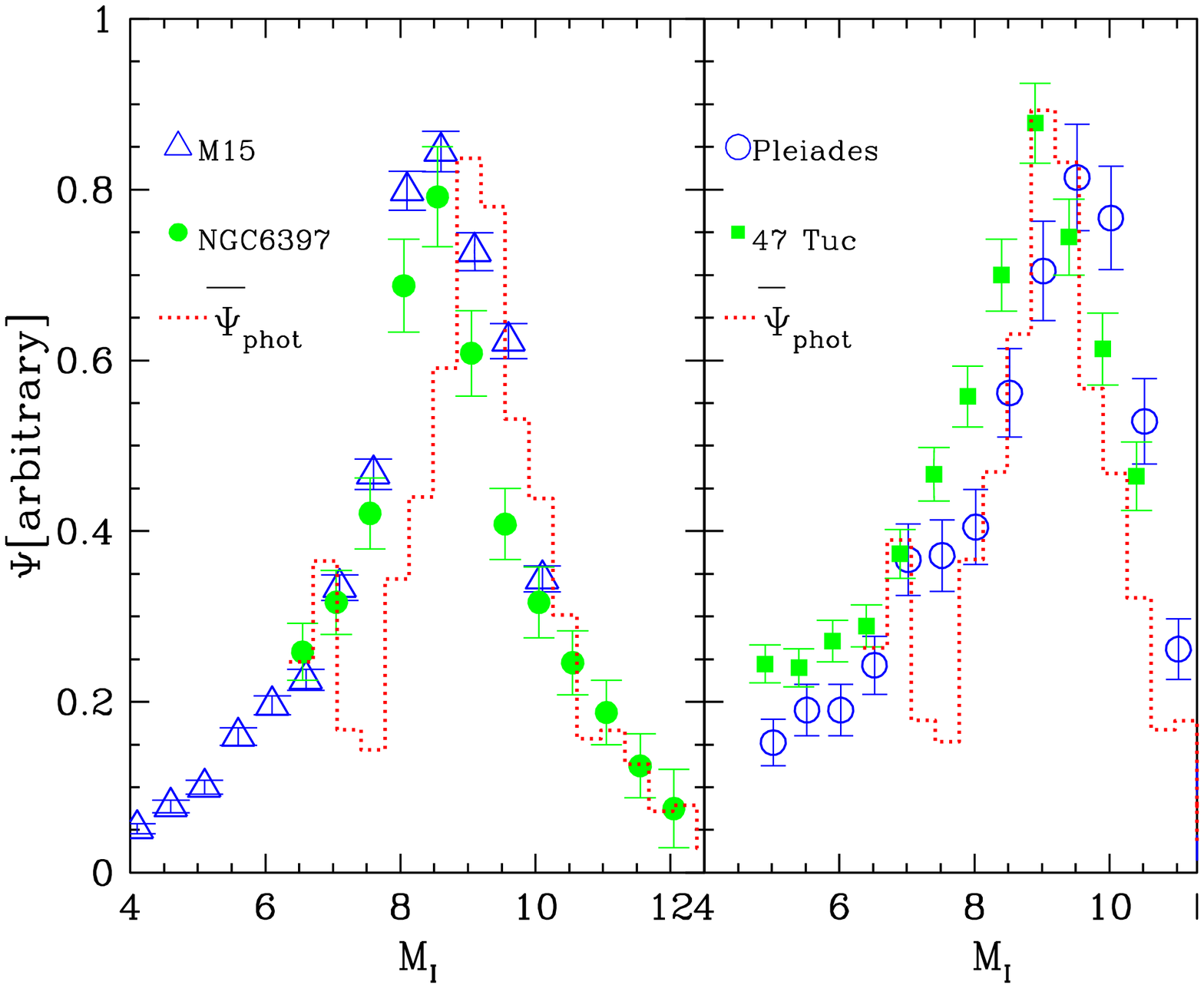}}}
\vskip -28mm
\caption[Stellar luminosity functions in star clusters]{\small{$I$-band LFs of stellar {\it systems} in four star
    clusters: globular cluster (GC) {\it M15} (\citealt{deMP95a},
    distance modulus $\Delta m=m-M=15.25$~mag); GC {\it NGC~6397}
    (\citealt{PdeMR95}, $\Delta m=12.2$); young Galactic cluster {\it
      Pleiades} (\citealt{HJH91}, $\Delta m=5.48$); GC {\it 47~Tuc}
    (\citealt{deMP95b}, $\Delta m=13.35$). The dotted histogram is
    $\overline{\Psi}_{\rm phot}(M_I)$ from Fig.~\ref{fig:MWlf},
    transformed to the $I$-band using the linear colour--magnitude
    relation $M_V=2.9+3.4\,(V-I)$ \citep{KTG93} and $\Psi_{\rm
      phot}(M_I) = (dM_V/dM_I) \, \Psi_{\rm phot}(M_V)$
    (Eq.~\ref{eq:passband}). The KTG peak is very pronounced in all
    LFs. It is due to an extremum in the derivative of the MLR
    (Fig.~\ref{fig:mlr}).
}} 
\label{fig:cllf}
\end{center}
\end{figure}

The local, {\it nearby LF} and the Malmquist-corrected deep {\it
  photometric LF} are displayed in Fig.~\ref{fig:MWlf}. They differ
significantly for stars fainter than $M_V\approx11.5$ which caused
some controversy in the past\footnote{\label{fn:cmr}This controversy
  achieved a maxiumum in 1995, as documented in \cite{K95a}.  The
  discrepancy evident in Fig.~\ref{fig:MWlf} between the nearby LF,
  $\Psi_{\rm near}$, and the photometric LF, $\Psi_{\rm phot}$,
  invoked a significant dispute as to the nature of this
  discrepancy. On the one hand \citep{K95a} the difference is thought
  to be due to unseen companions in the deep but low-resolution
  surveys used to construct $\Psi_{\rm phot}$, with the possibility
  that photometric calibration for VLMSs may remain problematical so
  that the exact shape of $\Psi_{\rm phot}$ for $M_V\simgreat 14$ is
  probably uncertain.  On the other hand \citep{RG97} the difference
  is thought to come from non-linearities in the $V-I, M_V$
  colour--magnitude relation used for photometric parallax. Taking
  into account such structure it can be shown that the photometric
  surveys underestimate stellar space densities so that $\Psi_{\rm
    phot}$ moves closer to the extended estimate of $\Psi_{\rm near}$
  using a sample of stars within 8~pc or further.  While this is an
  important point, the extended $\Psi_{\rm near}$ is incomplete (see
  footnote~\ref{fn:compl} on p.~\pageref{fn:compl}) and theoretical
  colour-magnitude relations do not have the required degree of
  non-linearity (e.g. fig.~7 in \citealt{Bochanski10}). The
  observational colour--magnitude data also do not conclusively
  suggest a feature with the required strength
  \citep{BCAH98}. Furthermore, $\Psi_{\rm phot}$ agrees almost
  perfectly with the LFs measured for star clusters of solar and
  population~II metallicity for which the colour-magnitude relation is
  not required (Fig.~\ref{fig:cllf}) so that non-linearities in the
  colour--magnitude relation cannot be the dominant source of the
  discrepancy.}.  That the local sample has a spurious but significant
over-abundance of low-mass stars can be ruled out by virtue of the
large velocity dispersion in the disk, $\approx30$~pc/Myr. Any
significant overabundance of stars within a sphere with a radius of
30~pc would disappear within one~Myr, and cannot be created by any
physically plausible mechanism from a population of stars with stellar
ages spanning the age of the Galactic disk. The shape of $\Psi_{\rm
  phot}(M_V)$ for $M_V\simgreat 12$ is confirmed by many independent
photometric surveys. That all of these could be making similar
mistakes, such as in colour transformations, becomes unlikely on
consideration of the LFs constructed for completely independent
stellar samples, namely star clusters (Fig.~\ref{fig:cllf}).

\subsection{The stellar mass--luminosity relation} 

The MF is related to the LF via the derivative of the stellar
mass--luminosity relation (Eq.~\ref{eq:mf_lf}). 

\begin{figure}
\begin{center}
\rotatebox{0}{\resizebox{0.9 \textwidth}{!}{\includegraphics{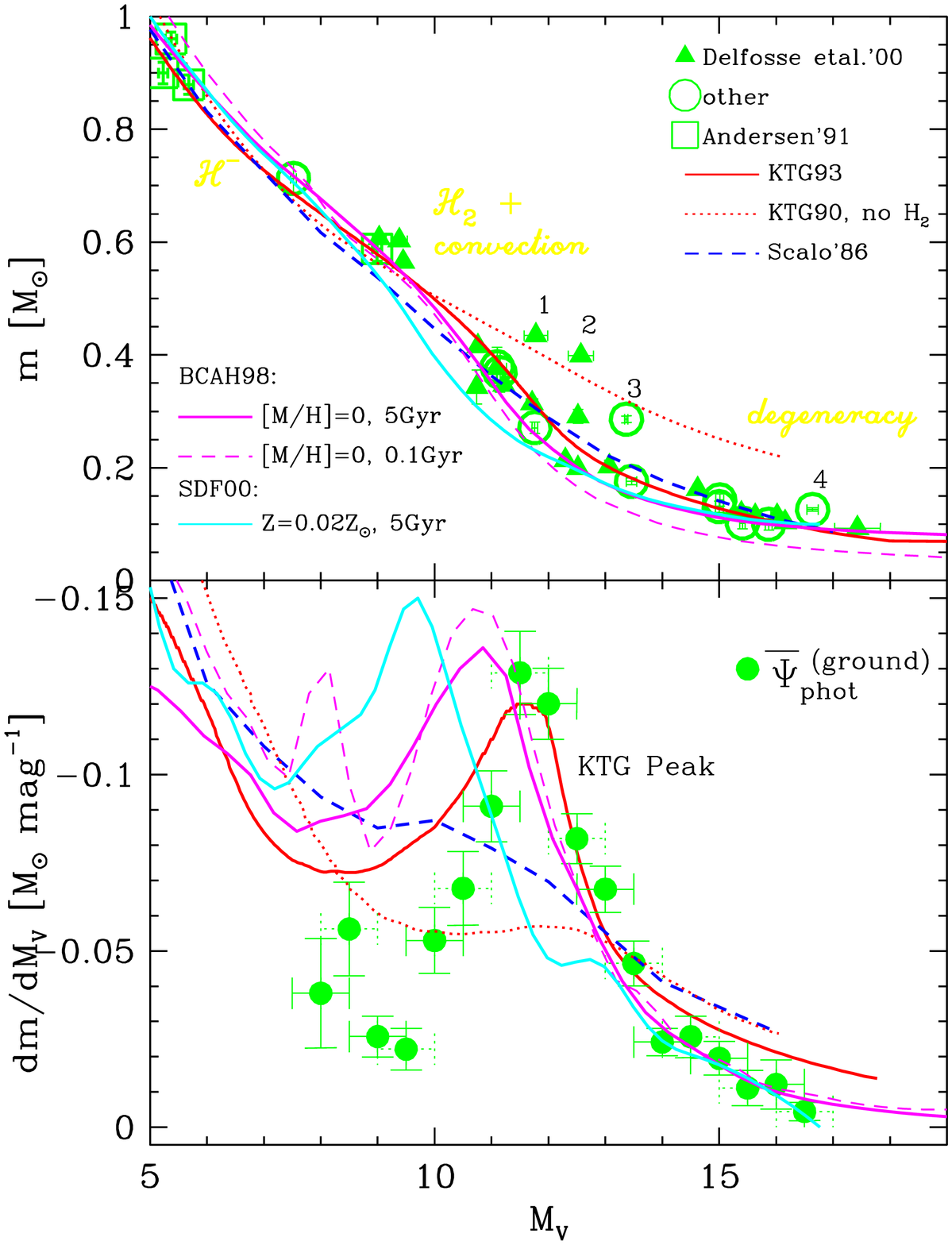}}}
\vskip -8mm
\caption[The mass-luminosity relation of stars]{\small{The
    mass-luminosity relation (MLR, upper panel) and its derivative
    (lower panel) for late-type stars. {\it Upper panel:} The
    observational data (solid triangles and open circles,
    \citealt{Detal00}; open squares, \citealt{A91}) are compared with
    the empirical MLR of \citet{Sc86} and the semi-empirical KTG93-MLR
    tabulated in \cite{KTG93}. The under-luminous data points~1,2 are
    GJ2069Aa,b and ~3,4 are Gl791.2A,B. All are probably metal-rich by
    $\approx0.5$~dex \citep{Detal00}.  Theoretical MLRs from
    \citet{BCAH98} (BCAH98) and \citet{SDF00} (SDF00) are also shown.
    The observational data \citep{A91} show that log$_{10}[m(M_V)]$ is
    approximately linear for $m>2\,M_\odot$. See also
    \cite{Malkov97}. {\it Lower panel:} The derivatives of the same
    relations plotted in the upper panel are compared with
    $\overline{\Psi}_{\rm phot}$ from Fig.~\ref{fig:MWlf}.  Note the
    good agreement between the location, amplitude and width of the
    KTG peak in the LF and the extremum in $dm/dM_V$.  }}
\label{fig:mlr}
\end{center}
\end{figure}

Eq.~\ref{eq:mf_lf} shows that any non-linear structure in the MLR is
mapped into observable structure in the LF, provided the MF does not
have a compensating structure. Such a conspiracy is implausible
because the MF is defined through the star-formation process, but the
MLR is a result of the internal constitution of stars.  The MLR and
its derivative are shown in Fig.~\ref{fig:mlr}.  It is apparent that
the slope is very small at faint luminosities leading to large
uncertainties in the MF near the hydrogen burning mass limit.

The physics underlying the non-linearities of the MLR are due to an
interplay of changing opacities, the internal stellar structure and
the equation of state of the matter deep inside the stars.  Starting
at high masses ($m\simgreat\,{\rm few}\,M_\odot$), as the mass of a
star is reduced H$^-$ opacity becomes increasingly important through
the short-lived capture of electrons by H-atoms resulting in reduced
stellar luminosities for intermediate and low-mass stars. The $m(M_V)$
relation becomes less steep in the broad interval $3<M_V<8$ leading to
the Wielen dip (Fig.~\ref{fig:MWlf}).  The $m(M_V)$ relation steepens
near $M_V=10$ because the formation of H$_2$ in the very outermost
layer of main-sequence stars causes the onset of convection up to and
above the photo-sphere leading to a flattening of the temperature
gradient and therefore to a larger effective temperature as opposed to
an artificial case without convection but the same central
temperature.  This leads to brighter luminosities and full convection
for $m\simless0.35\,M_\odot$.  The modern Delfosse data beautifully
confirm the steepening in the interval $10<M_V<13$ predicted in 1990.
In Fig.~\ref{fig:mlr} the dotted MLR demonstrates the effects of
suppressing the formation of the H$_2$ molecule by lowering it's
dissociation energy from 4.48~eV to 1~eV (\citealt{KTG90}, hereinafter
KTG).  The $m(M_V)$ relation flattens again for $M_V>14$,
$m<0.2\,M_\odot$ as degeneracy in the stellar core becomes
increasingly important for smaller masses limiting further contraction
\citep{HN63,ChB97}.  Therefore, owing to the changing conditions
within the stars with changing mass, a pronounced local maximum in
$-dm/dM_V$ results at $M_V\approx11.5$, postulated by KTG to be the
origin of the maximum in $\Psi_{\rm phot}$ near $M_V=12$.

\begin{figure}
\begin{center}
\rotatebox{0}{\resizebox{0.8
\textwidth}{!}{\includegraphics{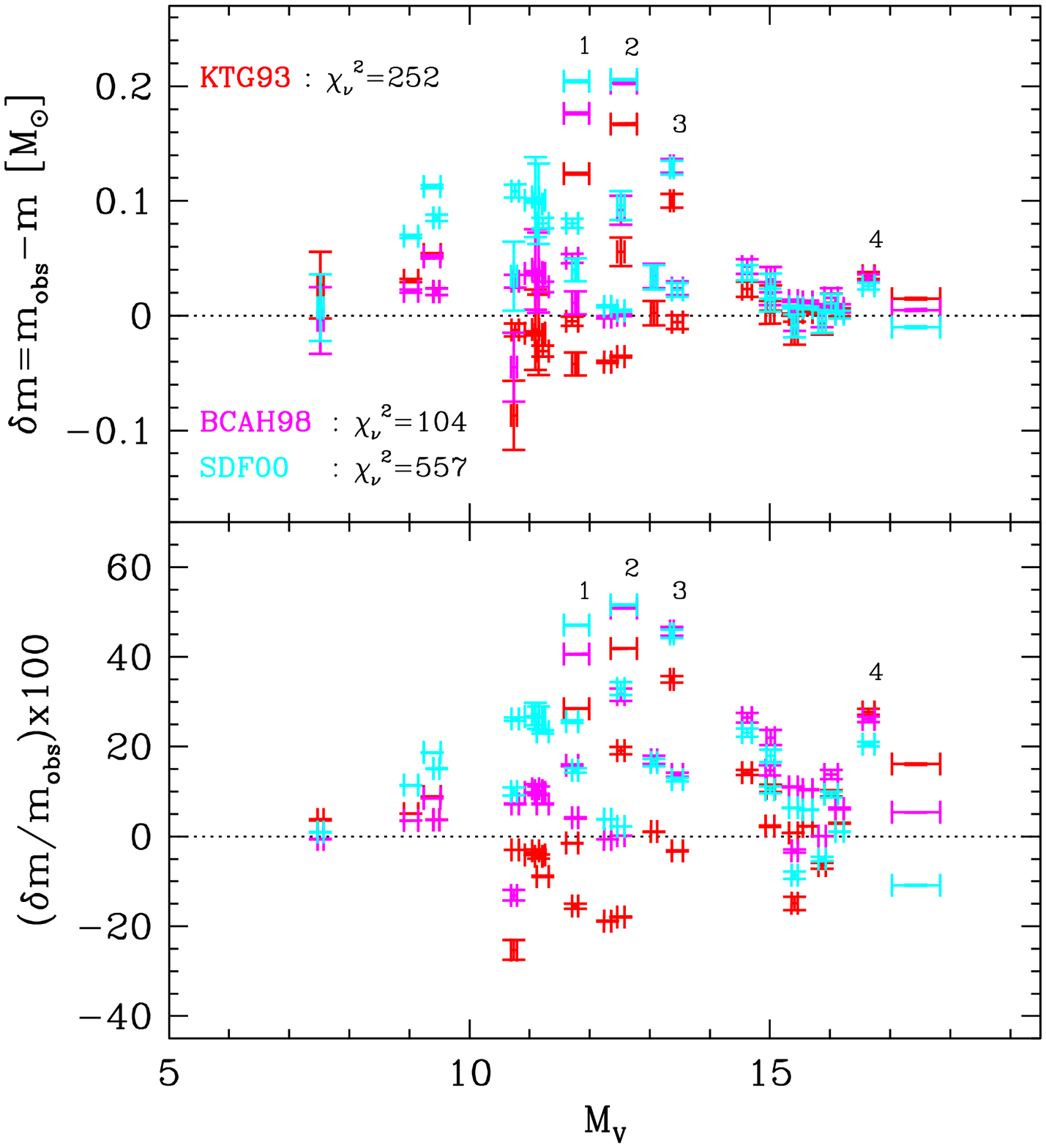}}}
\vskip -8mm
\caption[Deviations of the MLRs from empirical data]{\small{Deviations of the MLRs ($\delta m=m_{\rm
      obs}-m(M_V)$) from the empirical data, $m_{\rm obs}$, with
    errors $\epsilon_m$ shown in Fig.~\ref{fig:mlr} in $M_\odot$
    (upper panel) and in percent (lower panel with uncertainties
    $-(m(M_V)/m_{\rm obs}^2)\times\epsilon_m$). Colours refer to the
    models of Fig.~\ref{fig:mlr}.  Reduced $\chi_\nu^2$ ($\nu=26$ for
    31 data points, ignoring the four outliers) values indicate the
    formal goodness-of-fit. Formally, none of the MLRs available is an
    acceptable model for the data. This is not alarming though,
    because the models are for a single-metallicity, single-age
    population while the data span a range of metallicities and ages
    typical for the solar-neighbourhood stellar population, as
    signified by $\delta m \gg \epsilon_m$ in most cases.  The
    $\chi_\nu^2$ values confirm that the BCAH98 models \citep{BCAH98}
    and the semi-empirical KTG93 MLR \citep{KTG93} provide the
    best-matching MLRs. Note that the KTG93 MLR was derived from
    mass--luminosity data prior to 1980, but by using the shape of the
    peak in $\Psi_{\rm phot}(M_V)$ as an additional constraint the
    constructed MLR became robust.  The lower panel demonstrates that
    the deviations of observational data from the model MLRs are
    typically much smaller than 30~per cent, excluding the putatively
    metal-rich stars (1--4).  }}
\label{fig:dev}
\end{center}
\end{figure}

The implication that the LFs of all stellar populations should show
such a feature, although realistic metallicity-dependent stellar
models were not available yet, was noted \citep{KTG93}.  The
subsequent finding that all known stellar populations have the KTG
peak (Figs.~\ref{fig:MWlf} and~\ref{fig:cllf}) constitutes one of the
{\it most impressive achievements of stellar-structure theory}.
Different theoretical $m(M_V)$ relations have the maximum in
$-dm/dM_V$ at different $M_V$, suggesting the possibility of testing
stellar structure theory near the critical mass
$m\approx0.35\,M_\odot$, where stars become fully convective
\citep{KT97,BCC98}. But since the MF also defines the LF the shape and
location cannot be unambiguously used for this purpose unless it is
postulated that the IMF is invariant. Another approach to test stellar
models is by studying the deviations of observed $m(M_V)$ data from
the theoretical relations (Fig.~\ref{fig:dev}).

\begin{figure}
\begin{center}
\rotatebox{0}{\resizebox{0.8 \textwidth}{!}{\includegraphics{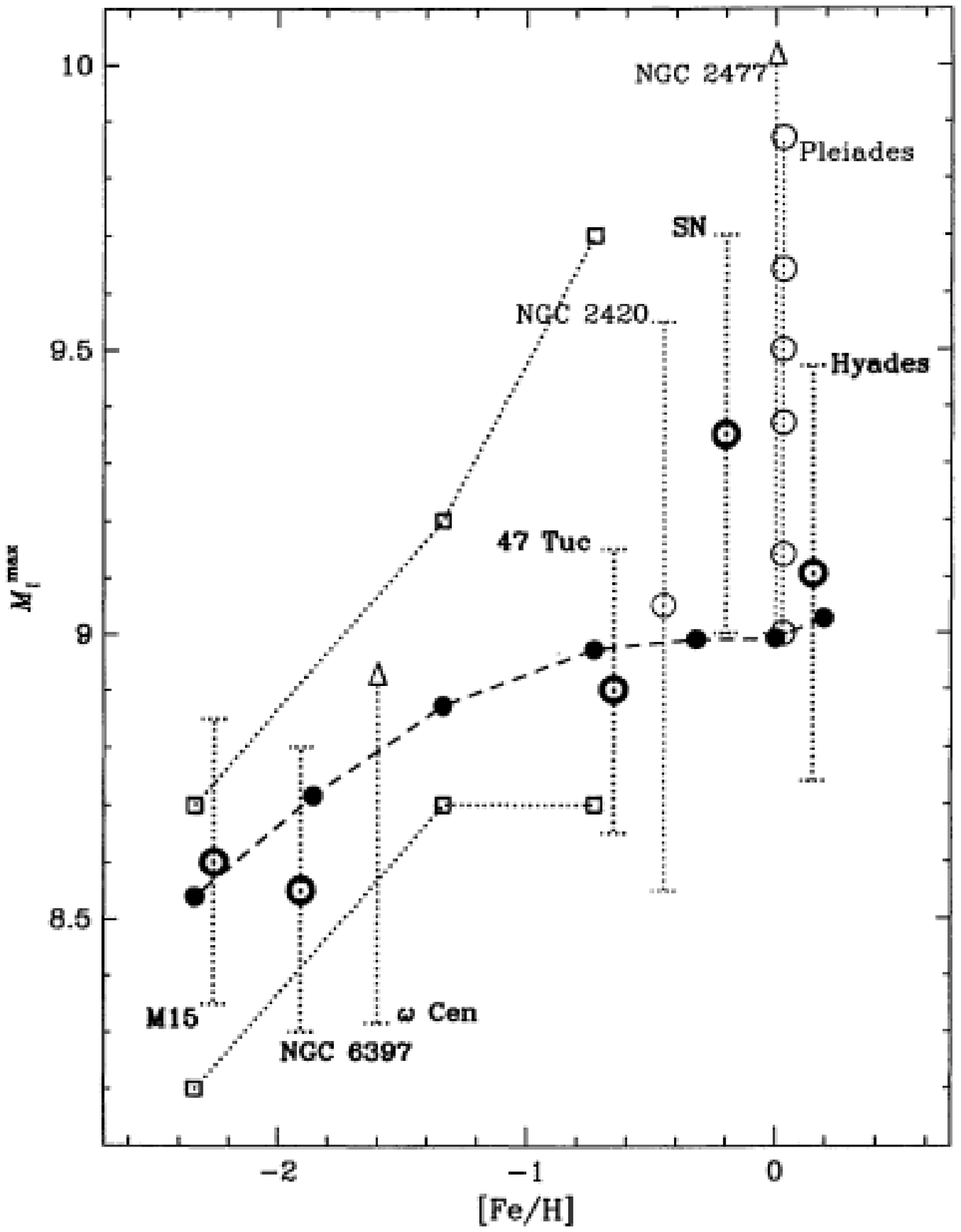}}}
\vskip 0mm
\caption[LF peak as a function of metallicity I.]{\small{ The position of the maximum in $-dm/dM_I$ as a
function of metallicity of the theoretical mass--luminosity data of
\citet{KT97} is shown as solid dots. The open squares represent bounds
by the stellar-structure models of \citet{dAM96}, and the open circles
are observational constraints for different populations (e.g. SN for
the composite solar-neighbourhood population, Pleiades for the simple
population of an intermediate-age cluster). Thick circles are more
certain than the thin circles, and for the Pleiades a sequence of
positions of the LF-maximum is given, from top to bottom, with the
following combinations of (distance modulus, age): (5.5, 70~Myr),
(5.5, 120~Myr), (5.5, main sequence), (6, 70~Myr), (6, 120~Myr), (6,
main sequence). For more details see \citet{KT97}.  }}
\label{fig:LFpeak}
\end{center}
\end{figure}

\begin{figure}
\begin{center}
\rotatebox{0}{\resizebox{0.9 \textwidth}{!}{\includegraphics{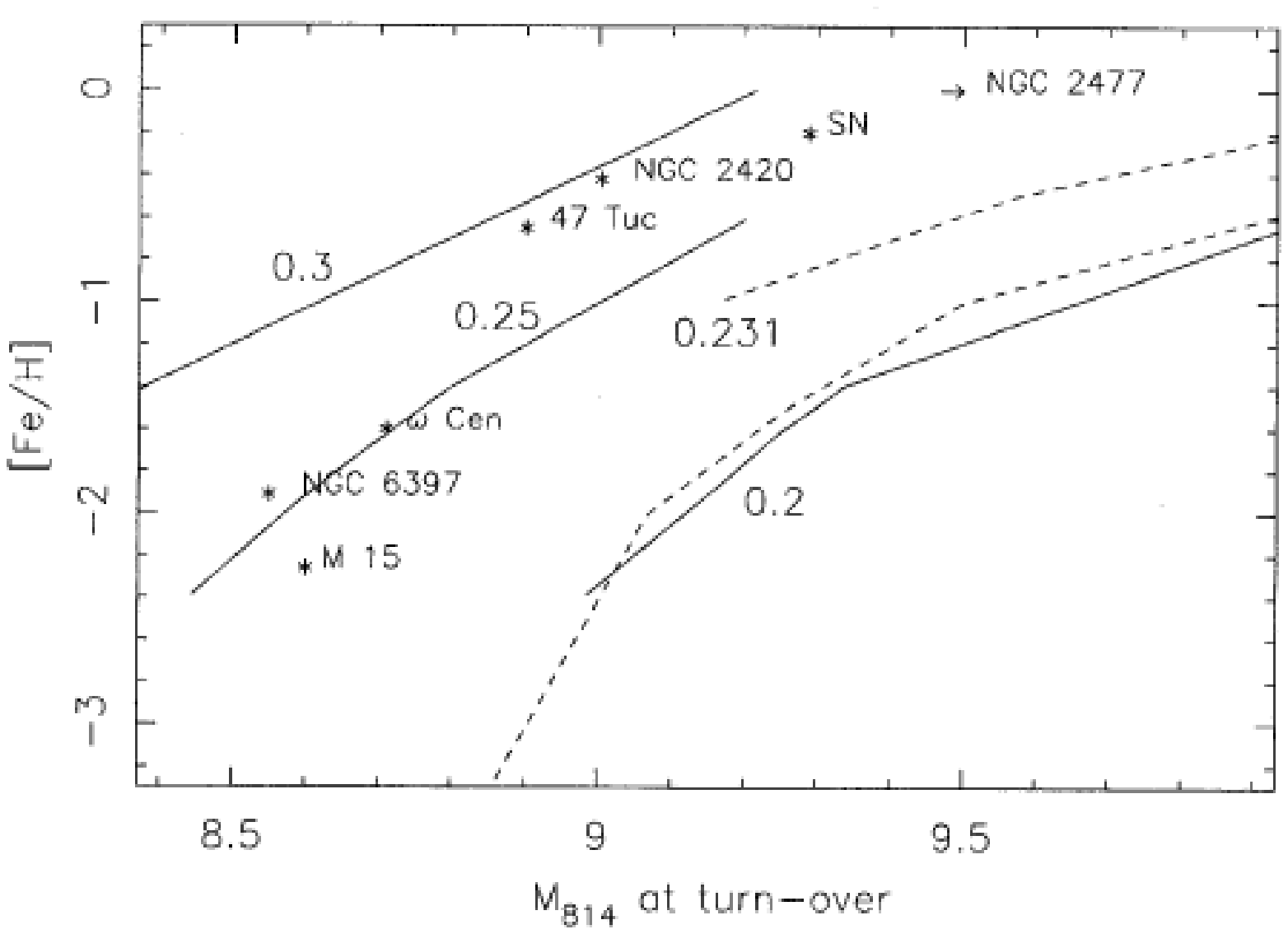}}}
\vskip 0mm
\caption[LF peak as a function of metallicity II.]{\small{Similar to Fig.~\ref{fig:LFpeak} but from
\citet{vonHippelGil96}, their fig.5: The absolute $I$-band-equivalent
magnitude of the maximum in the LF as a function of metallicity for
different populations.  The solid and dashed lines are loci of
constant mass ($0.2, 0.231, 0.3\,M_\odot$) according to theoretical
stellar structure calculations.  See \citet{vonHippelGil96} for more
details.  }}
\label{fig:LFpeak2}
\end{center}
\end{figure}

A study of the position of the maximum in the $I$-band LF has been
undertaken by \citet{vonHippelGil96} and \citet{KT97} finding that the
observed position of the maximum shifts to brighter magnitude with
decreasing metallicity, as expected from theory
(Figs.~\ref{fig:LFpeak} and~\ref{fig:LFpeak2}).


\subsection{Unresolved binary stars and the solar-neighbourhood IMF} 

In addition to the non-linearities in the $m(M_P)$ relation,
unresolved multiple systems affect the MF derived from the photometric
LF, in particular since no stellar population is known to exist that
has a binary proportion smaller than about 50~per cent, apart possibly
from dynamically highly evolved globular and open clusters
\citep{Sollima07,SCB09,MKO11}.

Suppose an observer sees 100~systems. Of these~40, 15~and 5~are
binary, triple and quadruple, respectively, these being realistic
proportions. There are thus 85~companion stars which the observer is
not aware of if none of the multiple systems are resolved. Since the
distribution of secondary masses is not uniform but typically
increases with decreasing mass for F-, G- and K-type primaries (it
decreases for M-type primaries, \citealt{MZ01, MKO11}), the bias is
such that low-mass stars are significantly underrepresented in any
survey that does not detect companions \citep{KTG91,MZ01}.  Also, if
the companion(s) are bright enough to affect the system luminosity
noticeably then the estimated photometric distance will be too small,
artificially enhancing inferred space densities which are, however,
mostly compensated for by the larger distances sampled by binary
systems in a flux-limited survey, together with an exponential density
fall-off perpendicular to the Galactic disk \citep{K01b}.  A faint
companion will also be missed if the system is formally resolved but
the companion lies below the flux limit of the survey.

\begin{figure}
\begin{center}
\rotatebox{0}{\resizebox{0.75
\textwidth}{!}{\includegraphics{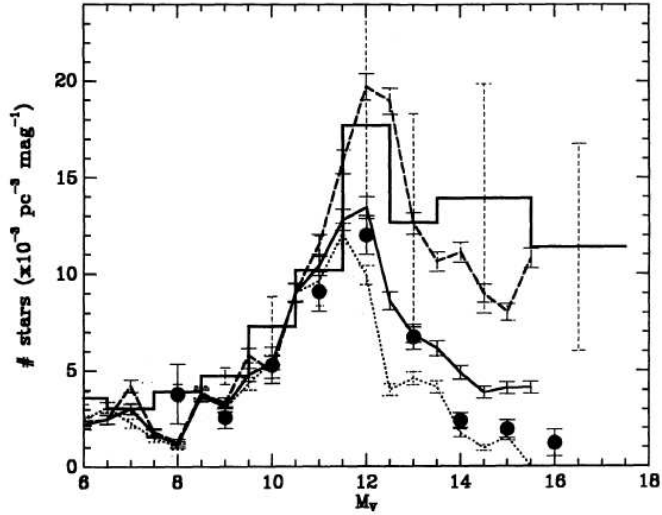}}}
\vskip 0cm
\caption[LF: model vs. observations I.]%
{\small { Comparison of the model field luminosity function (curves)
    of a single-metallicity and single-age population that is without
    measurement errors, with observations in the photometric V-band (a
    comparison of the corresponding LFs in bolometric magnitudes can
    be found in \citealt{K95e}). {\it The model} assumes the standard
    or canonical stellar IMF, Eq.~\ref{eq:imf} below.  The model
    single star luminosity function is normalised to the nearby
    luminosity function at $M_{\rm V}\approx 10, M_{\rm bol}\approx9$,
    giving the normalisation constant in the MF $k$
    (Eq.~\ref{eq:imf_mult}), and the plot shows $k\,\Psi_{\rm
      mod,sing}$ (long dashed curve), $k\,\Psi_{\rm mod,sys}(t=0)$
    (dotted curve, 100~per cent birth binary fraction in dynamically
    unevolved embedded clusters, see Fig.~\ref{fig:LFclmod}) without
    pre-main sequence brightening, and $k\,\Psi_{\rm mod,sys}(t=1\,{\rm
      Gyr})$ (solid curve, 48~per cent surviving binary fraction in
    dissolved clusters, see Fig.~\ref{fig:LFclmod}).  Note that the
    solid curve is the luminosity function for a realistic model of
    the Galactic field population of systems consisting of 48~per~cent
    binaries (which result from disruption of the 100~per cent binary
    birth population of Sec.~\ref{sec:bins} in the embedded clusters)
    which have a period distribution consistent with the observed G-,
    K-, and M-dwarf period distribution, the mass ratio distributions
    for G-dwarf systems as observed \citep{DM91}, and the overall
    mass-ratio distribution given by fig.~2 in \cite{Kr_etal03}, where
    a concise description of the ``standard star-formation model'' can
    be found.  {\it The observed} nearby stellar luminosity function,
    $\Psi_{\rm near}$, which is not corrected for Lutz-Kelker bias
    (\citealt{LK73}, tables~2 and~8 in \citealt{K95a}) and which is
    smoothed by using larger bin widths at the faint end, as detailed
    in section~4 of that paper, is plotted as the solid-line
    histogram.  The filled circles represent the best-estimate
    Malmquist corrected photometric luminosity function,
    ${\overline\Psi}_{\rm phot}$ (Fig.~\ref{fig:MWlf}).  By correcting
    for Malmquist bias \citep{SIP} the LF becomes that of a
    single-age, single-metallicity population.  Taken from
    \citet{K95e}.}}
\label{fig:lfmods}
\end{center}
\end{figure}

Comprehensive star-count modelling of the solar neighbourhood that
incorporates unresolved binary systems, metallicity and age spreads
and the density fall-off perpendicular to the Galactic disk with
appropriate treatment of Malmquist and Lutz-Kelker bias show that the
IMF, from which the solar-neighbourhood populations within a few~pc and
a few hundred~pc stem, can be {\it unified with one MF} which is a
two-part power-law with $\alpha_1=1.3\pm0.5, 0.07 <m/M_\odot \le 0.5$,
$\alpha_2\approx 2.2, 0.5<m/M_\odot \le 1$, a result obtained for two
different MLRs \citep{KTG93,K01b}. The index $\alpha_2$ is constrained
tightly owing to the well-constrained $\Psi_{\rm near}$, the
well-constrained empirical MLR in this mass range and because
unresolved binary systems do not significantly affect the
solar-neighbourhood LF in this mass range because primaries with
$m\simgreat1\,M_\odot$ are rare and are not sampled.  The stellar
sample in the mass range $0.5-1\,M_\odot$ is therefore complete.

Fig.~\ref{fig:lfmods} demonstrates models of the individual-star and
system LFs for the KTG93 MLR shown in Fig.~\ref{fig:mlr}.  The
significant difference between the individual-star and system LFs is
evident, being most of the explanation of the disputed discrepancy
between the observed $\Psi_{\rm near}$ and $\Psi_{\rm phot}$. Note
though that the observed photometric LF contains triple and quadruple
systems that are not accounted for by the model. Note also that the
photometric LF has been corrected for Malmquist bias and so
constitutes the system LF in which the broadening due to a metallicity
and age spread and photometric errors has been largely removed. It is
therefore directly comparable to the model system LF, and both indeed
show a very similar KTG peak.  The observed nearby LF, on the other
hand, has not been corrected for the metallicity and age spread nor
for trigonometric distance errors, and so it appears broadened.  The
model individual-star LF, in contrast, does not, by construction,
incorporate these, and thus appears with a more pronounced maximum.
Such observational effects can be incorporated rather easily into
full-scale star-count modelling \citep{KTG93}. The deviation of the
model system LF from the observed photometric LF for $M_V\simgreat 14$
may indicate a change of the pairing properties of the VLMS or BD
population~(Sec.~\ref{sec:bds}).

Since the nearby LF is badly defined statistically for $M_V\simgreat
13$, the resulting model, such as shown in Fig.~\ref{fig:lfmods}, is a
{\it prediction} of the true, individual-star LF that should become
apparent once the immediate solar-neighbourhood sample has been
enlarged significantly through the planned space-based astrometric
survey {\sc Gaia} \citep{Gil98}, followed by an intensive follow-up
imaging and radial-velocity observing programme scrutinising every
nearby candidate for unseen companions \citep{K01b}.  Despite such a
monumental effort, the structure in $\Psi_{\rm near}^{\rm GAIA}$ will
be smeared out due to the metallicity and age spread of the local
stellar sample, a factor to be considered in detail.

\vspace{2mm} \centerline{ \fbox{\parbox{\columnwidth}{ {\sc Main results}: {
        The universal structure in the stellar LF of main sequence
        stars (the Wielen dip and the KTG peak) is well understood. It
        is due to non-linearities in the stellar mass--luminosity
        relation. Binary systems have a highly significant effect on
        the LF of late-type stars. The solar-neighbourhood IMF
        which unifies $\Psi_{\rm near}$ and $\Psi_{\rm phot}$ has
        $\alpha_1=1.3\; (0.07-0.5\,M_\odot)$ and $\alpha_2=2.2\;
        (0.5 - 1\,M_\odot$).  }}}}
\vspace{2mm}

\subsection{Star clusters} 
\label{sec:stcl}

Star clusters less massive than about $M=10^5\,M_\odot$ and to a good
degree of approximation also those with $M>10^5\,M_\odot$ offer
populations that are co-eval, equidistant and that have the same
chemical composition. But, seemingly as a compensation of these
advantages the extraction of faint cluster members is very arduous
because of contamination by the background or foreground
Galactic-field population. The first step is to obtain photometry of
everything stellar in the vicinity of a cluster and to select only
those stars that lie near one isochrone, taking into account that
unresolved binaries are brighter than single stars. The next step is
to measure proper motions and radial velocities of all candidates to
select only those high-probability members that have coinciding space
motion with a dispersion consistent with the a priori unknown but
estimated internal kinematics of the cluster. Since nearby clusters
for which proper-motion measurements are possible appear large on the
sky, the observational effort is overwhelming. An excellent example of
such work is the 3D mapping of the Hyades cluster by
\cite{Roeseretal11}.  For clusters such as globulars that are isolated
the second step can be omitted, but in dense clusters stars missed due
to crowding need to be corrected for.

The stellar LFs in clusters turn out to have the same general shape as
$\Psi_{\rm phot}$ (Fig.~\ref{fig:cllf}), with the maximum being
slightly off-set depending on the metallicity of the population
(Figs.~\ref{fig:LFpeak} and~\ref{fig:LFpeak2}). A 100~Myr isochrone
(the age of the Pleiades) is also plotted in Fig.~\ref{fig:mlr} to
emphasise that for young clusters additional structure (in this case
another maximum near $M_V=8$ in the LF is expected via
Eq.~\ref{eq:mf_lf}). This is verified for the Pleiades cluster
\citep{Belikov98}, and is due to stars with $m<0.6\,M_\odot$ not
having reached the main-sequence yet \citep{ChB00}.

LFs for star clusters are, like $\Psi_{\rm phot}$, system LFs because
binary systems are not resolved in the typical star-count survey. The
binary-star population evolves due to encounters, and after a few
initial crossing times only those binary systems survive that have a
binding energy larger than the typical kinetic energy of stars in the
cluster \citep{Heggie75,MKO11}.

A further complication with cluster LFs is that star clusters
preferentially loose single low-mass stars across the tidal boundary
as a result of ever-continuing re-distribution of energy during
encounters while the retained population has an increasing binary
proportion and increasing average stellar mass. The global PDMF thus
flattens with time with a rate proportional to the fraction of the
cluster lifetime and, for highly evolved initially rich open clusters,
it evolves towards a delta function near the turnoff mass, the
mass-loss rate being a function of Galactocentric distance.  This is a
major issue for aged open clusters (initially $N<10^4$~stars) with
life-times of only a few~100~Myr.

\begin{figure}
\begin{center}
\rotatebox{0}{\resizebox{0.75 \textwidth}{!}{\includegraphics{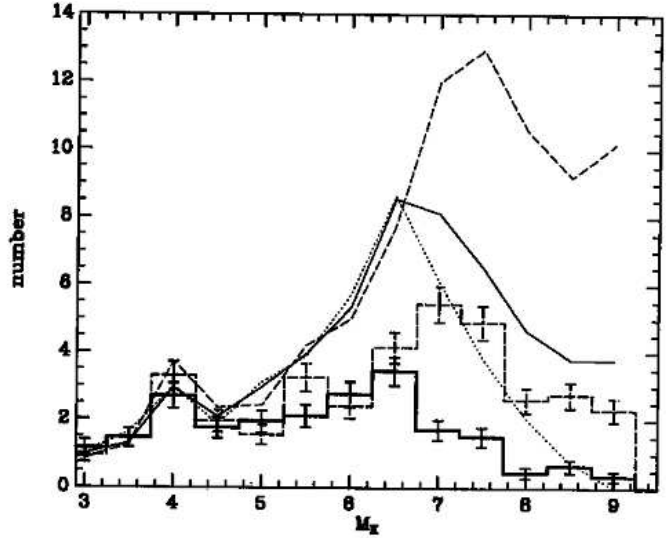}}}
\vskip -4mm
\caption[LF: model vs. observations II.]{\small{ Models of the $K$-band individual-star and system LFs
    in an ensemble of 20~dynamically highly evolved clusters (thin and
    thick histograms, respectively). An observer would deduce the
    thick histogram, which can only be transformed to the
    individual-star PDMF in the cluster if adequate correction for
    unresolved binaries is made. Such a correction leads to the upper
    thin histogram from which the PDMF can be inferred via
    Eq.~\ref{eq:mf_lf}.  Each cluster model consists initially of
    200~binaries with a half-mass radius of 0.85~pc, and the LFs are
    shown at an age of 480~Myr (44 initial crossing times) and count
    all stars and systems within the central sphere with a radius of
    2~pc. The clusters are random renditions from the same parent
    distributions (binary-star orbital parameters, IMF, stellar
    positions and velocities) and are initially in dynamical
    equilibrium. The upper dashed curve is the initial individual-star
    LF (KTG93 MLR, Fig.~\ref{fig:mlr}, and canonical IMF,
    Eq.~\ref{eq:imf} below) and the solid curve is the model
    Galactic-field LF of systems, also shown in
    Fig.~\ref{fig:lfmods}. This is an accurate representation of the
    Galactic-field population in terms of the IMF and mixture of
    single and binary stars, and is derived by stars forming in
    clusters such as shown here that dissolve with time.  Both of
    these LFs are identical to the ones shown in
    Fig.~\ref{fig:lfmods}. The dotted curve is the initial system LF
    (100~\% binaries).  From \citet{K95b}.  }}
\label{fig:LFclmod}
\end{center}
\end{figure}

These processes are now well quantified, and Fig.~\ref{fig:LFclmod}
shows that a dynamically very evolved cluster such as the Hyades has
been depleted significantly in low-mass stars. Even so, the
binary-star correction that needs to be applied to the LF in order to
arrive at the individual-star present-day LF is significant.

\begin{figure}
\begin{center}
\rotatebox{0}{\resizebox{0.65
\textwidth}{!}{\includegraphics{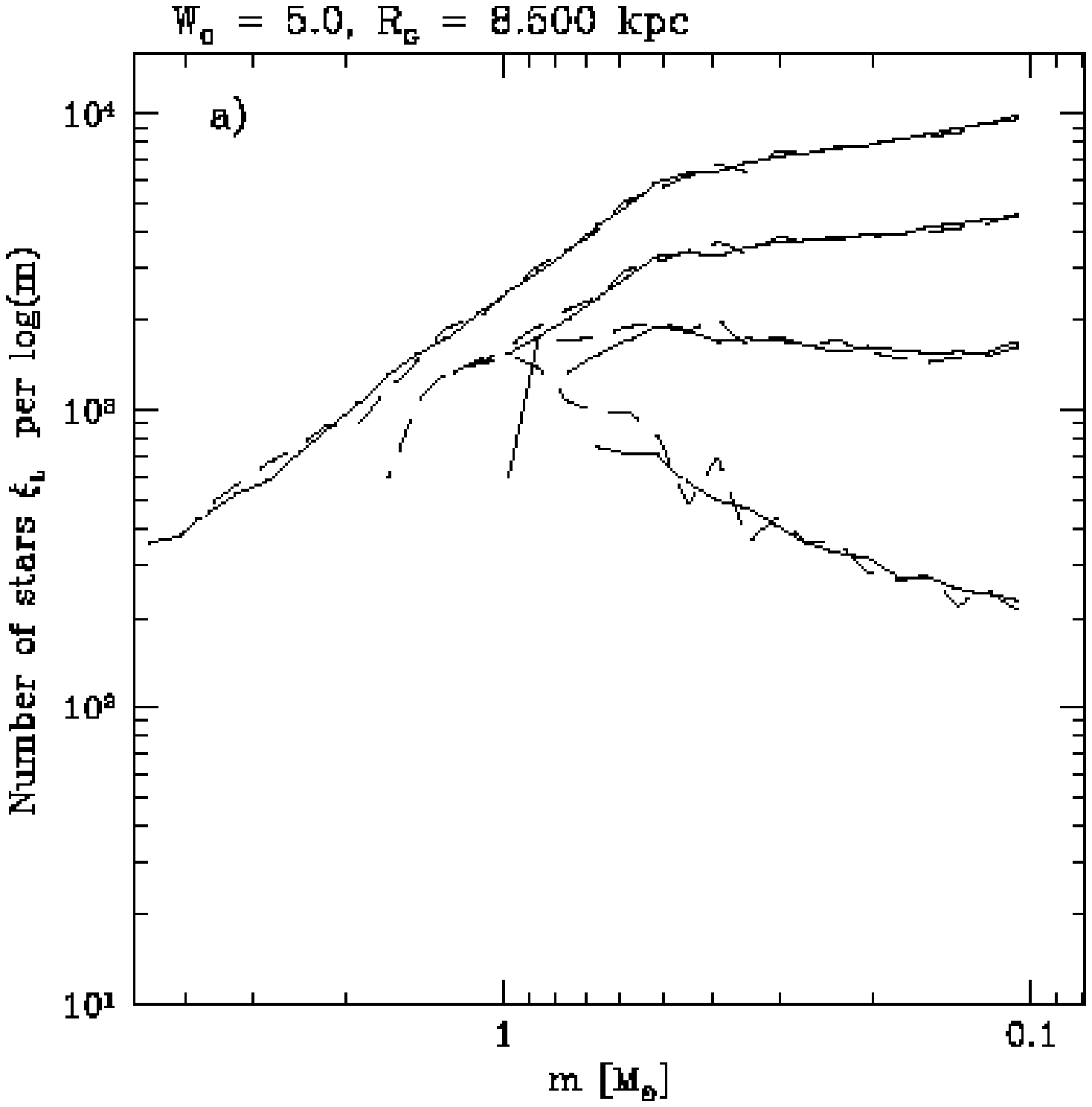}}}
\vskip -4mm
\caption[Dynamical evolution of PDMFs]{\small{PDMFs in a King-model cluster with concentration
    $W_0=5$ on a circular orbit about the MW centre with radius
    8.5~kpc. Shown are the MFs of all bound stars at ages
    corresponding to $\tau_f=\,$0~\%, 30~\%, 60~\% and 90~\% of the
    cluster life-time (from top to bottom). For each age the solid
    line represents one computation with $1.28\times 10^5$~stars, the
    dashed lines show the sum of four clusters each with 8000~stars
    (scaled to the same number of stars as the massive
    computation). Results for other circular and eccentric orbits and
    cluster concentrations are virtually indistinguishable. From
    \citet{BaumMakin03}. Note the progressive depletion of low-mass
    stars as the cluster ages.  }}
\label{fig:MFBaumg}
\end{center}
\end{figure}

A computationally challenging investigation of the systematic changes
of the MF in evolving clusters of different masses has been published
by \citet{BaumMakin03}. Baumgardt \& Makino quantify the depletion of
the clusters of low-mass stars through energy-equipartition-driven
evaporation and conclusively show that highly evolved clusters have a
very substantially skewed PDMF (Fig.~\ref{fig:MFBaumg}). If the
cluster ages are expressed in fractions, $\tau_f$, of the overall
cluster lifetime, which depends on the initial cluster mass, its
concentration and orbit, then different clusters on different orbits
lead to virtually the same PDMFs at the same $\tau_f$.  Their results
were obtained for clusters that are initially in dynamical equilibrium
and that do not contain binary stars (these are computationally highly
demanding), so that future analysis, including initially
non-virialised clusters and a high primordial binary fraction
(Sec.~\ref{sec:bins}), will be required to further refine these
results.

For the massive and long-lived globular clusters ($N\simgreat
10^5$~stars) theoretical stellar-dynamical work shows that the MF
measured for stars near the cluster's half-mass radius is
approximately similar to the global PDMF, while inwards and outwards
of this radius the MF is flatter (smaller $\alpha_1$) and steeper
(larger $\alpha_1$), respectively, owing to dynamical mass segregation
\citep{VH97}. However, mass loss from the cluster flattens the global
PDMF such that it no longer resembles the IMF anywhere
(Fig.~\ref{fig:MFBaumg}), for which evidence has been found in some
cases (\citealt{PZ99}, see also Sec.~\ref{sec:tophGC}). The MFs
measured for globular clusters must therefore generally be flatter
than the IMF, which is indeed born-out by observations
(Fig.~\ref{fig:apl} below). However, again the story is by no means
straightforward, because globular clusters have significantly smaller
binary fractions than population~II clusters \citep{Ivanova_etal05}.
The binary-star corrections are therefore smaller for globular cluster
MFs implying a larger difference between $\alpha_1$ for GCs and open
clusters for which the binary correction is very significant.

Therefore, and as already pointed out by \citet{K01a}, it appears
quite realistically possible that population~II IMFs were in fact
flatter (smaller $\alpha_1$) than population~I IMFs, as would be
qualitatively expected from simple fragmentation theory
(Sec.~\ref{sec:var_metal}). Clearly, this issue needs detailed
investigation which, however, is computationally highly demanding,
requiring the use of state-of-the art $N$-body codes and
special-purpose hardware.

\begin{figure}
\begin{center}
\rotatebox{0}{\resizebox{0.99 \textwidth}{!}{\includegraphics{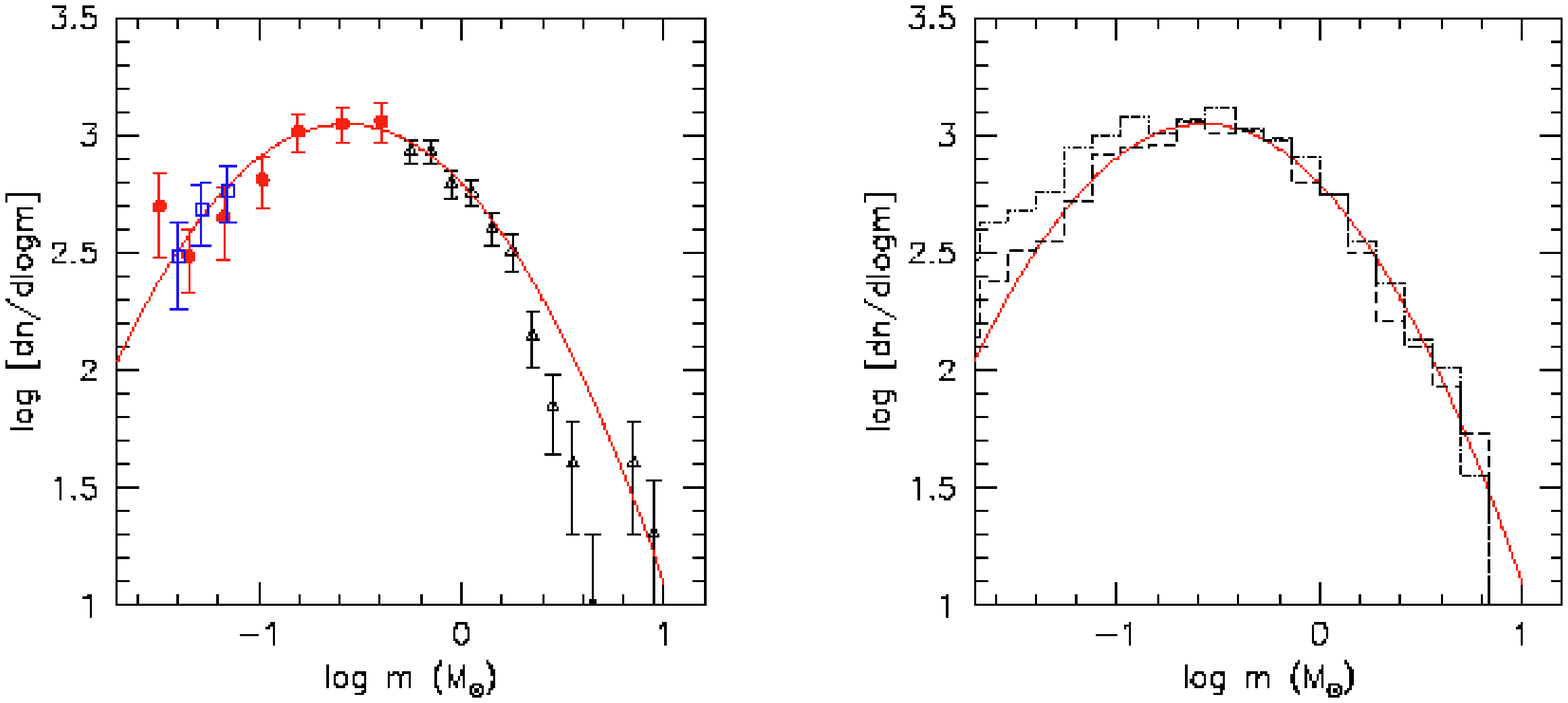}}}
\vskip -4mm
\caption[Mass function in the Pleiades]{\small{ The observationally
    deduced MF in the Pleiades cluster. {\it Left panel}: The symbols
    are observational data (for details see \citealt{MKB04}) and the
    curve is a log-normal fit. {\it Right panel}: The curve is the
    same log-normal fit.  Theoretical system MFs for two initial
    models of the Pleiades cluster according to \citet{KAH} are
    plotted at an age of 100~Myr.  These models assume the young
    cluster to be deeply embedded in gas, with a star-formation
    efficiency of 33~\%, a gas-expulsion time-scale shorter than the
    crossing time and to contain $10^4$ stars and BDs, whereby all
    stars and BDs are paired randomly to binary systems (i.e., in
    these models BDs are not treated as a separate population, see
    Sec.~\ref{sec:bds} for more realistic models). Model~A (dashed
    histogram) has an initial central number density $\rho_{\rm
      C}=10^{4.8}$~stars/pc$^3$, while model~B (dotted histogram) has
    $\rho_{\rm C}=10^{5.8}$~stars/pc$^3$.  The embedded phase lasts
    0.6~Myr, and during this time mass segregation develops in the
    initially denser cluster model~B. Note that these models are not a
    fit but a prediction of the Pleiades MF, assuming it had a
    canonical IMF (Eq.~\ref{eq:imf}).  Note that the initially denser
    cluster (upper histogram) retains more BDs as a result of these
    being ionised off their stellar primaries in the denser
    environment of model~B. Also note that the observational data
    suggest a deficit of early-type stars in the Pleiades (left panel)
    which is reminiscent of the deficit of massive stars noted for the
    ONC \citep{P-AK06a}.  }}
\label{fig:MorauxPl}
\end{center}
\end{figure}

The first realistic calculations of the formation of an open star
cluster such as the Pleiades demonstrate that the binary properties of
stars remaining in the cluster are comparable to those observed even
if all stars initially form in binary systems according to the BBP
(Eq.~\ref{eq:fPbirth}, \citealt{KAH, Kr_etal03}).  That work also
demonstrates the complex and counter-intuitive interplay between the
initial concentration, mass segregation at the time of residual gas
expulsion, and the final ratio of the number of BDs to stars
(Fig.~\ref{fig:MorauxPl}). Thus, this modelling shows that an
initially denser cluster evolves to significant mass segregation when
the gas explosively leaves the system. Contrary to naive expectation,
according to which a mass-segregated cluster should loose more of its
least massive members during expansion after gas expulsion, the
ensuing violent relaxation of the cluster retains more free-floating
BDs than the less-dense model. This comes about because BDs are split
from the stellar binaries more efficiently in the denser cluster.
This, however, depends on the BDs and stars following the same pairing
rules, which is now excluded (Sec.~\ref{sec:bds}).  During the long
term evolution of the mass function initially mass segregated star
clusters loose more low-mass stars when they are close to dissolution
as has been found by \citet{BaumDeMarchi08} after comparing star loss
from clusters in $N$-body calculations with the observed mass
functions of globular clusters. But additionally the gas expulsion
from embedded clusters has to be carefully taken into account to
explain the correlation between concentration of a globular cluster
and the slope of its PDMF (Sec.~\ref{sec:tophGC})

These issues remain an active area of research, because at least two
changes need to be made to the modelling: On the one hand, BDs need to
be treated as a population separate from the stellar one
(Sec.~\ref{sec:bds}) so that the free-floating BDs that result, in the
currently available models, from the disruption of star--BD binaries,
will not be available in reality. On the other hand some observations
suggest that star clusters may form highly mass-segregated.  The
mass-dependent loss of stars thus definitely remains an issue to be
studied.

The above work suggests that even clusters as young as the Pleiades
are significantly evolved because clusters of all masses form from
highly concentrated embedded morphologies \citep{Kr_paris05, MK10,
  Marks12a, Conroy12}.  Also, the low-mass stars in clusters as young
as the Pleiades or M35 (Fig.~\ref{fig:mfn1} below) have not yet
reached the main sequence, so that pre-main sequence stellar-evolution
calculations have to be resorted to when transforming measured
luminosities to stellar masses via the MLR.

For ages younger than a few~Myr this becomes a serious problem:
Classical pre-main sequence theory, which assumes hydrostatic
contraction of spherical non-, sometimes slowly-rotating stars from
idealised initial states breaks down because of the overlap with the
star formation processes that defies detailed treatment. Stars this
young remember their accretion history, invalidating the application
of classical pre-main sequence stellar evolution tracks, a point made
explicitly clear by \cite{Tout99,WK01} and \cite{WT03}, and are in any
case rotating rapidly and are non-spherical. Such realistic pre-main
sequence tracks are not available yet. The uncertainties due to such
processes have been partially discussed though \citep{Baraffe02,
  BCG09, Baraffe10}.

Research on the IMF in very young clusters benefits from spectroscopic
classification of individual stars to place them on a theoretical
isochrone of existing classical pre-main-sequence evolution theory to
estimate masses (e.g. \citealt{Meyer00, Luhman04, Barrado_etal04,
  Slesnick_etal04}). In such cases the deduced age spread becomes
comparable to the age of the cluster (Eq.~\ref{eq:imf_pdmf}). Binary
systems are mostly not resolved but can feign an apparent age spread
even if there is none in the underlying population as has been shown
by \citet{WKM09}. Differential reddening due to in-homogeneously
distributed remnant gas and dust has a significant effect on
estimating stellar masses \citep{Andersen09}.  But also episodic
accretion onto the proto-stars can mimic such age spreads
\citep{Tout99, BCG09, Baraffe10}. The reality of age spreads is an
important issue as it defines whether star clusters are almost coeval
or host prolonged star-formation. The finding of 10~to 30~Myr old
dwarfs by the Lithium depletion method in the $\approx1$~Myr ONC by
\citet{PRP07} seems to indicate the latter. But careful numerical
calculations have shown that a collapsing molecular cloud can trap a
corresponding amount of stars from a surrounding older OB~association
into the forming cluster \citep{PAK07}. The ONC is embedded in the
Orion OB1 association that has an age of 10 to 15 Myr which could
explain the older dwarfs in the ONC.

A few results are shown in Fig.~\ref{fig:mfn1}
and~\ref{fig:mfn2}. While the usual argument is for an invariant IMF,
as is apparent for most population~I stars (e.g. fig.5 in
\citealt{Chrev03}; fig.~3 in \citealt{Bastian2010}),
Fig.~\ref{fig:mfn1} shows that some appreciable differences in
measured MFs are evident. The M35 MF appears to be highly deficient in
low-mass stars. This clearly needs further study because M35 and the
Pleiades appear to be otherwise fairly similar in terms of age,
metallicity (M35 is somewhat less metal-rich than the Pleiades) and
the size of the survey volume.

Taking the ONC as the best-studied example of a very young and nearby
rich cluster (age$\;\approx 1$~Myr, distance$\;\approx 450$~pc;
$N\approx 5000-10000$~stars and BDs; \citealt{HC00,L00,MLL00, K00,
Slesnick_etal04}), Fig.~\ref{fig:mfn1} shows how the shape of the
deduced IMF varies with improving (but still classical) pre-main
sequence contraction tracks. This demonstrates that any sub-structure
cannot, at present, be relied upon to reflect possible underlying
physical mechanisms of star formation.

\begin{figure}
\begin{center}
\rotatebox{0}{\resizebox{0.99 \textwidth}{!}{\includegraphics{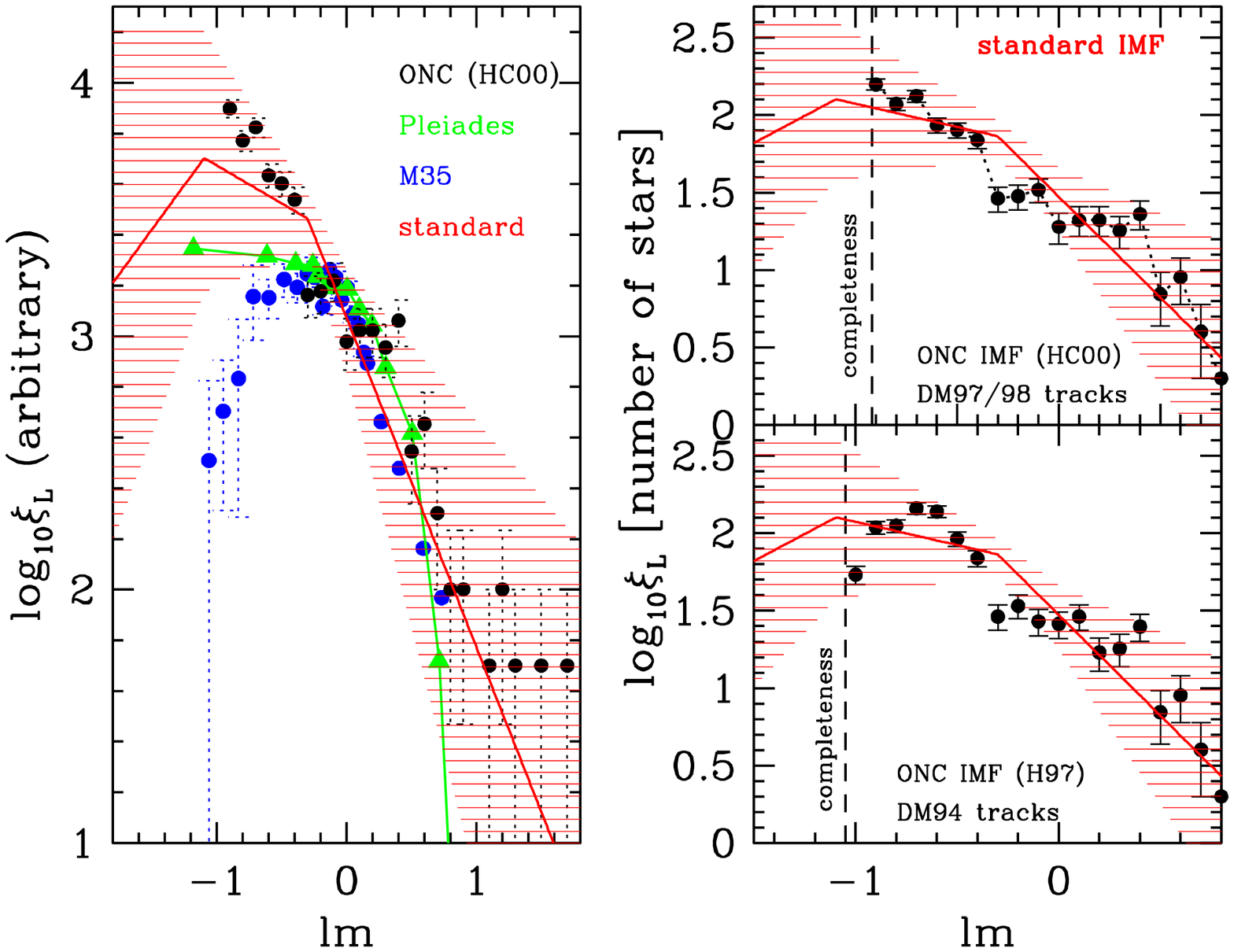}}}
\vskip -35mm
\caption[System mass function in the ONC]%
{\small{\it Left panel:} The observationally deduced system mass
  functions ($lm\equiv {\rm log}_{10}(m/M_\odot)$) (a)~in the ONC
  \citep{HC00}: optical data within $r\le2.5$~pc, $\tau_{\rm
    cl}<2$~Myr, [Fe/H]$=-0.02$ \citep{Estetal98} (b)~in the Pleiades
  \citep{Hambetal99}: $r\le6.7$~pc, $\tau_{\rm cl}\approx100$~Myr,
  [Fe/H]$=+0.01$, and (c)~in M35 \citep{Netal01}: $r\le4.1$~pc,
  $\tau_{\rm cl}\approx160$~Myr, [Fe/H]$=-0.21$, where $r$ is the
  approximate projected radius of the survey around the cluster centre
  and $\tau_{\rm cl}$ the nuclear age. The strong decrease of the M35
  MF below $m\approx0.5\,M_\odot$ remains present despite using
  different MLRs (e.g. DM97, as in the right panel).  None of these
  MFs are corrected for unresolved binary systems.  The canonical
  individual-star IMF (Eq.~\ref{eq:imf}) is plotted as the three
  straight lines assuming here continuity across the VLMS/BD mass
  range.  {\it Right panel:} The shape of the ONC IMF differs
  significantly for $m<0.22\,M_\odot$ if different pre-main sequence
  evolution tracks, and thus essentially different theoretical MLRs,
  are employed (DM stands for tracks calculated by D'Antona \&
  Mazzitelli, see \citealt{HC00} for details.)  }
\label{fig:mfn1}
\end{center}
\end{figure}

\begin{figure}
\begin{center}
\rotatebox{0}{\resizebox{0.99 \textwidth}{!}{\includegraphics{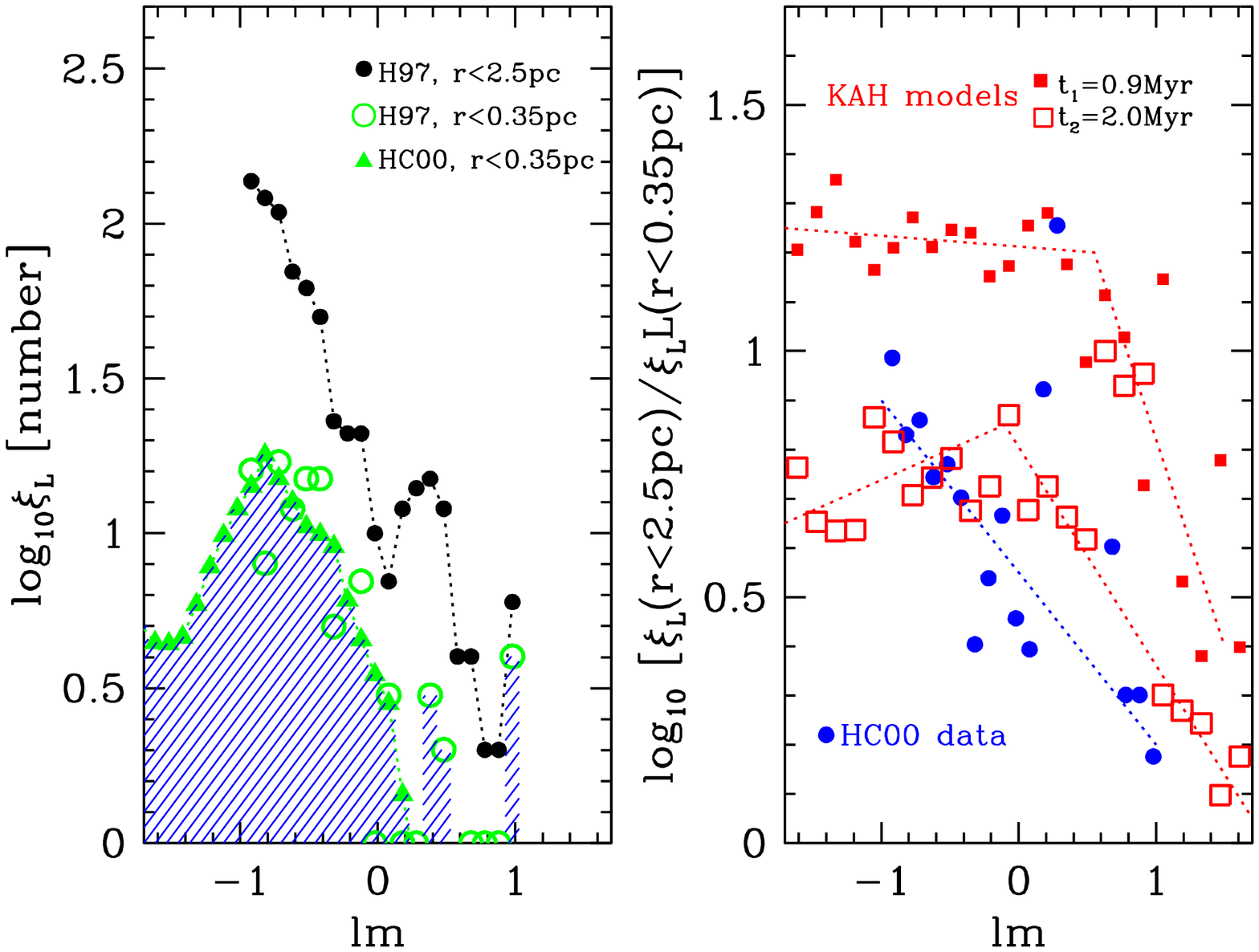}}}
\vskip -35mm
\caption[Mass segregation in the ONC]%
{\small{{\it Left panel:} Mass segregation is very pronounced in the
    ONC, as is evident by comparing the observationally deduced MF for
    all stars within $r=2.5$~pc with the MF for all stars with
    $r<0.35$~pc (\citealt{HC00}, HC00) ($lm\equiv {\rm
      log}_{10}(m/M_\odot)$). For both samples the reddening
    $A_V<2.5$~mag (\citealt{Hill97}, H97, is for an optical and
    spectroscopic survey, whereas HC00 is a near-infrared
    survey). {\it Right panel:} The ratio, $\xi_{\rm L}(r<2.5\,{\rm
      pc})/\xi_{\rm L}(r<0.35\,{\rm pc})$ (solid dots), of the MFs
    shown in the left panel increases significantly with decreasing
    mass, demonstrating the significant depletion of low-mass stars in
    the central region of the ONC. Stellar-dynamical models of the ONC
    \citep{KAH} approximately reproduce this trend at an age of 2~Myr
    for the canonical IMF (Eq.~\ref{eq:imf}, whereby the system masses
    of surviving binary systems are counted instead of the individual
    stars many of which are in unresolved binaries) even if no initial
    mass segregation is assumed (at $t=0$, $\xi_{\rm L}(r<2.5\,{\rm
      pc})/\xi_{\rm L}(r<0.35\,{\rm pc}) =\,$constant).  The model
    snapshots shown are from model~B in \cite{KAH} under the
    assumption that prior to gas-expulsion, the central stellar
    density was $\rho_{\rm C}=10^{5.8}$~stars/pc$^3$. The dotted lines
    are eye-ball fits to the plotted data. See also
    Fig.~\ref{fig:MorauxPl}). }}
\label{fig:mfn2}
\end{center}
\end{figure}


\vspace{2mm} \centerline{ \fbox{\parbox{\columnwidth}{\label{box:resonimf}
      {\sc Main results}: {Currently available evidence from resolved
        stellar populations largely points toward an IMF of late-type
        ($m\simless 1\,M_\odot$) stars which is independent of the
        environment and which can be described well by a power-law
        with an index of about $\alpha_2 = 2.3$ for $m\simgreat
        0.5\,M_\odot$ and $\alpha_1=1.3$ for $m\simless
        0.5\,M_\odot$. }}}} \vspace{2mm}

\section{The IMF of Very Low-Mass Stars (VLMSs) and of Brown Dwarfs
  (BDs)}
\label{sec:bds}

These are stars near to the hydrogen-burning mass limit (VLMS) or
objects below it (BDs). BDs are not massive enough to achieve
sufficiently high central pressures and temperatures to stabilise
against continued contraction by burning H and thus indefinitely cool
to unobservable luminosities and temperatures. The term ``brown
dwarf'' was coined by Jill Tarter in her 1975 PhD thesis, and was
later generally accepted as a name for sub-stellar
(i.e. non-hydrogen-burning) objects which presumably form like stars.

Observationally it is very difficult to distinguish between VLMSs and
BDs, because a sufficiently young BD may have colours and spectral
features corresponding to a VLMS.  BDs were studied as theoretical
objects in~1963 by \citet{HN63}, who performed the first truly
self-consistent estimate of the minimum hydrogen burning mass limit,
$m_{\rm H}$, by computing the luminosity at the surface and the energy
release rate by nuclear burning. Modern theory of the evolution and
internal constitution of BDs has advanced considerably owing to the
inclusion of an improved equation of state and realistic
model-atmospheres that take into account absorption by many molecular
species as well as dust allowing the identification of characteristic
photometric signatures \citep{ChB00}.  This work shows that the
critical mass below which an object cannot be stabilised by nuclear
fusion is $m_{\rm H}=0.075\,M_\odot$ for solar metallicity. For lower
metallicity $m_{\rm H}$ is larger since a larger luminosity (due to
the lower opacity) requires more efficient nuclear burning to reach
thermal equilibrium, and thus a larger mass.  The first BDs were
detected in~1995, and since then they have been found in the solar
neighbourhood and in young star clusters \citep{Basri00} allowing
increasingly sophisticated estimates of their mass distribution
\citep{Bouvier03}.

For the solar neighbourhood, near-infrared large-scale surveys have
now identified many dozens of BDs probably closer than 25~pc
(e.g. \citealt{Allen_etal04}). Since these objects do not have
reliable distance measurements an ambiguity exists between their ages
and distances, and only statistical analysis that relies on an assumed
star-formation history for the solar neighbourhood can presently
constrain the IMF \citep{Ch02}, finding a 60~\% confidence interval
$\alpha_0 = 0.3\pm0.6$ for $0.04-0.08\,M_\odot$ approximately for the
Galactic-field BD IMF \citep{Allen_etal04}.

Surveys of young star clusters have also discovered BDs by finding
objects that extend the colour--magnitude relation towards the faint
locus while being kinematical members. Given the great difficulty of
this endeavour, only a few clusters now possess constraints on the
MF. The Pleiades star cluster has proven especially useful, given its
proximity ($d\approx127$~pc) and young age ($\tau_{\rm
  cl}\approx100$~Myr). Results indicate
$\alpha_0\approx0.5-0.6$. Estimates for other clusters (ONC,
$\sigma$~Ori, IC~348, Cha~I) also indicate $\alpha_0\simless0.8$. In
their table~1, \citet{Allen_etal04} summarise the available
measurements for 11~populations finding that $\alpha_0\approx
0-1$. And \citet{Andersen08} find the low-mass IMF in~7 young
star-forming regions to be most consistent with being sampled from an
underlying log-normal (Chabrier) IMF.

However, while the log-normal (Chabrier) IMF is indistinguishable in
the stellar regime from the simpler canonical two-part power-law IMF
(see Fig.~\ref{fig:canIMF} below), it is to be noted that these and
other constraints on the IMF of BDs rely on assuming the IMF to be
continuous across the stellar/BD boundary. In the following it will
emerge that this assumption is not consistent with the binary
properties of stars and BDs. The IMF can therefore not be a continuous
log-normal across the VLMS/BD boundary.

\subsection{BD and VLMS binaries}
\label{sec:BDbins}

The above estimates of the BD IMF suffer under the same bias affecting
stars, namely from unseen companions which need to be taken into
account to infer the true BD IMF.

BD--BD binary systems are known to exist \citep{Basri00}, in the field
\citep{Bouy_etal03,Close_etal03} and in clusters
\citep{Martin_etal03}. Their frequency is not yet fully constrained
since detailed scrutiny of individual objects is time-intensive on
large telescopes but the data suggest a binary fraction of about 15~\%
only.  The results show conclusively that the semi-major axis
distribution of VLMSs and BDs is much more compact than that of
M~dwarfs, K~dwarfs and G~dwarfs. \citet{Bouy_etal03, Close_etal03,
  Martin_etal03} and \cite{PhanBao_etal05} all find that BD binaries
with semi-major axis $a\simgreat 15$~AU are very rare. Using
Monte-Carlo experiments on published multiple-epoch radial-velocity
data of VLMSs and BDs, \citet{MJ05} deduce an overall binary fraction
between~32 and~45~\% with a semi-major-axis distribution that peaks
near~4~AU and is truncated at about 20~AU.  In the Pleiades cluster
where their offset in the colour--magnitude diagram from the single-BD
locus makes them conspicuous, \citet{Pinfield_etal03} find the BD
binary fraction may be as high as 60~\%. This, however, appears
unlikely as a survey of more than six years UVES/VLT spectroscopy has
shown the BD binary fraction to be 10--30~per cent and that
incompleteness in the few~AU separation region is not significant
\citep{Joergens08}.  Using a very deep infrared survey of the
Pleiades, \cite{Lodieu07} suggest a BD binary fraction of~28 to 44~per
cent, consistent with the \cite{MJ05} result but only marginally so
with the \cite{Pinfield_etal03} value.

It has already been shown that the disruption in embedded clusters of
stellar binaries in a stellar population which initially consists of
100~per cent binary stars with periods (i.e. binding energies)
consistent with the observed pre-main sequence and proto-stellar
binary data (Eq.~\ref{eq:fPbirth}) leads to the observed main-sequence
binary population in the Galactic field \citep{K95c, GK05, MK11}.
Systems with BD companions have an even lower binding energy, and the
truncated semi-major-axis distribution of BDs may be a result of
binary-disruption in dense clusters of an initial stellar-like
distribution.

This notion is tested by setting-up the {\it star-like hypothesis}
\citep{Kr_etal03}:

\vspace{2mm} \centerline{ \fbox{\parbox{\columnwidth}{ {\sc Star-Like
        Hypothesis for BDs}:
BDs form as stars do from molecular-cloud cores. 
}}}
\vspace{2mm}

\noindent The {\sc Star-Like Hypothesis} implies that BDs form
according to the same binary pairing rules (the BBP,
Eq.~\ref{eq:fPbirth}) as stars do.

If BDs form as stars do then this hypothesis ought to be true since
objects with masses $0.04-0.07\,M_\odot$ should not have very
different pairing rules than stars that span a much larger range of
masses ($0.1-1\,M_\odot$) but show virtually the same
period-distribution function independently of primary mass (the M-, K-
and G-dwarf samples, \citealt{FiMa92,Mayor_etal92,DM91},
respectively). Thus, the hypothesis is motivated by observed orbital
distribution functions of stellar binaries not being sensitive to the
primary mass, which must come about if the overall physics of the
formation problem is similar.  Further arguments for a star-like
origin of BDs comes from the detection of accretion onto and disks
about very young BDs, and that the BDs and stars in Taurus-Auriga have
indistinguishable spatial and velocity distributions
\citep{WhiteBasri03}.

\begin{figure}
\begin{center}
\rotatebox{0}{\resizebox{0.95\textwidth}{!}{\includegraphics{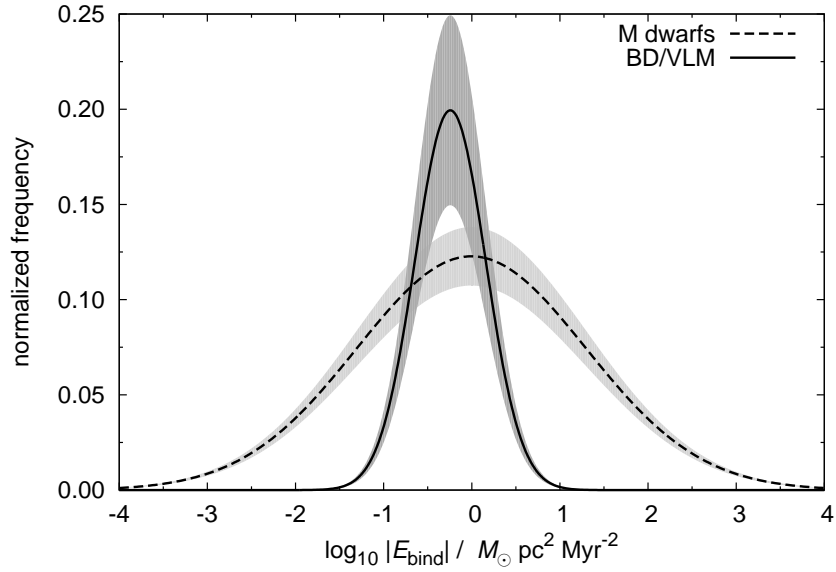}}}
\vskip -5mm
\caption[Binding energies of BD and M dwarf binaries]{\small{ The distribution of binding energies, $E_{\rm bin} =
    -G\,m_1\,m_2/(2\,a)$, of BDs (solid line) compared to those of
    M~dwarfs (MDs, dashed line). The BD distribution is computed as a
    Gaussian distribution based on BD/VLMS data from the Very Low Mass
    Binary Archive (http://vlmbinaries.org). Specifically, the
    Gaussian distribution has a mean semi-major axis log$_{10}(a_{\rm
      mean}/{\rm AU})=0.6$ with a half-width of
    log$_{10}(\sigma_a/{\rm AU})=0.4$.  The upper and lower envelopes
    correspond to BD binary fractions of $f_{\rm BD}=0.25$ and 0.15,
    respectively (area under the curves).  The MD energy distribution
    is computed by assuming the $a$-distribution from \citet{FiMa92}
    (which is practically identical to that of G~dwarfs) and choosing
    $10^7$ masses, $m_i\in(0.1-0.5\,M_\odot)$, from the canonical
    stellar IMF (Eq.~\ref{eq:imf}) and random pairing. For the MDs the
    Gaussian distribution has log$_{10}(a_{\rm mean}/{\rm AU})=1.5$
    and log$_{10}(\sigma_a/{\rm AU})=1.3$ such that $f_{\rm MD}=0.45$
    and~0.35 for the upper and lower envelopes, respectively.  }}
\label{fig:BDbindEn}
\end{center}
\end{figure}

Assuming the {\sc Star-Like Hypothesis} to hold, \citet{KAH,
  Kr_etal03} perform $N$-body calculations of ONC- and
Taurus-Auriga-like stellar aggregates to predict the semi-major-axis
distribution functions of BD--BD, star--BD and star--star binaries.
These calculations demonstrate that the binary proportion among BDs
becomes smaller than among low-mass stars after a few crossing times,
owing to their weaker binding energies.  The distribution of
separations, however, extends to similar distances as for stellar
systems (up to $a\approx 10^3$~AU), disagreeing completely with the
observed BD--BD binary distribution.  The {\sc Star-Like Hypothesis}
thus predicts far too many wide BD--BD binaries.  This can also be
seen from the distribution of binding energies of real BD binaries.
It is very different to that of stars by having a low-energy cutoff,
$^{\rm BD}E_{\rm bin, cut}\approx -10^{-0.9}\,M_\odot\,$(pc/Myr)$^2$,
that is much higher than that of the M~dwarfs, $^{\rm M}E_{\rm bin,
  cut}\approx -10^{-3}\,M_\odot\,$(pc/Myr)$^2$
(Fig.~\ref{fig:BDbindEn}).  This is a very strong indicator for some
fundamental difference in the dynamical history of BDs.

Furthermore, the $N$-body distributions contain a substantial number
of star--BD pairs, which also disagrees with the existence of very few
BD companions to nearby stars \citep{Basri00, PhanBao_etal05}.
Basically, if BDs form exactly like stars, then the number of star--BD
binaries would be significantly larger than is observed, since for
example G-dwarfs prefer to pair with M-dwarfs (why should BDs be any
different from M dwarfs in their pairing to G dwarfs?).  The observed
general absence of BD companions is referred to as the {\it BD desert}
\citep{ZuMaz01}, since stellar companions and planets are found at
such separations \citep{Halbw00,Vogt01}.  A few very wide star--BD
systems can form during the final stages of dissolution of a small
cluster \citep{Fuente98}, and three such common proper-motion pairs
have perhaps been found \citep{Gizis01}.

Finally, the {\sc Star-Like Hypothesis} also predicts far too few
star--star binaries in Taurus-Auriga, where binary disruption has not
been active.  This comes about because the large number of star--BD
systems in this model limits the number of star--star binaries given
the finite number of stellar primaries.

It is thus concluded that the observed BD population is incompatible
with the {\sc Star-Like Hypothesis}. Therefore BDs need to be treated
as a separate, or extra, population \citep{Kr_etal03}. This is
confirmed by \cite{PG11}, who constrain BD binary properties from the
observed ones given that their dynamical evolution in the birth
clusters needs to be corrected for.

\vspace{2mm} \centerline{ \fbox{\parbox{\columnwidth}{ \label{box:BDStarTh}
      {\sc BD/Star Population Synthesis Theorem}: When setting up a
      population of BDs and stars, BDs need to be algorithmically
      treated separately from the stars.  }}} \vspace{2mm}

\noindent {\sc Proof}: The BD dessert, see the rejection of
the {\sc Star-Like Hypothesis} above and/or \cite{Kr_etal03}.  End of
proof.  

A practical formulation of this theorem is posed as a {\sc Gedanken
  Experiment} on p.~\pageref{box:GE}.

\subsection{The number of BDs per star and  BD universality}
\label{sec:BDnr}

\citet{Briceno_etal02} report that Taurus-Auriga appears to form
significantly fewer BDs per star than the ONC. Both systems are very
different physically but have similar ages of about 1~Myr.  This
finding was interpreted to be the first possible direct evidence of a
variable IMF, being consistent qualitatively with the Jean-mass,
\begin{equation}
M_{\rm J}\propto \rho^{-1/2}T^{3/2},
\label{eq:MJeans}
\end{equation}
being larger in Taurus-Auriga than in the ONC because its gas density,
$\rho$, is smaller by one to two orders of magnitude, while the
temperatures, $T$, are similar to within a factor of a few
(Sec.~\ref{sec:introd_IMF}).

Given this potentially important finding, \citet{Kr_etal03} computed
$N$-body models of the stellar aggregates in Taurus-Auriga in order to
investigate the hypothesis that BDs form star-like. They find that the
same initial number of BDs per star in Taurus-Auriga and in the ONC
leads to different observed ratios because BD--BD and star--BD
binaries are disrupted more efficiently in the ONC; the observer thus
sees many more BDs there than in the comparatively dynamically
unevolved Taurus-Auriga groups. But, as already noted above, the {\sc
  Star-Like Hypothesis} must be discarded because it leads to too many
wide BD--BD binaries, and also it predicts too many star--BD
binaries. Given this problem, \citet{Kr_Bouv03b} study the production
rate of BDs per star assuming BDs are a separate population, such as
ejected embryos \citep{RC01}, or as gravitational instabilities in
extended circum-proto-stellar disks \citep{GW07, Thies10}.  Again they
find that both, the physically very different environments of
Taurus-Auriga and the ONC, can have produced the same ratios (about
one BD per 4~stars) if BDs are ejected embryos with a dispersion of
ejection velocities of about 2~km/s (this number is revised to
1.3~km/s below).

Based on some additional observations, \citet{Luhman04} revised the
\citet{Briceno_etal02} results by finding that the number of BDs per
star had been underestimated in Taurus-Auriga. Since the new
spectroscopic study of \citet{Slesnick_etal04} also revised the number
of BDs per star in the ONC downwards, \citet{Luhman04} retracts the
significance of the claimed difference of the ratio in Taurus-Auriga
and the ONC. Is a universal, invariant, BD production scenario still
consistent with the updated numbers?

Let the true ratio of the number of BDs per late-type star be
\begin{equation}
R \equiv { N(0.02-0.08\,M_\odot) \over N(0.15-1.0\,M_\odot)} 
  \equiv { N_{\rm BD,tot} \over N_{\rm st,tot}}.
\end{equation}
Note that here stars more massive than $1.0\,M_\odot$ are not counted
because Taurus-Auriga is mostly producing late-type stars given the
limited gas mass available (see also Fig.~\ref{fig:mmaxf}).  But the
observed ratio is
\begin{equation}
R_{\rm obs} = { N_{\rm BD,obs} \over N_{\rm st,obs}} = N_{\rm
  BD,tot}({\cal B}+{\cal U}) {(1+f) \over N_{\rm st,tot}} = R \,
\left( {\cal B} + {\cal U} \right)\, \left( 1+f \right),
\end{equation}
since the observed number of BDs, $N_{\rm BD,obs}$ is the total number
produced multiplied by the fraction of BDs that are gravitationally
bound to the population (${\cal B}$) plus the unbound fraction, ${\cal
  U}$, which did not yet have enough time to leave the survey
area. These fractions can be computed for dynamical models of the
Taurus-Auriga and ONC and depend on the mass of the Taurus-Auriga
sub-groups and of the ONC and on the dispersion of velocity of the
BDs.  This velocity dispersion can either be the same as that of the
stars if BDs form like stars, or larger if they are ejected embryos
\citep{RC01}.  The observed number of ``stars'' is actually the number
of systems such that the total number of individual stars is $N_{\rm
  st,tot} = (1+f)\,N_{\rm st,obs}$, where $f$ is the binary fraction
of stars (Eq.~\ref{eq:fbin}). Note that here no distinction is made
between single or binary BDs, which is reasonable given the low binary
fraction (about 15~\%) of BDs.  For Taurus-Auriga, \citet{Luhman04}
observes $R_{\rm TA, obs} = 0.25$ from which follows
\begin{equation}
R_{\rm TA} = 0.18 \;\; {\rm since} \;\;  f_{\rm TA}=1, {\cal B} + {\cal U} = 0.35+0.35
\end{equation}
\citep{Kr_etal03}.  According to \citet{Slesnick_etal04},
the revised ratio for the ONC is $R_{\rm ONC,obs} = 0.28$ so that
\begin{equation}
R_{\rm ONC}=0.19 \;\; {\rm because} \;\; f_{\rm ONC} = 0.5, {\cal B} + {\cal U} = 1+0
\end{equation}
\citep{Kr_etal03}. Note that the regions around the stellar groupings
in Taurus-Auriga not yet surveyed should contain about 30~\% of all
BDs, while all BDs are retained in the ONC. The ONC and TA thus appear
to be producing quite comparable BD/star number ratios.

Therefore, the updated numbers imply that about {\it one BD is
  produced per five late-type stars}, and that the dispersion of
ejection velocities is $\sigma_{\rm ej} \approx 1.3$~km/s. These
numbers are an update of those given in \citet{Kr_etal03}, but the
results have not changed much. Note that a BD with a mass of
$0.06\,M_\odot$ and a velocity of 1.3~km/s has a kinetic energy of
$10^{-1.29}\,M_\odot\,$(pc/Myr)$^2$ which is rather comparable to the
cut-off in BD--BD binding energies (Fig.~\ref{fig:BDbindEn}). {\it
  This supports the notion that most BDs may be mildly ejected
  embryos}, e.g. from the \cite{GW07} and \cite{Thies10}
circum-proto-stellar disks.

It appears thus that the different physical environments evident in
Taurus-Auriga and the ONC produce about the same number of BDs per
late-type star, so that there is no convincing evidence for
differences in the IMF among current nearby star-forming regions
across the hydrogen burning mass limit.

\begin{figure}
\begin{center}
\rotatebox{0}{\resizebox{0.78 \textwidth}{!}{\includegraphics{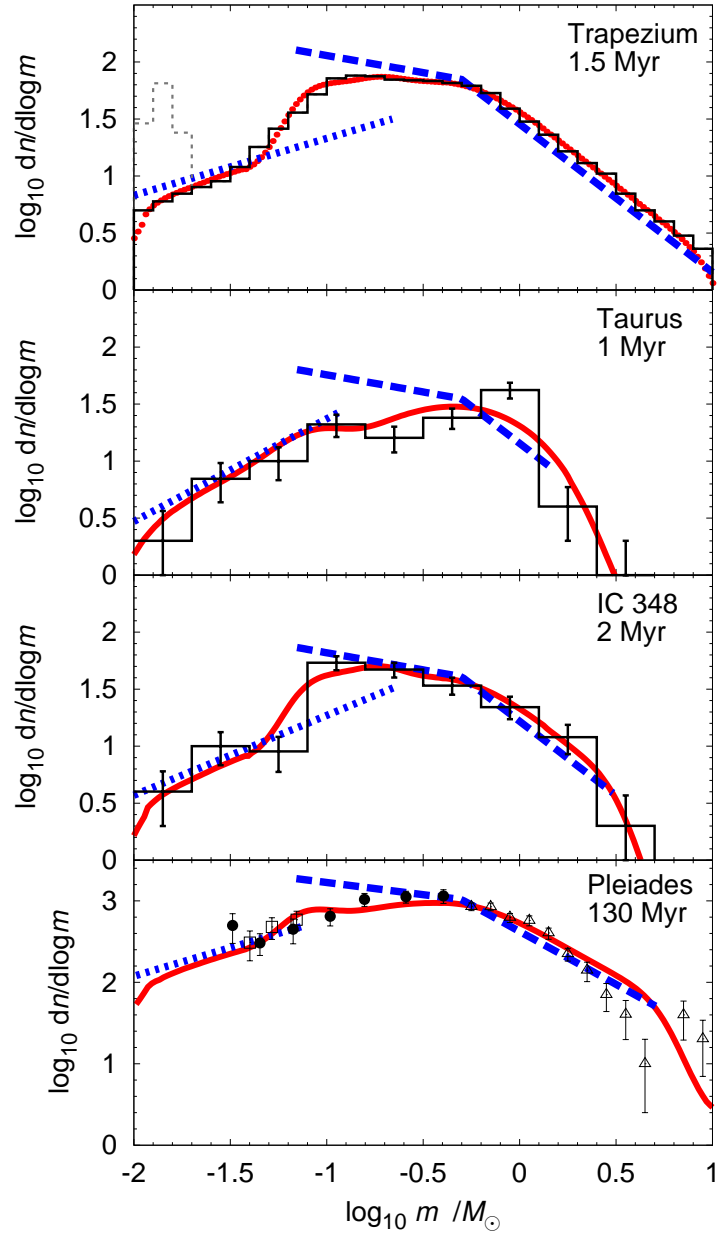}}}
\vskip0mm
\caption[Mass functions of BDs]{\small{The observationally determined MFs in four young
    clusters, the names and approximate ages of which are indicated in
    the panels, are shown as the histograms or data with error
    bars. The thick solid (red) curve is the model of unresolved
    systems by \cite{TK07}. In this model the stars have a binary
    fraction of~100~per cent in Taurus and 50~per cent in the other
    cases, with companion masses selected randomly from the canonical
    IMF (Eq.~\ref{eq:fPbirth}). The two-part power-law canonical IMF
    (Eq.~\ref{eq:imf}) is shown as the thick dashed line. The BDs are
    described by the additional power-law IMF shown as the dotted
    (blue) line. In the model they have a binary fraction of 15~per
    cent and do not mix with the stars (the brown dwarf desert). It
    can be seen that the model is essentially invariant and leads to
    an excellent description of the data despite the BD and stellar
    IMFs being discontinuous (Fig.~\ref{fig:imf_wsigma}). Further
    details are found in Sec.~\ref{sec:BD_IMF}.  }}
\label{fig:TAmfn}
\end{center}
\end{figure}

There is also no substantial evidence for a difference in the {\it
  stellar} IMF in these two star-forming regions, contrary to the
assertion by e.g. \citet{Luhman04}. Fig.~\ref{fig:TAmfn} shows the MF
in four young clusters. The caveat that the classical pre-main
sequence evolution tracks, upon which the observational mass
determinations rely, are not really applicable for such young ages
(Sec.~\ref{sec:stcl}) needs to be kept in mind though. Also, the
observationally derived MFs are typically obtained by treating the
luminous objects as single, while a majority are likely to be binary.

\subsection{BD flavours}
\label{sec:BDflavours}

BDs can come in different flavours depending on their formation
\citep{Kr_Bouv03b}: star-like BDs, ejected embryos, collisional BDs
and photo-evaporated BDs.  As seen above, star-like BDs appear to be
very rare, because BDs do not mix with stars in terms of pairing
properties. This is supported theoretically because in order to have a
cloud core produce only a BD or a BD--BD binary it needs to acquire an
extreme density to become gravitationally unstable. Such conditions
are very rare in typical turbulent molecular clouds that have Mach
numbers $\simless 6$ \citep{HC08, HC09}.

A more recently discussed possible channel of BD formation is via the
fragmentation of massive circum-stellar disks (FMCSDs) beyond about
100~AU and subsequent dynamical separation of the resulting weekly
bound BD companion \citep{GW07,SHW07,SW09a,Thies10,BV12}. This
mechanism for producing BDs can be considered a revised version of the
ejected BD scenario.  It appears to be the most promising physical
mechanism for producing the dominant flavour of BDs.

An environmental dependency of the VLMS and BD IMF is expected for the
FMCSD channel as shown by \cite{SWH11}. They point out that disk
fragmentation is enhanced when the star plus disk accretes
episodically as long as the accretion intervals are long enough to
allow the disk to cool. In dense environments episodic accretion would
occur too frequently thereby inhibiting fragmentation as the accreting
star's luminosity is kept as a high accretion luminosity thereby
heating the disk.

The collisional removal of accretion envelopes for the production of
unfinished stars needs to be discounted as a source for BDs because
this process is far too rare \citep{Kr_Bouv03b}. The removal of
accretion envelopes through photo-evaporation can occur, but only
within the immediate vicinity of an O~star and never in Taurus-Auriga
\citep{WhitZinn04}.  Even in the presence of some ionising stars, as
in the cluster NGC~6611 which ionises the Eagle Nebula,
\cite{Oliveira09} find no measurable effect of photo-evaporation on
the sub-stellar MF.  However, \citet{Kr_Bouv03b} show that the radius
within which photo-evaporation may be able to remove substantial
fractions of an accretion envelope within $10^5$~yr is comparable to
the cluster size in star-burst clusters that contain thousands of
O~stars. In such clusters photo-evaporated BDs may be very common.
Globular clusters (GCs) may then be full of BDs.

\vspace{2mm} \centerline{ \fbox{\parbox{\columnwidth}{ {\sc GC-BD Hypothesis}: 
In GCs the number ratio of BDs to stars may be very large ($\gg
1/5$). }}}
\vspace{2mm}

\subsection{The origin of BDs and their IMF}
\label{sec:BD_IMF}

It has thus emerged that in order to construct a realistic stellar and
BD population, BDs and VLMSs need to be treated according to a
different initialisation algorithm to that of stars. They need to be
separated when setting up the binary populations. This can be
visualised with the following Gedanken Experiment which is a practical
formulation of the {\sc BD/Star Population Theorem} on
p.~\pageref{box:BDStarTh}.

\vspace{2mm} \centerline{ \fbox{\parbox{\columnwidth}{\label{box:GE}
      {\sc Gedanken Experiment}: Imagine a box contains BDs, M-, K-
      and G-dwarf stars. In order to pair these to obtain the correct
      birth binary population (p.~\pageref{def_pk:bbp}), a distinction
      between M-, K- and G-dwarfs need not be made. But, nearly every
      time a BD is picked as a companion to a star, or a star is
      picked as a companion to a BD, the system needs to be discarded.
    }}} \vspace{2mm}

\noindent
The physical interpretation of this mathematical result is that BDs
form along with stars, just as planets do, but, just like planets,
they do not result from the same formation mechanism. Rather,
similarly as planets, BDs stem from gravitationally pre-processed
material.

Indeed, the conditions in a molecular cloud core very rarely are such
that a dense-enough core can collapse under self-gravity without
accreting too much material for it to not transcend the BD/star mass
boundary, which is why most BDs do not derive from a star-like origin
(see also Sec.~\ref{sec:cloudcore}).  But the outer regions of
circum-proto-stellar disks accumulate material which has time to loose
entropy (Sec.~\ref{sec:BDflavours}). These outer regions are
sufficiently dense to locally collapse under self gravity either
because they become unstable or because they are perturbed.  They are
not too dense and thus remain optically thin for a sufficiently long
time to allow the collapsing object to cool radiatively, and the
region around the collapsing object has a limited supply of local disk
material.  Fragmentation may occur in marginally stable disks upon an
external perturbation by the gravity of other stars
\citep{TKT05,Thies10} or by the gas-dynamical interaction of two disks
\citep{Shen_etal10}. Both such processes thus enlarge the parameter
space of circum-stellar disk conditions for BD formation since most
stars form in a clustered mode.

A circum-proto-stellar disk therefore sets the boundary conditions in
favour of BD formation. This is the same, but even more extreme, for
planet formation, for which a highly processed molecular cloud core is
required in the sense that gravo-hydrodynamical dynamics is augmented
significantly by the dynamics between solid particles in a
circum-stellar disk. Thus, while the formation of BDs is still a
purely gravo-hydrodynamical process within an extended disk, planet
formation is seeded by the coagulation of dust to larger solids which
may then induce gas accretion from the circum-stellar disk.

The differences in the binary properties of BDs and stars therefore
indicate that they are two different classes of objects with their own
separate mass distributions. The MF of planets is also never
considered a continuous extension of the stellar IMF \citep{Dominik11}.

The IMF of BDs needs to be derived from the observational star count
data by taking the above into account. \cite{TK07, TK08} have done
so. The results for Taurus-Auriga, the ONC, IC~348 and Pleiades
demonstrate that the IMF appears to be universal with $\alpha_0\approx
0.3$ albeit with a significant discontinuity near $0.1~M_\odot$
(Fig.~\ref{fig:TAmfn}). The data imply that there is an overlap
region: for the analysis to correctly account for the observed data
there must be VLMSs that form as BDs do, while there are massive BDs
that form as stars do (Fig.~\ref{fig:imf_wsigma}). Without account of
this overlap the IMF as well as the binary properties as a function of
mass may feign continuity (as suggested by \citealt{Kaplan12}).
Noteworthy is that this analysis re-derives the same BD-to-star
fraction as deduced above: about one BD forms per five stars.  The
same result on the number ratio has been inferred by \cite{Andersen08}
who however need to describe the BD MF as a decreasing log-normal form
as a result of insisting the IMF to be continuous across the
BD/stellar mass range. As stated above, this approach cannot account
for the BD-desert and the correct approach is to treat the BDs/VLMSs
as a distinct population from the stars implying two separate IMFs for
BDs and stars. This correct approach leads to a power-law solution for
the BD MF with $\alpha\approx 0.3$, i.e. a mildly rising MF towards
small masses, and separate BD and stellar IMFs.

The universality of the BD--stellar IMF is interesting, and may
suggest that the formation of BDs is mostly dependent on the
conditions prevalent in circum-proto-stellar disks. Indeed, in a large
ensemble of SPH simulations of circum-stellar disks in young star
clusters, \cite{Thies10} find that a theoretical BD and VLMS MF
emerges which is the same as the observationally deduced one.

The observational and theoretical result that the BD-branch becomes
insignificant (i.e. the the BD MF falls of steeply) above
$0.1-0.2\,M_\odot$ follows from a combination of two effects: the
limited amount of material around late-type pre-main sequence stars
that can be accreted onto a gravitational instability within the outer
region of a circum-proto-stellar mass and the rare occurrence of massive
proto-stars which are likely to have massive disks.

\vspace{2mm} \centerline{ \fbox{\parbox{\columnwidth}{ {\sc Main
        results}: Brown dwarfs are a separate population when compared
      to stars due to their distinct binary properties. The BD IMF has
      a power-law index $\alpha_0\approx 0.3$ and the IMF is
      discontinuous but with a significant overlap of masses between
      the sub-stellar and stellar regime.  This may hide the
      discontinuity unless a proper analysis is done.  About one BD
      forms per five stars.}}}  \vspace{2mm}

\section{The Shape of the IMF from Resolved Stellar Populations}

From the above discourse it thus becomes apparent that we have good
constraints on the stellar and BD IMF. These are valid only for the
regime of present-day (``normal'') star formation, i.e. star-formation
densities $\rho \simless 10^5\,M_\odot$pc$^{-3}$ and metallicities
[Fe/H]$\simgreat -2$. Since stars are formed as binary systems
(Sec.~\ref{sec:bins}) the system IMF is provided in
Sec.~\ref{sec:sysIMF} for binary fractions of 100~\% (the birth system
IMF) and for a binary fraction of 50~\% (the typical Galactic-field or
open star cluster system IMF). The Galactic-field IMF, which is the
IGIMF valid for the Milky Way, is provided in Sec.~\ref{sec:galIMF}.
In Sec.~\ref{sec:IMFvar} it is concluded that the IMF becomes
top-heavy for star formation under denser conditions and that it may
be bottom-light under metal-poor conditions.

The distribution of stars by mass in ``normal'' systems is a power-law
with exponent or index $\alpha_2\approx 2.3$ for stellar masses
$m\simgreat 0.5\,M_\odot$.  There exists a physical stellar mass
limit, $m_{\rm max *}\approx 150\,M_\odot$ such that $m\le m_{\rm max
  *}$ (Sec.~\ref{sec:maxlim}).  The distribution of stars below the
K/M~dwarf transition mass, $0.5\,M_\odot$, can also be described by a
power law, but with exponent $\alpha_1\approx 1.3$
(Sec.~\ref{sec:lmst}).  Given the latest results described in
Sec.~\ref{sec:bds}, the mass-distribution below the mass $m_1 \approx
0.1\,M_\odot$ is uncertain, but measurements indicate a power-law with
exponent $0 < \alpha_0 < 1$. Because the binary properties of VLMSs
and BDs differ substantially from those in the low-mass star regime,
it emerges stringently that BDs and some VLMSs need to be considered
as a separate population that is linked-to, but different from stars.
Fitting a functional description of the mass distribution with the
continuity constraint across $m_1$ would therefore be wrong.  It
follows that one single function such as the log-normal form, which
may be associated with the likelihood of occurrence of masses from the
fragmentation limit\footnote{\label{foot:fragmlimit} When a cloud
  collapses its density increases but its temperature remains constant
  as long as the opacity remains low enough to enable the contraction
  work to be radiated away. The Jeans mass (Eq.~\ref{eq:MJeans})
  consequently decreases and further fragments with smaller masses
  form. When, however, the density increases to a level such that the
  cloud core becomes optically thick, then the temperature increases,
  and the Jeans mass follows suit. Thus an opacity-limited minimum
  fragmentation mass of about $0.01\,M_\odot$ is arrived at
  \citep{LowLyndenBell76,Boss86, Kumar03, Bate2005b}.}, $m_0\approx
0.01\,M_\odot$, through to the physical stability limit, $m_{\rm
  max*}$, is not the correct description of the stellar and BD IMF

With these recent insights (power-law IMF over two orders of magnitude
in mass and discontinuity near the sub-stellar mass limit), little of
the argument for the advantages of a log-normal or any other
mathematical form (Table~\ref{tab:imfs} below) remains. Indeed, any
such other mathematical form has the disadvantage that the tails of
the distribution react to changes in the parametrisation in a way
perhaps not wanted when testing models.  To give an example: a single
log-normal form would change the slope of the IMF at large masses even
if only the LF for late-type stars is to be varied. The canonical
(Eq.~\ref{eq:imf}) two-part power-law stellar IMF, on the other hand,
would allow changes to the index at low masses without affecting the
high-mass end, and the addition of further power-law segments is
mathematically convenient. The canonical two-part power-law stellar
IMF also captures the essence of the physics of star formation, namely
a featureless power-law form for the largest range of stellar masses,
and a turnover near some fraction of a solar mass. This turnover
appears to be present already in the pre-stellar cloud-core MF and may
be due to the decreasing likelihood that low-mass cloud clumps
collapse under self-gravity (Sec.~\ref{sec:cloudcore}).

\subsection{The canonical, standard or average IMF}
\label{sec:canIMF}

The various constraints arrived at above are summarised by an IMF that
is a single power-law for BDs and a two-part power-law for stars
(Eq.~\ref{eq:imf}), using the notation from Eq.~\ref{eq:imf_bd}
and~\ref{eq:imf_mult}.

\vspace{2mm} \centerline{ \fbox{\parbox{\columnwidth}{ {\sc The Canonical
        IMF:} ($m$ is in units of $M_\odot$)
\[
\xi_{\rm BD} (m) = {k\over 3}
  \begin{array}{l@{\quad\quad\quad\quad\quad\quad\quad\quad,\quad}l}
   \left({m\over 0.07}\right)^{-0.3\pm0.4} &0.01 < m \simless 0.15,
  \end{array}
\]
\begin{equation}
\xi_\mathrm{star} (m) = k\left\{
  \begin{array}{l@{\quad,\quad}l@{\quad}}
   \left({m\over 0.07}\right)^{-1.3\pm0.3}  &0.07 < m \le 0.5,\\
   \left[\left({0.5\over 0.07}\right)^{-1.3\pm0.3}
       \right] \left({m\over 0.5}\right)^{-2.3\pm0.36} &0.5 < m \le 150\\
  \end{array}\right.
\label{eq:imf}
\end{equation}
}}}
\vspace{2mm}

%
%
%

\noindent
The uncertainties are discussed in Sec.~\ref{sec:apl}. This is the
individual-star/BD IMF which is corrected fully for multiple
companions. That is, in a star-formation event, the distribution of
stars and of BDs that form are given by this IMF. The constraint that
the population be optimally sampled (p.~\pageref{box:optdistr}) may be
invoked.  To formulate a realistic stellar and BD population would
then require these stars and BDs to be distributed into stellar and BD
binaries (Sec.~\ref{sec:bins}, \ref{sec:sysIMF}).

\begin{figure}
\begin{center}
\rotatebox{0}{\resizebox{0.9 \textwidth}{!}{\includegraphics{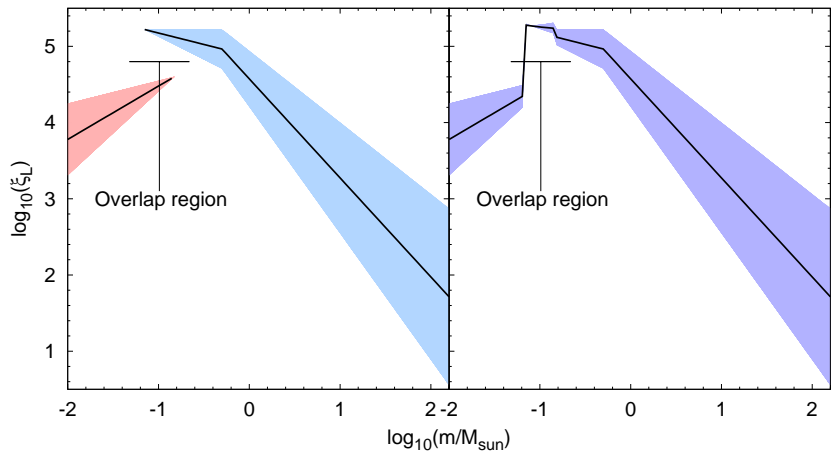}}}
\caption[BD discontinuity]{\small {\it Left frame:} The canonical IMF
  ({\it solid line}, Eq.~\ref{eq:imf}) with separate BD and stellar
  components. The upper and lower envelopes of the shaded regions are
  the minimum and maximum IMF when the uncertainties are taken into
  account.  Both components overlap between 0.07 and 0.15~$M_\odot$.
  {\it Right frame:} The sum of the BD and the stellar components as
  it would appear to an observer if perfect mass-measurements were
  available of all stars and BDs in all multiple systems. Note the
  ``bump'' that marks the overlap region. However, as demonstrated in
  Fig.~\ref{fig:TAmfn}, the real observational data wash out such
  features owing to the measurement uncertainties and the errors in
  transforming observed fluxes to masses. The model mass-histograms
  (thick red lines in Fig.~\ref{fig:TAmfn}) then fit the observed
  mass-histograms excellently for the young clusters for which such
  data are available.  }
\label{fig:imf_wsigma}
\end{center}
\end{figure}

Note that this form is a two-part power-law in the stellar regime, and
that BDs contribute about 4 per cent by mass only and need to be
treated as a separate population such that both IMFs overlap between
about 0.07~$M_\odot$ and 0.15$M_\odot$ (Fig.~\ref{fig:imf_wsigma}).
The gap or discontinuity between the BD and the stellar IMF can be
measured by the BD-to-star ratio at the classical BD-star border,
$m_{\rm H}\approx 0.075\,M_\odot$, with $\xi_\mathrm{BD}(m_{\rm
  H})/\xi_\mathrm{star}(m_{\rm H}) \approx \frac{1}{3}$, i.e. $k_{\rm
  BD}\approx1/3$ \citep{Kr_etal03,TK07,TK08}.

The canonical IMF can also be described as a log-normal function for low-mass
stars with a power-law extension to massive stars, yielding the {\it
  log-normal canonical IMF}. 

\vspace{2mm} \centerline{ \fbox{\parbox{\columnwidth}{ {\sc The Log-Normal
        Canonical IMF:} ($m$ is in units of $M_\odot$)
\[
\xi_{\rm BD} (m) = {k\,k_{\rm BD}}
  \begin{array}{l@{\quad\quad\quad\quad\quad\quad\quad,\quad}l}
   \left({m\over 0.07}\right)^{-0.3\pm0.4} &0.01 < m \simless 0.15,
  \end{array}
\]
\begin{equation}
\xi_\mathrm{star} (m) = k\left\{
  \begin{array}{l@{\quad\quad\quad\quad\quad,\quad}l@{\quad}}
{1\over m}\,{\rm exp}\left[ -{ (lm - lm_c)^2 \over 2\, \sigma_{lm}^2} \right]
&0.07 < m \le 1.0,\\
A \, \left({m\over 1.0}\right)^{-2.3\pm0.36} &1.0 < m \le 150\\
  \end{array}\right.
\label{eq:chabimf}
\end{equation}
}}}
\vspace{2mm}

\noindent
In Eq.~\ref{eq:chabimf} $lm_c\equiv {\rm log}_{10}m_c/M_\odot$,
continuity is assured at $1\,M_\odot$ and $\xi_\mathrm{BD}(m_{\rm
  H})/\xi_\mathrm{star}(m_{\rm H})$ $\approx \frac{1}{3}$, as in
Eq.~\ref{eq:imf}.

A least-squares fit of the log-normal canonical IMF to the two-part
power-law form (Eq.~\ref{eq:imf}), whereby
$\int_{0.07}^{150}m\,\xi(m)\,dm=1\,M_\odot$ for both ($m$ in Solar
units), yields $m_c=0.055\,M_\odot$ and $\sigma_{lm}=0.75$ with
$A=0.2440$ for continuity at~$1\,M_\odot$ and $k_{\rm BD}=4.46$ to
ensure $\xi_{\rm BD}(0.75\,M_\odot)/\xi_{\rm star}(0.75\,M_\odot)=1/3$
as being the best log-normal plus power-law representation of the
canonical IMF. Alternatively, the Chabrier parametrisation has
$m_c=0.079^{-0.016}_{+0.021}\,M_\odot$ and
$\sigma_{lm}=0.69^{-0.01}_{+0.05}$ (table~1 in \citealt{Chrev03}) with
$A=0.2791$ and $k_{\rm BD}=4.53$.

\begin{figure}
\begin{center}
\rotatebox{0}{\resizebox{0.8\textwidth}{!}{\includegraphics{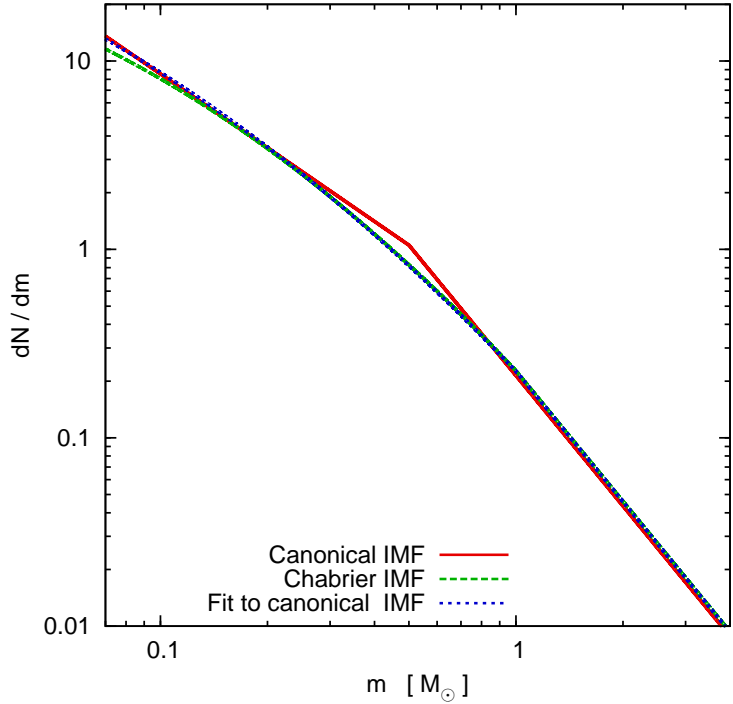}}}
\caption[Two-part power law vs. log-normal IMF]{\small{A comparison between the three canonical IMFs: the
    two-part power-law IMF (Eq.~\ref{eq:imf}, solid red curve) and the
    log-normal plus power-law IMF (Eq.~\ref{eq:chabimf}, the
    ``best-fit-canonical IMF'' and the ``Chabrier IMF'') in the
    interval from $0.07\,M_\odot$ to $4\,M_\odot$. Plotted is the
    number of stars per mass interval versus the stellar mass, both
    scales being logarithmic. The IMFs are normalised such that
    $\int^{150}_{0.07} m\,\xi(m)\, dm=1\,M_\odot$, where $m$ is the mass in
    Solar units. Note in particular that the three IMF forms are
    indistinguishable over the whole mass interval. They are identical
    above a mass of $1\,M_\odot$, except for a slightly different
    normalisation factor. }}
\label{fig:canIMF}
\end{center}
\end{figure}

The three forms of the canonical IMF are compared in
Fig.~\ref{fig:canIMF}.  The best-fit log-normal representation of the
two-part power-law canonical IMF is indistinguishable from the
Chabrier result, demonstrating the extreme robustness of the canonical
IMF.

The above canonical or standard forms have been derived from detailed
considerations of star-counts thereby representing an {\it average}
IMF: for low-mass stars it is a mixture of stellar populations
spanning a large range of ages ($0-10$~Gyr) and metallicities
([Fe/H]$\simgreat -2$).  For the massive stars it constitutes a
mixture of different metallicities ([Fe/H]$\simgreat -1.5$) and
star-forming conditions (OB associations to very dense star-burst
clusters: R136 in the LMC). Therefore the average IMF can be taken as
a canonical form, and the aim is to test the {\sc IMF Universality
  Hypothesis}:

\vspace{2mm}

\centerline{ \fbox{\parbox{\columnwidth}{{\sc IMF Universality Hypothesis}:
      the canonical IMF (Eq.~\ref{eq:imf},~\ref{eq:chabimf})
      constitutes the parent distribution of all stellar populations
      that form with densities $\rho\simless 10^5\,M_\odot/{\rm
        pc}^3$ and metallicities [Fe/H]$\simgreat
      -2$. \label{kroupa_hyp:univ}}}}

\vspace{2mm}

\noindent
Negation of this hypothesis would indeed imply a variable IMF. For
larger densities evidence has emerged that the IMF becomes top-heavy
(Eq.~\ref{eq:toph_imf_density}). Also, evidence has emerged that the
IMF may have a metallicity dependence (Eq.~\ref{eq:toph_imf_metal}).

\subsection{The IMF of systems and of primaries}
\label{sec:sysIMF}
\begin{figure}
\begin{center}
\rotatebox{0}{\resizebox{0.9\textwidth}{!}{\includegraphics{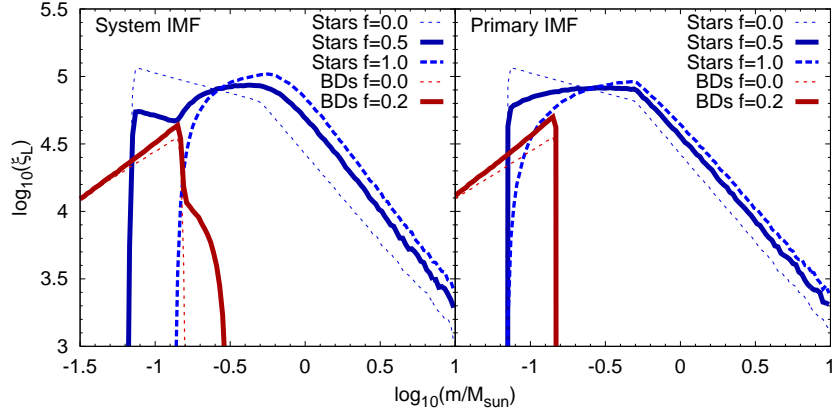}}}
\caption[BD mass function in dependence of binarity]{\small{{\it Left panel:} The system IMF is shown in blue and
    the system BD/VLMS IMF is in red. The thick curves are for a
    binary fraction among stars of $f=50\,$\% and for a BD binary
    fraction of $f_{\rm BD}=20\,$\%.  Random pairing of companion
    masses is assumed (Sec.~\ref{sec:bins}) from the canonical stellar
    (0.07~to $150\,M_\odot$) and BD (0.01~to $0.15\,M_\odot$)
    IMFs. These are shown here as the thin dotted curves for binary
    fraction$\,=0$ (Eq.~\ref{eq:imf}, also plotted in
    Fig.~\ref{fig:imf_wsigma}).  Note the flat extension at the
    lowest-mass end of the stellar system IMF. It comes from the fact
    that below $0.14\,M_\odot$ all stars are single since the minimum
    system mass is twice the minimum stellar mass ($0.07\,M_\odot$).
    The bump at the high-mass end of the BD system distribution comes
    from the most massive systems having masses larger than the most
    massive BD/VLMS ($0.15\,M_\odot$) in this model.  The medium
    dashed stellar system IMFs are for an initial stellar binary
    fraction of~100~\%, as would be found in dynamically not evolved
    star-forming regions. {\it Right panel:} the same but plotting
    only the IMF of primary masses. Note the significant difference
    between the canonical IMF and the system/primary-star IMFs in both
    panels. The system IMF and the IMF of primaries are approximated
    by Eq.~\ref{eq:sysimf} and~\ref{eq:primimf}, respectively. }}
\label{fig:sysimf}
\end{center}
\end{figure}

The canonical stellar IMF (Eq.~\ref{eq:imf}) is the distribution of
all stars to have formed together in one star-forming event. However,
since stars form as binary systems it is also useful to consider the
form of the system IMF which results from pairing stars chosen
randomly from the canonical stellar IMF. Random pairing is a good
description of late-type stellar systems in birth environments, but
massive stars tend to prefer more similar-mass companions
(Sec.~\ref{sec:bins}). The system IMF is given by Eq.~\ref{eq:sysimf}
by approximating the mass distribution by a power-law with index
$\alpha_1$ between $0.15$ and $0.5\,M_\odot$. It is assumed that all
stars are in binaries, i.e. $f=1$ and, for completeness, the system
IMF for a binary fraction of 50~\% is also evaluated.  The actual mass
distributions are plotted in Fig.~\ref{fig:sysimf}.

\vspace{2mm}  
\noindent {\sc The system IMF:}
\begin{equation}
          \begin{array}{l@{\quad\quad,\quad}ll@{\quad,\quad}l}
\alpha_1 = +0.66\pm0.3   &0.15 &\le m/M_\odot < 0.65 &\mbox{ ($f=0.5$)} \\
\alpha_1 \approx -0.22   &0.15  &\le m/M_\odot < 0.65  &\mbox{ ($f=1$)}.
          \end{array}
\label{eq:sysimf}
\end{equation}
\vspace{2mm}
Note that in each case $\alpha_0=0.2\pm0.4$ ($f_{\rm BD}=0.2$, random
pairing) and $\alpha_2=2.3 \; (m\simgreat 0.65\,M_\odot)$ as in the
canonical stellar IMF (Eq.~\ref{eq:imf}). 

Note that the binary effect contributes $\Delta \alpha \approx 0.64$
for $f=0.5$ and $\Delta \alpha \approx 1.5$ for $f=1$ where $\Delta
\alpha$ is the difference between the canonical $\alpha_1=1.3$ and the
above values.

\noindent
The IMF of primary stars, which may be closer to an observed IMF since
companions are usually faint and would be lost from the star-count, is
given by Eq.~\ref{eq:primimf} following the same procedure as for the
system IMF.

\vspace{2mm}
\noindent 
{\sc The IMF of primary stars:}
\begin{equation}
          \begin{array}{l@{\quad\quad,\quad}ll@{\quad,\quad}l}
\alpha_1 = +1.0\pm0.3   &0.15 &\le m/M_\odot < 0.50 &\mbox{ ($f=0.5$)} \\
\alpha_1 = +0.7\pm0.3   &0.15  &\le m/M_\odot < 0.50  &\mbox{ ($f=1$)}.
          \end{array}
\label{eq:primimf}
\end{equation}
\vspace{2mm}
Note that in each case $\alpha_0=0.2\pm0.4$ ($f_{\rm BD}=0.2$, random
pairing) and $\alpha_2=2.3 \; (m\simgreat 0.5\,M_\odot)$ as in the
canonical stellar IMF (Eq.~\ref{eq:imf}). 
%
%
%
%
%
%
%
%
%
%
%
%
%

\subsection{The Galactic-field IMF}
\label{sec:galIMF}

The Scalo power-law index $\alpha_3 = 2.7 \; (m\simgreat 1\,M_\odot)$ was
inferred by \citet{Sc86} from star counts for the MW disk population
of massive stars such that the three-part-power law KTG93 IMF needs to
be identified with the {\it composite IMF} (i.e. the IGIMF) introduced
in Sec.~\ref{sec:comppop}, arriving at the {\it KTG93 IMF}
\citep{KTG93}. It is updated in Eq.~\ref{eq:KTGimf} to account for the
separate BD and stellar populations.

\vspace{2mm}  
\noindent
{\sc The Galactic-field (KTG93) stellar IMF:}
\begin{equation}
          \begin{array}{l@{\quad\quad,\quad}ll@{\quad,\quad}l}
\alpha_0 = +0.3\pm0.4   &0.01 &\le m/M_\odot \simless 0.15 &n=0,\mbox{ (BDs)} \\ \hline
\alpha_1 = +1.3\pm0.3   &0.07 &\simless m/M_\odot < 0.50 &n=1,\mbox{ (stars)} \\
\alpha_2 = +2.3\pm0.3   &0.5  &\le m/M_\odot  <1    &n=2, \\
\alpha_3 = +2.7\pm0.4   &1    &\simless m/M_\odot        &n=3.
          \end{array}
\label{eq:KTGimf}
\end{equation}
\vspace{2mm} 

\noindent
The Galactic-field (KTG93) IMF is the IMF of all young stars within
the Galactic disk assuming all multiple stars can be resolved. It is
not equal to the canonical IMF for $m\simgreat 1\,M_\odot$.  In the
LMC the field-star IMF is known to be steep, $\alpha_3\approx 4.5$
\citep{Massey03}.

\subsection{The alpha plot}
\label{sec:apl}

\begin{figure}
\begin{center}
\rotatebox{0}{\resizebox{0.99 \textwidth}{!}{\includegraphics{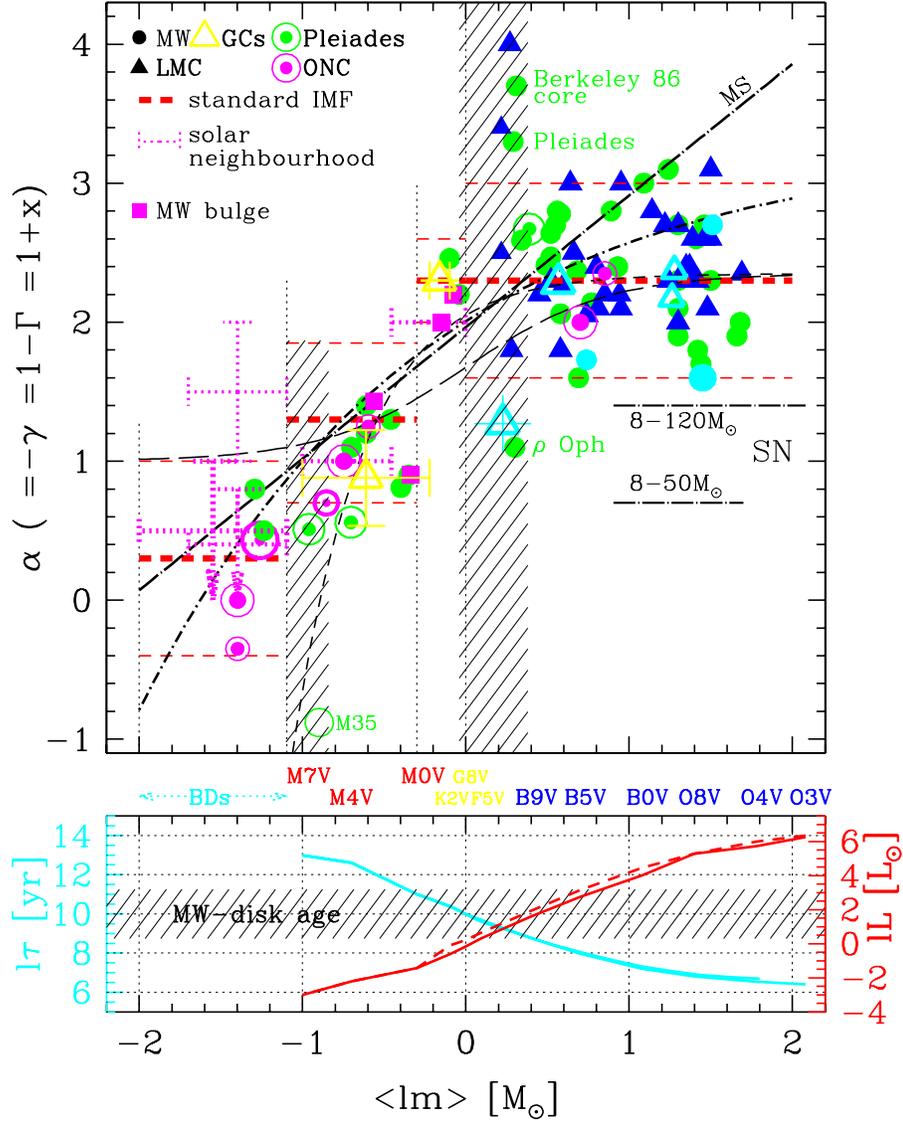}}}
\vskip -1mm
\caption[The $\alpha$ plot]{\small{ {\it Upper panel:} The alpha
    plot. The curves (e.g. labelled ``MS'') are various IMF slopes
    documented in Table~\ref{tab:imfs} and discussed in
    Sec.~\ref{sec:otherIMF}.  The plotted data are listed in the
    supplementary information of \cite{K02}.  The cyan open triangles
    are for R136 in the LMC. {\it Lower panel:} The bolometric stellar
    mass-luminosity relation, $lL(lm)$, shown by the solid and dashed
    curves, and the stellar main-sequence life-time (or turn-off
    masses at a given age), $l\tau$.  The possible range of Milky-Way
    (MW) disk ages are shown as the shaded region. The masses of
    stellar spectral types are indicated.  Notation: $lm\equiv {\rm
      log}_{10}(m/M_\odot)$, $l\tau\equiv {\rm log}_{10}(\tau/{\rm
      yr})$, $lL\equiv {\rm log}_{10}(L/L_\odot)$.  For more details
    see text. }}
\label{fig:apl}
\end{center}
\end{figure}

A convenient way for summarising various studies of the IMF is by
plotting independently-derived power-law indices in dependence of the
stellar mass over which they are fitted
\citep{Sc98,K01a,K02,Hillrev04}.  The upper panel of
Fig.~\ref{fig:apl} shows such data: The shape of the IMF is mapped by
plotting measurements of $\alpha$ at $<\!lm\!>=(lm_2-lm_1)/2$ obtained
by fitting power-laws, $\xi(m)\propto m^{-\alpha}$, to logarithmic
mass ranges $lm_1$ to $lm_2$ (not indicated here for clarity).
Circles and triangles are data compiled by \citet{Sc98} and
\cite{K01a} for MW and Large-Magellanic-Cloud (LMC) clusters and OB
associations, as well as newer data, some of which are emphasised
using different symbols (and colours).  Unresolved multiple systems
are not corrected for in all these data including the MW-bulge data.
The canonical stellar IMF (Eq.~\ref{eq:imf}), corrected for unseen
binary-star companions, is the two-part power-law (thick short-dashed
lines).  Other binary-star-corrected solar-neighbourhood-IMF
measurements are indicated as (magenta) dotted error-bars. 

In the lower panel of Fig.~\ref{fig:apl} are plotted the luminosities
of main-sequence stars and the stellar lifetimes as a function of mass
on the same mass-scale as the alpha plot shown in the upper panel.

For $m>1\,M_\odot$ correction for unseen companions does not affect
the IMF \citep{MA08, WKM09}. The M~dwarf ($0.1-0.5 \,M_\odot$) MFs for
the various clusters are systematically flatter (smaller $\alpha_1$)
than the canonical IMF, which is mostly due to unresolved multiple
systems in the observed values. This is verified by comparing to the
system IMF for a binary fraction of~50~\% (Eq.~\ref{eq:sysimf}).  Some
of the data do coincide with the canonical IMF though, and
\citet{K01a} argues that on correcting these for unresolved binaries
the underlying true individual-star IMF ought to have $\alpha_1\approx
1.8$. This may indicate a systematic variation of $\alpha_1$ with
metallicity because the data are young clusters that are typically
more metal-rich than the average Galactic field population for which
$\alpha_1=1.3$ (Eq.~\ref{eq:systemvar} below)

A power-law extension into the BD regime with a smaller index
($\alpha_0=+0.3$) is shown as a third thick short-dashed segment, but
this part of the mass distribution is not a continuous extension of
the stellar distribution, as noted in Sec.~\ref{sec:BD_IMF}.

The upper and lower thin short-dashed lines are the estimated 99~\%
confidence range on $\alpha_i$. The usual one-sigma uncertainties
adopted in Eq.~\ref{eq:imf} are, however, estimated from the
distribution of $\alpha$ values in Fig.~\ref{fig:apl} and
Fig.~\ref{fig:ahist}.

The long-dash-dotted horizontal lines in Fig.~\ref{fig:apl} labeled
``SN'' are those IMFs with $\alpha_3=0.70 (1.4)$ but $\alpha_0,
\alpha_1, \alpha_2$ as in Eq.~\ref{eq:imf}, for which 50~\% of the
stellar (including BD) mass is in stars with $8 - 50 (8 -
120)\,M_\odot$, respectively. It is noteworthy that none of the
available resolved clusters, not even including the Local Group
star-burst clusters, have such a top-heavy IMF.

The vertical dotted lines in Fig.~\ref{fig:apl} delineate the four
mass ranges over which the Galactic-field IMF is defined
(Eq.~\ref{eq:KTGimf}), and the shaded areas highlight those stellar
mass regions where the derivation of the IMF is additionally
complicated especially for Galactic field stars: for $0.07<m/M_\odot<
0.15$ long pre-main sequence contraction times \citep{ChB00} make the
conversion from an empirical LF to an IMF (Eq.~\ref{eq:mf_lf})
dependent on the precise knowledge of stellar ages and the
star-formation history and for $0.8< m/M_\odot<2.5$ uncertain
main-sequence evolution, Galactic-disk age and the star-formation
history of the MW disk do not allow accurate IMF determinations
\citep{Binney00, ES06}.

\subsection{The distribution of data points in the alpha-plot}
\label{sec:ahist}
The first thing to note about the data distribution in the alpha-plot
is that there is no readily discernible systematic difference in IMF
determinations neither with metallicity nor density of the population
(cf. Fig.~\ref{fig:kroupa_figmassey}).

In order to understand the origin and nature of the dispersion of
power-law indices evident in the alpha plot, \citet{K01a} investigates
the dispersion of $\alpha$ values for a given mass interval using
Aarseth-$N$-body models of evolving clusters. The result is that the
dispersion can be understood in terms of statistical sampling from a
universal IMF (as also found by \citealt{Elm97,Elm99}) together with
stellar-dynamical biases \footnote{Note that this does {\it not}
  constitute a proof of the stellar IMF being a probabilistic density
  distribution!}. {\sc Optimal Scampling} (Sec.~\ref{sec:optsamp}) was
not available in~2001 so that statistical variations that arise by
randomly sampling stars from the IMF could not then be separated from
variations of the IMF due to stellar-dynamical processes in young star
clusters.

\begin{figure}
\begin{center}
\rotatebox{0}{\resizebox{0.8\textwidth}{!}{\includegraphics{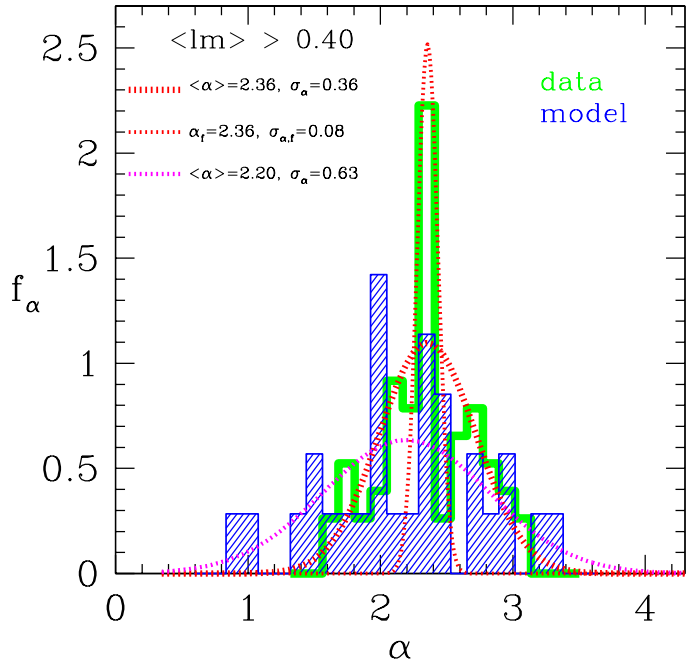}}}
\vskip -15mm
\caption[The distribution of high-mass star IMF
slopes]{\small{Distribution of $\alpha$ values for massive stars
    ($m>10^{0.4}\,M_\odot$). Note that the empirical data show a
    sharper (just at the Salpeter index) and more symmetrical
    distribution than the model, which is based on the Salpeter
    index. This is unexpected.  See {\sc Open Question~III} in
    Sec.~\ref{sec:ahist} for details.}}
\label{fig:ahist}
\end{center}
\end{figure}

Given the existing~2001 theoretical investigation, it is possible to
compare the theoretical distribution of $\alpha$ values for an
ensemble of star clusters with the observational data. This is done
for stars with $m>2.5\,M_\odot$ in Fig.~\ref{fig:ahist} where the open
(green) histogram shows the distribution of observational data from
Fig.~\ref{fig:apl}. The (blue) shaded histogram is the theoretical
ensemble of 12 star clusters containing initially 800 to $10^4$~stars
that are ``observed'' at 3 and 70~Myr: stellar companions are merged
to give the system MFs, which are used to measure $\alpha$, but the
input individual-star IMF is in all cases the canonical form
(Eq.~\ref{eq:imf}).  The dotted curves are Gaussians with the same
average $\alpha$ and standard deviation in $\alpha$ obtained from the
histograms. Fixing $\alpha_{\rm f}=<\!\!\alpha\!\!>$ and using only
$\!\mid \alpha \!\mid \le 2\,\sigma_\alpha$ for the observational data
gives the narrow thin (red) dotted Gaussian distribution which
describes the Salpeter peak extremely well (not by construction).

The interesting finding is thus that the observational data have a
very pronounced Salpeter/Massey peak, with broad near-symmetric
wings. This indicates that there are no significant biases that should
skew the distribution. For example, if the observational sample
contained clusters or associations within which the OB stars have a
low binary fraction compared to others that have a very high
multiplicity fraction, we would expect the binary-deficient cases to
deviate towards high $\alpha$ values since fainter companions are
hidden in the binary-rich cases. The absence of this effect is
consistent with the result obtained by \cite{MA08} and \cite{WKM09}
that multiple systems do not affect the derived power-law index of the
IMF for stars more massive than a few~$M_\odot$. For stars with mass
$m<1\,M_\odot$ unresolved multiples do have a significant effect, and
this has been corrected for in the canonical IMF (Eq.~\ref{eq:imf}).
Energetic dynamical ejections from cluster cores deplete the IMF of
the massive stars (\citealt{P-AK06a}, increasing the observed
$\alpha_3$, \citealt{BK12}) in the cluster while mass segregation has
the opposite effect.

In contrast to the observational data, the theoretical data show (i) a
distribution with a mean shifted to smaller $\alpha_2\approx 2.2$ that
has (ii) a larger width than the observational one. The input
canonical Salpeter/Massey index is not really evident in the
theoretical data, and if these were the observational data then it is
likely that the astronomical community would strongly argue for the
case that the IMF shows appreciable variations. This leads to the
following peculiar result:

\vspace{2mm} \centerline{ \label{quest:openIII}\fbox{\parbox{\columnwidth}{ {\sc
        Open Question III}: The empirical IMF power-law indices for
      massive stars are better behaved than the model data. This is
      unexpected because all the additional complications
      (observational uncertainties, uncertainties in transferring
      observed luminosities to masses such as stellar models,
      rotating/non-rotating stars) ought to deteriorate the empirical
      data, while the model has none of these uncertainties.  }}}
\vspace{2mm}

\noindent Clarifying the hitherto not understood difference between
the much more ``well-behaved'' observational data and the theoretical
data will need further theoretical work which will have to attempt to
re-produce the observational procedure as exactly as is possible (see
also the {\sc Sociological Hypothesis} on p.~\pageref{hyp:soc}).

\label{page:resolved}Note that {\sc Open question III} is naturally resolved if nature
follows {\sc optimal Sampling} (Sec.~\ref{sec:optsamp}) rather than
random sampling (Sec.~\ref{sec:massgen} from the
IMF.

{\it Is the scatter of data points in the alpha-plot
  (Fig.~\ref{fig:apl}) a result of IMF variations?} For this to be
conclusively convincing would require a measurement to lie further
from the canonical IMF than the conservative uncertainty range shown
in the figure. However, the adopted uncertainties on $\alpha_i$ in
Eq.~\ref{eq:imf} stem from the scatter in the alpha plot, so that this
argument is circular.

An independent indication of the uncertainties inherent to IMF
determinations can be obtained by comparing IMF estimates of the same
cluster by different authors. This is demonstrated for the
well-studied Pleiades, ONC and for 30~Dor
(Fig.~\ref{fig:apl}). Overall, the uncertainties in $\alpha$ are about
$\pm0.5$ which is also about the scatter evident in all the data, so
that there is no indication of significant outliers (except in the
shaded regions, see below).  Differences of $\Delta \alpha \approx
0.5$ for VLMSs and BDs are evident for the extremely young ONC
allowing an estimate of likely non-physical variations in the alpha
plot. Data reduction at these low masses is hampered by variability,
differential reddening, spurious detections \citep{Slesnick_etal04,
  Andersen09}.  It is clear that because the procedure of measuring
$\alpha(lm)$ is not standardised and because the IMF is not a single
power-law, author--author variations occur simply due to the use of
different mass ranges when fitting power-laws (see ``{\it Binning}''
bias in Sec.~\ref{sec:biases}).

Significant departures from the canonical IMF only occur in the shaded
areas of the alpha plot. These are, however, not reliable. The upper
mass range in the shaded area near $1\,M_\odot$ poses the problem that
the star-clusters investigated have evolved such that the turn-off
mass is near to this range.  Some clusters such as $\rho$~Oph are so
sparse that more massive stars did not form. In these cases the shaded
range is close to the upper mass limit leading to possible stochastic
stellar-dynamical biases since the most massive stars meet near the
core of a cluster due to mass segregation, but three-body or
higher-order encounters there can cause expulsion from the
cluster. Furthermore, $\rho$~Oph is still forming, leading to unknown
effects that are likely to enhance variations in the first derivative
of the IMF (i.e. in $\alpha$ values).  

The shaded area near $0.1\,M_\odot$ poses the problem that the VLMSs
are not on the main sequence for most of the clusters studied, and are
again prone to bias through mass-segregation by being underrepresented
within the central cluster area that is easiest to study
observationally. Especially the latter is probably biasing the M35
datum, but some effect with metallicity may be operating, especially
so since M35 appears to have a smaller $\alpha$ near the H-burning
mass limit than the Pleiades cluster which has a similar age but has a
larger abundance of metals (Fig.~\ref{fig:mfn1}). The M35 cluster
ought to be looked at again with telescopes.

Two other well-studied massive star-burst clusters have $\alpha
\approx\,${\it Salpeter/Massey} (30~Dor and NGC~3603) implying no
clear evidence for a bias that resolved star-burst clusters prefer
smaller $\alpha$ and thus more massive stars relatively to the number
of low-mass stars. Low-mass stars are known to form in 30~Dor
\citep{Sirianni_etal00}, although their MF has not been measured yet
due to the large distance of about 55~kpc. From the ONC we know that
the entire mass spectrum $0.05\simless m/M_\odot\simless 40$ is
represented but that it has a deficit of massive stars \citep{P-AK06a}
(Fig.~\ref{fig:mfn1}). The Pleiades appear to have had an IMF very
similar to the canonical one, although for massive stars a steeper IMF
with $\alpha_3\approx 2.7$ may also be suggested by theoretical work
(Fig.~\ref{fig:MorauxPl}).

But it remains an unsolved issue (see {\sc Open Question III} on
p.~\pageref{quest:openIII}) as to why the theoretical data have a larger
dispersion of $\alpha$ values than the empirical ones
(Fig.~\ref{fig:ahist}). The models discussed above and plotted in
Fig.~\ref{fig:ahist} had as the input IMF the Salpeter index, whereas
the empirical data suffer under all the biases associated with
transforming measured fluxes to stellar masses (Sec.~\ref{sec:biases})
and have been collated from various sources published by different
authors.  In particular the very narrow empirical distribution exactly
around the Salpeter index is remarkable. This leads to the following
Sociological Hypothesis:

\vspace{2mm} \centerline{ \fbox{\parbox{\columnwidth}{ \label{hyp:soc}{\sc
        Sociological Hypothesis}: {The measured $\alpha$ indices for
        massive stars are affected by sociological predispositions.
}}}}
\vspace{2mm} 

\noindent However, the excellent documented efforts of the teams
working on measuring $\alpha$ for massive stars would not necessarily
lend support to this hypothesis. Actually, as already observed on
p.~\pageref{page:resolved}, {\sc Optimal Sampling} would automatically
resolve {\sc Open Question III} therewith negating the {\sc
  Sociological Hypothesis}.

The available evidence is thus that low-mass stars and massive stars
form together even in extreme environments. This is also supported by
an impressive observational study \citep{L00,Luhman04} of many
close-by star-forming regions using one consistent methodology to
avoid author--author variations.  The result is that the IMF does not
show conclusive differences from low-density star-forming regions in
small molecular clouds ($n= 0.2-1$~stars/pc$^3$ in $\rho$~Oph) to
high-density cases in giant molecular clouds ($n= (1-5)\times
10^4$~stars/pc$^3$ in the ONC). This result extends to the populations
in the truly exotic ancient and metal-poor dwarf-spheroidal satellite
galaxies which are speculated to be dominated by dark matter but
definitely constitute star-forming conditions very different from
present-day events. Two such close companions to the Milky Way have
been observed \citep{Grill98,Felt99} finding the same MF as in
globular clusters for $0.5\simless m/M_\odot \simless 0.9$ and thus no
evident differences to the canonical IMF. However, evidence for
top-heavy IMFs in pc-scale starbursts and for bottom-heavy IMFs in
elliptical galaxies has emerged (Sec.~\ref{sec:currentstate}
and~\ref{sec:cosmoev}, respectively).


\vspace{2mm} \centerline{ \fbox{\parbox{\columnwidth}{ {\sc Main results}:
      Within the observational uncertainties the IMF of all known
      resolved stellar populations is well described by the canonical,
      standard or average IMF. It is given by Eq.~\ref{eq:imf} and has
      been corrected for unresolved multiple systems. The only
      structure evident in the {\it stellar} IMF is thus a turnover
      near $0.5\,M_\odot$ and a rapid turn-down below
      $0.075\,M_\odot$. The BD IMF is a separate single power-law such
      that about one BD forms per five stars. {\sc Open Question III}
      leads to the {\sc Sociological Hypothesis} which may naturally
      be negated by {\sc Optimal Sampling}.  }}} \vspace{2mm}


\section{Comparisons and Some Numbers}

In this Section some useful numbers are provided, and a comparison
between various IMF forms is made with cumulative functions in the
number of stars and stellar mass being plotted for general use.

\subsection{The solar-neighborhood mass density and some other numbers}

Given the reasonably well constrained shape of the stellar IMF
(Eq.~\ref{eq:imf}), it is of interest to investigate the implied
number and mass density in the Galactic disk. Here we consider the
solar neighbourhood.  In order to normalise the IMF to the
solar-neighbourhood stellar number density the observed stellar LF is
conveniently used.

The nearby Hipparcos LF, $\Psi_{\rm near}({\rm Hipp})$
(Fig.~\ref{fig:MWlf}), has
$\rho=(5.9\pm0.3)\times10^{-3}$~stars/pc$^3$ in the interval
$M_V=5.5-7.5$ corresponding to the mass interval $[m_2, m_1]=[0.891,
  0.687]\,M_\odot$ \citep{K01b} using the KTG93 MLR
(Fig.~\ref{fig:mlr}). $\int_{m_1}^{m_2}\xi(m)\,dm=\rho$ yields
$k=0.877\pm0.045$~stars/(pc$^3\,M_\odot$).  The number fractions, mass
fractions and Galactic-field mass densities contributed by stars in
different mass-ranges are summarised in Table~\ref{tab:frac}.

The local mass density made up of interstellar matter is $\rho^{\rm
  gas}\approx0.04\pm0.02\,M_\odot$/pc$^3$. In stellar remnants it is
$\rho^{\rm rem}\approx0.003\,M_\odot$/pc$^3$ \citep{Weide} or
$\rho^{\rm rem}\approx0.005\,M_\odot$/pc$^3$ (\citealt{Chrev03} and
references therein).  Giant stars contribute about $0.6\times
10^{-3}\,M_\odot$pc$^{-3}$ \citep{Haywood_etal97}, so that
main-sequence stars make up about half of the baryonic matter density
in the local Galactic disk (Table~\ref{tab:frac}).  BDs, which for
some time were regarded as candidates for contributing to the
dark-matter problem, do not constitute a dynamically important mass
component of the Galaxy, contributing not more than~5~\% in mass.
This is corroborated by dynamical analysis of local stellar space
motions that imply there is no need for dark matter in the Milky-Way
disk \citep{FF94}, and the revision of the thick-disk mass density to
larger values \citep{Fuhrmann04,Soubiran_etal03} further reduces the
need for dark matter within the solar circle.

Table~\ref{tab:frac} also shows that a star cluster looses about
12~per cent of its mass through stellar evolution within 10~Myr if
$\alpha_3=2.3$ (turnoff-mass $m_{\rm to}\approx20\,M_\odot$), or
within 300~Myr if $\alpha_3=2.7$ (turnoff-mass $m_{\rm
  to}\approx3\,M_\odot$). After 5~Gyr the mass loss through stellar
evolution alone amounts to about 45~per cent if $\alpha_3=2.3$ or
30~per cent if $\alpha_3=2.7$.  Mass loss through stellar evolution
therefore poses no risk for the survival of star clusters for the IMFs
discussed here, since the mass-loss rate is slow enough for the
cluster to adjust adiabatically. A star-cluster would be threatened
through mass loss from supernova explosions if $\alpha\simless 1.4$
for $8<m/\,M_\odot\le 120$ which would mean a mass-loss of $\simgreat
50$~per cent within about 40~Myr when the last supernova explodes. It
is remarkable that none of the measurements in resolved populations
has found such a low $\alpha$ for massive stars
(Fig.~\ref{fig:apl}). Mass loss due to stellar evolution might pose a
threat to the survival of post-gas-expulsion clusters when the system
is mass-segregated as has been shown by \citet{Vesperini09}.

\begin{landscape}
\begin{table}
\vspace{-2cm}
\begin{center}
\begin{tabular}{l|ccc|ccc|c|c}

\hline\hline

mass range     
&\multicolumn{3}{c|}{$\eta_N$}
&\multicolumn{3}{c|}{$\eta_M$}    
&$\rho^{\rm st}$
&$\Sigma^{\rm st}$  \\

[$M_\odot$]    
&\multicolumn{3}{c|}{[per cent]}
&\multicolumn{3}{c|}{[per cent]}
&[$M_\odot/{\rm pc}^3$]  
&[$M_\odot/{\rm pc}^2$] \\

&\multicolumn{3}{c|}{$\alpha_3$}
&\multicolumn{3}{c|}{$\alpha_3$}    
&$\alpha_3$
&$\alpha_3$  \\

&2.3 &2.7 &4.5 &2.3 &2.7 &4.5 &4.5 &4.5\\

\hline






0.07--0.5
&77.71
&79.39
&82.38

&28.42
&37.96
&52.90

&$2.17\times10^{-2}$
&9.73

\\ 

0.5--1
&13.25
&13.54
&14.05

&16.66
&22.24
&31.00

&$1.27\times10^{-2}$
&5.70

\\ 

1 -- 8
&8.45
&6.87
&3.57

&33.44
&31.62
&16.01

&$6.56\times10^{-3}$
&2.95

\\ 

8 -- 120 
&0.59
&0.20
&0.00

&21.48
&8.18
&0.09

&$3.69\times10^{-5}$
&$1.66\times10^{-2}$

\\ 

\hline

$\overline{m}/M_\odot=$
&$0.545$
&$0.417$
&$0.310$

&
&
&

&$\rho_{\rm tot}^{\rm st}=0.041$
&$\Sigma_{\rm tot}^{\rm st}=18.4$

\\

\hline \hline

&\multicolumn{1}{r|}{}
&\multicolumn{2}{c|}{$\alpha_3=2.3$} 
&\multicolumn{2}{c||}{$\alpha_3=2.7$}
&
&\multicolumn{2}{c}{$\Delta M_{\rm cl}/M_{\rm cl}$}
\\ 

&\multicolumn{1}{r|}{$m_{\rm max}$} 
&$N_{\rm cl}$  &$M_{\rm cl}$  
&$N_{\rm cl}$  &\multicolumn{1}{r||}{$M_{\rm cl}$}
&$m_{\rm to}$  &\multicolumn{2}{c}{[per cent]}
\\

&\multicolumn{1}{r|}{[$M_\odot$]} & &[$M_\odot$] & 
&\multicolumn{1}{r||}{[$M_\odot$]} &[$M_\odot$] 
&$\alpha_3=2.3$ &$\alpha_3=2.7$\\

\cline{2-9}

&\multicolumn{1}{r|}{1}      &13    &3.2     &15                 
&\multicolumn{1}{r||}{3.8}
&80 &2.2 &0.5
\\
                                                                     
&\multicolumn{1}{r|}{8}      &173   &82     &489                
&\multicolumn{1}{r||}{187}
&60 &4.0 &0.9
\\
                                                                      
&\multicolumn{1}{r|}{20}     &601   &307     &2415               
&\multicolumn{1}{r||}{970}
&40 &6.7 &1.7
\\
                                                                      
&\multicolumn{1}{r|}{40}     &1756  &893     &8839    
&\multicolumn{1}{r||}{3623}
&20 &12.2 &3.6
\\

&\multicolumn{1}{r|}{60}     &3803  &1993     &21509    
&\multicolumn{1}{r||}{8885}
&8 &21.5 &8.2
\\

&\multicolumn{1}{r|}{80}     &8023  &4275     &48750    
&\multicolumn{1}{r||}{20224}
&3 &32.1 &15.8
\\
  
&\multicolumn{1}{r|}{100}    &20820 &11236     &133129    
&\multicolumn{1}{r||}{55385}
&1 &45.2 &29.9
\\

&\multicolumn{1}{r|}{119}  &509319   &277416 &$3.38\times10^6$ 
&\multicolumn{1}{r||}{$1.41\times10^6$}
&0.7 &48.6 &34.4
\\

\hline\hline

\end{tabular}
\end{center}
\caption[Stellar number and mass fractions]%
{\small{The number fraction $\eta_N=100\,\int_{m_1}^{m_2}
    \xi(m)\,dm/ \int_{m_l}^{m_u}\xi(m)\,dm$, and the mass fraction
    $\eta_M=100\,\int_{m_1}^{m_2} m\,\xi(m)\,dm/ M_{\rm cl}$, $M_{\rm
      cl}= \int_{m_l}^{m_u} m\,\xi(m)\,dm$, in per cent of
    main-sequence stars in the mass interval $m_1$ to $m_2$, and the
    stellar contribution, $\rho^{\rm st}$, to the Oort limit and to
    the Galactic-disk surface mass-density, $\Sigma^{\rm
      st}=2\,h\rho^{\rm st}$, near to the Sun, taking
    $m_l=0.07\,M_\odot$, $m_u=120\,M_\odot$ and the Galactic-disk
    scale-height $h=250$~pc ($m<1\,M_\odot$ \citealt{KTG93}) and
    $h=90$~pc ($m>1\,M_\odot$, \citealt{Sc86}). Results are shown for
    the canonical IMF (Eq.~\ref{eq:imf}), for the high-mass-star
    Galactic-field IMF ($\alpha_3=2.7, m>1\,M_\odot$), and for the
    PDMF ($\alpha_3=4.5$, \citealt{Sc86,KTG93}) which describes the
    distribution of stellar masses now populating the Galactic
    disk. For gas $\Sigma^{\rm gas}=13\pm3\,M_\odot$/pc$^2$ and
    remnants $\Sigma^{\rm rem}\approx3\,M_\odot$/pc$^2$ \citep{Weide}.
    The average stellar mass is $\overline{m}= \int_{m_l}^{m_u}
    m\,\xi(m)\,dm/ \int_{m_l}^{m_u}\xi(m)\,dm$.  $m_{\rm max}$ and
    $M_{\rm ecl}$ are calculated using eqs~\ref{eq:mm} and~\ref{eq:Mecl}
    with $m_{\rm max}=120\,M_\odot$. $N_{\rm cl}$ is the
    number of stars that have to form in a star cluster such that the
    most massive star in the population has the mass $m_{\rm
      max}$. The mass of this population is $M_{\rm cl}$, and the
    condition is $\int_{m_{\rm max}}^{120\,M_\odot}\xi(m)\,dm=1$ with
    $\int_{0.07}^{m_{\rm max}} \xi(m)\,dm = N_{\rm cl}-1$.  $\Delta
    M_{\rm cl}/M_{\rm cl}$ is the fraction of mass lost from the
    cluster due to stellar evolution, assuming that for
    $m\ge8\,M_\odot$ all neutron stars and black holes are kicked out
    due to an asymmetrical supernova explosion, but that white dwarfs
    are retained \citep{Wetal92}. The masses of the white dwarfs are
    estimated as $m_{\rm WD} = 0.109\,m_{\rm ini}+0.394\,[M_\odot]$,
    which is a linear fit to the masses of observed white dwarfs
    \citep{Kalirai2008}. The evolution time for a star of mass $m_{\rm
      to}$ to reach the turn-off age is available in
    Fig.~\ref{fig:apl}. Note that brown dwarfs are not considered for
    any of the numbers listed in this table. }}
\label{tab:frac}
\end{table}
\end{landscape}

\subsection{Other IMF  forms and cumulative functions}
\label{sec:otherIMF}

The standard or canonical power-law IMF (Eq.~\ref{eq:imf}) provides a
good description of the data combined with mathematical ease and
physical meaning. Integrating the IMF is a frequently required task,
and it is especially here that the two- or even multi-part power-law
description of the canonical IMF shows its strength.  For example,
if the IMF were a probabilistic density distribution function then
writing down a mass-generating function (Sec.~\ref{sec:massgen}) is
practically trivial allowing perfectly efficient (each random deviate
$X$ yields a usable mass $m$) discretisation of the stellar population
into individual stellar masses.  A further strong advantage of this
parametrisation is that each section can be changed without affecting
another part of the IMF, as stated above. As an explicit example,
should the BD MF be revised, $\alpha_0$ can be adopted accordingly
without affecting the rest of the mass distribution in the
well-constrained stellar regime.  That the two- or multi-part
power-law form has a discontinuity in the derivative at $0.5\,M_\odot$
(Fig.~\ref{fig:apl}) has no implications for stellar-populations and
is therefore not a disadvantage, especially so since there exists no
theoretically derived IMF form of significance. It is, however, true
that the real IMF must be differentiable, but it must have a rapid
change in slope near $0.5\,M_\odot$ and below about $0.08\,M_\odot$
where the stellar IMF decays rapidly. The Chabrier-formulation of the
log-normal canonical IMF (Eq.~\ref{eq:chabimf}, Fig.~\ref{fig:canIMF})
also has a discontinuity in its derivative but at $1\,M_\odot$, and
does not reproduce the rapid falloff of the stellar IMF below about
$0.08\,M_\odot$ unless it is cutoff there, nor is it easily
integrable.

Often a single Salpeter power law IMF is applied ($\alpha=2.35$ for
$0.1\,M_\odot \simless m$). Table~\ref{tab:Salpetererror} gives the
stellar masses (Eq.~\ref{eq:Mecl1}) for the single-power-law Salpeter
IMF in comparison to the canonical IMF. For example, using a single
power-law ``Salpeter'' IMF with $\alpha=2.3$ for a stellar population
with stars in the mass range $0.1-0.8\,M_\odot$ instead of the
canonical IMF would lead to an overestimate of the stellar mass
by~$0.66/0.37=78$~\% and to an overestimate of the number, $N$, of
stars by $1.687/0.702=2.4$.
\begin{table}
\begin{center}
\begin{tabular}{lllll}

\hline\hline
$m_1/M_\odot$  &$m_2/M_\odot$  &$N$  &$M_{\rm ecl}/M_\odot$   &IMF used\\
\hline
0.07     &150                &1.000     &1.000  &{\rm canonical IMF}\\
0.1      &150                &0.823     &0.973  &{\rm canonical IMF}\\
0.1      &0.8                &0.702     &0.370  &{\rm canonical IMF}\\
0.5      &150                &0.223     &0.719  &{\rm canonical IMF}\\
\hline 
0.07     &150                &2.874     &1.424  &{\rm Salpeter IMF,} $\alpha=2.30$\\
0.1      &150                &1.808     &1.264  &{\rm Salpeter IMF}\\
0.1      &0.8                &1.687     &0.660  &{\rm Salpeter IMF}\\
0.5      &150                &0.223     &0.719  &{\rm Salpeter IMF}\\
\hline 
0.07     &150                &3.378     &1.543  &{\rm Salpeter IMF,} $\alpha=2.35$\\
0.1      &150                &2.087     &1.348  &{\rm Salpeter IMF}\\
0.1      &0.8                &1.961     &0.756  &{\rm Salpeter IMF}\\
0.5      &150                &0.238     &0.719  &{\rm Salpeter IMF}\\
\hline\hline
\end{tabular}
\end{center}
\caption[Canonical vs. Salpeter IMF]{\small{
    The mass in stars, $M_{\rm ecl}$, normalised to
    $1\,M_\odot$ for the canonical IMF between $0.07\,M_\odot$ and
    $150\,M_\odot$ in comparison to the mass for other mass ranges
    ($m_1$ to $m_2$) and
    the commonly used single-power-law Salpeter IMF taken here to have 
    $\alpha=2.3$ and~2.35. All IMFs are normalised to have
    identical $M_{\rm ecl}$ for $m>0.5\,M_\odot$ and no stellar remnants are
    included. $N$ is the correspondingly normalised number of stars. }}
\label{tab:Salpetererror}
\end{table}

\begin{figure}
\begin{center}
\rotatebox{0}{\resizebox{0.9\textwidth}{!}{\includegraphics{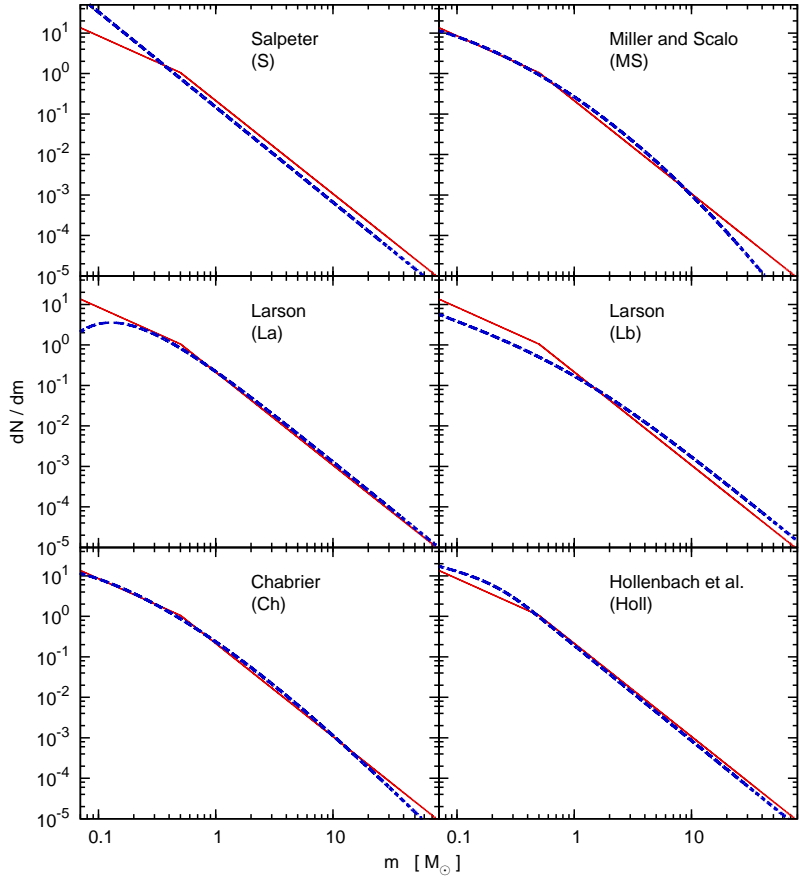}}}
\caption[Comparison of IMFs]{\small{A comparison between the canonical IMF
    (Eq.~\ref{eq:imf}, solid red curve) and the IMFs of
    Tab.~\ref{tab:imfs} in the interval from $0.07\,M_\odot$ to
    $80\,M_\odot$. Plotted is the number of stars per mass interval
    versus the stellar mass, both scales being logarithmic. All IMFs
    are normalised such that $\int^{150}_{0.07} m \, \xi(m) \,
    dm=1\,M_\odot$, where $m$ is the mass in Solar units. Note that
    the ``Chabrier'' IMF in the lower left panel is the earlier
    Chabrier IMF listed in Table~\ref{tab:imfs}, and is not the later
    log-normal plus power-law extension plotted in
    Fig.~\ref{fig:canIMF}. See also Fig.~\ref{fig:cumN}
    and~\ref{fig:cumM} for a comparison of the cumulative functions.}}
\label{fig:compIMFs}
\end{center}
\end{figure}

\begin{figure}
\begin{center}
\rotatebox{0}{\resizebox{0.9\textwidth}{!}{\includegraphics{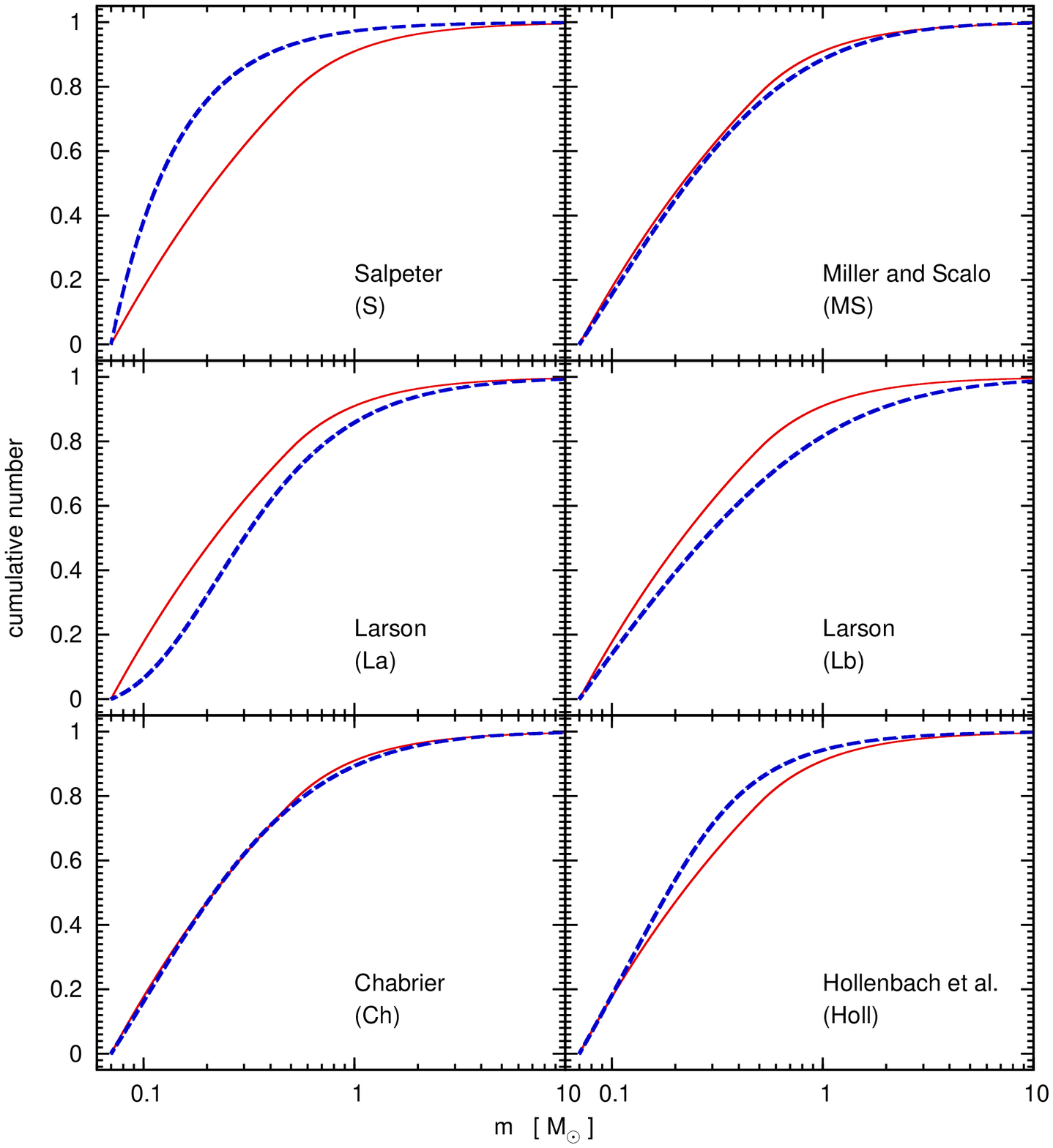}}}
\caption[Cumulative number plots for different IMFs]{\small{Cumulative
    function of the number of stars in the stellar mass range
    $0.07-150\,M_\odot$, plotted here over the range
    $0.07-10\,M_\odot$, for the alternative IMF forms listed in
    Tab.~\ref{tab:imfs} and plotted in Fig.~\ref{fig:compIMFs}.  In
    all the panels, the canonical IMF (Eq.~\ref{eq:imf}) is shown as
    the solid red curve, while the alternative IMF forms are plotted
    with the dashed blue curve. Note that half of a saturated
    canonical stellar population has $m<0.21\,M_\odot$.  }}
\label{fig:cumN}
\end{center}
\end{figure}

\begin{figure}
\begin{center}
\rotatebox{0}{\resizebox{0.9\textwidth}{!}{\includegraphics{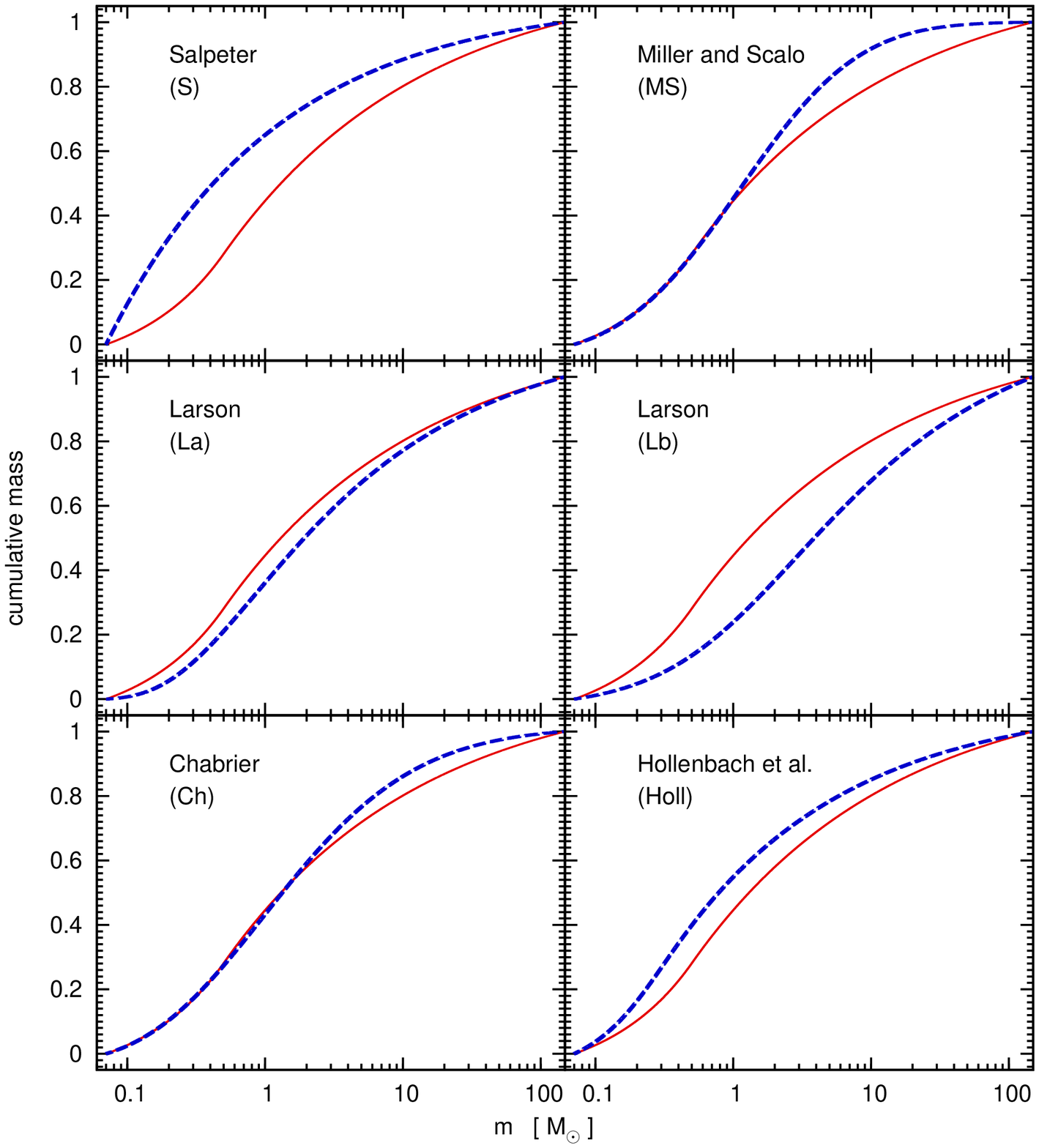}}}
\caption[Cumulative mass plots for different IMFs]{\small{ Cumulative
    function of the mass in stars in the stellar mass range
    $0.07-150\,M_\odot$ for the alternative IMF forms listed in
    Tab.~\ref{tab:imfs} and plotted in Fig.~\ref{fig:compIMFs}.  In
    all the panels, the canonical IMF (Eq.~\ref{eq:imf}) is shown as
    the solid red curve, while the alternative IMF forms are plotted
    with the dashed blue curve.  Note that half the mass of a
    saturated canonical stellar population has $m<1\,M_\odot$.  }}
\label{fig:cumM}
\end{center}
\end{figure}

Additional forms are in use and are preferred for some investigations:
In Fig.~\ref{fig:apl} the quasi-diagonal (black) lines are alternative
analytical forms summarised in Table~\ref{tab:imfs}. They are compared
to the canonical IMF in Fig.~\ref{fig:compIMFs} and their cumulative
number and cumulative mass functions are presented in
Fig.~\ref{fig:cumN} and Fig.~\ref{fig:cumM}, respectively.

Of the other IMF forms sometimes in use, the generalised Rosin-Rammler
function (Eq.~{\it Ch} in Table~\ref{tab:imfs}, thick
short-dash-dotted curve) best represents the data, apart from a
deviation for $m\simgreat 10\,M_\odot$, which can be fixed by adopting
a Salpeter/\-Mas\-sey power-law extension for $m>1\,M_\odot$
(Eq.~\ref{eq:chabimf}). Interpretation of $m_o$ in terms of a
characteristic stellar mass poses difficulties. As can be seen in
Fig.~\ref{fig:canIMF}, the difference between a Chabrier IMF and a
canonical IMF is negligible. The popular Miller-Scalo log-normal IMF
(Eq.~{\it MS} in Table~\ref{tab:imfs}) deviates strongly from the
empirical data at high masses. Larson's Eq.~{\it Lb} in
Table~\ref{tab:imfs} fits rather well, except that it may predict too
many BDs.  Finally, the {\it effective initial mass function for
  galactic disks} proposed by \citet{Hollenbach_etal05, Parravano11}
(Eq.~{\it Holl}) reproduces the data in the alpha-plot quite well
(fig.~1 in \citealt{Hollenbach_etal05, Parravano11}), and is not
incorporated into Fig.~\ref{fig:apl} here. Note however that a {\it
  composite IMF} (i.e. the IGIMF and the local IGIMF, LIGIMF), which
would be the correct IMF for a whole disk galaxy or parts thereof,
respectively, ought to be steeper (have a larger $\alpha$) at high
masses which is precisely what \cite{Sc86} deduced for the MW disk
($\alpha_3\approx 2.7$, Eq.~\ref{eq:KTGimf}, Sec.~\ref{sec:comppop}).

The closed functional IMF formulations (Eqs.~{\it MS, La, Lb, Ch,
  Holl}) have the advantage that possible variations of the IMF with
physical conditions can be studied more naturally than with a
multi-power-law form, because they typically have a characteristic
mass that can be varied explicitly. However, they cannot readily
capture the variation of the stellar IMF with $\rho$ and [Fe/H]
discovered recently (Eq.~\ref{eq:toph_master}).

\begin{landscape}
 
\begin{table}
\begin{center}
\vspace{-1cm}
\hskip-0.5cm
{\small
\begin{tabular}{l|l|l|l}

\hline\hline
general
&\multicolumn{2}{l}{$dN = \xi(m)\,dm = \xi_{\rm L}(m)dlm$}\\
&\multicolumn{2}{l}{$\xi_{\rm L}(m) = (m\,{\rm ln}10)\,\xi(m)$}
&{\it gen}\\

Scalo's IMF index \citep{Sc86} 
&\multicolumn{2}{l}{$\Gamma(m)\equiv {d\over dlm}
\left({\rm log}_{10}\xi_{\rm L}(lm)\right)$}
&{\it Gam}\\

&\multicolumn{2}{l}{$\Gamma = -x = 1+\gamma = 1-\alpha$}
&{\it ind}\\
&e.g.  for power-law form:
&$\xi_{\rm L} = A\,m^\Gamma = A\,m^{-x}$\\
&&$\xi = A'\,m^{-\alpha} = A'\,m^{+\gamma}$\\
&&$A' = A/{\rm ln}10$\\

\hline\hline

\citet{S55} 
&$\xi_{\rm L}(lm) = A\,m^\Gamma$
&$\Gamma=-1.35 \, (\alpha=2.35)$
&{\it S}\\

&\multicolumn{2}{l}{$A = 0.03\,{\rm pc}^{-3}\,{\rm
log}_{10}^{-1}M_\odot; \quad 0.4\le m/M_\odot \le 10$}\\

\hline

\citet{MS79} 
&$\xi_{\rm L}(lm) = A\,{\rm exp}\left[-{\left(lm-lm_o\right)^2 \over 
2\,\sigma_{lm}^2 } \right]$
&$\Gamma(lm) = -{\left(lm-lm_o\right) \over \sigma_{lm}^2}\,{\rm log}_{10}e$
&{\it MS}\\

{\it thick long-dash-dotted line}
&\multicolumn{2}{l}{$A = 
106\,{\rm pc}^{-2}\,{\rm log}_{10}^{-1}M_\odot;
\quad lm_o=-1.02; \quad \sigma_{lm}=0.68$}\\

\hline

\citet{L98}
&$\xi_{\rm L}(lm) = A\, m^{-1.35} {\rm exp}\left[-{m_o\over
                    m}\right]$
&$\Gamma(lm) = -1.35  + {m_o\over m}$
&{\it La}\\
{\it thin short-dashed line}
&\multicolumn{2}{l}{$A=-\,; \quad\quad m_o=0.3\,M_\odot$}\\

\hline

\citet{L98}
&$\xi_{\rm L}(lm) = A\,\left[1 + {m\over m_o}\right]^{-1.35}$ 
&$\Gamma(lm) = -1.35\left(1 + {m_o\over m}\right)^{-1}$
&{\it Lb}\\
{\it thin long-dashed line}
&\multicolumn{2}{l}{$A=-\,; \quad\quad m_o=1\,M_\odot$}\\

\hline

\citet{Ch01,Ch02}
&$\xi(m) = A\,m^{-\delta}\,{\rm exp}
         \left[-\left({m_o\over m}\right)^\beta \right]$
&$\Gamma(lm) = 1 - \delta + \beta\left(m_o\over m\right)^\beta$
&{\it Ch}\\

{\it thick short-dash-dotted line}
&\multicolumn{2}{l}{$A=3.0\,{\rm pc}^{-3}\,M_\odot^{-1};
\quad m_o=716.4\,M_\odot; \quad \delta=3.3;  \quad
\beta=0.25 $}\\

\hline

\citet{Hollenbach_etal05,Parravano11}
&$\xi_{\rm L}(m) = k\,m^{-\Gamma} ( 1-{\rm exp} [-(m/m_{\rm ch})^{\gamma+\Gamma}] )$
&
&{\it Holl}\\

{\it not plotted in Fig.~\ref{fig:apl}}

&\multicolumn{2}{l}{$\gamma = 0.4, \Gamma=1.35, m_{\rm ch}=0.18\,M_\odot$}\\

\hline\hline

\end{tabular}}
\end{center}
\caption[Summary of published IMFs]{\small{Summary of different proposed analytical IMF forms
    discussed in Sec.~\ref{sec:otherIMF} (the modern power-law form,
    the canonical IMF, is presented in Eq.~\ref{eq:imf} and its
    log-normal equivalent is given by Eq.~\ref{eq:chabimf}).  Notation:
    $lm\equiv{\rm log}_{10}(m/M_\odot)={\rm ln}(m/M_\odot)/{\rm
      ln}10$; $dN$ is the number of all stars in the mass interval
    $m$ to $m+dm$ and in the logarithmic-mass interval $lm$ to
    $lm+dlm$.  The mass-dependent IMF indices, $\Gamma(m)$ (Eq.~{\it
      Gam}), are plotted in Fig.~\ref{fig:apl} using the line-types
    defined here.  Eq.~{\it MS} was derived by Miller\&Scalo assuming
    a constant star-formation rate and a Galactic disk age of 12~Gyr
    (the uncertainty of which is indicated in the lower panel of
    Fig.~\ref{fig:apl}). \citet{L98} does not fit his forms
    (Eq.~{\it La} and~{\it Lb}) to solar-neighbourhood star-count data
    but rather uses these to discuss general aspects of likely
    systematic IMF evolution; the $m_o$ in Eq.~{\it La} and~{\it Lb}
    given here are approximate eye-ball fits to the canonical IMF.  }}
\label{tab:imfs}
\end{table}
\end{landscape}

\section{The Origin of the IMF}

The two fundamental theoretical ansatzes for understanding the form of
the IMF are discussed in Sec.~\ref{sec:introd_IMF}, and here a few
more detailed aspects of the theoretical problem are
raised. Observationally it appears that the form of the stellar IMF is
already established in the pre-stellar cloud core MF. Open questions
concerning our understanding of the IMF remain.

\subsection{Theoretical notions}
\label{sec:theor}

The Jeans mass scale (Eq.~\ref{eq:MJeans}, see also
Sec.~\ref{sec:introd_IMF}) is useful as a general indication of the
rough mass scale where fragmentation of a contracting gas cloud
occurs. But the concept breaks down when considering the stellar
masses that form in star clusters. The central regions of these are
denser, formally leading to smaller Jeans masses which is the opposite
of the observed trend, where even in very young clusters massive stars
tend to be located in the inner regions. More complex physics is
clearly involved ({\sc Self-Regulatory Ansatz} in
Sec.~\ref{sec:introd_IMF}).  \cite{MurrayLin96} develop an $N$-body model
for the formation of stars in star clusters by considering the
formation and dynamical evolution and interactions of cloudlets which
can form stars which heat their surroundings thereby quenching further
star formation. They naturally arrive at mass-segregated clusters with
power-law IMFs.  Interestingly, \citet{EKW08} show that the Jeans-mass
depends only weakly on the ambient radiation field, temperature,
metallicity and density for a large range of initial conditions,
therewith possibly explaining a largely invariant IMF. The
insensitivity of the IMF to the initial kinetic structure of the gas
is also explicitly demonstrated by \cite{Bate2009} in a
high-resolution SPH computation of a large-scale molecular cloud.  The
inclusion of radiation and magnetic fields is now becoming possible
and is touched upon further down in this section.

The impressive agreement between the theoretical $m_{\rm max}-M_{\rm
  ecl}$ relation Eq.~\ref{eq:mmax_pred}), which results from SPH and
FLASH computations of star formation with and without feedback, and
the observational data (Fig.~\ref{fig:mmaxf}), together with these
simulations also reproducing the general form of the IMF quite well,
suggests that the essential physics of star-formation has been
understood. Another success of star formation computations is the
excellent reproduction of the observed mass distribution of BDs and
VLMSs (Sec.~\ref{sec:BD_IMF}).

But, as discussed in Sec.~\ref{sec:bins} this theoretical work has not
yet explained the birth binary properties. Also, as stated in
Sec.~\ref{sec:introd_IMF}, this theoretical work predicts a variation
of the IMF with the temperature and the density of the gas. The
consensus in the community has been until now that such variation has
not been found (more on this in Sec.~\ref{sec:currentstate}).

Stars almost certainly regulate their own mass by feedback (winds,
radiation, outflows) limiting to a certain degree the amount of mass
that can be accreted ({\sc Self-Regulatory Ansatz} in
Sec.~\ref{sec:introd_IMF}). Indeed, this is an attractive proposition
in-line with the structure in the $m_{\rm max}-M_{\rm ecl}$ relation
at $m_{\rm max}\simgreat 10\,M_\odot$
(Sec.~\ref{sec:stmass_clmass}). The $m_{\rm max}-M_{\rm ecl}$ relation
suggests a self-regulated balance between the amount of matter that
can accrete onto a star and the feedback and local fragmentation which
modulates the accretion. In particular noteworthy is the result
(Sec.~\ref{sec:mmaxInterp}) that feedback from the central massive
stars may reduce the accretion onto the surrounding less massive stars
leading to an improved agreement with the observational $m_{\rm
  max}-M_{\rm ecl}$ data.  Self-regulatory behaviour may be a reason
why the shape of the IMF appears to be so invariant to metallicity
\citep{Krumholz10, Myers11}.

The coagulation of proto-stars probably plays a significant role in
the densest regions where the cloud-core collapse time, $\tau_{\rm
  coll}$, is longer than the fragment collision time-scale, $\tau_{\rm
  cr}$. The collapse of a fragment to a proto-star, with about 95~per
cent of its final mass, takes no longer than $\tau_{\rm coll}\approx
0.1$~Myr \citep{WK01}, so that the core crossing time
\begin{equation}
\tau_{\rm cr}/{\rm Myr} \approx 42\,\left( (R/{\rm pc})^3 \over
(M_{\rm ecl}/M_\odot) \right)^{1\over2} < 0.1\;{\rm Myr},
\label{eq:collapset}
\end{equation}
where $R$ is the half-mass radius of a Plummer-sphere model, implies
$M/R^3 > 10^5\, M_\odot$/pc$^3$. Such densities correspond to
star-formation-rate densitities 
\begin{equation}
SFRD>0.1\,M_\odot\,{\rm yr}^{-1}\,{\rm pc}^{-3}
\end{equation}
\noindent if the embedded cluster forms over a time-scale of
$1\,$Myr. It is under such conditions that proto-stellar interactions
are expected to affect the emerging stellar mass-spectrum. These are
found in the centres of very dense and rich embedded star clusters
before they expand as a result of gas expulsion (see also
\citealt{ElmSchad03} for a more detailed discussion). Thus, only for
massive stars ($m\simgreat 10\,M_\odot$) is the form of the IMF likely
affected by coagulation \citep{Bonn98,Kl01,ZH07}. It is indeed
remarkable that evidence for a top-heavy IMF has only emerged in
star-forming systems in which $\rho_{\rm gas+stars} >
10^5\,M_\odot\,{\rm pc}^{-3}$ (Sec.~\ref{sec:tophUCD}).

But the observed mass segregation in very young clusters cannot as yet
be taken as conclusive evidence for primordial mass segregation that
results naturally from competitive accretion/fragmentation induced
starvation (p.~\pageref{sec:mmaxInterp}) and coagulation, unless
precise $N$-body computations of the embedded cluster are performed
for each case in question. For example, models of the ONC show that
the degree of observed mass segregation can be established dynamically
within about 2~Myr (see Fig.~\ref{fig:mfn2}) despite the embedded and
much denser configuration having no initial mass segregation. The
notion behind the ``no initial mass-segregation'' assumption is that
star clusters fragment heavily sub-clustered
\citep{Meg,Bont01,Kl01,MCB10,MC11}, and each sub-cluster may form a
few OB stars with a few hundred associated lower-mass stars
(Table~\ref{tab:frac}), so that the overall morphology may resemble a
system without significant initial mass segregation
\citep{McMillan07,Fellhauer09} or even with ``inverse mass
segregation'' \citep{Parker11}.

The theoretical time-scale, $t_2-t_1$ in Fig.~\ref{fig:mfn2}, for mass
segregation to be established even in an initially not sub-clustered
cluster can be shortened by decreasing the relaxation time. This can
be achieved by reducing the number of stars and sub-clusters or by
increasing the pre-gas expulsion density in the model. But it may
prove impossible to find agreement at the same time with the density
profile and kinematics because the ONC is probably expanding rapidly
now given that it is virtually void of gas \citep{KAH}. Clearly the
issue of initial mass segregation requires more study.

An interesting approach to explain why the stellar IMF has a
Salpeter/\-Massey slope above $1\,M_\odot$ has been proposed by
\cite{Oey11} by applying the concepts of the IGIMF theory
(Sec.~\ref{sec:comppop}) to the notion that a star-cluster forms as a
hierarchical fractal structure \citep{Elm97, Elm99}. Sally Oey is able
to demonstrate that the true stellar IMF of the stars that form within
each sub-clump (here referred to as the ``{\it elementary IMF}'',
EIMF) has $\alpha_{\rm scl}\approx2$ in agreement with the MF of
embedded clusters, $\beta\approx2$ (Eq.~\ref{eq:ECMF}). If the
least-massive sub-clump has a mass comparable to the lowest-mass star,
then the stellar IMF of the whole cluster steepens to the canonical
Salpeter/Massey value, $\alpha=2.3$ (Eq.~\ref{eq:imf}), upon addition
of all the sub-clumps. Notable is that this effect had already been
discovered in SPH simulations of star formation \citep{MCB10}.  The
reason for this is that low-mass sub-clumps cannot form massive stars,
which therewith become underrepresented in the whole cluster. A
slightly top-heavy IMF with $\alpha=\alpha_{\rm scl}$ may result in a
star-burst cluster if all sub-clumps are more massive than the most
massive star can be, since then each sub-clump can form the whole
range of stellar masses, i.e. if each sub-clump were saturated. This
is essentially the exact same reason why the IGIMF may become top
heavy if the least massive cluster is saturated
(p.~\pageref{box:saturated}).

The Oey model raises a number of questions: on the one hand side, it
remains to be seen how the other well studied effects of coagulation,
competitive accretion and stellar feedback would affect the resultant
stellar IMF. And, on the other hand side, it leaves the question open
why the pre-stellar cloud core MF already has the shape of the stellar
IMF (Sec.~\ref{sec:cloudcore}). Last not least, observations, while
very difficult, would need to confirm that the stellar IMF in
sub-clumps has $\alpha_{\rm sc}\approx 2$ rather than the canonical
$\alpha=2.3$.

But what about radiation feedback and magnetic fields? The results
discussed above are based nearly exclusively on pure
gravo-hydrodynamical computations only, and radiation and magnetic
fields are likely to significantly affect the fragmentation and
accretion behaviour of a gas cloud. Such calculations are highly
demanding (e.g. \citealt{Stamatellos07}) and not many results allowing
statistically significant statements exist yet. The fragmentation of a
magnetised cluster-forming cloud clump and the properties of the
emerging bound cores such as their masses, radii, mean densities,
angular momenta, spins, magnetisations and mass-to-flux ratios are
documented and are found to be comparable to observational properties.
\cite{PB09} present SPH calculations of a $50\,M_\odot$ gas cloud with
radiative feedback and a magnetic field being incorporated together
for the first time. They show that the magnetic field supports the gas
cloud on large scales while the radiation field limits fragmentation
on a small scale through heating of the gas nearby proto stars. The
overall result is a similar mass distribution of proto stars but a
significantly reduced star-formation efficiency. \cite{LiWang10} find
that, besides reducing the overall star formation rate, both magnetic
fields and outflow feedback reduce the characteristic mass of the
cluster stars.  The influence of a magnetic field and radiation
feedback on the formation of massive stars in a rotating
$1000\,M_\odot$ cloud is studied for the first time computationally by
\cite{Peters11} using the adaptive-mesh FLASH code. In comparison to
the otherwise equal purely gravo-hydrodynamical computations it is
found that by yielding large-scale support the magnetic field limits
secondary fragmentation and by carrying angular momentum outwards it
enhances accretion onto the central massive proto-star. The number of
fragments is only reduced by a factor of about two though.

At extreme star-formation rate densities (SFRDs) the gas can be heated
by supernova(SN)-generated cosmic rays (CRs) which propagate into the
dense cloud region heating it if the SN frequency is sufficiently high
to sustain a CR flux.  Many SNe in confined places may also contribute
to turbulent heating of the gas.  Such a very excited gas phase has
been revealed by high-J CO line observations of ultra-luminous
infrared galaxies (Papadopoulos, private communication).  These
physical processes raise the Jeans mass possibly leading to a
top-heavy IMF (Sec.~\ref{sec:CR}).

\subsection{The IMF from the cloud-core mass function?}
\label{sec:cloudcore}

The possible origin of most stellar masses is indicated by a
remarkable discovery by the team around Philippe Andr{\'e} in Saclay
for the low-mass $\rho$~Oph cluster in which star-formation is
on-going. Here the pre-stellar and proto-star MF are indistinguishable
and both are startlingly similar to the canonical IMF, even in showing
the same flattening of the power-law near $0.5\,M_\odot$
\citep{Motte,Bont01}. The pre-stellar cores have sizes and densities
that can be interpreted as Jeans-instabilities for the conditions in
the $\rho$~Oph cloud, so that cloud-fragmentation due to the collapse
of density fluctuations in a dissipating turbulent interstellar medium
\citep{NordlundPadoan02,PadoanNordlund02,
  MacLowKlessen04,TilleyPudritz04,PadoanNordlund04,HC08,HC09,El11}
augmented with mass-dependent accretion and accretion-truncation
prescription (\citealt{Veltchev11, Myers11a}, see also
Sec.~\ref{sec:introd_IMF}) appears to be the most-important mechanism
shaping the stellar IMF for masses $0.1\simless
m/M_\odot\simless\,$a~few $\,M_\odot$.

If this result can be generalised then it may indicate that the shape
of the IMF describing the vast population of stars would be mostly
determined by the spectrum of density fluctuations in molecular
clouds. Similar results have indeed been obtained for the Serpens
clouds and for the clouds in Taurus--Auriga
(\citealt{TestiSargent98,Onishi_etal02}, respectively).

More recently \cite{Alves07} used a novel technique by mapping the
extinction through the Pipe Nebula to derive the pre-stellar
cloud-core MF. The result is a cloud-core MF with the same shape as
the canonical IMF but shifted to larger masses by a factor of about
three.  And, the massive effort within the Gould Belt Survey with the
Herschel telescope towards the Aquila rift and Polaris Flare regions
yields virtually the same result \citep{Andre10}.  This is interpreted
to mean that the star-formation efficiency (SFE) is about 30~per cent
independently of the core mass.  It is interesting that this SFE is
similar to the efficiencies deduced for embedded clusters on a scale
of about one~pc \citep{Lada_Lada03}.  Such a SFE also arises from
magneto-hydrodynamical computations of the formation of a proto-star
from a cloud core: as shown by \cite{MachidaMatsumoto11} 20--50~per
cent of the infalling material is expelled pole-wards by a
magnetically-driven wind.  Why the more recent results indicate a SFE
shift in mass between the cloud-core MF and the stellar IMF which was
not seen in the earlier work will need to be understood.

The majority of stellar masses making up the canonical IMF thus do not
appear to suffer significant subsequent modifications such as
competitive accretion, proto-stellar mergers or even self-limitation
through feedback processes, assuming there is a one-to-one map between
a pre-stellar core mass and the stellar mass.  A keynote result
supporting the ``no-interaction'' conjecture is presented by
\cite{Andreetal07} using kinematical measurements in $\rho$~Oph: The
relative velocities of the pre-stellar and proto-stellar cloud cores
are too small for the individual condensations to interact with each
other before evolving into pre-main sequence stars. This implies the
following conjecture:

\vspace{2mm} \centerline{ \fbox{\parbox{\columnwidth}{\label{conj:IMForigin}
      {\sc The IMF Origin Conjecture}: The IMF is mostly determined by
      cloud fragmentation at the pre-stellar stage. Competitive
      accretion is not the dominant mechanism at the proto-stellar
      stage. Competitive accretion may govern the growth of starless,
      self-gravitating condensations initially produced by
      gravo-turbulent fragmentation towards an IMF-like mass spectrum
      \citep{Andreetal07}.  }}} \vspace{2mm}

The work of \citet{PadoanNordlund02} with modifications by \cite{El11}
demonstrates that, under certain reasonable assumptions, the mass
function of gravitationally unstable cloud cores deriving from the
power-spectrum of a super-sonic turbulent medium leads to the observed
canonical IMF above $1\,M_\odot$. The flattening at lower masses is a
result of a reduction of the SFE because at small masses only the
densest cores can survive sufficiently long to collapse, this being
the reason why BDs do not, as a rule, form as stars do directly from
self-induced gravitational collapse of a core
(Sec.~\ref{sec:BD_IMF}). \cite{TilleyPudritz07} apply their
magneto-hydrodynamical simulations to test the Padoan-Nordlund model
of turbulent fragmentation finding disagreements in the position of
the peak mass of the core mass spectrum. 

If this holds for more massive star-forming regions is not
clear. \citet{BMC09} show by observation that massive young dense
cores in Cygnus X tend to be more fragmented than expected from
turbulence regulated monolithic collapse but that they are less
fragmented than predicted by gravo-turbulent scenarios and that the
fragments do not show a canonical IMF distribution. The
magneto-hydrodynamical plus radiation feedback simulations discussed
above would be relevant in understanding this observational result.
\cite{HC08, HC09} calculate the mass spectrum of the self-gravitating
cores based on an extension of the Press-Shechter statistical
formalism thereby also reproducing well the observed IMF.  But it
depends on the turbulence spectrum and the IMF ought to therefore be
variable.  The generation of turbulence in the ISM through the passage
of curved shock waves is investigated by
\cite{KevlahanPudritz09}. They conclude that ``the composite nature of
the IMF -—- a log-normal plus power-law distribution -—- is shown to be
a natural consequence of shock interaction and feedback from the most
massive stars that form in most regions of star formation in the
galaxy''.  A three-component IMF was also suggested by \cite{El04} to
account for the apparently seen different IMFs in different
star-forming regimes.

\vspace{2mm} \centerline{ \fbox{\parbox{\columnwidth}{ {\sc The Similarity
        Statement}: The intriguing observational result is that the
      stellar, proto-stellar and pre-stellar clump mass spectra are
      similar in shape to the stellar IMF.  }}} \vspace{2mm}

\noindent
This is consistent with the independent finding that the properties of
binary systems in the Galactic field can be understood if most stars
formed in modest, $\rho$~Oph-type clusters with primordial binary
properties as observed in Taurus-Auriga (Eq.~\ref{eq:domode},
Sec.~\ref{sec:bins}), and with the independent result derived from an
analysis of the distribution of local star clusters that most stars
appear to stem from such modest clusters \citep{AdamsMyers01}.
However, the canonical IMF is also similar to the IMF in the ONC
(Fig.~\ref{fig:mfn1}) implying that fragmentation of the pre-cluster
cloud proceeded similarly there.

The above {\sc Similarity Statement} may not be entirely true because
each cloud core typically forms a multiple stellar system
(Sec.~\ref{sec:bins}, \citealt{Getal08,SLB09}).  In the computer
simulations, the similarity of the clump mass function and the IMF
does not necessarily imply a direct one-to-one mapping of clumps to
stars. The temporal evolution of the clump and stellar MF in
SPH-models shows that stars come from a broad range of clump masses
despite the similar shape of the MF.

The SPH computations by \citet{BonnellBate02} and collaborators of the
formation of dense clusters indeed not only predict the observed
$m_{\rm max}-M_{\rm ecl}$ relation (Fig.~\ref{fig:mmaxf}), but they
also show that a Salpeter/Massey power-law IMF is obtained as a result
of competitive accretion and the merging of proto-stars near the
cluster core driven by gas-accretion onto it, independently of
metallicity as long as $Z/Z_\odot$ is larger than $10^{-5}$
\citep{CGB09}. The reason as to why the IMF is so invariant above a
few $M_\odot$ may thus be that the various physical processes all
conspire to give the same overall scale-free result (see also
Sec.~\ref{sec:theor}).

{\sc Open Question IV} emerges here: Various theories of the IMF as
resulting from the pre-stellar cloud core MF account for the observed
shape of the canonical IMF. This is also the case for theories based
on competitive accretion and on coagulation. However, if star
formation is intrinsically hierarchical through first the emergence of
sub-clumps of stars that merge dynamically, then as discussed in
Sec.~\ref{sec:theor}, the stellar IMF in each sub-clump must be
flatter with $\alpha_{\rm sc}\approx 2$ than the Salpeter/\-Massey
index. Why does theory not predict this flatter {\it elementary IMF}
(EIMF)?

\vspace{2mm} \centerline{ \label{quest:openIV}\fbox{\parbox{\columnwidth}{
      {\sc Open Question IV}: The hierarchical model of star-cluster
      formation implies the EIMF to be flatter than the canonical
      IMF. Why has theory never predicted this EIMF?  }}} \vspace{2mm}

\noindent A word of caution is advisable in view of the modelling of
star-formation and the resulting IMF.  An excellent example of how
state-of-the art modelling may be somewhat misleading is as follows:
Observations found a top-heavy PDMF in the Arches cluster with an
apparent lack of stars below $6\,M_\odot$ \citep{Stolteetal02} and
this was often interpreted as a top-heavy IMF because of the youth of
the object. \citet{KSJ07} therefore presented a state-of-the-art
SPH-model of star-formation from warm gas in the Galactic Centre which
produces a top-heavy IMF with a down-turn below about
$6\,M_\odot$. But \citet{Kimetal06} showed that the observed Arches
PDMF is in fact readily explained by strong stellar dynamical
evolution due to the extreme environment with no need for a
non-canonical IMF.  And, stars less massive than $6\,M_\odot$ have
formed in the cluster without a sign of a deficit.

\vspace{2mm} \centerline{ \fbox{\parbox{\columnwidth}{ {\sc Main result}:
      Observations have led to the understanding that the pre-stellar
      cloud-core MF is very similar to the proto-stellar MF and to the
      stellar IMF suggesting that gravitationally-driven instabilities
      in a turbulent medium may be the primary physical mechanism
      setting the shape of the IMF for stars in the mass range
      $0.1\simless m/M_\odot \simless \; {\rm few}$. Theoretical work
      has progressed significantly, but remains too inconsistent with
      observations to allow the conclusion that a theory of the IMF
      exists.  }}}  \vspace{2mm}

\section{Variation of the IMF}
\label{sec:IMFvar}

From the previous discussion it has emerged that the IMF appears to be
universal in resolved star forming systems as are found largely in the
vicinity of the Sun and in very nearby extra-galactic systems (LMC,
SMC, dSph satellites).

But the stellar IMF has been predicted theoretically to systematically
vary with star-forming conditions. This has been shown with Jeans-mass
arguments including SPH simulations and for self-regulated mass-growth
physics (Sec.~\ref{sec:introd_IMF}). Stellar populations formed from
triggered star formation in expanding shells have also been suggested
to be significantly top-heavy \citep{Dale11}.

In Sec.~\ref{sec:introd_IMF} the {\sc Variable IMF Prediction} is
emphasised as a robust result of IMF theory. Is there evidence
supporting this prediction?  Next some recently emerging observational
evidence for a dependency of the IMF on star-forming conditions is
presented which may be part of the long-expected violation of the {\sc
  Invariant IMF Hypothesis} (Sec.~\ref{sec:introd_IMF}).

\subsection{Trivial IMF variation through the $m_{\rm max}-M_{\rm
    ecl}$ relation}
\label{sec:mmaxIMFvar}

The existence of the $m_{\rm max}-M_{\rm ecl}$ relation
(Sec.~\ref{sec:stmass_clmass}) trivially implies that the stellar IMF
varies with increasing stellar mass, $M_{\rm ecl}$, of the population
formed in the star-formation event. This is best seen by the increase
of the average stellar mass with increasing $M_{\rm ecl}$ in contrast
to what is expected if the IMF were merely a probabilistic
distribution function (Fig.~\ref{fig:avmass}).

Note that the relatively small scatter of the observational $m_{\rm
  max}-M_{\rm ecl}$ data (Fig.~\ref{fig:mmaxf}) and the sharpness of
the distribution of IMF power-law indices (Fig.~\ref{fig:ahist}) may
be taken to imply that pure random sampling from the IMF is excluded
as a viable model for stellar populations
(cf. Figs~\ref{fig:optIMF}--\ref{fig:avmass}). This is further
supported by the lack of evidence for massive stars forming in
isolation (Sec.~\ref{sec:isolated}) and the lack of stars more massive
than a few$\,M_\odot$ in the Taurus-Auriga and Orion Nebula star
forming regions (p.~\pageref{box:igimfpred}).

\subsection{Variation with metallicity}
\label{sec:var_metal}

Differences in the metallicity, $Z$, of the populations sampled in the
nearby Local Group do not lead to discernible variations of the IMF
for massive stars as has been shown using star counts with
spectroscopic classification (Fig.~\ref{fig:kroupa_figmassey}).  Thus
the distribution of masses for massive stars has been interpreted to
not be significantly affected by the metallicity of the star-forming
gas.

That low-mass stars are forming together with massive stars in the
low-metallicity environment within the Magellanic Clouds with a MF
similar to the canonical IMF has been demonstrated through the deep
photometric surveying effort by Dimitrios Gouliermis and collaborators
(e.g. \citealt{GBH06, DGH09}). Detailed studies of star-formation in
the low-metallicity environment of the Small Magellanic Cloud is being
pushed by this team (e.g. \citealt{Gouliermis10}), but the work is
very challenging due to biases through crowding, resolution and mass
estimation from photometric data. Further, \cite{YKT08} find there to
be no measurable difference in system MFs down to $0.1\,M_\odot$
between the extreme outer Galactic disk and the inner, more metal-rich
regions. More metals lead to a dustier gas cloud and how the
characteristic dust grain size may affect the final characteristic
stellar mass has been studied by \cite{CB11}.

A metallicity effect for low-mass stars may however be uncoverable
from a detailed analysis of Milky Way star clusters. Present-day
star-forming clouds typically have somewhat higher metal-abundances
([Fe/H]$\approx+0.2$) compared to 5~Gyr ago ([Fe/H]$\approx-0.3$)
\citep{BM98} which is the mean age of the population defining the
canonical IMF. The data in the empirical alpha-plot
(Fig.~\ref{fig:apl}) indicate that some of the younger clusters may
have an individual-star MF that is somewhat steeper (larger
$\alpha_1$) than the canonical IMF {\it when unresolved binary-stars
  are corrected for}. This may mean that clouds with a larger [Fe/H]
produce relatively more low-mass stars which is tentatively supported
by the typically but not significantly flatter MFs in globular
clusters \citep{PZ99} that have [Fe/H]$\,\approx -1.5$, and the
suggestion that the old and metal-poor ([Fe/H]$=-0.6$) thick-disk
population also has a flatter MF below $0.3\,M_\odot$ with
$\alpha_1\approx0.5$ \citep{RR01}.  If such a systematic effect is
present, then for $m\simless 0.7\,M_\odot$ and to first order,
\begin{equation}
\alpha \approx 1.3 + \Delta\alpha\, {\rm[Fe/H]},
\label{eq:systemvar}
\end{equation}
with $\Delta\alpha \approx0.5$ \citep{K01a}.

Is this evidence supporting {\sc The Variable IMF Prediction}
(Sec.~\ref{sec:introd_IMF})? At the present Eq.~\ref{eq:systemvar}
needs to be taken as suggestive rather than conclusive
evidence. Measuring the stellar IMF for low-mass stars in metal-poor
environments, such as in young star-clusters in the Small Magellanic
Cloud \citep{GBH05,GBH06,SGD08,DGH09}, would thus be an important
goal.  In Sec.~\ref{sec:tophGC} we will uncover somewhat more robust
evidence for a variation of the stellar IMF towards top-heaviness with
decreasing metallicity but coupled to increasing density, possibly
yielding a for the first time uncovered variation of the overall form
of the IMF with metallicity in Eq.~\ref{eq:toph_imf_metal} \citep{Marks12}.

That metallicity does play a role is becoming increasingly evident in
the planetary-mass regime in that the detected exo-planets appear to
occur mostly around stars that are more metal-rich than the Sun
\citep{ZuMaz01,Vogt01}.

\subsection{Cosmological evidence for IMF variation}
\label{sec:cosmoev}

\citet{L98} invoked a \emph{bottom-light} IMF that is
increasingly deficient in low-mass stars the earlier the stellar
population formed, while for high-mass stars it is equal to the
canonical IMF at all times. This theoretical suggestion is motivated
by the decreasing ambient temperatures due to the expansion of the
Universe, implying a decrease of the Jeans mass in a star-forming gas
cloud and thus a decrease of the average mass of the stars with
decreasing redshift.  Such an IMF could explain the relative paucity
of metal-poor G-dwarfs in the Solar Neighbourhood (the G-dwarf
problem), compared to the predictions of a closed-box model for the
chemical evolution of galaxies. This is because, in this model,
low-mass stars form less frequently at times when the self-enrichment
of galaxies has not yet reached the current level.

Empirical evidence for a bottom-light IMF-variation was suggested by
\citet{VD08} in order to explain how the integrated colours of massive
elliptical (E) galaxies change with red-shift. The stellar populations
of E~galaxies are old, so that they have evolved passively most of the
time until the present day. Corrected for red-shift, the stellar
populations of E~galaxies are therefore bluer the more distant
(i.e. younger) they are.  \citet{VD08} finds however that the observed
reddening is less with decreasing redshift than can be accounted for
by stellar evolution.  This trend may be understood if the IMF were a
bottom light power-law with a characteristic mass $m_C\approx
2\,M_\odot$ and an index near $1\,M_\odot$ of $\alpha_2\approx1.3$
when E~galaxies formed, rather than the canonical $m_C\approx
0.1\,M_\odot, \alpha_2=2.3$.

But such an IMF appears to be in contradiction with the observed
near-canonical PDMFs of globular clusters (GCs). Just like E~galaxies,
GCs have formed their stellar populations at high
red-shifts. Nevertheless, GCs still have a large population of stars
that should not have formed in them, according to the model proposed
in \citet{VD08}. Thus, the IMF either did not have the time-dependency
suggested in \citet{VD08}, or the IMF in GCs was considerably
different from the one in E~galaxies (see the review by
\citealt{Bastian2010}). By studying various integrated
gravity-sensitive features in luminous E~galaxies, \cite{Cenarro03}
and \cite{VD10} on the other hand find evidence for a very
bottom-heavy IMF. \cite{Cenarro03} propose the IMF to vary according
to $\alpha=3.41 + 2.78 [{\rm Fe/H}] - 3.79 [{\rm Fe/H}]^2$.

At present it is unclear how these discrepant results may be brought
into agreement. The detailed dynamical analysis by \cite{Deasonetal11}
demonstrates that~E galaxies are consistent with a canonical IMF
rather than with a bottom-light IMF over a large range of masses. This
modelling assumes dark matter to provide the mass deficit and it is
unclear whether an entirely different approach not relying on the
existence of dynamically relevant dark matter would lead to a
different conclusion. A contradicting result is obtained by
\cite{GrilloGobat10} who show that for a sample of~13~E galaxies a
match is found between the dynamical masses and photometric (plus dark
matter) masses if the IMF was a Salpeter power-law, i.e. if it was
bottom-heavy relative to a canonical IMF
(Table~\ref{tab:Salpetererror}). Chemo-evolutionary population
synthesis models \citep{Vazdekis96, Vazdekis97} need an initially
top-heavy IMF which, after a short burst of star formation, becomes
bottom-heavy to explain the optical and near-infrared colours and line
indices of the most metal-rich E~galaxies.  It is unclear at this
stage why the dynamical and stellar population modelling of~E~galaxies
leads to such diverging results.

Other empirical evidence for a time-variability of the IMF was
proposed by \citet{Baugh2005} and \citet{Nagashima2005}.
\citet{Baugh2005} modelled the abundances of Lyman-break galaxies and
submillimetre galaxies. Both types of galaxies are considered to be
distant star-forming galaxies, but the latter are thought to be
obscured by dust that transforms the ultraviolet radiation from
massive stars into infrared radiation. The model from
\citet{Baugh2005} is based on galaxy formation and evolution via
accretion and merging according to $\Lambda$CDM-cosmology. They
include a detailed treatment of how the radiation from stars is
converted to dust emission. Their model only returns the correct
abundances of Lyman-break galaxies and sub-millimetre galaxies if they
assume two modes of star formation.  One of the two modes suggested by
\cite{Baugh2005} occurs with the canonical IMF and the other one is
with a \emph{top-heavy} IMF.  The mode with the top-heavy IMF is
thought to be active during star bursts, which are triggered by galaxy
mergers. Quiescent star formation in the time between mergers is
thought to be in the mode with the canonical
IMF. \citet{Nagashima2005} used the same model for galaxy evolution in
order to explain the element abundances in the intra-cluster medium of
galaxy clusters. They also need the two mentioned modes of star
formation in order to succeed. Noteworthy is that this two-mode IMF
model is qualitatively in natural agreement with the IGIMF theory
(Sec.~\ref{sec:comppop}). In the IGIMF theory the IGIMF becomes
top-heavy only in star bursts.

\cite{Wilkins08} note that the stellar mass density observed in the
local universe is significantly smaller than what would be expected by
integrating the cosmic star formation history within the standard
(dark-matter dominated) cosmological model.  They show that a single
top-heavy power-law IMF with high-mass star index $\alpha = 2.15$,
i.e. a smaller number of long-lived low-mass stars per massive star,
can reproduce the observed present-day stellar mass density.

This evidence for IMF-variations stands and falls with the validity of
$\Lambda$CDM cosmology. Modelling galaxy formation and evolution to be
consistent with observations is in fact a major problem of
$\Lambda$CDM cosmology. There are discrepancies in the Local Group
\citep{Kroupa10, Kroupa12} and in the Local Volume of galaxies
\citep{PN10} that already now shed major doubt as to the physical
applicability of the concordance cosmological model to the real world.
For instance, the actual galaxy population is less diverse than the
one $\Lambda$CDM theory predicts \citep{Disney2008}. It thus remains
to be seen whether the results by \citet{Baugh2005},
\citet{Nagashima2005} and \cite{Wilkins08} can be confirmed using a
refined model for galaxy evolution, or if instead a fundamentally
different cosmological model, which would yield different
redshift-time-distance relations, is required. It is noteworthy in
this context that evidence for top-heavy IMFs in high-star
formation-rate density environments has emerged, as is discussed in
the following sections.

\subsection{Top-heavy IMF in starbursting gas}
\label{sec:CR}

We have seen above that the IMF is largely insensitive to star-forming
conditions as found in the present-day Local Group. The observed
embedded clusters in these ``normal'' photon-dominated star-forming
regions have gas plus star densities $\rho\simless 10^5\,M_\odot\,{\rm
  pc}^{-3}$ ($M_{\rm ecl}\simless 10^{5.5}\,M_\odot$ within a
half-mass radius $r_{\rm h}\approx 0.5\,$pc). The objects form within
about~1~Myr leading to a star-formation rate density $SFRD \simless
0.1\,M_\odot\,{\rm pc}^{-3}\,{\rm yr}^{-1}$ for which no significant
IMF variation has been found, apart from the trivial variation through
the $m_{\rm max}-M_{\rm ecl}$ relation (Sec.~\ref{sec:mmaxIMFvar}). 

When the SFRD becomes very high, understanding the formation of individual
proto-stars becomes a challenge, as they are likely to coalesce before
they can individually collapse (Eq.~\ref{eq:collapset}) such that
probably a top-heavy IMF may emerge \citep{Dab10}. 

For high ambient SFRDs and a sufficiently long duration of the star
burst, \cite{Papadopoulos10} has shown that the cosmic rays (CRs)
generated by exploding supernovae of type~II (SNII) heat the clouds
which are too optically thick to cool therewith becoming CR dominated
regions within which the conditions for star formation are
significantly altered when compared to normal photon-dominated
star-forming regions.  This Papadopoulos-CR-mechanism raises the Jeans
mass and must lead to top-heavy IMFs in star-bursts
\citep{Papadopoulos11}.

This is also found by \cite{HocukSpaans11} who compute the IMF which
arises from star formation in a $800\,M_\odot$ cloud being irradiated
with X-rays and CRs as well as UV photons from a population of massive
stars and an accreting super-massive black hole (SMBH) at 10~pc
distance. CRs penetrate deep into the cloud therewith heating it,
while the ambient X-rays lead to a thermal compression of the cloud.
Their adaptive-mesh refinement computations with the FLASH code
include shear which opposes gravitational contraction through self
gravity of the cloud.  The turnover mass in the IMF increases by a
factor of~2.3 and the high-mass index becomes $\alpha\simless 2$ such
that the resultant IMF is top-heavy. Shear lessens the effects of the
CRs and X-rays by the IMF becoming bottom heavy, but only for the most
massive SMBHs. Low-mass SMBHs ($<10^6\,M_\odot$) or star-bursts
without massive BHs would thus lead to top-heavy IMFs as long as the
CR flux is significant.

This may well be the dominant physical mechanism why star-bursts have
a top-heavy IMF. Indeed, \cite{KSJ07} and \cite{HocukSpaans11}
demonstrate, respectively, with SPH and FLASH simulations, that under
warmer conditions a stellar mass-spectrum forms which is dominated by
massive stars. The Jeans mass can also be raised through turbulent
heating through expanding supernova shells if the explosion rate of
SNII is large enough, but details need to be worked out (Papadopoulos,
2011, private communication).

Observational evidence for top-heavy IMFs in regions of high SFRD have
now emerged for GCs (Sec.~\ref{sec:tophGC}) and UCDs
(Sec.~\ref{sec:tophUCD}). Top-heavy IMFs are also reported in the
central regions of Arp~220 and Arp~299 assuming SFRs of duration
longer than at least a few 0.1~Myr (the evidence becomes less
significant for a top-heavy IMF if the objects experienced a
star-burst of $\simless {\rm few}\,10^5\,$yr duration). The central
region of Arp~220 has a star-formation rate of $100\,M_\odot$/yr and
the large number of observed type~II supernovae requires a top-heavy
IMF (\citealt{Lonsdale06,Parraetal07}, see also
Sec.~\ref{sec:tophUCD}).  Noteworthy is that the Arp~220 central
region consists of what may be UCD-sized sub-structures
(Sec.~\ref{sec:tophUCD}). The high frequency and spatially confined
occurrence of supernovae in Arp~299 indicates that the star-formation
occurs in highly sub-clustered regions with dimensions less
than~0.4~pc and within larger structures of~30~pc scale
\citep{Ulvestad09}. This is reminiscent of the cluster-complex model
for the origin of some UCDs \citep{FK02, Bruens11}.  Likewise,
\cite{Perezetal09} and \cite{Anderson11} deduce a top-heavy IMF in
Arp~299 from the relative frequency of different supernova types.

\subsection{Top-heavy IMF in the Galactic centre}
\label{sec:tophGalCen}

Observations of the Galactic Centre revealed one or two disks of about
6~Myr old stars orbiting the central super-massive black hole (SMBH)
at a distance between 0.04~pc and 0.4~pc.  \cite{NS05} argued that the
X-ray luminosity of the Sgr A$^*$ field is too low to account for the
number of young $<3\,M_\odot$ stars expected from a canonical IMF,
considering the large number of O stars observed in the disks. The low
X-ray luminosity may be explained by a low-mass cut-off near
$1\,M_\odot$.  A more recent systematic search of OB stars in the
central parsec revealed a significant deficit of B-type stars in the
regime of the young disks, suggesting a strongly top-heavy IMF for
these disks with a best-fit power law of $\alpha_3 = 0.45\pm0.3$.
\citep{BMT09}. This appears to be the very best evidence for a truly
top-heavy IMF.

Using SPH simulations of star formation in fragmenting gas accretion
disks, \cite{BR08} find that the IMF of the disk stars can be bimodal
(and thus top-heavy) if the infalling gas cloud is massive enough
($\approx 10^5\,M_\odot$) and the impact parameter of the encounter
with the SMBH is as small as $\approx 0.1\,$pc. Thus, only under quite
extreme conditions will a top-heavy IMF emerge, whereby the strong
tidal forces and the rotational shear seem to be the dominant physical
mechanisms shaping the IMF since only cloud cores massive enough can
collapse to a star. This is explicitly calculated by
\cite{HocukSpaans11} for a range of SMBH masses (Sec.~\ref{sec:CR}).

Observations of the central parsec of the Milky Way show that this
region is dominated by a dense population of old stars with a total
mass of $\approx 1.5 \times 10^6\,M_\odot$.  This stellar cluster
around the SMBH has also been probed for evidence for a non-canonical
IMF.  By means of stellar evolution models using different codes,
\cite{LBK10} show that the observed luminosity in the central parsec
is too high to be explained by a long-standing top-heavy IMF as
suggested by other authors, considering the limited amount of mass
inferred from stellar kinematics in this region. In contrast,
continuous star formation over the Galaxy's lifetime following a
canonical IMF results in a mass-to-light ratio and a total mass of
stellar black holes (SBHs) consistent with the
observations. Furthermore, these SBHs migrate towards the centre due
to dynamical friction, turning the cusp of visible stars into a core
as observed in the Galactic Centre. It is thus possible to
simultaneously explain the luminosity and dynamical mass of the
central cluster and both the presence and extent of the observed core,
since the number of SBHs expected from a canonical IMF is just enough
to make up for the missing luminous mass.

In conclusion, observations of the Galactic Centre are well consistent
with continuous star formation following the canonical IMF.  Only the
centre-most young stellar disks between about~0.04 and 0.4~pc from the
SMBH show a highly top-heavy IMF, but the circumstances that led to
their formation must be very rare, since these have not affected most
of the central cluster.

\subsection{Top-heavy IMF in some star-burst clusters}
\label{sec:sbcl}

There are indications of top-heavy IMFs such as in some massive
star-burst clusters in the M82 galaxy. Using spectroscopy of the
unresolved M82-F cluster, \cite{SmGa01} derive, via the inferred
velocity dispersion, a mass and from the luminosity a mass-to-light
ratio that is significantly smaller than the ratio expected from the
canonical IMF for a 60~Myr population. The implication is that the
M82-F population may be significantly depleted in low-mass stars, or
equivalently it may have a top-heavy IMF, provided the velocity
dispersion is representative of the entire cluster. A possibility that
will have to be addressed using stellar-dynamical modelling of forming
star clusters is that M82-F may have lost low-mass stars due to tidal
shocking \citep{SmGa01}. Highly pronounced mass segregation which
leads to a dynamically decoupled central core of OB stars is an
important mechanism for reducing the measured mass-to-light ratio
\citep{Boily_etal05}, while rapid expulsion of residual gas from
forming clusters enhances the measured mass-to-light ratios
\citep{GB06}. But also a younger age would reduce the inferred
depletion in low-mass stars and some hints exist that M82-F might be
as young as 15 Myr \citep{MGV05}.

In an extensive literature study of Galactic and extragalactic
observations \cite{El05} concluded that dense star-forming regions
like starbursts might have a slightly shallower IMF, a view shared by
\cite{Eisenhauer01}.

\subsection{Top-heavy IMF in some globular clusters (GCs)}
\label{sec:tophGC}

Observations of~17 globular clusters (GCs) for which PDMFs were
measurable over the mass range $0.5-0.8\,M_\odot$, showed the
higher-metallicity GCs to have flatter PDMFs
\citep{Djorgovski93}. This lacked an explanation until now. In
particular, this trend is difficult to reconcile with standard
dynamical evolution scenarios as it is unclear how dynamics could
possibly know about the metal content of a cluster.

\cite{deMarchi2007} performed a deep homogeneous star-count survey
of~20 Milky Way GCs using the HST and VLT and measured the global
PDMFs in the mass range $0.3-0.8\,M_\odot$. They discovered the least
concentrated GCs to have a bottom-light PDMF, while the other GCs show
a canonical MF.  $N$-body calculations predict that two-body-encounter
driven dynamical evolution preferentially removes low-mass stars from
a star cluster \citep{VH97, BaumMakin03}. This occurs because two-body
relaxation drives the cluster into core collapse and energy
conservation leads to the expansion and thus evaporation of low-mass
stars.  Thus, the most concentrated star clusters are expected to show
the strongest depletion of low-mass stars, in disagreement with the
observations.

A possibility would be that the low-concentration GCs were formed with
a bottom-light IMF. However, there is no known theory of star
formation which could account for this: the low-concentration clusters
typically have a higher metallicity than the high-concentration
clusters (see below in this Sec.), and so the data would imply that
the IMF ought to have been bottom-light in the higher-metallicity GCs.
This however contradicts basic star-formation theory
(Sec.~\ref{sec:introd_IMF}).

This apparent disagreement between theory and observation can be
resolved if GCs formed mass-segregated and with the canonical IMF
(Eq.~\ref{eq:imf}) for stars less massive than about $1\,M_\odot$.
\cite{BaumDeMarchi08} performed $N$-body models using the Aarseth code
and demonstrate that the range of PDMFs observed by
\cite{deMarchi2007} can be arrived at if GCs were born mass segregated
and filling their tidal radii such that they do not need to first
evolve into core collapse and so they immediately begin loosing
low-mass stars.  However, the stellar-dynamically induced trend of
PDMF with concentration is not able to account for the observed
metallicity dependence. Furthermore, it is not clear why GCs ought to
be formed mass segregated but filling their tidal radii since all
known young clusters are well within their tidal radii.

An alternative scenario by \cite{Marks08} also assumes the young
compact GCs to be formed mass segregated with a canonical IMF
below~$1\,M_\odot$. But after formation the expulsion of residual gas
unbinds the low-mass stars that typically reside near the outer region
of the clusters, leading to flattening of the MF. This ansatz allows
for a metal-dependency, since the process of residual-gas expulsion is
expected to be enhanced for metal-richer gas which has a stronger
coupling to radiation than metal-poor gas, similar to the
metallicity-dependent stellar winds.  For each of the~20 GCs in the
\cite{deMarchi2007} sample, the best-fitting tidal-field strength,
radius, star-formation efficiency and gas-expulsion time-scale is
obtained. This uncovers remarkable correlations between the
gas-expulsion quantities, the tidal field strength and the
metallicities, allowing a very detailed reconstruction of the first
collapse phase of the Milky Way about 12~Gyr ago \citep{MK10}. The
correlations for example confirm the expectation that gas expulsion is
more efficient and thus dynamically more damaging in metal-richer gas,
and also that metal-poorer GCs were denser than their metal-richer
slightly younger counterparts which were subject to stronger tidal
fields.

This in-turn suggests that in order to provide enough feedback energy
to blow out the residual gas, the IMF had to be top-heavy in
dependence of the initial density of the GC \citep{Marks12}.  Assuming
the gas leaves a cluster with the velocity of the sound speed of
about~10~km/s the gas-expulsion time-scales for clusters with radii
between 0.5~and 1~pc would lie between~$\tau_{\rm gas}=0.05$
and~0.1~Myr.  The resultant high-mass IMF slopes derived for the GCs
from their individual PDMFs assuming $0.05\simless \tau_{\rm gas}/{\rm
  Myr}\simless 0.15$ cover a wide range, $0.9\simless \alpha_3
\simless 2.3$, where $\alpha_3$ is the slope for $m\simgreat
1\,M_{\odot}$, for an IMF that is canonical otherwise
(Fig.~\ref{fig:topheavy}).

The calculated IMF slopes also correlate with the metallicity of the
GCs, such that the PDMFs show the observed correlation after the
metallicity-depend\-ent gas expulsion process ends and the remaining GC
revirialises.  This correlation with metallicity, quantified in
Eq.~\ref{eq:toph_imf_metal} (Fig.~\ref{fig:a3_metal}), is an important
clue, as it implies that GCs formed from metal-poorer gas were more
compact and had a more top-heavy IMF, just as is indeed expected from
star-formation theory (Sec.~\ref{sec:introd_IMF} and~\ref{sec:theor}).

\vspace{2mm} 
\centerline{ \fbox{\parbox{\columnwidth}{ {\sc The Top-Heavy
        Stellar IMF / Metallicity Dependence}:  The suggested dependence of
      $\alpha_3$ on globular cluster-forming cloud metallicity can
      be parametrised as
\begin{equation}
          \begin{array}{l@{\;,\;}ll@{\;\;}l}
\alpha_3 = \alpha_2  
      &m > 1\, M_\odot  \quad \wedge        &{\rm [Fe/H]} \ge -0.5, \\
\alpha_3 = 0.66\,{\rm [Fe/H]} + 2.63
      &m> 1\, M_\odot \quad \wedge         &{\rm Fe/H]} < -0.5. 
          \end{array}
\label{eq:toph_imf_metal}
\end{equation}
}}}
\vspace{2mm}

\noindent
\cite{Strader11} observed high-resolution spectra of~200 GCs of the
Andromeda galaxy and discovered that the near-infrared M/L ratios
decrease significantly with increasing metallicity of the GCs. This
cannot be explained by secular dynamical evolution but follows from a
PDMF which is systematically more bottom-light for more metal rich
GCs. This is thus the same finding as discussed above for the~20 GCs
of the MW, and a possible explanation put forward by \cite{Strader11}
is metallicity-dependent gas expulsion.

\cite{Marks12} find that $\alpha_3$ decreases with increasing
pre-globular cluster cloud-core density, $\rho$
(Eq.~\ref{eq:toph_imf_density}, Fig.~\ref{fig:topheavy}). Such a trend
is to be expected theoretically if the massive and initially dense
GCs, each of which was a star-burst (star-formation rate density
SFRD$\simgreat 0.1\,M_\odot$/(yr~pc$^3$) for an initial half-mass
radius of about 0.5~pc and formation time-scale of about 1~Myr),
involves the merging of proto-stellar cores since the collision
probability is higher in denser systems (Eq.~\ref{eq:collapset}).

The high-mass IMF index $\alpha_3$ thus depends on $\rho$ and on
[Fe/H]. The increasing top-heaviness, deduced with a
principle-component-type analysis by \cite{Marks12}, with decreasing
[Fe/H] and increasing $\rho$ is quantified here for the first time in
Eq.~\ref{eq:toph_master} and Fig.~\ref{fig:toph_master}. It is
remarkable that the theoretically expected trend of the IMF with
$\rho$ and [Fe/H] has now emerged from elaborate stellar-dynamical
analysis of deep observations of GCs.

The extremely top-heavy IMFs for some of the GCs raise the question
whether they could survive the strong mass loss these IMFs imply due
to stellar evolution. This is especially an issue if clusters start
mass-segregated \citep{Vesperini09}. 

\subsection{Top-heavy IMF in UCDs}
\label{sec:tophUCD}

Further independently obtained evidence for top-heavy IMFs at high
SFRDs comes from ultra compact dwarf galaxies (UCDs) which have been
observed in nearby galaxy clusters. They typically have effective
radii of a dozen~pc, and are understood to have formed as a star-burst
(SFRD$\,\approx 1-100\,M_\odot$/(yr~pc$^3$) for an initial radius of
about 1~pc and formation time scale of 1~Myr, \citealt{DKB09}). The
Papadopoulos-CR-heating process (Sec.~\ref{sec:CR}) may be a factor
during the formation of UCDs. They are, like galaxies, collisionless
stellar-dynamical systems such that two-body relaxation driven
evaporation of low-mass stars is insignificant, contrary to the case
for GCs \citep{Anders09,MH11,FK11}.

For a significant sample of UCDs high resolution spectra are available
allowing estimates of their stellar velocity dispersions. These
velocity dispersions and the effective radii imply dynamical masses
between~$10^6$ and~$10^8 M_{\odot}$. The dynamical masses of UCDs are
thus similar to the values of the much more extended dwarf spheroidal
(dSph) galaxies. Combining the dynamical masses of UCDs with their
luminosities leads to estimates for their dynamical mass-to-light
(M/L) ratios. While the uncertainties of these estimates are large,
the most likely values for these M/L ratios are systematically higher
than compared to the expectation for a stellar population that formed
with the canonical IMF.

Dark matter is not a viable explanation for the enhanced M/L ratios of
UCDs, given the current understanding of how UCDs are formed. One such
idea is that UCDs evolve from star cluster complexes as are observed
in interacting systems like the Antennae (NGC 4038 and NGC 4039,
\citealt{Bruens11}).  The $300-500$~Myr old ultra-massive ``star
cluster'' W3 in the merger remnant galaxy NGC~7252 is indeed an object
that supports such a model \citep{Fel2005} since it is too young for
it to be a stripped nucleus of a nucleated dwarf galaxy.  Another idea
is that UCDs are extremely massive GCs
(e.g. \citealt{Mieske2002}). This notion is motivated by the
similarities that UCDs share with GCs, e.g. their seemingly continuous
mass-radius relation. Thus, there is supportive evidence for both
concepts, but they also imply that UCDs are essentially free of dark
matter. Also note that dSph galaxies are usually thought to populate
the least massive and therefore the most dense dark-matter
halos. However, the dark matter densities inferred from the dynamics
of dSph galaxies are still about two orders of magnitude too low to
influence the dynamics of UCDs to the required
extent\footnote{Adiabatic contraction \citep{Blu1986} in UCDs may
  however alleviate this problem but unlikely sufficiently so
  \citep{Mur2009}.}. This leaves a variation of the IMF as the most
natural explanation for the high dynamical M/L ratios of UCDs
\citep{DKB09}.

In an old stellar system like a UCD, both a \emph{bottom-heavy} IMF
and a \emph{top-heavy} IMF lead to a M/L ratio that exceeds the
expectation for the canonical IMF. In the case of a top-heavy IMF, the
M/L ratio of the stellar population is high only if it is old. This is
because in an old population, the massive (and therefore bright) stars
have turned into essentially non-luminous remnants. In the case of a
bottom-heavy IMF, the M/L ratio of the stellar population is enhanced
by the high M/L ratios of low-mass stars. The two cases are not easy
to distinguish by observations, simply because in either case a
population that is characterised by its low luminosity would have to
be detected. 

If the high M/L ratios of UCDs are caused by a bottom-heavy IMF then
this should be traceable for the highest M/L ratios by a
characteristic absorption feature in the spectra of low-mass stars
\citep{MieskeKr08}.

If, in contrast, the high M/L ratios of UCDs are the consequence of a
top-heavy IMF then this may be noticeable by the number of UCDs with
bright X-ray sources.  A system can form low-mass X-ray binaries
(LMXBs) by stellar-dynamical encounters between black holes and
neutron stars on the one hand, and main-sequence stars on the
other. When the main sequence star evolves its companion remnant may
accrete its envelope therewith becoming visible as an LMXB. Assuming a
canonical IMF it has been shown that the incidence of LMXBs in GCs
generally follows the expected correlation with GC mass.  UCDs however
turn out to be far over-abundant as X-ray sources. This overabundance
of UCDs as X-ray sources can be accounted for with the same top-heavy
IMF dependence on UCD mass as is obtained independently from matching
the M/L ratios. This constitutes a strong indication that UCDs formed
with top-heavy IMFs \citep{Dabringhausen11}.

\begin{figure}
\begin{center}
\rotatebox{0}{\resizebox{0.85 \textwidth}{!}{\includegraphics{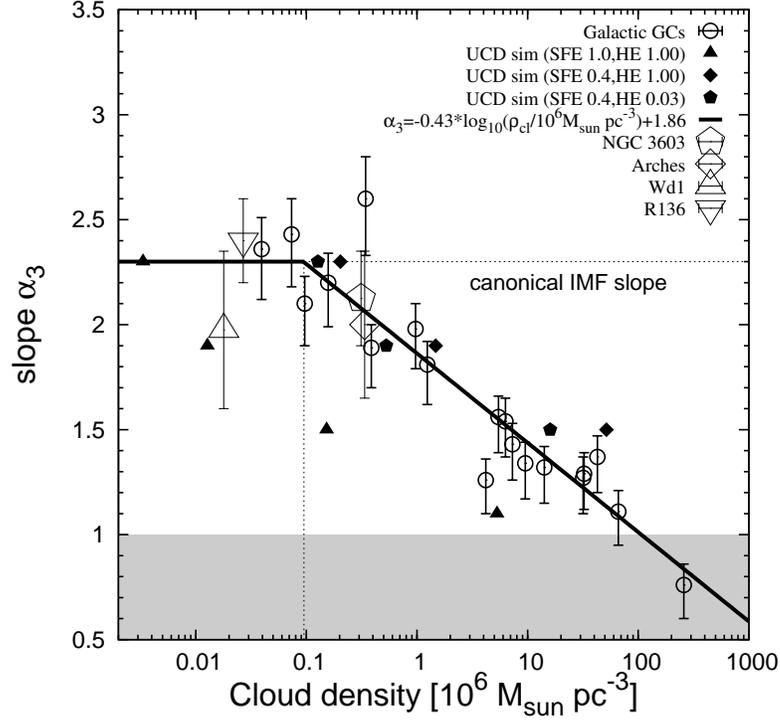}}}
\vskip 0mm
\caption[IMF of massive stars ($\alpha_3$) as a function of cloud
density]{\small{ The high-mass IMF slope, $\alpha_3$, deduced for
    Galactic GCs (open circles) as a function of the pre-GC cloud-core
    density within their initial half-mass radii
    (Sec.~\ref{sec:tophGC}).  Upper limits are for a gas-expulsion
    time-scale of $\tau_{\rm gas}=0.15\,$Myr, open circles are for
    $\tau_{\rm gas}=0.1\,$Myr and the lower limits are for $\tau_{\rm
      gas}=0.05\,$Myr.  Also plotted are MF slope values found in the
    literature for the massive, young clusters NGC~3603, Arches, Wd1
    and R136. Their corresponding cloud-core densities were calculated
    using their PD half masses within their PD half-mass radii
    ($0.2\,$pc for NGC~3603, $0.24\,$pc for Arches, $1\,$pc for Wd1,
    $1.1\,$pc for R136) assuming that the clusters have formed with a
    SFE of $1/3$, i.e. their gaseous progenitors were three times more
    massive and that their sizes did not change. The filled symbols
    correspond to simulations devoted to finding the most probable IMF
    slopes in systems of different density that lead, after residual
    gas expulsion and supernova driven evolution, to objects that
    resemble the properties of UCDs today
    (Sec.~\ref{sec:tophUCD}). Different symbols correspond to
    different input parameters (star formation efficiency SFE and
    heating efficiency HE; see Sec.~\ref{sec:tophUCD}). The overall
    trend for GCs and UCDs is consistent in the sense that denser
    systems form flatter IMFs.  The solid line is a fit to the GC data
    (Eq.~\ref{eq:toph_imf_density}). Below 1~$M_{\odot}$, the assumed
    IMFs are equal to the canonical IMF (Eqs.~\ref{eq:imf_mult}
    and~\ref{eq:imf}).  The gray shaded region at $\alpha_3 < 1$ are
    IMFs which contain more than 99~per cent mass in stars with
    $m>1\,M_\odot$ making cluster survival after supernova explosions
    unlikely.  Note that if the GCs and UCDs form on a time scale of
    1~Myr then their star-formation rate densities would be
    $0.1-100\,M_\odot$/(yr~pc$^3$). From \cite{Marks12}.}}
\label{fig:topheavy}
\end{center}
\end{figure}
\begin{figure}
\begin{center}
\rotatebox{0}{\resizebox{0.85 \textwidth}{!}{\includegraphics{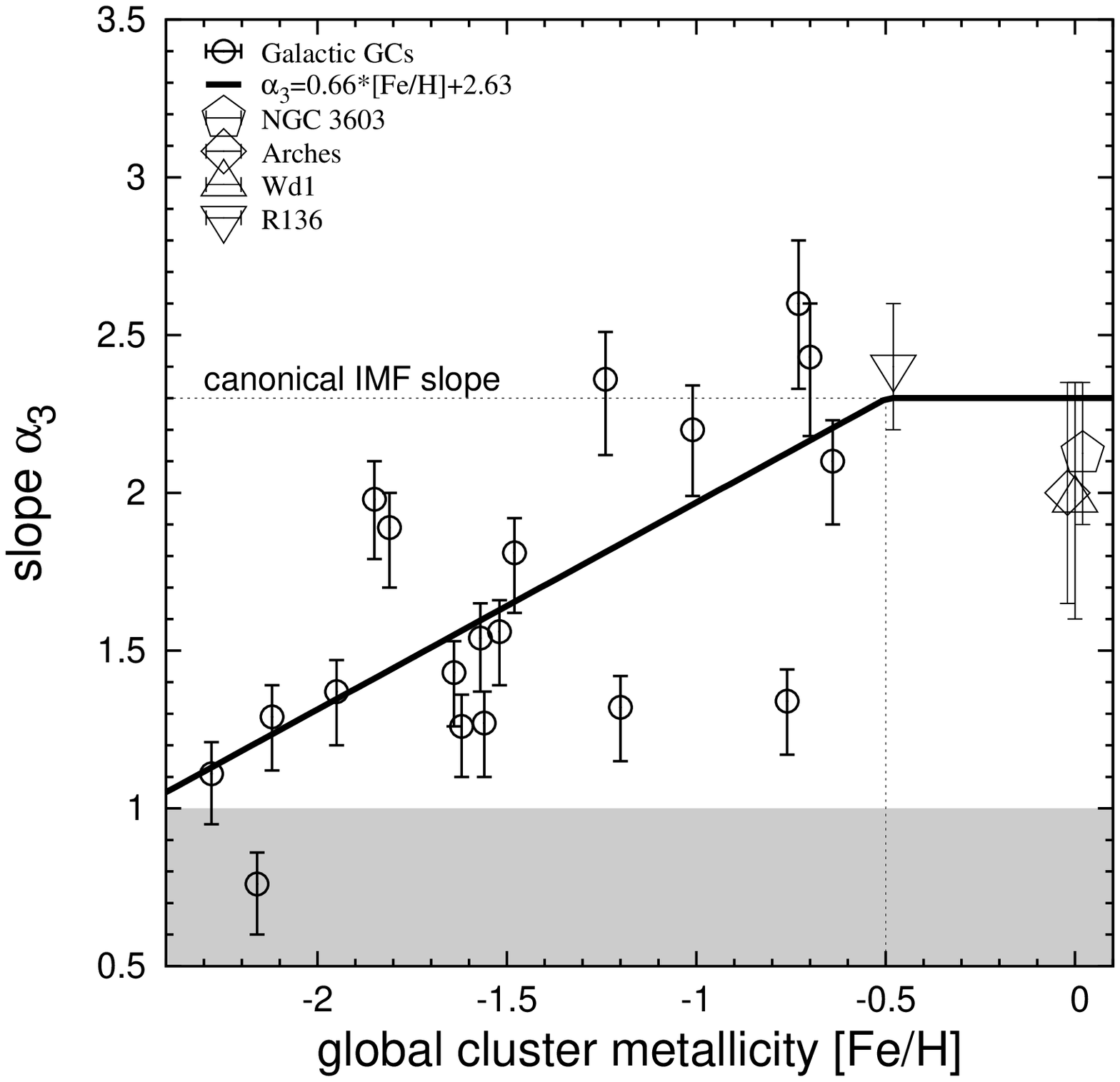}}}
\vskip 0mm
\caption[$\alpha_3$ as a function of metallicity]{\small{ The
    high-mass IMF slope, $\alpha_3$, deduced for Galactic GCs (open
    circles) as a function of the GC metallicity
    (Sec.~\ref{sec:tophGC}).  Otherwise as Fig.~\ref{fig:topheavy}.
  }}
\label{fig:a3_metal}
\end{center}
\end{figure}

The results reported in Sec.~\ref{sec:tophGC} on GCs and here on UCDs
are compared in the $\alpha_3$ vs birth-density diagramme
(Fig.~\ref{fig:topheavy}, Eq.~\ref{eq:toph_imf_density}). A remarkable
agreement of how $\alpha_3$ varies with birth density for GCs and UCDs
emerges.

Would UCDs with the deduced top-heavy IMF survive mass loss through
stellar evolution?  \citet{Dab10} calculate a set of numerical models
for the early dynamical evolution of UCDs with the canonical IMF and
top-heavy IMFs, using the particle-mesh code {\sc Superbox}.  They
assume that UCDs are hyper-massive star clusters, i.e. that their
stellar population formed in a starburst that took place in a dense
molecular cloud. A short formation time-scale of the UCDs is indeed
suggested by the enhanced $\alpha$-element abundances that
\cite{Evstigneeva2007} report for them. This implies that all massive
stars of a UCD evolve over a time-span of approximately 40~Myr,
i.e. the life-time of the least massive stars that become a type~II
supernova.

The main driver for the dynamical evolution of a UCD during this epoch
is the mass-loss through gas expulsion and supernova explosions. The
rate of this mass loss depends on a number of parameters: the rate at
which energy is deposited into the interstellar gas of the UCD, the
rate at which the interstellar gas is replenished by the ejecta from
type~II supernovae, the star formation efficiency (SFE, which sets how
much interstellar gas can be expelled aside from supernova ejecta) and
the heating efficiency (HE, which is the fraction of the energy
inserted into the interstellar medium that is not radiated away but
used up by removing gas from the UCD). This mass loss and the
consequences are quantified in \citet{Dab10} by comparing the energy
input by type~II supernova explosions and the radiation of stars over
a small time interval with the binding energy of the gas that is bound
to the UCD at that time, leading to individual mass-loss histories for
each of the considered UCD models. These mass-loss histories are
implemented into the code that is used to calculate their dynamical
evolution.

The mass loss is more pronounced the more top-heavy the IMF is because
more matter is set free by stars that evolve fast and the energy
deposition rates are high. As a result, UCDs with a very top-heavy IMF
dissolve because of heavy mass-loss, except for very high SFEs. Such
high effective SFEs may be realistic because the matter lost from
stars will accumulate within the UCD and may form new stars
\citep{PAK09, Wuensch11}.  In that case, the UCD models with a
high-mass IMF slope close to $\alpha_3=1$ evolve into objects that
resemble an observed UCD at the end of the calculation. For moderate
SFEs, the models in \citet{Dab10} evolve into UCD-like objects if they
have either the canonical IMF or a moderately top-heavy IMF with a
high-mass slope in between~1.5 and~2. However, as shown above, the
case of the canonical IMF can be excluded due to the high M/L ratios
of and high LMXB-occurrences in UCDs.  The UCD models that are
consistent with the constraints set by observed UCDs also have IMFs
and initial densities that are remarkably close to the high-mass IMF
slope versus initial cloud densities derived for Galactic GCs
(Sec.~\ref{sec:tophGC}) based on their PDMFs, as can be seen in
Fig.~\ref{fig:topheavy}.

The variation of the IMF with cluster-forming cloud density can be
summarised as follows:

\vspace{2mm} \centerline{ \fbox{\parbox{\columnwidth}{ {\sc The Top-Heavy
        Stellar IMF / Density Dependence}: Resolved stellar
      populations show an invariant IMF (Eq.~\ref{eq:imf}), but for
      $SFRD \simgreat 0.1\,M_\odot$/(yr~pc$^3$) the IMF becomes
      top-heavy, as inferred from UCDs and some GCs. The dependence of
      $\alpha_3$ on cluster-forming cloud density $\rho$ (stars plus
      gas) can be parametrised for $m>1\,M_\odot$ as \citep{Marks12}
\begin{equation}
          \begin{array}{l@{\;\;}l@{\;\;}l}
\alpha_3 = \alpha_2,  
  \quad        &\rho < 9.5\times 10^4\,M_\odot{\rm /pc}^3, \\
\alpha_3 = 1.86-0.43\,{\rm log}_{10}\left(\rho_{\rm cl}/
                                               (10^6\,M_\odot\,{\rm pc}^{-3})\right),
  \quad       &\rho \ge 9.5\times 10^4\,M_\odot{\rm /pc}^3. 
          \end{array}
\label{eq:toph_imf_density}
\end{equation}
}}}
\vspace{2mm}

\noindent This IMF is in good agreement with the supernova rate
observed in Arp~220 and Arp~299 (Sec.~\ref{sec:CR}). Note that for $m
\le 1\,M_\odot$ the IMF is canonical (Eq.~\ref{eq:imf}). Note also
that the top-heavy IMF is a fit to the GC constraints, and that this
provides a good description of the independent constraints arrived at
by the UCDs.

\subsection{The current state of affairs concerning IMF variation with
  density and metallicity and concerning theory}
\label{sec:currentstate}

The results achieved over the past decade in self-consistent
gravo-hydro\-dynami\-cal modelling of star formation in turbulent
clouds with the SPH and FLASH methods have been very successful
(Sec.~\ref{sec:stmass_clmass}). This is evident in the overall
reproduction of the stellar IMF as well as of the $m_{\rm max}-M_{\rm
  ecl}$ relation allowing detailed insights into the physics driving
the growth of an ensemble of stars forming together in one CSFE (see
{\sc Definitions} on p.~\pageref{box:definitions}) with and without
feedback. Also, computations with these same techniques of the
fragmentation of circum-proto-stellar disks lead to an excellent
agreement with the BD IMF.  These simulations have not yet been
able to reproduce the birth binary-star properties
(Sec.~\ref{sec:bins}) which may be due to as yet necessarily
inadequate inclusion of the various feedback processes but also
because the smoothing length and sink-particle radius required in
every SPH simulation limits the binary-orbital resolution to at least
a few~100$\,$AU.  Computational star-formation has also ventured into
the difficult terrain of including magnetic fields and radiative
feedback finding a certain degree of compensating effects in terms of
the emerging stellar masses with a significantly reduced
star-formation efficiency (Sec.~\ref{sec:theor}).

Following on from above, the current situation of our understanding of
IMF variations may be described as follows
(Sec.~\ref{sec:introd_IMF},~\ref{sec:theor}): Theory has, over
decades, robustly predicted the IMF to vary with star-forming
conditions such that metal poor environments and/or warmer gas ought
to lead to top-heavy IMFs.  Observations and their interpretation
including corrections for biases have, on the other hand, been
indicating the IMF to be invariant, this being the consensus reached
by the community as mitigated in most reviews. The suggestions for
top-heavy IMFs in star bursts (e.g. \citealt{El05, Eisenhauer01}) were
typically taken to be very uncertain due to the evidence stemming from
distant unresolved and hard-to-observe star-forming systems.

Now theoretical work has set itself the task of explaining the
invariance, and various suggestions have been made: \cite{Bate2005b,
  BCB06, EKW08, Bate2009, Krumholz10, Myers11}. But evidence in favour
of IMF variations has proceeded to come forth.

Perhaps the first tentative indication from resolved stellar
populations for a possible change of the IMF towards a bottom-heavy
form with increasing metallicity of the star-forming gas has emerged
through the analysis of recent star forming events
(Eq.~\ref{eq:systemvar}). This evidence is still suggestive rather
than conclusive. And even if true it is very difficult to extract any
IMF-variation because of combinations of the following issues that
mask IMF variations as they typically act randomisingly:

\vspace{2mm} \centerline{ \fbox{\parbox{\columnwidth}{ {\sc Masking IMF
        variations}:
      \begin{itemize} \item Major uncertainties in pre-main sequence
        stellar evolution tracks (footnote~\ref{foot:pms} on
        p.~\pageref{foot:pms}). \item The loss of low-mass stars from
        young, intermediate and old open clusters through residual gas
        expulsion and secular evolution. \item Different evolutionary
        tracks of initially similar star clusters subject to different
        tidal fields. \item By observing the outcome of current or
        recent star formation we are restricted to it occurring under
        very similar physical conditions. \item By using the canonical
        IMF as a bench mark, IMF variations become more difficult to
        unearth: the variation about the mean is smaller than the
        difference between the extrema. \item The fossils of
        star-forming events that were very different to our currently
        observationally accessible ones are given by Galactic GCs and
        dSph satellite galaxies. But given their typical distances,
        the PDMFs were not reliably measurable below about
        $0.5\,M_\odot$. The evidence for or against variations of the
        MF has thus been limited to the mass range $0.5-0.8\,M_\odot$.
\end{itemize}
\label{box:masking} }}} \vspace{2mm}

The landmark paper by \cite{deMarchi2007} (Sec.~\ref{sec:tophGC}) for
the first time provided unambiguous evidence for a systematically
changing global PDMF in Galactic GCs. This break-through became
possible because the PDMF could be measured down to about
$0.3\,M_\odot$ with the HST and VLT for a homogeneous sample of~20 GCs
giving us a greater leverage on the PDMF and the dynamical history of
the GCs. It emerges that the only model able to account for the
observed variation and its correlation with metallicity is one in
which the IMF becomes increasingly top-heavy with increasing density
(Eq.~\ref{eq:toph_imf_density}) and decreasing metallicity
(Eq.~\ref{eq:toph_imf_metal}) in a gas-expulsion scenario.  A
gas-expulsion origin is independently also suggested as an explanation
for the metallicity--M/L anti-correlation which follows from
a high-resolution analysis of 200~GCs of the Andromeda galaxy
\citep{Strader11}.

Furthermore, the modern observations of UCDs have lead to the
discovery that they typically have somewhat elevated $M/L$ ratia which
can be explained by a top-heavy IMF systematically changing with UCD
mass. The enhanced frequency of UCDs with LMXBs can also be explained
with a top-heavy IMF systematically changing with UCD mass. And, both
independent results on how the IMF varies with UCD mass agree
(Sec.~\ref{sec:tophUCD}). Last not least, the deduced variation of the
IMF in UCDs is in good agreement with the variation of the IMF as
deduced from the de Marchi, Paresce \& Pulone GCs, when expressed in
terms of the density of the star-forming cloud. Notwithstanding this
agreement, the arrived at variation of the IMF is also consistent with
the top-heavy IMF suggested in star-bursting systems such as Arp~220
and Arp~299 (Sec.~\ref{sec:CR}, \ref{sec:tophUCD}). The evidence for
top-heavy IMFs thus comes from GCs and UCDs that had SFRDs higher than
any other known stellar-dynamical system including elliptical galaxies
which had global $SFRD \simless 4\times 10^{-9}\,M_\odot\,{\rm
  pc}^{-3}\,{\rm yr}^{-1}$ for radii of about 5~kpc and formation
time-scales of about~0.5~Gyr.

\begin{figure}
\begin{center}
\rotatebox{0}{\resizebox{0.85 \textwidth}{!}{\includegraphics{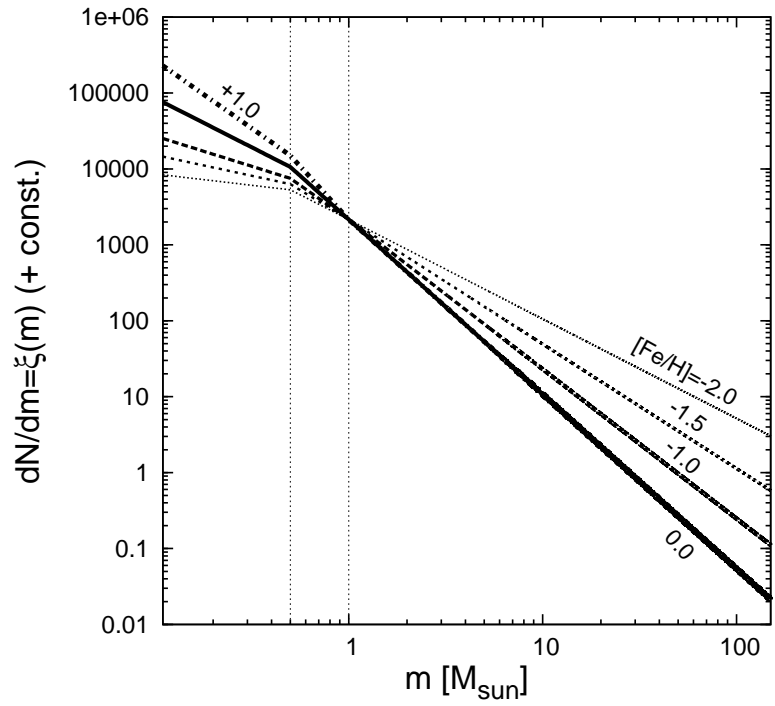}}}
\vskip 0mm
\caption[Variation of the IMF in dependence of metallicity]{\small{
    The variation of the stellar IMF with metallicity
    (Eq.~\ref{eq:systemvar} and~\ref{eq:toph_imf_metal}).  The IMFs
    are normalised to agree at $1\,M_\odot$. Note that the figure is
    based for $m<1\,M_\odot$ on an extrapolation of
    Eq.~\ref{eq:systemvar} below about [Fe/H]$\approx -0.5$.  Is this
    the long sought after systematic variation of the stellar IMF with
    metallicity?  From \cite{Marks12}.}}
\label{fig:varIMFmetal}
\end{center}
\end{figure}

The thus inferred systematic variation of the stellar IMF with
metallicity is documented in Fig.~\ref{fig:varIMFmetal}. 
Note that there is an implicit dependence of the IMF on the density of
the star-forming cloud (Eq.~\ref{eq:toph_imf_density}).  The
metallicity and density dependencies are correlated because metal
poorer gas clouds may collapse to larger pre-cluster densities than
metal rich clouds which fragment earlier and into smaller masses.

\begin{figure}
\begin{center}
\rotatebox{0}{\resizebox{0.85 \textwidth}{!}{\includegraphics{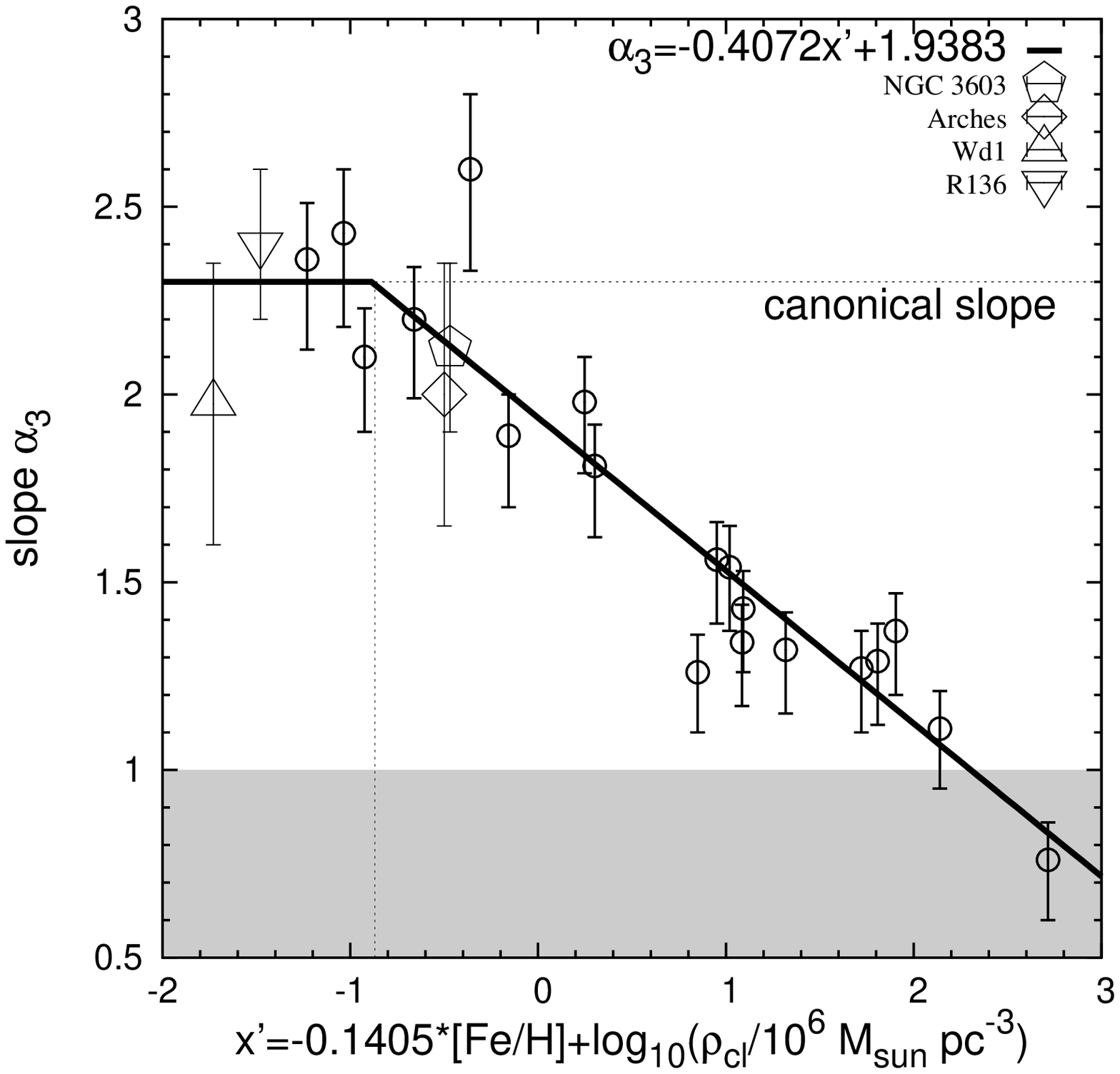}}}
\vskip 0mm
\caption[$\alpha_3$ as a function of density and metallicity]{\small{
    The variation of the stellar IMF with metallicity and cloud
    density (Eq.~\ref{eq:toph_master}) as deduced from deep
    observations of MW GCs using a principal-component-type analysis.
    From \cite{Marks12}. Otherwise as Fig.~\ref{fig:topheavy}.  }}
\label{fig:toph_master}
\end{center}
\end{figure}

This correlation is evident in the dependency of $\alpha_3$ on $\rho$
and [Fe/H] calculated by \cite{Marks12}
using a principle component-type analysis of the GCs
discussed in Sec.~\ref{sec:tophGC}. It is formulated in
Eq.~\ref{eq:toph_master} and plotted in Fig.~\ref{fig:toph_master}. 

\vspace{2mm} \centerline{ \fbox{\parbox{\columnwidth}{ {\sc The Stellar IMF
        Dependence on Density and Metallicity}: Resolved stellar
      populations show an invariant IMF (Eq.~\ref{eq:imf}), but for
      $SFRD \simgreat 0.1\,M_\odot$/(yr~pc$^3$) the IMF becomes
      top-heavy, as inferred from deep observations of GCs. The
      dependence of $\alpha_3$ on cluster-forming cloud density,
      $\rho$, (stars plus gas) and metallicity, [Fe/H],can be
      parametrised as
\begin{equation}
          \begin{array}{l@{\;\;}ll@{\;\;}l}
\alpha_3 = \alpha_2,  
      &m> 1\,M_\odot  \quad \wedge        &x < -0.89, \\
\alpha_3 = -0.41 \times x + 1.94,
      &m> 1\,M_\odot \quad \wedge         &x\ge -0.89,\\[3mm]
x = -0.14 \,{\rm [Fe/H]} + 0.99\, {\rm log}_{10}\left( \rho/ \left(10^6
    \, M_\odot \, {\rm pc}^{-3}\right) \right).
          \end{array}
\label{eq:toph_master}
\end{equation}
}}}
\vspace{2mm}

How does this suggested dependence of the IMF on metallicity compare
with the observational exclusion of any metallicity dependence in the
Local Group (Fig.~\ref{fig:kroupa_figmassey})? The SMC has
[Fe/H]$\approx -0.6$ such that on average $\alpha_3\approx 2.1$ (using
Eq.~\ref{eq:toph_imf_metal}) which is consistent with the SMC datum in
Fig.~\ref{fig:kroupa_figmassey}. For the LMC [Fe/H]$\approx -0.4$ such
that on average $\alpha_3\approx 2.3$ which is consistent with the LMC
data in Fig.~\ref{fig:kroupa_figmassey}.  Thus, the IMF variation
deduced from GCs and UCDs are easily accommodated by the Local Group
data.

On galaxy scales, tentative evidence for IMF variations has begun to
emerge in 2003 with the prediction that the IMF in a whole galaxy
ought to be steeper (top-light) if the stellar IMF were invariant
(Sec.~\ref{sec:comppop}). This (unwantingly) resolved the
long-standing but mostly ignored discrepancy between the canonical IMF
index $\alpha=2.3$ and the Scalo IMF index for the MW field
($\alpha\approx 2.7$), and generalisation to galaxies of different
type lead to an immediate understanding of the
metallicity--galaxy-mass relation and other systematic effects.

The first observational evidence for a systematic change of the
galaxy-wide IMF indeed appeared in~2008 and~2009
(Sec.~\ref{sec:igimf_pred} below).

\vspace{2mm} \centerline{ \fbox{\parbox{\columnwidth}{ {\sc In Conclusion} of
      this Sec.~\ref{sec:IMFvar}: Recent research has begun to uncover
      increasing evidence for a possible systematic variation of the
      IMF in agreement with the theoretical expectations, but this
      evidence requires verification by further observational work.
      At present the here quantified systematic variation of the IMF
      is more of a suggestion than established fact, but this
      suggestion may give a framework and a target for future work.
    }}} \vspace{2mm}

\section{Composite Stellar Populations - the IGIMF}
\label{sec:comppop} 

The IMF in individual pc(cluster)-scale star-forming events
ranging in mass from a few~$M_\odot$ up to $10^8\,M_\odot$ is
reasonably well constrained, as the previous sections have shown.
Integrated galaxy-wide properties, on the other hand, depend on the
galaxy-wide content of all newly formed stars, i.e. on the composition
of all collective star formation events (CSFEs. i.e. embedded star
clusters, Sec.~\ref{sec:intro}) in a galaxy. 

Following philosophical {\sc Approach~A} (Sec.~\ref{sec:philosophy}),
it has usually been assumed that the galaxy-wide IMF is identical to
the canonical stellar IMF which is established on the pc-scale events.
This is based on the simplest-of-all assumptions that the stellar
distribution is sampled purely randomly from the invariant IMF,
i.e. that the IMF is a probabilistic density distribution function
(e.g. \citealt{El04}).  Thus, for example, $10^5$ clusters, each
of~$10\,M_\odot$, would have the same composite (i.e. combined) IMF as
one cluster with mass $10^6\,M_\odot$. This assumption, while being
simple, has important implications for the astrophysics of
galaxies. For example, luminosities such as the H$\alpha$ flux would
scale linearly with the SFR leading to the much discussed
Kennicutt-Schmidt star-formation law, $\Sigma_{\rm SFR} \propto
\Sigma_{\rm gas}^N$ with $N\approx 1.5$, where $\Sigma_{\rm SFR}$ and
$\Sigma_{\rm gas}$ are the SFR surface density and gas-mass surface
density, respectively \citep{Kennicutt1994, Kennicutt08}.

The existence of the physical $m_{\rm max} - M_{\rm ecl}$ relation
(Sec.~\ref{sec:stmass_clmass}) has, on the other hand, profound
consequences for {\it composite populations}.  It immediately implies,
for example, that $10^{5}$ clusters, each weighing $10\,M_\odot$, {\it
  cannot} have the same composite (i.e. combined) IMF as one cluster
with $10^6\,M_\odot$, because such small clusters can never make stars
more massive than about $2.5\,M_\odot$ (Fig.~\ref{fig:mmaxf}).  And
since low-mass clusters are far more numerous than massive clusters,
galaxies would have steeper composite, or integrated galactic IMFs
(IGIMFs), than the stellar IMF in each individual cluster
(\citealt{KW03}, also hinted at independently by
\citealt{Vanbev82}). Furthermore, massive-star-sensitive galaxy
luminosities would not scale linearly with the SFR leading to a
significant revision of the H$\alpha$--SFR relation with corresponding
major implication for the galaxy-wide star-formation law ($N=1$
instead of~1.5, Eq.~\ref{eq:sflaws} below).

This is indeed supported to be the case by the theory of star
formation (see the {\sc BVB Conjecture} on p.~\pageref{quote:bvb})
which implies {\sc Optimal Sampling} to possibly be closer to reality
than the purely probabilistic IMF approach.

\subsection{IGIMF basics}

The galaxy-wide IMF, the {\it integrated galactic IMF}, is the sum of
all the stellar IMFs in all CSFEs formed over a time span $\delta
t$. While this is a next to trivial concept, it turns out to be
extremely powerful in particular when its foundation is sought in {\sc
  Approach~B} (Sec.~\ref{sec:philosophy}), i.e. in {\it Optimal
  Sampling} (Sec.~\ref{sec:optsamp}).  The IGIMF is therefore the
integral over the embedded cluster MF (ECMF, $\xi_{\rm ecl}$):

\vspace{2mm}
\noindent\fbox{\parbox{\columnwidth}{
{\sc Definition}: The IGIMF is an integral over all
star-formation events in a given star-formation ``epoch'' $t, t+\delta t$,

\begin{equation} 
\xi_{\rm IGIMF}(m;t) = 
\int_{M_{\rm ecl,min}}^{M_{\rm
ecl,max}(SFR(t))} \xi\left(m\le m_{\rm max}\left(M_{\rm
ecl}\right)\right)~\xi_{\rm ecl}(M_{\rm ecl})~dM_{\rm ecl},
\label{eq:igimf_t}
\end{equation}
with the normalisation conditions eqs.~\ref{eq:maxMecl} and~\ref{eq:SFR}
\[
M_{\rm ecl}-m_{\rm max}(M_{\rm ecl}) = \int_{0.07\,M_\odot}^{m_{\rm
    max}(M_{\rm ecl})} m'\,\xi(m')\,dm',
\]
\[
1=\int_{m_{\rm max}(M_{\rm ecl})}^{m_{\rm max *}}\,\xi(m')\,dm',
\]
which together yield the $m_{\rm max}-M_{\rm ecl}$ relation
(Eq.~\ref{eq:Mecl}).  }}

\vspace{2mm}

\noindent Here $\xi(m\le m_{\rm max})~\xi_{\rm ecl}(M_{\rm
  ecl})~dM_{\rm ecl}$ is the stellar IMF contributed by $\xi_{\rm
  ecl}~dM_{\rm ecl}$ CSFEs with stellar mass in the interval $M_{\rm
  ecl}, M_{\rm ecl}+dM_{\rm ecl}$.  The ECMF is often taken to be a
power-law,
\begin{equation}
\xi_{\rm ecl}(M_{\rm ecl}) \propto M_{\rm ecl}^{-\beta},
\label{eq:ECMF}
\end{equation}
with $\beta \approx 2$ \citep{Lada_Lada03}, whereby an ``embedded
cluster'' is taken here to be a CSFE and not a gravitationally bound
star cluster (see {\sc Definitions} on
p.~\pageref{box:definitions}). $M_{\rm ecl,max}$ follows from the
maximum star-cluster-mass {\it vs}
global-star-formation-rate-of-the-galaxy relation,
\begin{equation}
M_{\rm ecl,max}=8.5\times 10^4\;\left({ {\rm SFR} \over M_\odot/{\rm
      yr} }\right)^{0.75},
\label{eq:maxcl_sfr}
\end{equation}
(Eq.~1 in \citealt{WK05b}, as derived by \citealt{WKL04} using
observed maximum star cluster masses). A relation between $M_{\rm
  ecl,max}$ and $SFR$, which is a good description of the empirical
data, can also be arrived at by resorting to {\sc Optimal
  Sampling}. It follows by stating that when a galaxy has, at a time
$t$, a $SFR(t)$ over a time span $\delta t$ over which an optimally
sampled embedded star cluster distribution builds up with total mass
$M_{\rm tot}(t)$, then there is one most massive CSFE,
\begin{equation}
1 = \int_{M_{\rm ecl, max}(t)}^{M_{\rm U}} \xi_{\rm ecl}(M_{\rm
  ecl}')\,dM_{\rm ecl}',
\label{eq:maxMecl}
\end{equation}
with $M_{\rm U}$ being the physical maximum star cluster than can form
(for practical purposes $M_{\rm U}>10^8\,M_\odot$), and
\begin{equation}
SFR(t) = {M_{\rm tot}(t) \over \delta t} = {1\over \delta t}
\int_{M_{\rm ecl, min}}^{M_{\rm ecl,max}(t)} M_{\rm ecl}'\,\xi_{\rm
  ecl}(M_{\rm ecl}')\,dM_{\rm ecl}'.
\label{eq:SFR}
\end{equation}
$M_{\rm ecl,min}\,=\,5\,M_{\odot}$ is adopted in the standard
modelling and corresponds to the smallest ``star-cluster'' units
observed (the low-mass sub-clusters in Taurus-Auriga in
Fig.~\ref{fig:mmaxf}).

\citet{WKL04} define $\delta t$ to be a ``star-formation epoch'',
within which the ECMF is sampled optimally, given a SFR. This
formulation leads naturally to the observed $M_{\rm ecl,max}(SFR)$
correlation if the ECMF is invariant, $\beta\approx2.35$ and if the
``epoch'' lasts about $\delta t=10$~Myr.  Thus, the embedded cluster
mass function is optimally sampled in about 10~Myr intervals,
independently of the SFR. This time-scale is nicely consistent with
the star-formation time-scale in normal galactic disks measured by
\citet{Egusa_etal04} using an entirely independent method, namely from
the offset of HII regions from the molecular clouds in spiral-wave
patterns. In this view, the ISM takes about 10~Myr to transform via
molecular cloud formation to a gas-free population of dispersing young
simple stellar populations.  

\begin{figure}
\begin{center}
\rotatebox{0}{\resizebox{0.9\textwidth}{!}{\includegraphics{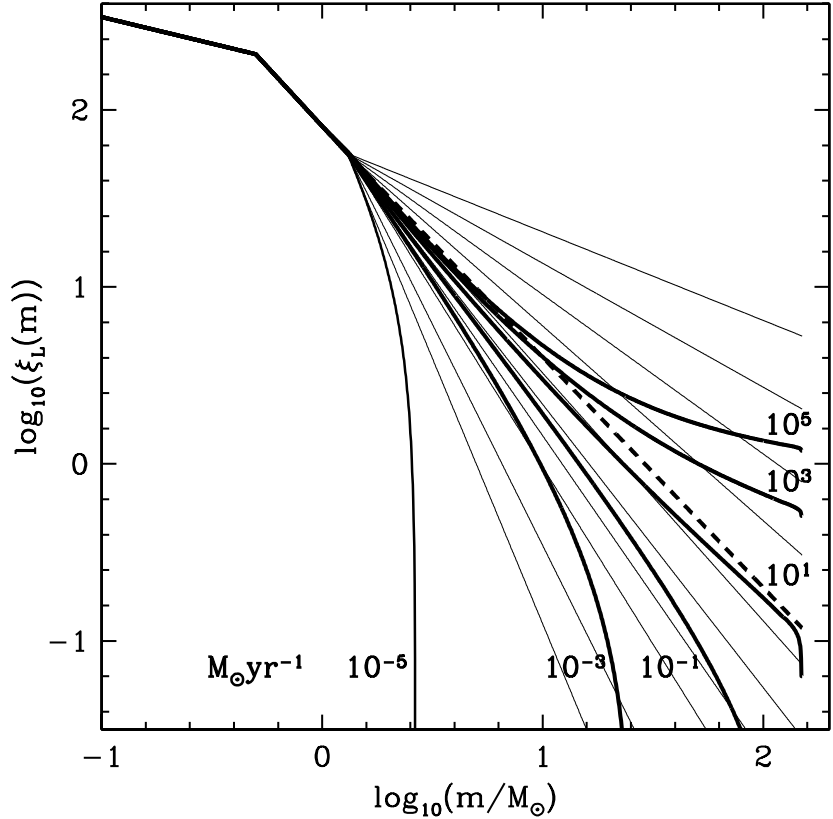}}}
\vskip 0mm
\caption[IGIMF as function of the SFR]{\small{The dependence of the
    logarithmic IGIMF (Eq.~\ref{eq:igimf_t}) on the SFR of a galaxy.
    The IGIMF is normalised by the total number of stars such that it
    does not change visibly at low stellar masses in this plot.  This
    IGIMF has been computed by adopting the canonical IMF which
    becomes top-heavy at embedded-star-cluster densities $\rho >
    10^5\,M_\odot/({\rm yr}\,{\rm pc}^3)$
    (Eq.~\ref{eq:toph_imf_density}), an ECMF with $\beta=2, M_{\rm
      ecl,min}=5\,M_\odot$ and the semi-analytical $m_{\rm max}-M_{\rm
      ecl}$ relation (Eq.~\ref{eq:mmax}).  For a given $M_{\rm ecl}$,
    $\rho= ({1\over2}\,M_{\rm ecl}/SFE) \, 3/(4\,\pi\,r_{0.5}^3)$ is
    the cloud (stellar plus gas) density, whereby a star-formation
    efficiency of $SFE=1/3$ and initial half-mass radius
    $r_{0.5}=0.5\,$pc are assumed \citep{MK10,Dab10,Marks12a}.  The
    thin lines are IMFs with different power-law indices, $\alpha'$,
    for $m>1.3\,M_\odot$ (the IGIMF is identical to the canonical IMF,
    Eq.~\ref{eq:imf}, below this mass). $\alpha'=1.5, 1.7, 1.9, 2.1,
    2.3, 2.4, 2.6, 2.8, 3.0, 3.5, 4.0$ (top to bottom), whereby the
    canonical value $\alpha'=2.3=\alpha_3$ is shown as the thick
    dashed line. Thus, for example, the IGIMF has $1.9 < \alpha' <
    2.1$ when $SFR=10\,M_\odot/{\rm yr}$. }}
\label{fig:igimf}
\end{center}
\end{figure}

The time-integrated IGIMF then follows from
\begin{equation} 
\label{eq:igimf}
\xi_{\rm IGIMF}(m) = \int_0^{\tau_{\rm G}} \frac{\xi_{\rm IGIMF}(m;t)}{\delta t}\,dt,
\end{equation}
where $\tau_{\rm G}$ is the age of the galaxy under scrutiny.  The
time-integrated IGIMF, $\xi_{\rm IGIMF}(m)$, is the stellar IMF of all
stars ever to have formed in a galaxy, and can be used to estimate the
total number of supernovae ever to have occurred, for
example. $\xi_{\rm IGIMF}(m;t)$, on the other hand, includes the
time-dependence through a dependency on $SFR(t)$ of a galaxy and
allows one to compute the time-dependent evolution of a stellar
population over the life-time of a galaxy, e.g. its instantaneous
population of massive stars (Fig.~\ref{fig:igimf}).  Note that
\begin{equation}
\xi_{\rm IGIMF}(m)=\left(\tau_{\rm G} \over \delta t \right)\,
\xi_{\rm IGIMF}(m;t) \quad {\rm if}\; SFR(t)=\;{\rm const}.
\end{equation}

\subsection{IGIMF applications, predictions and observational verification}
\label{sec:igimf_pred}

Since stellar clusters with larger masses are observed to form at
higher SFRs (Eq.~\ref{eq:maxcl_sfr}), the ECMF is sampled to larger
masses in galaxies that are experiencing high SFRs, leading to IGIMFs
that are flatter than for low-mass galaxies that have only a low-level
of star-formation activity.  \citet{WK05b} show that the sensitivity
of the IGIMF power-law index for $m\simgreat 1\,M_\odot$ towards $SFR$
variations increases with decreasing $SFR$. In starbursts extending to
$SFR\approx 10^4\,M_\odot$/yr the IGIMF can become top-heavy
\citep{WKP11} because the ultra-massive star clusters that enter the
IGIMF integral have top-heavy IMFs (Eq.~\ref{eq:toph_imf_density}).
The dependence of the IGIMF on the SFR of a galaxy is shown in
Fig.~\ref{fig:igimf}).

Thus, galaxies with a small mass in stars can either form with a very
low continuous SFR (appearing today as low-surface-brightness but
gas-rich galaxies) or with a brief initial SF burst (dE or dSph
galaxies). It is also possible for a dwarf galaxy to evolve through
multiple bursts such that its IGIMF varies. Thus the IGIMF ought to
vary significantly among dwarf galaxies (Fig.~\ref{fig:igimf}).  In
all cases, however, the IGIMFs are invariant for $m\simless
1.3\,M_\odot$ which is the maximal stellar mass in $5\,M_\odot$
``clusters'' (Fig.~\ref{fig:mmaxf}), assuming $M_{\rm ecl,
  min}=5\,M_\odot$ to be the invariant lower-mass limit of CSFEs.

An interesting application of the IGIMF theory to a particular system,
our MW Bulge, is as follows: By studying the metallicity distribution
of Bulge stars, (\citealt{Bollero07a, Bollero07b} among others) deduce
the Bulge of the MW to have formed rapidly on a time scale of
$\simless 1\,$Gyr with a top-heavy IMF with $\alpha_3\simless
2.1$. Given that the mass of the MW Bulge amounts to about
$10^{10}\,M_\odot$, it follows that the Bulge would have formed with
$SFR>10\,M_\odot/$yr. From Fig.~\ref{fig:igimf} it can be seen that
the resulting IGIMF has an equivalent power-law index $1.9 < \alpha' <
2.1$ for $m>1.3\,M_\odot$, in excellent agreement with the IMF
constraints based on the metallicity distribution. The IGIMF theory
therefore naturally accounts for the Bulge IMF, that is, no parameters
have to be adjusted apart from specifying the SFR.

\begin{figure}
\begin{center}
\rotatebox{0}{\resizebox{0.9 \textwidth}{!}{\includegraphics{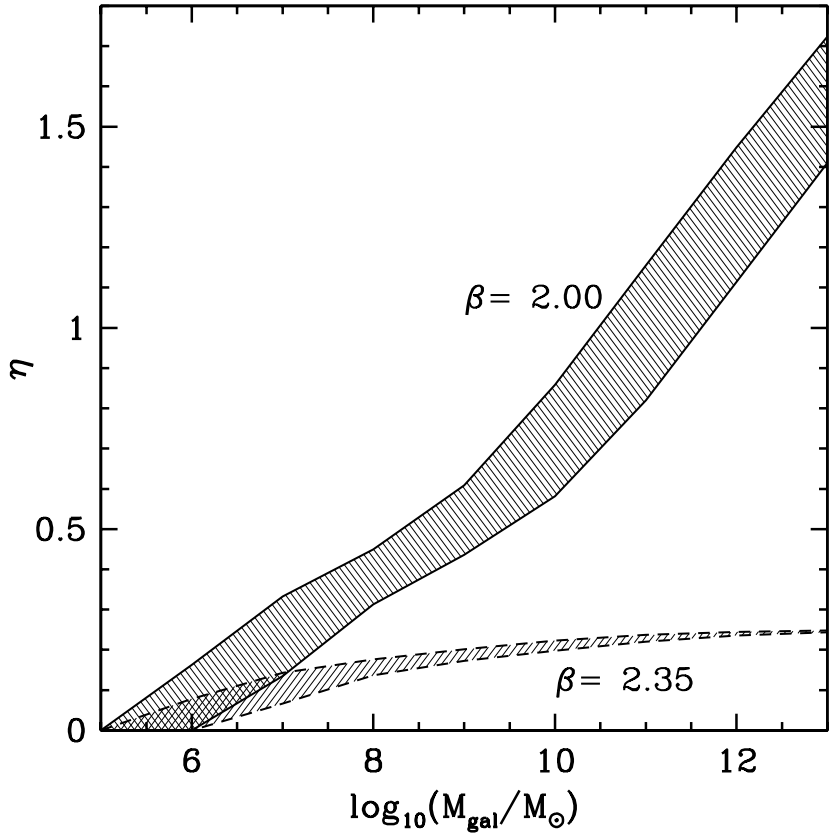}}}
\vskip 0mm
\caption[IGIMF and number of SNII]{\small{ The number of supernovae of type~II (SNII) per star
    in the IGIMF divided by the number of SNII per star in the
    canonical IMF, $\eta$, as a function of the stellar galaxy mass,
    $M_{\rm gal}$. The upper shaded area is for an ECMF with
    $\beta=2$, while the lower shaded area assumes $\beta=2.35$, both
    with $M_{\rm ecl, min}=5\,M_\odot$.  The upper bound for each
    shaded region is for an initial SF burst model of $1\,$Gyr
    duration ($SFR=M_{\rm gal}/1\,{\rm Gyr}$), while the lower bounds
    are for a constant SFR over a Hubble time ($SFR=M_{\rm
      gal}/13.7\,{\rm Gyr}$).  For details see \citet{WK05b}, but the
    here plotted $\eta$ is computed from the IGIMFs shown in
    Fig.~\ref{fig:igimf}. For most galaxies the SNII rate is expected
    to be smaller than the expected rate for an invariant IMF
    \citep{GP05}. But it can be seen that the number of SNII per star
    becomes larger than for a canonical IMF when $\beta=2.0$ and
    $SFR>10\,M_\odot/{\rm yr}$ because the IGIMF becomes top-heavy
    when CSFEs with cloud density $\rho>10^5\,M_\odot/{\rm pc}^3$ are
    included. Star-bursting galaxies therewith come with an
    overabundance of type~II SN {\it if} the ECMF has $\beta=2$ and
    clusters with mass down to $5\,M_\odot$ form in star
    bursts. Similar results are obtained for $\beta=2.35$ and an ECMF
    without low-mass clusters (not plotted here). Such and similar
    calculations allow observational testing of the IGIMF theory
    e.g. through constraining the mass of the extreme SN progenitors
    in dependence of the SFR of the host galaxy
    \citep{Neilletal11}. }}
\label{fig:SN}
\end{center}
\end{figure}

Because the IGIMF steepens above about $1.3\,M_\odot$ with decreasing
SFR, this being the {\it IGIMF-effect}, all galaxy-wide applications
based on a constant IMF require a critical consideration and possibly
a complete revision. A few studies on the outcome when a galaxy-wide
constant IMF is replaced by a SFR-dependent IGIMF do already exist:
Low-surface-brightness galaxies would appear chemically young, while
the dispersion in chemical properties ought to be larger for dwarf
galaxies than for more massive galaxies \citep{GP05,WK05b}. The
observed mass-metallicity relation of galaxies can be naturally
explained quantitatively in the IGIMF context \citep{KWK07}. The
[$\alpha$/Fe] element abundance ratios of early-type galaxies, which
decreases with decreasing stellar velocity dispersion, can be
understood as an IGIMF-effect \citep{RCK09} with the associated
reduction of the need for downsizing. And indeed, the chemical
evolution modelling of the Fornax dwarf-spheroidal satellite galaxy
demonstrates that this system must have produced stars up to at most
about $25\,M_\odot$ in agreement with the prediction of the IGIMF
theory given the low $SFR \approx 3\times 10^{-3}\,M_\odot$/yr deduced
for this system when it was forming stars in the past
\citep{Tsujimoto11}.  Another interesting implication is that the
number of supernovae per star would be significantly smaller over
cosmological times than predicted by an invariant Salpeter IMF
(\citealt{GP05}, Fig.~\ref{fig:SN}), except in phases when the average
cosmological SFR is higher than about $10\,M_\odot$/yr.

The relation between the produced total H$\alpha$ luminosity and the
underlying SFR is linear in the classical Kennicutt picture, i.e. in
the context of a constant galaxy-wide IMF.  It turns out that this
relation becomes strongly non-linear at SFRs comparable to the SMC and
smaller \citep{PWK07}.  The implication of the revised
$L_\mathrm{H\alpha}$-SFR relation is fundamental: In the classical
picture the calculated gas depletion times ($\tau$, the ratio of
available neutral gas mass and current SFR) of dwarf irregular
galaxies are much larger than those of large disk galaxies. This has
been taken to mean that dwarf galaxies have lower ``star-formation
efficiencies'', $1/\tau$, than massive galaxies.  But the
IGIMF-revised $L_\mathrm{H\alpha}-SFR$ relation reveals a fundamental
constant gas depletion time scale of about~$\tau=3$~Gyr over almost
five orders in magnitude in total galaxy neutral gas mass
\citep{PAK09b}. Dwarf galaxies thus have ``star formation
efficiencies'' comparable to those of massive galaxies.  

Furthermore, it is possible to formulate the IGIMF on local scales and
not only on global (galaxy-wide) ones.  This can be achieved
straightforwardly by replacing all galaxy-wide quantities in the
IGIMF-theory (Eq.~\ref{eq:igimf_t}) by their corresponding surface
densities.  This local IGIMF (LIGIMF) theory \citep{PAK08} readily
explains the observed radial H$\alpha$ cut-off in disk galaxies as
well as the different radial profiles in H$\alpha$ and FUV observed by
\citet{BOISSIER07}.

Summarising the two star-formation laws of galaxies which emerge from
the IGIMF theory \citep{PAK07, PAK08, PAK09b},
\[
{SFR \over M_\odot \, {\rm yr}^{-1}} =  {1 \over 2.8\,{\rm Gyr}} \,
  {M_{\rm gas} \over M_\odot},
\]
\begin{equation}
{\Sigma_{\rm SFR} \over M_\odot\,{\rm pc}^{-2} \, {\rm yr}^{-1}}
= {1\over 2.8\, {\rm Gyr}} \, {\Sigma_{\rm gas} \over M_\odot\,{\rm pc}^{-2}},
\label{eq:sflaws}
\end{equation}
where $SFR$ is the global star-formation rate of the galaxy with mass
in neutral gas mass of $M_{\rm gas}$, and $\Sigma_{\rm SFR}$ and
$\Sigma_{\rm gas}$ are the surface star formation rate and surface gas
densities, respectively.

A compilation of a number of issues relating to the astrophysics of
galaxies that are naturally resolved within the IGIMF theory are
listed in the box {\sc IGIMF Successes} (p.~\pageref{box:igimfsucc}).

The IGIMF concept has allowed a number of predictions: Based on the
IGIMF-theory a decreasing galaxy-wide H$\alpha$/FUV-flux ratio with
decreasing total SFR has been predicted \citep{PWK09}. This prediction
has been confirmed qualitatively \citep{MWK09} and quantitatively
\citep{LGT09}. Additionally, \citet{HG08} found, in the integrated
properties of over 50000 SDSS galaxies, that galaxies of lower mass
seem to have steeper IMFs than more massive ones, as would be expected
from the IGIMF. A direct confirmation of the IGIMF would be to measure
the IGIMF effect. A few predictions and tests are compiled in the box
{\sc IGIMF Predictions/Tests} (p.~\pageref{box:igimfpred}).

\vspace{2mm} \centerline{ \fbox{\parbox{\columnwidth}{ \label{box:igimfsucc}
      {\sc IGIMF Successes}:
\begin{description}
\item The mass--metallicity relation of galaxies emerges naturally \citep{KWK07};
\item The [$\alpha$/Fe] element abundance ratios of early-type galaxies emerge naturally
 \citep{RCK09};
\item The observed radial H$\alpha$ cut-off in disk galaxies as well
  as the different radial profiles in H$\alpha$ and FUV emerge
  naturally \citep{PAK08};
\item The SFR of a galaxy is proportional to its mass in neutral gas
  (Eq.~\ref{eq:sflaws}).
\item The gas depletion time-scales of dwarf irregular and large disk
  galaxies are about~2.8~Gyr, implying that dwarf galaxies do not have
  lower star formation efficiencies than large disk galaxies
  \citep{PAK09b};
\item The stellar-mass buildup times of dwarf and large galaxies are
  only in agreement with downsizing in the IGIMF context, but
  contradict downsizing within the traditional framework that assumes
  a constant galaxy-wide IMF.  The stellar-mass build-up times in
  dwarf galaxies become shorter than a Hubble time and therewith
  naturally solve the hitherto unsolved problem that the times are
  significantly longer than a Hubble time if an invariant IMF is
  assumed \citep{PAK09b}.
\item The IGIMF solution for the IMF of the Galactic Bulge is in
  excellent agreement with the top-heavy IMF derived from
  chemical-evolution studies of the Bulge.
\item  For a SFR$=119\,M_\odot$/yr the IGIMF has $\alpha\approx2$ 
(Fig.~\ref{fig:igimf}). This is in good agreement with the constraint 
$\alpha=1.9\pm 0.15$ observed for the $z \approx 2.5$ lensed galaxy SMM J163554.2+661225 
with Herschel by \cite{Finkelstein11}, who adopted a maximum stellar mass
of~$100\,M_\odot$ whereas the IGIMF theory adopts $m_{\rm max*} =
150\,M_\odot$ therewith biasing their IMF solution to slightly
steeper indices.
\end{description}
}}}
\vspace{2mm}

\vspace{2mm} \centerline{ \fbox{\parbox{\columnwidth}{ {\sc IGIMF
        Predictions/Tests}: \begin{description} \label{box:igimfpred} 
\item If the IMF were a stochastic or probabilistic distribution
        function then a population of e.g.~1500 very young stars
        should contain~9 stars more massive than~$8\,M_\odot$
        (Table~\ref{tab:frac}). The finding by \cite{Hsu12} that the
        L1641 cloud is deficient in~O and early~B stars to a
        3--4~sigma significance level constitutes a direct
        observational verification of the LIGIMF effect (many
        low-density star-forming clumps).
      \item In the young star-forming region Taurus-Auriga, random
        sampling from the IMF predicts~9 stars more massive than
        $3.25\,M_\odot$ while none are observed. However, such a low
        number of stars above $3.25\,M_\odot$ is to be expected if one
        assumes a local IGIMF effect from the sub-clusters in that
        region. The closest star-forming region is thus fully
        consistent with the IGIMF theory while being in conflict with
        the hypothesis that star formation is equivalent to randomly
        sampling stars from the IMF;
\item The fraction among all stars of massive stars in a galaxy with a
  low SFR is smaller than in a galaxy with a larger SFR. E.g. for
  $SFR=10^{-3}\,M_\odot/$yr no star more massive than $18\,M_\odot$
  ought to be seen in the galaxy while~30 are expected for an
  invariant canonical stellar IMF (Weidner et al, in~prep.);
\item The number of type~II supernovae is smaller in all dwarf and
  normal galaxies than hitherto thought assuming an invariant stellar
  IMF (Fig.~\ref{fig:SN}).
\end{description}
}}}
\vspace{2mm}

\noindent
These new insights should lead to a revision of theoretical work on
galaxy formation that typically until now relied on an invariant IMF.
Empirical evidence in favour of or against the notion of a
galaxy-variable IGIMF is being studied (e.g. \citealt{Corbelli09,
  Calzetti10, Fumagalli11, Neilletal11, Weiszetal11,
  Roy11}) \label{igimftests} and will ultimately lead to a refinement
of the ideas. Important for workers to realise here is that
stellar-dynamical processes are a central physics ingredient when
analysing populations of unresolved star clusters and the spatial
distribution of massive stars.  Also, care needs to be exercised in
testing hypotheses self-consistently (Sec.~\ref{sec:hyptest}).  At a
fundamental level, the IGIMF theory is correct, since a galaxy is
trivially the sum of all star-formation events. The astrophysical
constraints over many orders of magnitude of galaxy mass allow one to
constrain the fundamental parameters that define the particularly
valid IGIMF. These parameters are the ECMF (e.g., do dwarf and massive
galaxies form only massive clusters when they experience a star
burst?), the exact form of the $m_{\rm max}-M_{\rm ecl}$ relation, and
the variation of the stellar IMF with star-formation rate density.

\vspace{2mm} \centerline{ \fbox{\parbox{\columnwidth}{ {\sc Main results}: Due
      to the clustered nature of star-formation the composite or
      integrated IMF of galaxies or of parts thereof is steeper than
      the canonical IMF for low to modest SFRs ($SFR\simless
      1\,M_\odot/$yr) or SFR surface densities, respectively, and it
      can be top-heavy for larger SFRs. The behaviour of the IGIMF
      depends on the variation of the ECMF with the SFR. The IGIMF
      theory for the first time links in a computationally accessible
      way empirically calibrated star-formation processes on a
      pc~scale with astrophysical behaviour of systems on galactic and
      cosmological scales.  }}} \vspace{2mm}

\section{The Universal Mass Function}
\label{sec:barMF}

How does the stellar IMF fit in with the mass distribution of all
condensed objects, from planets to massive galaxy clusters?

\begin{figure}
\begin{center}
\rotatebox{0}{\resizebox{0.9 \textwidth}{!}{\includegraphics{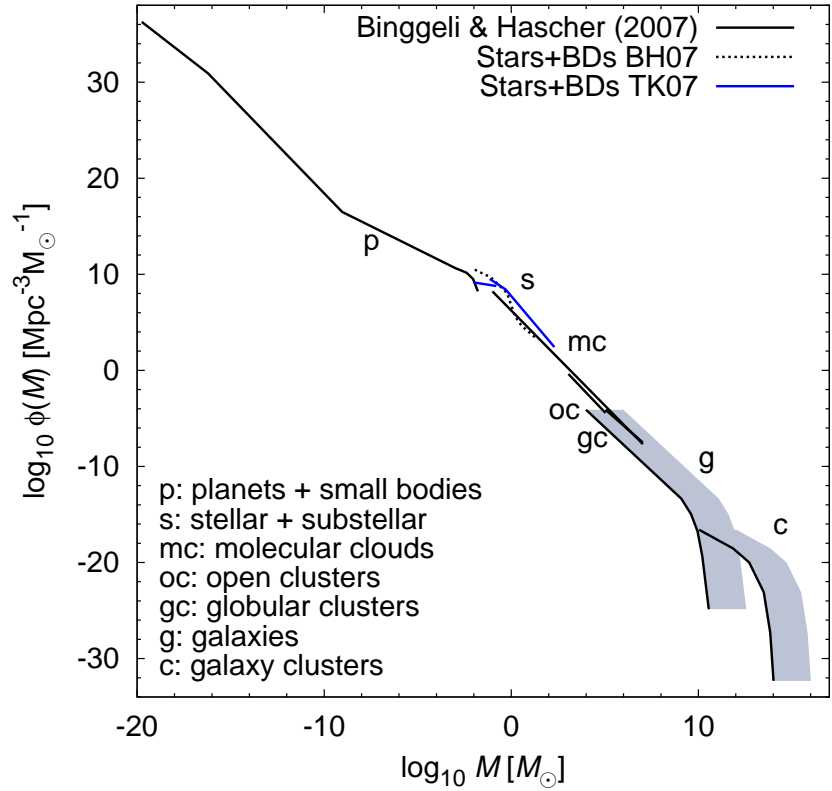}}}
\vskip 0mm
\caption[The universal mass function of condensed structures]{\small{
    The universal mass function (UMF) of condensed structures
    according to \cite{BinggeliHascher07}. BH07 estimated the absolute
    normalisation based on the observed occurrence of the objects
    ranging from planets to galaxy clusters. The canonical IMF are the
    TK07 lines (Eq.~\ref{eq:imf}, \citealt{TK07}). Note that if dark
    matter does not exist then the galaxy and galaxy cluster MFs are
    given by the black line (lower limit to the shaded region), while
    the dark matter plus baryonic masses of galaxies are given by the
    upper envelope of this shaded region.  }}
\label{fig:umf1}
\end{center}
\end{figure}

\cite{BinggeliHascher07} deduce a universal MF (UMF) for all
astronomical objects over a mass range of 36~decimal orders of
magnitude, divided into seven groups: 1.~Planets and small bodies
(meteoroids, asteroids), 2.~stars, BDs and stellar remnants,
3.~molecular clouds, 4.~open clusters, 5.~globular clusters,
6.~galaxies (dark matter plus baryonic masses), and 7.~galaxy clusters
(dark matter plus baryonic masses).  Their result is reproduced in
Fig.~\ref{fig:umf1}. The shape as well as the normalisation are based
on observational data, however, using a present-day mass function
rather than the IMF for stars, with a slightly different low-mass
stellar slope and BDs as being continuously connected to stars (dotted
curve). The canonical IMF is shown by the blue curves. The UMF follows
an approximate $\propto m^{-2}$ power law which is remarkably similar
to the Salpeter one ($\alpha=2.35$) as well as to a logarithmic mass
equi-distribution ($\alpha=2$).
 
\cite{BinggeliHascher07} write ``It is gratifying that the two halves
almost perfectly connect to each other around one solar mass. Remember
that this normalization was achieved on the basis of the mean
universal mass density carried by the stars, embodied in the galaxies
on the one hand, and comprised by the stars themselves on the other.''
As pointed out by \cite{BinggeliHascher07}, a possible reason for this
fairly continuous UMF may be that gravitation is the dominant agent
for creating the structures. 

\section{Concluding Comments}
\label{sec:concs}

Spectacular advances have been achieved over the past two decades in
the field of IMF research and this affects a vast area of
astrophysics.  The discrepant observed nearby and distant luminosity
functions of stars in the local Galactic disk have been unified with
one IMF, and the strong peak in the luminosity function near $M_V=12$
is well understood as a result of the changing internal constitution
of stars as a function of their mass.  The unification of the
luminosity functions also ultimately lead to a unification of the
observed discrepant binary populations in different environments
through dynamical processing in star-forming regions.

The observationally determined stellar IMF is described by
Eq.~\ref{eq:imf} and/\-or~Eq.~\ref{eq:chabimf}). Its form has
traditionally been understood to be invariant in contradiction to the
{\sc Variable IMF Prediction} (p.~\pageref{box:varIMFpred}).  The IMF
of stellar systems and of the Galactic field is provided by
Eqs.~\ref{eq:sysimf} and~\ref{eq:KTGimf}, respectively.

By studying the IMF star-formation theory is being tested.  According
to the {\sc IMF Origin Conjecture} (p.~\pageref{conj:IMForigin}) the
vast majority of stellar masses ($0.1 \simless m/M_\odot$ $\simless$
${\rm few}\,$) do not appear to be affected by competitive accretion
or proto-stellar interactions (Sec.~\ref{sec:cloudcore}).
According to the computational star-formation research the most
massive star in a star-formation event correlates physically with the
mass of the event.  This $m_{\rm max}-M_{\rm ecl}$ relation is a
natural outcome of the competitive accretion-growth and the
fragmentation-induced starvation of stellar masses, and the
computations reproduce the general shape of the IMF. Different ideas
(competitive accretion, coagulation and simply the distribution of
gravitationally unstable regions in turbulent clouds) all lead to
virtually the same type of theoretical power-law IMF. The IMF appears
to be mostly form invariant within space-time correlated
star-formation events (CSFEs, i.e. in individual embedded star
clusters) for star-formation rate densities $SFRD \simless
0.1\,M_\odot\,{\rm pc}^{-3}\,{\rm yr}^{-1}$ within a spatial scale of
about a~pc. But, the IMF varies (trivially) among such individual
CSFEs of stellar mass $M_{\rm ecl}$ through the $m_{\rm max}-M_{\rm
  ecl}$ relation. This relation follows from the notion of {\sc
  Optimal Sampling}.  Computational star-formation has achieved a
remarkable degree of realism, although the binary population has not
yet emerged properly and the computations cannot yet reach the birth
of populous star clusters.

That {\sc Optimal Sampling} appears to be describing a freshly born
stellar population would invalidate the concept of a stellar
population being a purely random representation of the IMF, leading to
{\sc Open Question II} (p.~\pageref{quest:openII}). The remarkable
similarity of observationally determined IMF power-law indices ({\sc
  Open Question III}, p.~\pageref{quest:openIII}) may well be an
indication that nature follows {\sc Optimal Sampling}.  Two other {\sc
  Open Questions} related to star formation and the IMF are
furthermore stated (p.~\pageref{quest:openI}, and
\pageref{quest:openIV}).

Evidence for a variation of the shape of the IMF has emerged, being
consistent with the long-previously predicted IMF variation.  The
evidence for top-heavy IMFs for $SFRD \simgreat 0.1\,M_\odot\,{\rm
  pc}^{-3}\,{\rm yr}^{-1}$ (Eq.~\ref{eq:toph_imf_density}) comes from
either unresolved clusters or from populations that are very difficult
to observe but appears to be increasingly established. The long-sought
after evidence for a systematically varying stellar IMF with
metallicity (Fig.~\ref{fig:varIMFmetal}) may have emerged.  A
principal-component-type analysis of the PDMFs of GCs has now for the
first time yielded a formulation of a systematically varying IMF with
star-forming cloud density and metallicity (Eq.~\ref{eq:toph_master}).

Among other intriguing recent results are that BDs appear to be a
distinct population from that of low-mass stars; their pairing
properties have a different energy scale. This is well reproduced by
computational star formation.  BDs and stars thus follow different
mass distributions which do not join.  A continuous log-normal
function across the VLMS/BD mass scale does not therefore correctly
describe the IMF.

Furthermore, the IMF does appear to have a physical maximum stellar
mass that has now been found empirically. Stars with
$m\simgreat150\,M_\odot$ do not appear to form, unless they implode
invisibly shortly after being formed. Populous star clusters with
super-canonical ($m>150\,M_\odot$) stars are termed to be super
saturated (p.~\pageref{box:saturated}). There is no statistically
meaningful observational evidence for the formation of massive stars
in isolation, in agreement with star-formation computations. The
invariance of the observationally derived IMF precludes the exotic IMF
associated with isolated massive star formation.

By realising that CSFEs are the true fundamental building blocks of a
galaxy, all such events with their IMFs need to be added up to arrive
at the integrated galactic initial mass function. This IGIMF varies in
dependence of the SFR of the galaxy. According to the IGIMF theory,
galaxies with low SFRs have a smaller ratio between the number of
massive stars and low-mass stars than galaxies with high SFRs. This is
the ``IGIMF effect''.  This formulation allows computation of the
IGIMF as a function of time for galaxies with different SFRs.  One
implication of this is that equally-old galaxies can have very
different chemical compositions ranging from unevolved to evolved, and
that the cosmological supernova~type~II rate per star would be
significantly different and dependent on galaxy type than if an
invariant stellar IMF is assumed. Galaxy-formation and evolution
computations with this latter assumption are not likely to be correct.
Indeed, it transpires that only with the IGIMF theory is a fundamental
time-scale of about~3~Gyr uncovered on which all late-type galaxies
consume their current gas supply. Very simple star-formation laws for
galaxies emerge (Eq.~\ref{eq:sflaws}).  Also, only with the IGIMF
theory are the stellar-mass buildup times of dwarf galaxies consistent
with the Hubble time. The top-heavy IMF of the Galactic Bulge deduced
from chemical evolution research follows immediately from the IGIMF
theory.

Thus, with the IGIMF theory it has now become possible to calculate
how the observationally well-constrained star-formation on pc~scales
propagates through to galactic scales of cosmological relevance. A
unification of scales has therewith been achieved which was quite
unthinkable only a few years ago.

Many details still need to be worked out though. For example, direct
verification of the IGIMF effect is needed. This can be achieved by
directly counting the number of massive stars in dwarf galaxies with
low SFRs, or, by counting the number of massive stars in a giant
molecular cloud and comparing this number with the number of late-type
stars formed there.  Also, the existence of and exact form of the
$m_{\rm max}-M_{\rm ecl}$ relation is important not only for the
calculation of the IGIMF but also for understanding to which degree
star-formation is self-regulated on a pc~scale. Such work will
establish which of the two {\sc Philosophical Approaches} of
Sec.~\ref{sec:philosophy} describe the astrophysics of star-formation
and of galaxies.  The fact that the Orion South Cloud has formed a
significant deficit of massive stars with half the cloud producing
thousands of stars without a single massive star \citep{Hsu12} and the
fact that the $m_{\rm max}-M_{\rm ecl}$ relation is excellently mapped
with a small scatter at the lowest masses \citep{KM11} already
constitute major evidence that the IGIMF theory and its fundamental
assumptions appear to be given by nature.

Concerning the wider picture, the stellar and BD IMFs fit-in quite
continuously into the universal mass function of condensed structures
as pointed out by \cite{BinggeliHascher07}.

\vspace{5mm}

{\bf Acknowledgements:} {\small PK would like to thank Christopher Tout and
  Gerry Gilmore for very stimulating and important contributions
  without which much of this material would not have become available.
  PK is especially indebted to Sverre Aarseth whose friendly tutoring
  (against ``payments'' in the form of many bottles of {\it
    lieblichen} German white wine) eased the numerical dynamics work
  in 1993/94.  Douglas Heggie be thanked for fruitful discussions
  with PK on {\sc Optimal Sampling} in Heidelberg in Sept.~2011.  We
  thank Sambaran Banerjee for very helpful comments on the manuscript.
  PK would also like to thank M.R.S.~Hawkins who had introduced him to
  this field in~1987 whilst PK visited the Siding-Spring Observatory
  as a summer vacation scholar at the ANU. Mike gave PK a delightful
  lecture on the low-mass LF one night when visiting his observing run
  to learn about modern, state-of-the-art Schmidt-telescope surveying
  with photographic plates {\it before} PK embarked on post-graduate
  work.  This research was much later supported through DFG grants
  KR1635/2, KR1635/3 and a Heisenberg fellowship, KR1635/4, KR1635/12,
  KR1635/13 and currently KR1635/25. MM acknowledges the Bonn/Cologne
  International Max-Planck Research School for support.}

\clearpage
\lhead{}
\chead{}
\bibliographystyle{aa}
\bibliography{ref} 
\end{document}